%% file: main.tex
\documentclass[PhD]{dukethesis2006}

\usepackage{ifthen}
\usepackage[cmex10]{amsmath} 
\usepackage{longtable}
\usepackage{makecell}
\usepackage{algorithm,algorithmic}
\usepackage{color}
\usepackage[normalem]{ulem}
                             
\usepackage{amssymb,amsthm}
\usepackage{mathtools,bbm}  
\usepackage{IEEEtrantools}                           
\usepackage{xifthen,xparse}
\usepackage{stmaryrd}
\usepackage[hidelinks]{hyperref}  
\usepackage{float}
\usepackage{hhline}

\usepackage{blkarray, bigstrut}
\usepackage{booktabs}
\usepackage{supertabular}

\usepackage{rotating}

\usepackage[table]{xcolor}
\usepackage{listings}

\definecolor{dkgreen}{rgb}{0,0.6,0}
\definecolor{gray}{rgb}{0.5,0.5,0.5}
\definecolor{mauve}{rgb}{0.58,0,0.82}
\definecolor{darkgreen}{rgb}{0,0.5,0}
\definecolor{magenta}{rgb}{1,0,1}
\definecolor{frankfurt_blue_body}{rgb}{0.2,0.2,0.698}

\usepackage{tikz,pgfplots,tkz-graph,color}
\pgfplotsset{compat=newest}
\pgfplotsset{plot coordinates/math parser=false}
\usetikzlibrary{decorations.markings,calc,positioning,matrix}

\usetikzlibrary{quantikz}

\usepackage{etoolbox}
\let\bbordermatrix\bordermatrix
\patchcmd{\bbordermatrix}{8.75}{4.75}{}{}

\allowdisplaybreaks

\newtheorem{theorem}{Theorem}
\newtheorem{corollary}[theorem]{Corollary}
\newtheorem{remark}[theorem]{Remark}
\newtheorem{lemma}[theorem]{Lemma}
\newtheorem{definition}[theorem]{Definition}
\newtheorem{fact}[theorem]{Fact}
\newtheorem{example}{Example}

\newenvironment{example*}
  {\addtocounter{example}{-1}\example}
  {\endexample}

\newcommand{\MZ}{\mathbb{Z}}

\newcommand{\MCC}{\mathcal{C}}
\newcommand{\MCCd}{\mathcal{C}^{\perp}}
\newcommand{\MCQ}{\mathcal{Q}}
\newcommand{\vecnot}[1]{\underline{#1}}
\newcommand{\bg}{\boldsymbol{g}}
\newcommand{\llbr}{[\![}
\newcommand{\rrbr}{]\!]}
\newcommand{\cnot}[2]{{\text{CX}}_{#1 \rightarrow #2}}
\newcommand{\lcnot}[2]{\overline{{\text{CX}}}_{#1 \rightarrow #2}}
\newcommand{\cz}[2]{{\text{CZ}}_{#1#2}}
\newcommand{\lcz}[2]{\overline{{\text{CZ}}}_{#1#2}}
\newcommand{\lX}{\bar{X}}
\newcommand{\lZ}{\bar{Z}}
\newcommand{\lY}{\bar{Y}}
\newcommand{\lH}{\bar{H}}
\newcommand{\lP}{\bar{P}}

\newcommand{\Fn}{\mathbb{Z}_2^n}

\newcommand{\syminn}[2]{\langle #1, #2 \rangle_{\text{s}}}

\newcommand{\norm}[1]{\left\| #1 \right\|}	
\newcommand{\dket}[1]{\left\lvert #1 \right\rangle}
\newcommand{\dbra}[1]{\left\langle #1 \right\rvert}
\NewDocumentCommand\dketbra{+m+g}{%
  \IfNoValueTF{#2}
    {\left\lvert #1 \right\rangle \left\langle #1 \right\vert}
  {\left\lvert #1 \right\rangle \left\langle #2 \right\rvert}%
}
\NewDocumentCommand\dbraket{+m+g}{%
  \IfNoValueTF{#2}
    {\left\langle #1 \vert #1 \right\rangle}
  {\left\langle #1 \vert #2 \right\rangle}%
}

\makeatletter
\renewcommand*\env@matrix[1][\arraystretch]{%
  \edef\arraystretch{#1}%
  \hskip -\arraycolsep
  \let\@ifnextchar\new@ifnextchar
  \array{*\c@MaxMatrixCols c}}
\makeatother

\makeatletter
\newcommand{\longdash}[1][2em]{%
  \makebox[#1]{$\m@th\smash-\mkern-7mu\cleaders\hbox{$\mkern-2mu\smash-\mkern-2mu$}\hfill\mkern-7mu\smash-$}}
\makeatother
\newcommand{\omitskip}{\kern-\arraycolsep}
\newcommand{\llongdash}[1][2em]{\longdash[#1]\omitskip}
\newcommand{\rlongdash}[1][2em]{\omitskip\longdash[#1]}

\author{Narayanan Rengaswamy}

\advisor{Henry Pfister}
\member{Arthur Robert Calderbank, Co-Advisor}
\member{Kenneth Brown}
\member{Iman Marvian}
\member{Jianfeng Lu}

\department{Electrical and Computer Engineering}

\title{Classical Coding Approaches to Quantum Applications}

\newcommand{\singlespacing}{%
  \let\CS=\small\renewcommand{\baselinestretch}{1.0}\CS}
\newcommand{\doublespacing}{%
  \let\CS=\small\renewcommand{\baselinestretch}{1.6}\CS}

\begin{document}

\maketitle

\makeabstract

\Copyright

\abstract
\vspace{-0.1in}

Quantum information science strives to leverage the quantum-mechanical nature of our universe in order to achieve large improvements in certain information processing tasks.
Such tasks include quantum communications and fault-tolerant quantum computation. 
In this dissertation, we make contributions to both of these applications.

In deep-space optical communications, the mathematical abstraction of the binary phase shift keying (BPSK) modulated pure-loss optical channel is called the pure-state channel. 
It takes classical inputs and delivers quantum outputs that are pure (qubit) states.
To achieve optimal information transmission, if classical error-correcting codes are employed over this channel, then one needs to develop receivers that collectively measure all output qubits in order to optimally identify the transmitted message.
In general, it is hard to determine these optimal collective measurements and even harder to realize them in practice.
So, current receivers first measure each qubit channel output and then classically post-process the measurements. 
This approach is sub-optimal.
We investigate a recently proposed quantum algorithm for this task, which is inspired by classical belief-propagation algorithms, and analyze its performance on a simple $5$-bit code.
We show that the algorithm makes optimal decisions for the value of each bit and it appears to achieve optimal performance when deciding the full transmitted message.
We also provide explicit circuits for the algorithm in terms of standard gates.
For deep-space optical communications, this suggests a near-term quantum advantage over the aforementioned sub-optimal scheme.
Such a communication advantage appears to be the first of its kind.

Quantum error correction is vital to building a universal fault-tolerant quantum computer.
An $\llbr n,k,d \rrbr$ quantum error-correcting code (QECC) protects $k$ information (or logical) qubits by encoding them into quantum states of $n > k$ physical qubits such that any undetectable error must affect at least $d$ physical qubits.
In this dissertation we focus on stabilizer QECCs, which are the most widely used type of QECCs. 
Since we would like to perform universal (i.e., arbitrary) quantum computation on the $k$ logical qubits, an important problem is to determine fault-tolerant $n$-qubit physical operations that induce the desired logical operations.
Our first contribution here is a systematic algorithm that can translate a given logical Clifford operation on a stabilizer QECC into all (equivalence classes of) physical Clifford circuits that realize that operation.
We exploit binary symplectic matrices to make this translation efficient and call this procedure the Logical Clifford Synthesis (LCS) algorithm.

In order to achieve universality, a quantum computer also needs to implement at least one non-Clifford logical operation.
We develop a mathematical framework for a large subset of diagonal (unitary) operations in the Clifford hierarchy, and we refer to these as Quadratic Form Diagonal (QFD) gates.
We show that all $1$- and $2$-local diagonal gates in the hierarchy are QFD, and we rigorously derive their action on Pauli matrices.
This framework of QFD gates includes many non-Clifford gates and could be of independent interest.
Subsequently, we use the QFD formalism to characterize all $\llbr n,k,d \rrbr$ stabilizer codes whose code subspaces are preserved under the transversal action of $T$ and $T^{-1}$ gates on the $n$ physical qubits.
The $T$ and $T^{-1}$ gates are among the simplest non-Clifford gates to engineer in the lab.
By employing a ``reverse LCS'' strategy, we also discuss the logical operations induced by these physical gates.
We discuss some important corollaries related to triorthogonal codes and the optimality of CSS codes with respect to $T$ and $T^{-1}$ gates.
We also describe a few purely-classical coding problems motivated by physical constraints arising from fault-tolerance.
Finally, we discuss several examples of codes and determine the logical operation induced by physical $Z$-rotations on a family of quantum Reed-Muller codes.
A conscious effort has been made to keep this dissertation self-contained, by including necessary background material on quantum information and computation.

\acknowledgements

First and foremost, I extend my deepest gratitude to my supervisors Henry Pfister and Robert Calderbank for their continued guidance and enthusiasm.
Their mentorship and professional support are primarily responsible for my current status as an early career researcher.
I had enjoyed working with Henry for my Master's thesis at Texas A{\&}M University, so I was excited to apply to Duke and continue working with him.
Right from the beginning, Henry has provided me a great amount of freedom in choosing the topic I wanted to work on.
Thanks to this, I spent my first year continuing with my Master's work on polar codes and learning about the vibrant research area of compressed sensing.
While I was not making much progress at a brisk pace, I had several interactions with Henry that helped me get a handle on finding the right questions to ask.
It is often said that asking the right questions solves half of the problem, and it was exciting to learn this by experience at an early stage.
Since Henry has an impressive breadth of knowledge, from physics to communications to mathematics, all of my conversations with him leave me with new associations between a variety of problems.
I think developing such connections across topics is a key skill for a successful researcher, and I am grateful to Henry for being an accessible resource in this regard.
Also, he has always been available whenever I had questions and knocked on his door, and this has proven to be a delightful luxury for me.

Later, my course project on compressed sensing developed into the report for my Ph.D. Qualifying Examination.
It was during this time that I began to interact with Robert, since I was working on his construction of compressed sensing matrices using Kerdock codes.
Robert is an extremely humble and kind human being, besides being the brilliant mathematician and researcher that he is well-known for.
Even after his long and successful professional career, his enthusiasm for research and discovery is astounding to say the least.
When I slowly moved from compressed sensing to learning about quantum information science, thanks to a small reading group to which Henry invited me, I started learning more from Robert about the basics of quantum computation.
Then I got an opportunity to serve as the teaching assistant for a coding theory course co-taught by Robert and Henry in Fall 2017.
At this time I was developing some notes for my own purposes regarding the construction of logical Pauli operators for CSS codes, the ``C'' being Calderbank.
Robert encouraged me to give a lecture in the course based on my notes, and at the end of the lecture he hinted at a method for extending my approach to all logical Clifford gates.
It was also at this time when my friend Swanand Kadhe, from Texas A{\&}M, was visiting Duke and we collaborated on this work.
We both had a great time during that project.
This was the turning point in my Ph.D. and it kickstarted my contributions to synthesizing logical operators for stabilizer codes.
From then on, it has been an incredible journey in quantum computing, and I am very thankful to Robert for guiding me all along.
There are several such exciting incidents with both Henry and Robert, but I will leave them for another occasion in the interest of space.
I am also grateful to my supervisors for encouraging me to attend workshops, present in several conferences, and visit research groups.
This has provided me great exposure to the works of peers in the field, and help establish new friendships that I hope will last long.

I would like to thank Kenneth Brown, Iman Marvian, and Jianfeng Lu for serving on my dissertation committee.
Ken is a delightful person to interact with and I learned a great deal from his course on quantum error correction.
He has also been very kind to introduce me to senior researchers during conferences.
When I first developed the main result on transversal $T$ gates, he took the time to read my draft and together we began to develop the first new code using the theorem.
That meeting was a wonderful source of encouragement for me.
Eventually, it led to more results and landed me a contributed talk at the 2020 Quantum Information Processing (QIP) conference, where most paper submissions get accepted only as posters. 

I enjoyed taking Iman's courses on quantum information science and quantum information theory.
In terms of research, back in 2017 I had started investigating Joseph Renes' belief-propagation algorithm that passes quantum messages, when I had just learned the fundamentals of quantum information.
However, my progress was slow in understanding the workings of the algorithm.
After Iman joined Duke in early 2018, I had several interactions with him regarding this algorithm.
Those conversations were very helpful to navigate the subtleties of quantum information that manifest in this algorithm.
Iman is an expert in quantum information and he is always enthusiastic about listening to new ideas and sharing his insightful thoughts.

Jianfeng is a soft-spoken applied mathematician whose research interests span optimization, probability, differential equations and quantum chemistry algorithms.
He is very energetic to interact with and also helped me understand the aforementioned quantum belief-propagation algorithm through his inquisitive nature.
Besides that, during my work on logical Clifford gates for stabilizer codes, I had finished constructing an algorithm to map a sequence of binary vectors to their pairs using symplectic transvections.
But I needed to find a way to determine all symplectic solutions that satisfy the sequence of mappings.
While I was stuck in this pursuit, an afternoon conversation with Jianfeng helped me solve this problem completely, and that led to my first paper on quantum computing, with Swanand, in early 2018.

I am grateful to all Duke staff and professors at Duke from whom I learned a lot, especially Galen Reeves, Cynthia Rudin, Jungsang Kim, and David Brown.
I would like to thank Kathy Peterson at the Rhodes Information Initiative at Duke (iiD) for all the administrative help and putting a smile on everyone's face through her charming personality.
She is industrious and systematic in her work, and finds time to address people's problems spontaneously even when she is taking care of her family.
I have had the opportunity to interact with her regularly regarding the Kavli Quantum Coffee Hour meetings, and I would like to thank her on behalf of all of us for keeping us fed during talks.
I would also like to thank Ariel Dawn at iiD for administrative and organizational help.
Kathy and Ariel helped me a lot while I was a volunteer for organizing the 2016 North American School of Information Theory (NASIT) here at Duke.

It was a pleasure to have Mengke Lian as a lab mate, and I have learned a great deal from him about several topics.
He helped me with hints to solve tricky problems in the course on compressed sensing and the interdisciplinary course on information theory, algorithms and statistical mechanics.
Mengke's enthusiasm to simultaneously develop computer programs as well as solve challenging theoretical problems was inspiring and it enriched my graduate experience greatly.
I also enjoyed developing an information theory crossword puzzle with him for NASIT 2016.
I would also like to thank Christian H{\"a}ger for many fun lunch conversations about research and other things.
Sarah Brandsen has been another wonderful colleague in the group, and I would like to thank her for engaging me in conversations about quantum state discrimination.
This led to a very nice paper in collaboration with Mengke, Kevin Stubbs, and Henry.
Kevin is a calm graduate student in Gross Hall who works with Jianfeng, and he is always enthusiastic about solving math problems.
He was a persevering force in making the above collaboration come true.

I had a wonderful time interacting with Ken Brown's group -- Michael Newman, Dripto Debroy, Muyuan Li, Natalie Brown, Leonardo Andreta de Castro, Swarnadeep Majumder, and Shilin Huang.
In particular, conversations with Mike have helped me learn a lot and one of them was responsible for kickstarting my project on transversal $T$ gates.
I was also fortunate to meet and know Anirudh Krishna and Pavithran Iyer, fellow Chennaites from the University of Sherbrooke. 
I have already learned a great deal from both of them in our short span of friendship.

I also enjoyed my interactions with visitors to our group -- Elia Santi, Fabrizio Carpi, and Mustafa Cemil Co{\c s}kun.
In particular, Mustafa is such an energetic character that he can liven up any conversation.
I had great interactions with him and also learned several things about polar codes and list decoders.
It was a lot of fun to chat over food with him and other Gross Hall colleagues -- Emina Hodzic, Marianna Pepona, Elchanan Solomon, Shan Shan, Matthias Sachs, and Deborshee Sen.

I would like to thank my wonderful ``Today!'' group friends Vaishakhi Mayya, Moritz Binder, Panchali Nag, Soudipta Chakraborty, Hersh Singh, and Sravya Chelikani for the great fun during lunches and dinners in different cuisines.
Each one of them added a distinct character to the group and it helped keep conversations very alive. 
I will never forget the ``Pie Chart'' tradition after dinners, especially the ``Moritz Factor''!
We also had great times during our road trips to Kitty Hawk, Savannah and the Smokies.
It was also a delight to chat with Dhanasekar Sundararaman, Ashwin Shankar, and Manoj Kumar Jana about varied things, and we had a lot of fun playing table tennis and badminton.
During my first two years, I thoroughly enjoyed the Saturday volleyball sessions with the wonderful group of Chinese postdocs and two Russians, Mikhail Dobrikov and Elena Dobrikova.
I thank Ming-Feng Hsueh for enthusiastically organizing those sessions.

I am always thankful to my wonderful friends Vaidhyanathan Sabesan, Nagaraj Janakiraman, Vasudevan Ramachandran, Shravan Sriram, Ramprasad Ramesh, Subramanian Ramalingam, Abhay Shankar Anand, Jagadish Prasad, Aditya Rajagopalan and Raghav Thanigaivel for all the delightful moments we shared together.
I am thankful to Venkatasubramanian Sankararaman, Janani Venkatasubramanian, and Bharath Kashyap for the enjoyable 2018 December in Arizona.
I also have great memories with Divya Sekar and Srikanth Rohini from my wonderful 2017 summer in San Jose.
I thank Karthik Thothathri for the hospitality and memorable trips during my week in Singapore in summer 2017, when I attended the ``Beyond I.I.D. in Information Theory'' workshop at the National University of Singapore.
I had one of the best times with my college buddy Vishnu Anirudh during my visit to England in summer 2019 for the ``Quantum Error Correction (QEC)'' conference.

It is a boon to have family close by when you are in a foreign country.
I owe an immeasurable amount of love and gratitude to my aunt, Vasanthi Murali Krishnan, and uncle, Murali Krishnan, and their adorable children, Shrey and Dhruv.
I had amazing times with them, who lived only a few hours away, whenever I got a chance to spend time with them.
I would like to thank all my family with whom I got a chance to spend time with, but I had a lot of cherishable moments specifically with my cousins, Vaikunth Raghavan, Srikanth and Parthasarathy Gomadam, and Vedha and Srinath Sampathkumar.
I also had a great day with Jeyashree Krishnan and her parents, my aunt Uma and uncle Krishnan, when I met them in Paris while I was there for the 2019 ``IEEE International Symposium on Information Theory''.

Last but certainly not the least, I am extremely blessed and fortunate to have Sudha Rengasami and Rengaswamy Purushothaman as my parents.
It is truly their unconditional love and support that has made my graduate journey possible.
I am eternally indebted to them for whatever good characteristics I possess, and their sincerity, dedication, and perseverance have served as my guiding light all along.
I hope my graduate school accomplishments at least repay a tiny portion of their sacrifices for me in these past 4.5 years.
I love you both so much, Amma and Appa!
Finally, I extend my humble gratitude to the Almighty for providing me with everything, specifically strength and perseverance, to pursue my passion for research, despite my suffering from the evergreen impostor syndrome many a time.

\tableofcontents

\listoffigures

\listoftables

\doublespacing
\chapter{Introduction}
\pagenumbering{arabic}

\input{ch1_intro.tex}

\chapter{Essentials of QIQC}
\input{ch2_background.tex}

\chapter{A Quantum Communication Advantage}
\input{ch3_bpqm.tex}

\chapter{Binary Symplectic Geometry in QC}
\input{ch4_groups.tex}

\chapter{Logical Clifford Synthesis (LCS) for Stabilizer Codes}
\input{ch5_lcs_algorithm.tex}

\chapter{Quadratic Form Diagonal (QFD) Gates}
\input{ch6_qfd_gates.tex}

\chapter{Stabilizer Codes that Support QFD Gates}
\input{ch7_stabilizer_codes_qfd.tex}


\chapter{Conclusions}
\input{conclusion}

\clearpage
\appendix

\chapter{LCS Algorithm: Code and Solutions}
\input{ch5_appendix.tex}

\clearpage
\phantomsection

\singlespacing

\newcommand{\etalchar}[1]{$^{#1}$}

\biography
\doublespacing

Narayanan Rengaswamy received his Bachelor of Technology (B.Tech.) degree with distinction in Electronics and Communication Engineering (ECE) from Amrita University, Coimbatore, in 2013.
He ranked first in the Coimbatore campus and third across all 3 engineering campuses of Amrita University in the B.Tech. ECE program.
Then he completed his Master of Science (M.S.) in Electrical Engineering at Texas A{\&}M University, USA, in 2015, where he worked on ``Cyclic Polar Codes'' for his thesis under the supervision of Henry Pfister.
This led to his first conference paper in the 2015 IEEE International Symposium on Information Theory (ISIT)~\cite{Rengaswamy-isit15}, the flagship conference of his field.
He spent the summer of 2015 as a graduate research intern in Alcatel-Lucent Bell Labs, Stuttgart, Germany.
During this time, he worked on ``Spatially-Coupled LDPC Codes on Burst Erasure Channels'' under the supervision of Laurent Schmalen and Vahid Aref.
This later led to two conference papers~\cite{Rengaswamy-izsc16,Aref-istc16} and his first journal publication in the prestigious IEEE Transactions on Information Theory~\cite{Aref-it18}.

Subsequently he moved to Duke University in January 2016 in pursuit of a Doctor of Philosophy (Ph.D.) in Electrical Engineering.
His supervisors are Henry Pfister and Robert Calderbank, and the focus of his dissertation is fault-tolerant quantum computation and an application in quantum communications.
He has written several papers during his Ph.D.~\cite{Rengaswamy-arxiv18*2,Rengaswamy-arxiv19b,Rengaswamy-pra19,Can-arxiv19,Rengaswamy-arxiv19c,Rengaswamy-arxiv20} and presented them as posters and talks in conferences~\cite{Rengaswamy-isit18,Can-isit19}.
His work on transversal $T$ gates~\cite{Rengaswamy-arxiv19c} was one of the 73 out of 283 papers that were presented as talks in the prestigious 2020 Quantum Information Processing (QIP) conference.
His first external collaboration is with Saikat Guha and Kaushik Seshadreesan from the University of Arizona~\cite{Rengaswamy-arxiv20}.
He presented this work as a Graduation Day talk and poster in the 2020 Information Theory and Applications (ITA) workshop.
He has reviewed papers for IEEE Transactions on Information Theory, Quantum, IEEE Transactions on Vehicular Technology, and the conferences ISIT 2018, 2020, and ITW (Information Theory Workshop) 2018.

\end{document}

%% file: ch1_intro.tex




\label{ch:ch1_intro}

\section{The Interplay of Revolutionary Theories}

Twentieth century witnessed the development of three revolutionary scientific theories: computer science, information theory, and quantum mechanics.
Devices that perform computations have a long history, starting between 2700-2300 BCE with the Sumerian Abacus and continuing in the 1800s CE with the pioneering contributions of Charles Babbage and Ada Lovelace for developing programmable machines.
However, it was the contributions of scientists such as Gottfried Leibniz, George Boole, Akira Nakashima, Claude Shannon, Alan Turing, and John von Neumann that gave computer science the theoretical rigor required for generality and scalability.
Computer science concerns with studying computational tasks, constructing deterministic or randomized algorithms for solving them, analyzing the time and memory requirements for these algorithms, and optimizing the algorithms for such resource \emph{complexity}.
These algorithms represent the underlying task using binary digits (\emph{bits}) at the most fundamental level, and all operations (or manipulations) involve binary arithmetic.
Computer scientists have classified computational tasks into \emph{complexity classes} such that tasks in the same class share similar resource requirements.
Two prominent complexity classes are ``P'' and ``NP''.
``P'' is the set of all decision problems that can be solved by a deterministic machine (algorithm) in polynomial time, i.e., in time that is polynomial in the size of the problem.
``NP'' is the set of all decision problems whose solutions can be (deterministically) \emph{verified} in polynomial time, or equivalently that can be solved by a non-deterministic machine (algorithm) in polynomial time.
Although ``P'' is contained in ``NP'', it remains unknown if they are equal.
Since problems in ``NP'' (outside ``P'') do not \emph{yet} have an efficient (polynomial-time deterministic) algorithm, computer scientists are always interested in more powerful (faster) ways to solve different tasks.

While computer science studies the manipulation of information to infer quantities of interest, the realm of information theory concerns with the representation, storage and reliable communication of information.
Claude Shannon had studied the application of Boolean algebra for designing efficient electrical circuits, consisting of relay contacts and switch blades, that possess certain desired characteristics.
As he himself recognized, his 1937 Master's thesis on this work~\cite{Shannon-aiee38} had immediate practical ramifications since automatic telephone exchanges and industrial motor-control equipment often involved building such complex electrical circuits.
Later, during World War II, he studied the problem of reliable communication amidst noise.
This led to his celebrated 1948 papers on ``A Mathematical Theory of Communication''~\cite{Shannon-bstj48,Shannon-bstj48b}, where he laid the foundation for information theory.
He established two coding theorems, one that characterized the limit of data compression such that the data can be retrieved reliably, and the other that characterized the maximum rate at which information can be transmitted reliably over a given noisy channel.
The former limit is the \emph{entropy} of the data source, and the latter rate is the \emph{capacity} of the channel.
He embraced the notion of a \emph{bit} of information, and established both these quantities to be measured in that unit.
This single work forms the bedrock of the design of all modern communication and data storage systems.
Over these decades, information theorists have been studying the fundamental limits in several communication settings, including in secure communications under the presence of eavesdroppers or adversaries.

In his coding theorems, Shannon had established that one would have to effectively encode and decode information in order to achieve the limits of data compression and reliable communication.
Hence, coding theorists have been working on explicit constructions of \emph{codes} that achieve these fundamental limits asymptotically.
From a practical perspective, they have also designed codes that enable retrievable compression of data and reliable communication over noise \emph{with tolerable resource overheads}.
There are many computer scientists who have made vital contributions to coding theory as well.
Therefore, information and coding theory is an area that weaves together probability theory, statistics, discrete mathematics, linear algebra and iterative algorithms in the pursuit of reliable storage and communication of information.
Since the representation of information is fundamental to any information-theoretic analysis of it, alternative notions of the ``bit'' would give rise to their own information theory.
This realization firmly ties information theory to the physical laws that govern the materials used to store information.
Succinctly, it reinforces the fact that \emph{information is physical}.

Quantum mechanics is the most fundamental physical theory that explains the workings of our world, and it has withstood (experimental) tests performed time and again in the past century.
The pioneers of this theory include Max Planck, Albert Einstein, Niels Bohr, Erwin Schr{\"o}dinger, C.V. Raman, Max Born, Werner Heisenberg, and Wolfgang Pauli.
According to a postulate of quantum mechanics~\cite{Nielsen-2010}, a quantum system in a deterministic state can be mathematically described by a (normalized) complex vector in a Hilbert space, the so-called \emph{wave function}.
If there is uncertainty in the state of the system, then it is represented as a ``bag of states'' that can be encapsulated into a \emph{density matrix}.
By the other postulates, such a system undergoes a \emph{unitary} evolution until being measured by the environment or an observer.
The evolution is governed by the Schr{\"o}dinger equation and the measurement is characterized by an \emph{observable}, which is a self-adjoint (Hermitian) operator.
Most importantly, prior to the measurement the system might be in a \emph{superposition} of several states in the basis defined by the observable, but the measurement forces (or collapses) the system into one of these basis states according to a probabilistic law.
This aspect of the theory is particularly counter-intuitive since in daily life we measure quantities of macroscopic objects without tangibly changing its (macroscopic) state.
In fact, Einstein was very much bothered by this postulate and it led to one of his famous statements: ``God does not play dice''.
However, several experiments have verified the quantification of this measurement postulate which is called the \emph{Born rule}.
Thus, the process of measurement connects the underlying quantum world to the perceived classical world.

By the postulates of quantum mechanics, the dimension of the Hilbert space grows exponentially in the number of quantum particles, or \emph{qubits} (for ``quantum-bits''), at hand. 
Therefore, it is quite computationally challenging to study non-trivial quantum systems, such as moderately large molecules, via classical computations.
The ability to study such systems could enable, among other things, discovery of better drugs for diseases and design of better materials.
So the acclaimed physicist Richard Feynman suggested in the 1980s that perhaps a \emph{quantum-mechanical} computer can be built and used to ``simulate'' other quantum systems of interest.
In order to be a general purpose machine, i.e., a \emph{universal quantum computer}, such a computer must be capable of preparing arbitrary quantum states, applying any unitary operation, and measuring any observable on the final system state.
Such a machine would differ from a conventional computer because its fundamental building block would be a qubit instead of a bit.
While a bit strictly takes only two values $0$ or $1$, the superposition property of quantum objects allows a qubit to be a mixture of $0$ ($\dket{0}$) and $1$ ($\dket{1}$).
Furthermore, using quantum laws it is possible to \emph{entangle} two qubits, which is a way of introducing correlation between them beyond what is possible classically between two bits.
Such quantum properties have been exploited by scientists since the late 1980s to (theoretically) demonstrate computational speedups beyond classical computers.
Although it is hard to exactly pin down the source of the computational power, superposition and entanglement are thought to be vital for this purpose.
Finally, the notion of the qubit produces a new representation of information.
Hence, scientists have also studied how these quantum properties can be used to achieve higher rates of communication over quantum channels than classically possible, i.e., \emph{quantum} information theory.

This dissertation lies at the intersection of these three revolutionary theories. 
In the communication setting, we demonstrate that a recently proposed quantum algorithm based on classical belief-propagation can be used to achieve a communication advantage in deep-space optical communications.
In the computation setting, we address several questions concerning reliable universal quantum computation under the presence of noise in a quantum-mechanical computer.
We will first discuss some challenges in both these settings before summarizing the contributions of this dissertation.

\section{Challenges in QIQC}

The foundations of quantum information and quantum computation (QIQC) have been laid out by appropriately adapting concepts from computer science and information theory to quantum-mechanical laws.
It must be emphasized that this is not merely a technical exercise but requires a fundamental change in the way we think of storing, processing, and communicating information.
However, bringing quantum technologies to reality introduces several theoretical and practical challenges.
While the QIQC community has been continuously addressing these challenges with ingenious ideas, several questions remain partly or fully unsolved, which makes the field exciting to work in at this point of time.
In particular, we will show later that some of these quantum-motivated questions give rise to new self-contained coding problems that provide new avenues for classical coding theorists.

It is important to note that there have already been impressive practical demonstrations of quantum technologies such as IBM's publicly accessible quantum computers and Google's recent ``quantum supremacy'' experiment.
In the latter, Google (and NASA) performed a real experiment on their 53-qubit quantum computer and demonstrated a computational advantage over classical computers~\cite{Arute-nature19}. 
The chosen task is related to sampling from the output distribution of certain random circuits~\cite{Aaronson-arxiv19}.
While the extent of the advantage has been disputed by IBM~\cite{Pednault-arxiv19}, this is widely considered to be a milestone hardware demonstration.
Hence, many people now believe that there are no fundamental \emph{scientific} obstructions to building a reliable universal quantum computer.
In other words, the challenges faced by the leading technologies today, e.g., trapped-ions and superconducting circuits, require advances in the way these systems are \emph{engineered}, so that overheads are minimized while reliability is ensured.
We will briefly discuss a few of these challenges that relate to this dissertation.

Building a quantum computer involves an intensive interdisciplinary effort between physicists, engineers, computer architects, material scientists, and software developers.
From a device physics and engineering perspective, the components used to make the computer remain quite noisy.
For example, the technology of trapping ions in vacuum and using their energy levels to form qubits is one of the leading strategies today for building a reliable universal quantum computer.
The operations on these qubits involve shining lasers at them, individually and also in groups, to realize single-qubit and multi-qubit unitary operations.
However, due to imperfections in the control of the laser, there is always some residual noise in these operations.
This is just once source of error and there are many such practical issues causing imperfections.
The best error rates reported for this technology today are about $10^{-4}$ for single-qubit gate errors and $10^{-3}$ for two-qubit gate errors~\cite{Blume-Kohout-ncomm17,Leung-prl18}, where error rates usually represent the \emph{diamond norm} between the ideal and actual gates.
Measurement error rates usually lie between single-qubit and two-qubit error rates.
Such error rates limit the length of reliable computation that can be performed in today's systems.
Moreover, these error rates are usually reported when operations are performed in isolation with only those qubits involved in the gates being present.
In other words, when the systems are scaled and one attempts to perform gates on different qubits and different pairs of qubits, the overall average gate error rates are significantly higher.
Experimentalists continue to address such issues in scaling, but the advantage in ion-traps is that so far the qubits have all-to-all connectivity unlike technologies such as superconducting circuits~\cite{Linke-nas17}.
This means that when circuits are mapped from paper to the actual hardware, all gates are directly implementable and do not require sophisticated mapping strategies that take into account limited qubit connectivity.
As a result, gate counts remain under control.
On the other hand, operations in superconducting circuits are much faster while qubit lifetimes (``coherence times'') are shorter.
Moreover, they are able to leverage the techniques developed for the highly mature silicon industry.
Hence, it remains to be seen which of these technologies prove to be preferable for truly scalable and reliable quantum computing.

A computer that can implement arbitrarily long quantum circuits composed of arbitrary state preparations, unitary operations and measurements, while being tolerant to errors in all these components, is called a \emph{universal fault-tolerant quantum computer (UFTQC)}.
In order to build a UFTQC, it appears inevitable that the initial data need to be encoded in a \emph{quantum error-correcting code (QECC)}, and all operations need to be performed on the QECC \emph{fault-tolerantly}.
The basic idea in classical and quantum ECC is to add redundancy to the raw data in order to protect against some of the many possible errors.
An $\llbr n,k,d \rrbr$ QECC encodes $k$ logical (information) qubits into $n$ physical qubits, and the smallest undetectable (or troublesome) error acts on at least $d$ of the $n$ qubits.
\emph{Stabilizer} QECCs, which are the quantum analogues of classical linear codes, and \emph{subsystem} QECCs, which generalize stabilizer codes, are the commonly used QECCs both in theory and practice.
In this dissertation, we mainly consider stabilizer codes.

For quantum computation, the celebrated \emph{threshold theorem} states that if the noise in all components of a quantum computer is below a certain threshold, then it is possible to make a UFTQC out of the system by employing QECCs~\cite{Knill-arxiv96,Aharonov-phdthesis99}.
The exact threshold depends on the specific QECC strategy, and the leading approach is to use the \emph{surface code} which has a $\approx 1\%$ threshold under circuit-level noise~\cite{Wang-arxiv09,Wang-pra11,Fowler-arxiv12}.
While the error rates quoted above for ion-trap systems might appear well below this threshold, recollect that scalability of the operations remains an important issue to be fully addressed.
The family of surface codes is defined on $d \times d$ 2D (square) lattices with nearest-neighbor connectivity, which is convenient for practical realizations in limited connectivity superconducting systems.
However, surface codes form a $\llbr 2d^2, 1, d \rrbr$ family (or $\llbr 2d^2, 2, d \rrbr$ depending on the construction), which means their rate is $1/2d^2$.
This introduces a lot of overhead physically per logical qubit.
Moreover, since a UFTQC is not just a memory device, one has to devise fault-tolerant schemes to perform a universal set of quantum operations on the $k$ logical qubits.
While there is a large literature on fault-tolerant logical operations for surface codes, the resulting resource overheads are too demanding for near-term devices~\cite{Fowler-arxiv12,Litinski-quantum19}.
Therefore, a broad and interesting open problem is to devise a high-rate QECC family with growing distance, and then demonstrate universal fault-tolerant quantum computation on it.

While we will not discuss the surface code much more in this dissertation, we will describe results that relate to systematically realizing logical \emph{Clifford} operations on \emph{stabilizer} QECCs.
This can algorithmically aid in translating the search for fault-tolerant logical operations on arbitrary stabilizer QECCs into a systematic algorithmic procedure.
In this process, we will show that \emph{symplectic matrices} form a natural and convenient control plane for quantum computation.
We will emphasize that further research into the structure of these matrices might prove very useful for algorithmic developments in compilers for UFTQCs.
We will also discuss some important results that we obtain for realizing logical \emph{non-Clifford} operations, which are harder to realize fault-tolerantly than logical Clifford operations.
Fault-tolerant realization of a generating set of logical Clifford operations and any one logical non-Clifford operation enables universal fault-tolerant quantum computation.

On the communication front, the reproduction of a transmitted message at the receiver requires the ability to reliably distinguish the received signals corresponding to different messages.
In classical communications, one implements detectors to convert the received waveform into either a bit-stream (``hard information'') or a sequence of probabilities (``soft information''), and subsequently processes them appropriately.
For coded transmission, to estimate the transmitted codeword (or message), soft decoders use the structure of the error-correcting code to combine the soft information under the guidance of Bayes inference.
Most importantly, the received waveform is in a definite state that corresponds to, say, a bit of information, and the detector essentially acts as a transducer that converts the representation of the information.
As a simple example, if ``0'' is encoded into a 0 Volt signal over a time interval and ``1'' is encoded into a 5 Volts signal over a time interval, then the detector simply converts 0V to ``0'' and 5V to ``1''.
However, for quantum communications the received ``waveform'' could be a qubit that is in one of two \emph{non-orthogonal} states, and measurement collapses it into one of two definite states.
Therefore, for non-trivial channels the measurement itself introduces uncertainty. 
Hence, measuring each output qubit might not be the best strategy for reliable communication.

Indeed, Carl Helstrom first calculated the optimal \emph{joint} measurement on all received qubits in order to optimally distinguish between two candidate messages~\cite{Helstrom-jsp69,Helstrom-ieee70}.
In the case of more than two hypotheses states, the Yuen-Kennedy-Lax conditions~\cite{Yuen-it75} provide a way to calculate such an optimal measurement that minimizes the probability of error.
However, none of these methods provide a systematic approach to decompose the optimal measurement into a sequence of operations that can be performed in the lab.
Therefore, many schemes proposed in the literature for capacity-approaching communication or minimum probability of error communication have decoders that remain as theoretical constructs.
In this dissertation we will discuss in length about a recently proposed quantum algorithm that is inspired by the classical belief-propagation algorithm~\cite{Renes-njp17}. 
This algorithm appears to provide optimal decoding strategies for deep-space optical communications, while also having a systematic decomposition into lab operations.
The advantage here is that one might not need a UFTQC to construct this receiver. 
Instead, this is an avenue for experimentalists where they might focus on reliably realizing a particularly structured circuit to obtain a quantum advantage over state-of-the-art receivers.
Hence, this points to a near-term application for non-universal quantum computers that appears to be the first instance of a practical quantum \emph{communication} advantage.
Other near-term applications that have been explored in the literature include variational optimization for logistics, metrology for high-precision measurements, quantum simulation for drug discovery, and chemistry for nitrogen fixation.

\section{Contributions of this Dissertation}

The dissertation can be broadly split into two parts. 
The first part describes our results for quantum communications and the second part discusses contributions to quantum computation.
Now, we will briefly summarize each of these contributions.

The abstract model for the channel in deep-space optical communications, under the binary phase shift keying (BPSK) modulation, is a classical-quantum (CQ) channel called the \emph{pure-state} channel.
Renes recently proposed a quantum algorithm that generalizes the classical belief-propagation (BP) algorithm to classical codes transmitted over the pure-state channel~\cite{Renes-njp17}.
This algorithm uses qubits as messages and appears to be the first instance where the messages passed in a BP algorithm are not classical probabilities.
This makes the algorithm inherently quantum and distinguishes itself from past applications of BP to quantum problems.
However, simulation and analysis of this algorithm have remained unexplored.
In Chapter~\ref{ch:ch3_bpqm}, we analyze this algorithm for a simple $5$-bit code and examine the density matrices involved in the process.
Using these density matrices we show that the algorithm provides optimal decisions for each bit value. 
It also appears to be optimal in terms of decoding the entire transmitted codeword.
To establish the latter, we compared our calculated performance with the fundamental limit for this setting which can only be calculated numerically~\cite{Krovi-pra15}.
Since there are some small numerical differences between this limit and the algorithm's performance, we are still looking for a mathematical proof to verify that the two are exactly the same (for deciding the transmitted codeword).
Currently, we verify our analysis using Monte-Carlo simulations.

The contributions to quantum computing are four-fold.
First, synthesizing fault-tolerant logical operations for stabilizer QECCs is a fundamental problem.
The usual strategy is to construct a universal set of logical operations via few gates that generate the (logical) Clifford group and at least one non-Clifford gate.
Several researchers have constructed fault-tolerant logical Clifford gates for specific codes (or code families) by focusing on specific properties of the code.
For any stabilizer code, we instead devise a systematic algorithm that can enumerate all physical Clifford circuits, up to an equivalence class, that realize a given logical Clifford circuit~\cite{Rengaswamy-arxiv18*2,Rengaswamy-arxiv19b}.
Our use of the binary symplectic formalism for the Clifford group makes this algorithm efficient.
We refer to this as the \emph{logical Clifford synthesis (LCS)} algorithm.
The key advantage here is the generality of the algorithm. 
This could make it useful in compilers for quantum computers, especially for those that dynamically choose codes to adapt to the current qubit environment.
While the resulting circuits are not always fault-tolerant, we show that a better understanding of the structure of symplectic matrices will lead us in that direction.
Also, an implementation of the algorithm is available at: \url{https://github.com/nrenga/symplectic-arxiv18a}.

Second, an important obstacle in generalizing the LCS algorithm to synthesize logical non-Clifford gates is the lack of a symplectic formalism for such operations.
Thus, we also extend the binary symplectic formalism for Clifford gates to an integer symplectic framework that unifies a large class of diagonal unitaries~\cite{Rengaswamy-pra19}.
These diagonal unitaries have entries that are power-of-$2$ roots of unity raised to a quadratic form in the variables indexing the rows (or columns) of the unitary.
We refer to these operators as \emph{quadratic form diagonal (QFD)} gates.
The \emph{Clifford hierarchy} of unitary operators was defined to demonstrate universal quantum computation via quantum teleportation, given access to some standard resources.
We show that all $1$- and $2$-local diagonal gates in the hierarchy are QFD.
Since most operations performed in the lab are $1$- or $2$-local in some Pauli basis, this allows us to unify them in a single framework, up to basis change operations such as the Hadamard gate.
Moreover, we characterize QFD gates by their action on Pauli matrices under conjugation, since Pauli matrices form an orthonormal basis for all square matrices.
This characterization provides explicit formulae that might find several applications within QIQC and we discuss a few in Section~\ref{sec:discuss}.

Third, it remains non-trivial to translate logical non-Clifford operations to physical (non-Clifford) operations even by applying the QFD framework via (an extension of) the strategy used in the LCS algorithm. 
Hence, we start by reversing our approach and trying to understand when a physical operation preserves the code space of a given stabilizer code. 
This is a necessary condition for such an operation to induce a non-trivial logical operation on the encoded (logical) qubits.
More specifically, utilizing the QFD framework, we devise a systematic procedure for identifying when a given physical QFD (non-Clifford) gate preserves the code space of a stabilizer code.
In general, it appears difficult to characterize the structure required in the stabilizer code for it to support an arbitrary fixed QFD gate.
However, when the QFD gate is a transversal pattern of $T, T^{\dagger}$ and identity on the physical qubits of a stabilizer code, we have solved this problem completely~\cite{Rengaswamy-arxiv19c}.
This is useful because the $T$ gate is a standard choice for the non-Clifford gate that provides universality when combined with all Clifford gates.
From this solution, we also deduce new self-contained classical coding problems, related to self-dual codes, that enable the construction of stabilizer codes that support $T$ gates~\cite{Rengaswamy-arxiv20}.
We also prove useful corollaries related to \emph{triorthogonal codes} and the optimality of \emph{Calderbank-Shor-Steane (CSS)} type stabilizer codes for $T$ gates.
These results address some important open questions in quantum error correction.
A partial extension to finer angle $Z$-rotations is also provided and this leads to some interesting trigonometric problems.
We also devise a family of quantum Reed-Muller codes and characterize the logical operation induced by \emph{transversal} finer angle $Z$-rotations.
Here, transversal means that the same $Z$-rotation is applied individually on all physical qubits.

Lastly, we consider a popular approach to assess the quality of a real quantum computer called \emph{randomized benchmarking}~\cite{Magesan-physreva12}.
It involves applying a sequence of randomly chosen gates to ``twirl'' the underlying noise channel into a \emph{depolarizing channel}, which can be thought of as the quantum analogue of the \emph{binary symmetric channel}.
For this to work, the gates have to be chosen from a statistical ensemble of unitaries called a \emph{unitary $2$-design}.
Such a finite-sized ensemble matches the Haar measure on all unitaries up to the second moment.
There are many known randomized constructions for unitary $2$-designs and one systematic algberaic construction by Cleve et al.~\cite{Cleve-arxiv16}.
We examine the symmetries of the classical $\mathbb{Z}_4$-linear Kerdock code by mapping it bijectively to a collection of \emph{graph states}, which form a special type of quantum \emph{stabilizer states}.
We use binary symplectic matrices to show that the symmetries of the Kerdock code form a subgroup of the Clifford group that is \emph{Pauli mixing}~\cite{Webb-arxiv16}.
This means that if any Pauli matrix is conjugated by a randomly chosen symmetry element from the aforementioned group, then the result is uniformly distributed over all Pauli matrices.
Since Pauli mixing ensembles are known to form unitary $2$-designs, we effectively prove that the (Clifford) symmetries of the $\mathbb{Z}_4$-linear Kerdock code form a unitary $2$-design~\cite{Can-arxiv19}.
Moreover, this design has an almost optimal size and is isomorphic to the design of Cleve et al. while having a simpler description.
The connection to classical codes is new and we think that this might provide opportunities to construct deterministic ensembles that form unitary $t$-designs for $t \geq 4$, since the Clifford group itself forms a unitary $3$-design.
As unitary $t$-designs are widely used in QIQC, such constructions have several applications.
Furthermore, by combining this with the LCS algorithm, we can systematically translate the physical unitary $2$-design into a \emph{logical} unitary $2$-design. 
This provides a concrete approach for \emph{logical} randomized benchmarking protocol~\cite{Combes-arxiv17}.
A computer implementation of this design is available at: \url{https://github.com/nrenga/symplectic-arxiv18a}.
Due to space considerations, we do not include this last contribution in the dissertation and instead refer the reader to our detailed paper on this construction~\cite{Can-arxiv19}.

\section{From ECCs to QECCs and back: A Prelude}

Until the 1990s, it was thought that noise would be the fundamental obstruction to building a reliable quantum computer.
This was because, just like for analog computers, one expects that it is impossible to protect information against a continuous noise model.
However, Calderbank and Shor~\cite{Calderbank-physreva96}, and independently Steane~\cite{Steane-physreva96}, devised a framework for constructing quantum codes from classical error-correcting codes, and this construction is now referred to as \emph{CSS codes} (for their initials).
CSS codes form an interesting subclass of stabilizer codes~\cite{Gottesman-phd97,Calderbank-it98*2}.
The ingredients for the CSS construction are two classical binary linear codes $C_1$ and $C_2$ with parameters $[n,k_1,d_1]$ and $[n,k_2,d_2]$, respectively, such that $C_2 \subset C_1$.
Let the dual codes of $C_1$ and $C_2$ be denoted by $C_1^{\perp}$ and $C_2^{\perp}$, respectively.
If $C_2^{\perp}$ has minimum distance $d_2^{\perp}$, then the construction yields a $\llbr n,k_1-k_2, d \rrbr$ stabilizer code CSS($C_1,C_2$), where $d \geq \min\{ d_1,d_2^{\perp} \}$.
The key difference here is that the distance of the code is defined with respect to a \emph{discretized} set of Pauli errors.
Ideally, this can be justified by arguing that when syndromes are measured, the measurement collapses the state into a Pauli ``frame''.
In other words, the true noise operation can be decomposed into a linear combination of Paulis, which form an orthonormal basis, and the measurement projects the state onto \emph{one} of these terms.
This is, however, not always true since we could still be left with a residual error that is a sum of (fewer) Pauli terms.
But it has been shown that if the code can correct a discrete set of errors $\{ E_i \}$ then it can protect against any linear combination of these errors~\cite[Section 2.6]{LB-2013}.
Thus, for many reasonable circumstances, the CSS construction provides a systematic way to translate ``good'' classical codes (ECCs) into ``good'' quantum codes (QECCs).

In classical error correction, ``good'' code families are usually characterized by high rates, growing distance, and efficient close-to-optimal decoders.
For quantum error correction, an additional requirement is the existence of fault-tolerant realizations of a universal set of logical operators.
This feature makes designing good QECCs more challenging and introduces interesting trade-offs.
It is well-known that CSS codes constructed out of classical self-orthogonal codes have a \emph{transversal} implementation of the logical Clifford operators.
Here, transversal means that the physical operation splits into individual operations on the code qubits or on pairs of corresponding code qubits in two different code blocks.
In general, implementing multi-qubit logical operations fault-tolerantly within a single code block is challenging.
It is also difficult to find fault-tolerant implementations of logical non-Clifford operations.
Inspired by this challenge, we will now describe a self-contained new classical coding problem based on CSS codes.
This is intimately related to realizing non-Clifford logical operations fault-tolerantly via transversal $T$ gates, as we explain later.

A \emph{CSS-T} code is a CSS($C_1,C_2$) code such that the following hold:
\begin{enumerate}
    \item $C_2$ is an even code, i.e., $w_H(x) \equiv 0$ (mod $2$) for all $x \in C_2$, where $w_H(x)$ is the Hamming weight of $x$.
    
    \item For each $x \in C_2$, there exists a dimension $w_H(x)/2$ self-dual code in $C_1^{\perp}$ that is supported on $x$, i.e., there exists $ C_x \subseteq C_1^{\perp}$ s.t. $|C_x| = 2^{w_H(x)/2}, C_x = C_x^{\perp}$, and $z \in C_x \Rightarrow z \preceq x\ (\text{supp}(z) \subseteq \text{supp}(x))$, where $\text{supp}(x)$ denotes the support of $x$.
\end{enumerate}
A simple example that clarifies these conditions is as follows.
Define $C_2$ to be the $[6,1,6]$ repetition code and $C_1$ to be the concatenation of three $[2,1,2]$ repetition codes.
Then it is easy to see that the only non-trivial codeword in $C_2$ is $x = [1,1,1,1,1,1]$, and $C_1 = C_1^{\perp}$ is generated by $z_1 = [1,1,0,0,0,0], z_2 = [0,0,1,1,0,0], z_3 = [0,0,0,0,1,1]$.
Hence, both of the above conditions are satisfied and this determines a $\llbr 6,2,2 \rrbr$ CSS (or CSS-T) code.

A more non-trivial construction using Reed-Muller codes is as follows, which we elaborate on later.
Set $C_1 = \text{RM}(r,m)$ and $C_2 = \text{RM}(r-1,m)$ where $\frac{m-1}{3} < r \leq \frac{m}{3}$.
As an example, let $m = 6, r = 2$.
Since RM codes are defined by monomials, this means we have $6$ binary variables $x_1,x_2,\ldots,x_6$, $C_2$ is generated by all monomials of degree $0$ or $1$, and $C_1$ has additional generators corresponding to degree $2$ monomials.
Consider, for example, the codeword in $C_2$ that is the evaluation of the monomial $x_1$.
In this case we know that $C_1^{\perp} = \text{RM}(m-r-1,m) = \text{RM}(3,6)$.
Now consider the codewords in $C_1^{\perp}$ corresponding to monomials $x_1, x_1 x_i, x_1 x_i x_j$ for $i,j \geq 2$ and $j \neq i$.
Since these are in the support of $x_1$, we can drop $x_1$ and look at the monomial inside that support.
This gives us monomials of the form $1, \tilde{x}_i = x_{i-1}, \tilde{x}_i \tilde{x}_j = x_{i-1} x_{j-1}$ that correspond to $m-1 = 5$ variables.
But we immediately recognize that these monomials exactly generate the code RM($2,5$) that is self-dual.
Similar arguments can be made for all other codewords in $C_2$ and hence this construction gives a $\llbr 2^m, \binom{m}{r}, 2^r \rrbr$ family of CSS-T quantum Reed-Muller codes.

Although we have a large CSS-T family with growing distance, we see that the rate vanishes asymptotically.
Hence, a major open problem here is to construct a ${\llbr n,k_1 - k_2,d \rrbr}$ CSS-T family with constant rate, i.e., $(k_1 - k_2)/n = \Omega(1)$, and growing distance that is ideally asymptotically linear as well, i.e., $d/n = \Omega(1)$.
We think that there is much to be gained here by leveraging the rich classical literature on self-dual codes~\cite{Rains-arxiv02,Nebe-2006}.
If such a family is found, then it would translate into tremendous resource savings for an important subroutine called \emph{magic state distillation}~\cite{Bravyi-pra12}.
This protocol uses such codes to distill ``magic'' states to perform fault-tolerant logical non-Clifford gates.
Even outside this protocol, a constant-rate CSS-T family might directly translate to an efficient UFTQC.

\section{Organization of this Dissertation}

Chapter~\ref{ch:ch2_background} provides the essential background to describe our contributions to quantum communications in Chapter~\ref{ch:ch3_bpqm}.
Chapter~\ref{ch:ch4_groups} expands on Chapter~\ref{ch:ch2_background} and provides a detailed discussion of the mathematical framework for quantum error correction.
Chapters~\ref{ch:ch5_lcs_algorithm},~\ref{ch:ch6_qfd_gates}, and~\ref{ch:ch7_stabilizer_codes_qfd} discuss our contributions to quantum computation in the order described above.

%% file: ch2_background.tex

\label{ch:ch2_background}

\section{Qubits and Density Matrices}

A single qubit is represented by a unit vector in $\mathbb{C}^2$, the $2$-dimensional complex vector space~\cite{Nielsen-2010}.
The single-qubit \emph{computational basis} is defined to be the standard basis:
\begin{align}
\dket{0} \coloneqq e_0 = 
\begin{bmatrix}
1 \\ 0
\end{bmatrix}, \quad
\dket{1} \coloneqq e_1 = 
\begin{bmatrix}
0 \\ 1
\end{bmatrix}.
\end{align}
So a single qubit can be written as $\dket{\psi} = \alpha \dket{0} + \beta \dket{1}$, where $\alpha, \beta \in \mathbb{C}$ satisfy $|\alpha|^2 + |\beta|^2 = 1$.
A system composed of $n > 1$ qubits resides in the Hilbert space $\mathbb{C}^N$, where $N \coloneqq 2^n$.
The computational basis for $\mathbb{C}^N$ is $\{ \dket{v} = \dket{v_1} \otimes \dket{v_2} \otimes \cdots \otimes \dket{v_n} \in \mathbb{C}^N, \ v_i \in \{0,1\} \}$, where $\otimes$ denotes the \emph{Kronecker product}.
Observe that if each entry of the vector $\dket{v}$ is indexed by a binary vector $x \in \mathbb{Z}_2^n$, then $\dket{v} \in \mathbb{C}^N$ is a standard basis vector with a $1$ in the entry indexed by $x = v$ and zeros elsewhere.
So an $n$-qubit system in state $\dket{\psi}$ can be written as
\begin{align}
\dket{\psi} = \sum_{v \in \mathbb{Z}_2^n} \alpha_v \ket{v}, \quad \alpha_v \in \mathbb{C}, \quad |\dbraket{\psi}|^2 = \sum_{v \in \mathbb{Z}_2^n} |\alpha_v|^2 = 1,
\end{align}
where $\dbra{\psi} \coloneqq \dket{\psi}^{\dagger}$ is the Hermitian (conjugate) transpose of the state $\dket{\psi}$, and $\dbraket{\psi}{\phi}$ represents the complex inner product between two states $\dket{\psi}$ and $\dket{\phi}$, i.e, $\dbraket{\psi}{\phi} = \dbraket{\phi}{\psi}^*$.
Using this inner product, the computational basis forms an orthonormal basis for $\mathbb{C}^N$.
A state is \emph{entangled} if it does not decompose under the Kronecker product, e.g., $(\dket{00} + \dket{11})/\sqrt{2}$.

States such as above are commonly referred to as \emph{pure states} because they represent a system that is in a definitive state.
When there is uncertainty in the state of a system, it is generally represented as a ``bag of states'' $\{ p_x, \dket{\psi_x} \}$, where $\dket{\psi_x}$ are pure states that are not necessarily orthogonal and $p_x$ is the probability that the system is in the state $\dket{\psi_x}$.
It is convenient to think of pure states as rank-$1$ matrices $\rho_x \coloneqq \dketbra{\psi_x}$.
With this representation, the \emph{density matrix} of the \emph{mixed state} corresponding to the bag of states is
\begin{align}
\rho \coloneqq \sum_x p_x \dketbra{\psi_x} \in \mathbb{C}^{N \times N}.
\end{align}
It is easy to verify that $\rho$ is a Hermitian (or self-adjoint) operator with unit trace, because the trace is linear and $\text{Tr}(\dketbra{\psi_x}) = \text{Tr}(\dbraket{\psi_x}) = \dbraket{\psi_x} = 1$.
It can be shown that $\rho$ is pure, i.e., $\rho = \dketbra{\psi_x}$ for exactly one $x$, if and only if $\text{Tr}(\rho^2) = \text{Tr}(\rho) = 1$.

A useful interpretation of $\rho$ is the following.
Let $X$ be a random variable that takes values in $\{ \dketbra{\psi_x} \}$.
Then $\rho$ is the \emph{statistical expectation} of this random variable $X$.
Hence, $\rho$ does not correspond to a physical state but is the representation of our knowledge about the system.
So if an algorithm is analyzed by examining its action on the density matrix of a system, then this is implicitly an ``average case analysis''.
Therefore, care must be taken to differentiate this from the analysis for a given instance of the system, i.e., for a given $\dket{\psi_x}$.
This observation will be useful when we study the BPQM algorithm.
Also note that $\rho$ does not uniquely represent a system. 
If $\rho$ is not diagonal, then diagonalizing $\rho$ yields a new bag of states, consisting of its eigenstates, that gives rise to the same density matrix.

\section{Basic Unitaries}

Any reversible operation on a quantum system can be described by a unitary matrix~\cite{Nielsen-2010}.
The \emph{unitary group}, $\mathbb{U}_N$, on $n = \log_2(N)$ qubits is defined as
\begin{align}
\mathbb{U}_N \coloneqq \{ U \in \mathbb{C}^{N \times N} \colon UU^{\dagger} = U^{\dagger}U = I_N \},
\end{align}
where $U^{\dagger}$ is the conjugate transpose of $U$ and $I_N$ is the $N \times N$ identity matrix\footnote{Henceforth, we might drop the subscript so that $I$ represents the identity matrix of the appropriate size in context.}.
When a unitary matrix $U \in \mathbb{U}_N$ acts on a system described as a bag of states, the system evolves as $\{ p_x, U \dketbra{\psi_x} U^{\dagger} \}$.
Hence, the density matrix evolves as $\rho \mapsto U \rho U^{\dagger}$.
All unitaries $U$ can be decomposed into a sequence of standard single- and multi-qubit gates which we will describe here.
For more details on these operations, see~\cite{Nielsen-2010}.

\subsection{Single-Qubit Gates}

The single-qubit \emph{Pauli} operators are given by the Hermitian unitary matrices
\begin{align}
I_2 \coloneqq 
\begin{bmatrix}
1 & 0 \\ 0 & 1
\end{bmatrix}, \ 
X \coloneqq 
\begin{bmatrix}
0 & 1 \\ 1 & 0
\end{bmatrix}, \ 
Z \coloneqq 
\begin{bmatrix}
1 & 0 \\ 0 & -1
\end{bmatrix}, \ 
Y \coloneqq \imath X Z = 
\begin{bmatrix}
0 & -\imath \\ \imath & 0
\end{bmatrix},
\end{align}
where $\imath \coloneqq \sqrt{-1}$.
The eigenbasis of $Z$ is the computational basis $\{ \dket{0}, \dket{1} \}$, the eigenbasis of $X$ is the \emph{conjugate basis} $\{ \dket{+}, \dket{-} \}$, and the eigenbasis of $Y$ is $\left\{ \frac{\dket{0} + \imath \dket{1}}{\sqrt{2}}, \frac{\dket{0} - \imath \dket{1}}{\sqrt{2}} \right\}$, where
\begin{align}
\dket{+} \coloneqq \frac{\dket{0} + \dket{1}}{\sqrt{2}}, \quad \dket{-} \coloneqq \frac{\dket{0} - \dket{1}}{\sqrt{2}}.
\end{align}
The normalized Pauli matrices, $\frac{1}{\sqrt{2}} \{ I_2, X, Z, Y \}$, form an orthonormal basis for all $2 \times 2$ matrices under the \emph{Frobenius inner product} 
$\langle A, B \rangle_F \coloneqq \text{Tr}(A^{\dagger} B)$.

The \emph{Hadamard} gate, $H$ (or sometimes $H_2$ to emphasize that it is on a single qubit), is the change-of-basis operator on $\mathbb{C}^2$ between the computational and conjugate bases, i.e., $HXH^{\dagger} = Z,\ HZH^{\dagger} = X$ and $H = H^{\dagger}$.
The \emph{Phase} gate, $P$, fixes the computational basis and maps $X \mapsto Y$ under conjugation.
The ``$T$'' gate, also called the ``$\pi/8$'' gate, also fixes the computational basis but maps $X \mapsto (X + Y)/\sqrt{2}$.
These are defined by the matrices
\begin{align}
H \coloneqq \frac{1}{\sqrt{2}}
\begin{bmatrix}
1 & 1 \\ 1 & -1
\end{bmatrix}, \quad
P \coloneqq 
\begin{bmatrix}
1 & 0 \\ 0 & \imath
\end{bmatrix} = \sqrt{Z}, \quad
T \coloneqq 
\begin{bmatrix}
1 & 0 \\ 0 & e^{\imath\pi/4}
\end{bmatrix} \equiv \exp\left( \frac{-\imath\pi}{8} Z \right),
\end{align}
where for the last equivalence we have ignored the global phase $e^{-\imath\pi/8}$ that is undetectable by any measurement in quantum mechanics.
As claimed above, these matrices satisfy
\begin{align}
HXH^{\dagger} = Z,\ HZH^{\dagger} = X,\ PXP^{\dagger} = Y,\ PZP^{\dagger} = Z = TZT^{\dagger},\ TXT^{\dagger} = \frac{X+Y}{\sqrt{2}} = e^{\frac{-\imath\pi}{4}} YP.
\end{align}
We will derive the relations for the $T$ gate more systematically in Chapter~\ref{ch:ch6_qfd_gates}.
It is easily seen that $H$ and $P$ can be used to derive all Pauli matrices: $P^2 = Z, HZH^{\dagger} = X, PXP^{\dagger} = Y$.
In fact the \emph{Clifford group} on a single qubit is generated by $H$ and $P$, and up to global phases it is a finite (quotient) group with $24$ elements.
Moreover, $H$ and $T$ can be used to approximate any single-qubit unitary $U$ up to any precision.
More precisely, given any unitary $U \in \mathbb{U}_2$ and precision $\epsilon > 0$, there exists a finite sequence of unitaries $U_1, U_2, \ldots, U_m$ with $U_i \in \{ H,T \}$ such that the error, in terms of the operator $2$-norm, is
\begin{align}
E\left( U, U_m \cdots U_1 \right) \coloneqq \max_{\dket{\psi}} \left\| \left( U - U_m \cdots U_1 \right)  \dket{\psi} \right\|_2 = \left\| U - U_m \cdots U_1 \right\|_{2} < \epsilon.
\end{align}
Using this definition of error it can be shown that both $U$ and $U_m U_{m-1} \cdots U_1$ approximately produce the same measurement statistics after operating on any input state~\cite{Nielsen-2010}.
Furthermore, if we intend to apply a sequence of unitaries $V_1,V_2,\ldots,V_m$ but only perform their respective approximations $U_1,U_2,\ldots,U_m$ in practice, then the error grows at most additively~\cite[Section 4.5.3]{Nielsen-2010}:
\begin{align}
E(V_m V_{m-1} \cdots V_1, U_m U_{m-1} \cdots U_1) \leq \sum_{i=1}^{m} E(V_i,U_i).
\end{align}
Note here that, when the sequence of operations is $V_1, \ldots, V_m$ in that order, the overall unitary matrix is calculated as $V_m V_{m-1} \cdots V_1$ since the matrix acts to its right on the input state.
These observations about the error also generalize to the multi-qubit case.

\subsection{Multi-Qubit Gates}

For $n$ qubits, the \emph{Pauli group} consists of Kronecker products of $n$ single-qubit Paulis with overall phases $\imath^{\kappa}$, where $\kappa \in \{0,1,2,3\}$.
As in the $n=1$ case, the (normalized) Hermitian elements in the group form an orthonormal basis for all $N \times N$ matrices under the Frobenius inner product.
Similarly, the Hadamard, Phase, and $T$ gates on different qubits are constructed via Kronecker products, and these qubit indices are denoted in subscripts, e.g., $H_1, P_2, T_3$.
The $n$-qubit \emph{Clifford group} is formally defined as the \emph{normalizer} of the Pauli group inside the group of all unitaries, $\mathbb{U}_N$.
We will delve into more details about these groups in Chapter~\ref{ch:ch4_groups}.
This Clifford group is generated by $H, P$ and either the \emph{Controlled-$X$ (CX)} or \emph{Controlled-$Z$ (CZ)} gates on arbitrary (pairs of) qubits.
They are defined as
\begin{align}
\text{CX}_{1 \rightarrow 2} \coloneqq \dketbra{0}_1 \otimes I_2 + \dketbra{1}_1 \otimes X_2, \quad \text{CZ}_{12} \coloneqq \dketbra{0}_1 \otimes I_2 + \dketbra{1}_1 \otimes Z_2.
\end{align}
Let us examine the action of these gates on the $2$-qubit computational basis.
If $a,b \in \{0,1\}$,
\begin{align}
\text{CX}_{1 \rightarrow 2}\, (\dket{a}_1 \otimes \dket{b}_2) = \dket{a}_1 \otimes \dket{a \oplus b}_2, \quad 
\text{CZ}_{12}\, (\dket{a}_1 \otimes \dket{b}_2) = (-1)^{ab} \dket{a}_1 \otimes \dket{b}_2.
\end{align}
Here, $\oplus$ denotes addition modulo $2$.
Therefore, the CX gate is more commonly called the \emph{Controlled-NOT (CNOT)} gate.
In this form, the first qubit is called the ``control'' and the second qubit is called the ``target''.
Observe that swapping the control and target for CZ introduces no difference but swapping them alters the action of the CX gate non-trivially, i.e., $\text{CX}_{2 \rightarrow 1} \neq \text{CX}_{1 \rightarrow 2}$ while $\text{CZ}_{12} = \text{CZ}_{21}$.
When we have an input state that is a superposition of computational basis states (of the form $\dket{a} \otimes \dket{b}$), then the aforementioned CNOT rule is applied linearly to each basis state in the superposition.

For brevity, we will use the following equivalent notations for states and operators:
\begin{align}
\dket{u} \otimes \dket{v} = \dket{u}_1 \dket{v}_2 = \dket{u,v} = \dket{uv}, \quad
A_i \otimes B_j \otimes C_k = A_i B_j C_k,
\end{align}
where the subscripts $1,2$ represent qubit indices and each of $i,j,k$ potentially represents a subset of qubits.
For example, we usually write $X_1 \otimes \text{CZ}_{23} \otimes P_4 = X_1\, \text{CZ}_{23}\, P_4$.

It can be verified that CX and CZ satisfy the identity $(I_1 \otimes H_2) \, \text{CX}_{1 \rightarrow 2}\, (I_1 \otimes H_2) = \text{CZ}_{12}$.
The CX and CZ gates can also be defined by their action on Pauli matrices as follows.
\begin{align}
\text{CX}_{1 \rightarrow 2} \, (X_1 \otimes I_2) \, \text{CX}_{1 \rightarrow 2}^{\dagger} = X_1 \otimes X_2, & \quad \text{CZ}_{12} \, (X_1 \otimes I_2) \, \text{CZ}_{12}^{\dagger} = X_1 \otimes Z_2 \\
\text{CX}_{1 \rightarrow 2} \, (I_1 \otimes X_2) \, \text{CX}_{1 \rightarrow 2}^{\dagger} = I_1 \otimes X_2, & \quad \text{CZ}_{12} \, (I_1 \otimes X_2) \, \text{CZ}_{12}^{\dagger} = Z_1 \otimes X_2 \\
\text{CX}_{1 \rightarrow 2} \, (Z_1 \otimes I_2) \, \text{CX}_{1 \rightarrow 2}^{\dagger} = Z_1 \otimes I_2, & \quad \text{CZ}_{12} \, (Z_1 \otimes I_2) \, \text{CZ}_{12}^{\dagger} = Z_1 \otimes I_2 \\
\text{CX}_{1 \rightarrow 2} \, (I_1 \otimes Z_2) \, \text{CX}_{1 \rightarrow 2}^{\dagger} = Z_1 \otimes Z_2, & \quad \text{CZ}_{12} \, (X_1 \otimes I_2) \, \text{CZ}_{12}^{\dagger} = I_1 \otimes Z_2.
\end{align}
In circuit notation, these standard single- and two-qubit gates and their identities are given in Table~\ref{tab:circuit_identities}.
It is also convenient to expand CX and CZ into their full matrix form:
\begin{align}
\text{CX}_{1 \rightarrow 2} \coloneqq
\begin{bmatrix}
1 & 0 & 0 & 0 \\
0 & 1 & 0 & 0 \\
0 & 0 & 0 & 1 \\
0 & 0 & 1 & 0
\end{bmatrix}, \quad
\text{CZ}_{12} \coloneqq
\begin{bmatrix}
1 & 0 & 0 & 0 \\
0 & 1 & 0 & 0 \\
0 & 0 & 1 & 0 \\
0 & 0 & 0 & -1
\end{bmatrix}.
\end{align}
Henceforth, just as we mentioned for the identity earlier, we might drop the subsystem indices when using CX and CZ unless, e.g., if the control and target are swapped for CX.
The gate set $\{ H, T, \text{CX} \}$, or equivalently $\{ H, T, \text{CZ} \}$, can be used to approximate any $n$-qubit unitary with arbitrary precision $\epsilon$ in the operator norm, similar to the $n=1$ case.

Two standard $3$-qubit gates that are sometimes convenient to use are:
\begin{align}
\text{CCX} \coloneqq
\begin{bmatrix}
1 & 0 & 0 & 0  &  0 & 0 & 0 & 0 \\
0 & 1 & 0 & 0  &  0 & 0 & 0 & 0 \\
0 & 0 & 1 & 0  &  0 & 0 & 0 & 0 \\
0 & 0 & 0 & 1  &  0 & 0 & 0 & 0 \\
0 & 0 & 0 & 0  &  1 & 0 & 0 & 0 \\
0 & 0 & 0 & 0  &  0 & 1 & 0 & 0 \\
0 & 0 & 0 & 0  &  0 & 0 & 0 & 1 \\
0 & 0 & 0 & 0  &  0 & 0 & 1 & 0
\end{bmatrix}, \quad
\text{CCZ} \coloneqq
\begin{bmatrix}
1 & 0 & 0 & 0  &  0 & 0 & 0 & 0 \\
0 & 1 & 0 & 0  &  0 & 0 & 0 & 0 \\
0 & 0 & 1 & 0  &  0 & 0 & 0 & 0 \\
0 & 0 & 0 & 1  &  0 & 0 & 0 & 0 \\
0 & 0 & 0 & 0  &  1 & 0 & 0 & 0 \\
0 & 0 & 0 & 0  &  0 & 1 & 0 & 0 \\
0 & 0 & 0 & 0  &  0 & 0 & 1 & 0 \\
0 & 0 & 0 & 0  &  0 & 0 & 0 & -1
\end{bmatrix}.
\end{align}
These are called the \emph{Controlled-Controlled-$X$ (CCX)} and \emph{Controlled-Controlled-$Z$ (CCZ)} gates, and CCX is commonly called the \emph{Toffoli} gate.
Their actions can be written as 
\begin{align}
\text{CCX} \, \dket{a,b,c} = \dket{a,b,c \oplus ab}, \quad \text{CCZ}\, \dket{a,b,c} = (-1)^{abc} \dket{a,b,c}.
\end{align}
Hence, CCX flips the third qubit if and only if the first two qubits are $1$.
As mentioned before for CX and CZ, when the input is a superposition of computational basis states, the gates act linearly on each basis state, e.g., CCX $(\dket{000} + \dket{111})/\sqrt{2} = (\dket{000} + \dket{110})/\sqrt{2}$.
It is well-known that the CCX gate itself is universal for classical computation.
In circuit notation these are denoted as CCX $\equiv$ %
\begin{tikzcd}
\qw & \ctrl{1} & \qw \\
\qw & \ctrl{1} & \qw \\
\qw & \targ{} & \qw
\end{tikzcd}
and CCZ $\equiv$ %
\begin{tikzcd}
\qw & \ctrl{1} & \qw \\
\qw & \ctrl{1} & \qw \\
\qw & \control{} & \qw
\end{tikzcd}.

Finally, note that any controlled unitary or inversely controlled unitary is written as
\begin{align}
\text{C-}U \coloneqq \dketbra{0} \otimes I + \dketbra{1} \otimes U, \ \ \text{Ci-}U \coloneqq \dketbra{0} \otimes U + \dketbra{1} \otimes I = X_1\, \text{C-}U\, X_1.
\end{align}
Extending this, a coherently controlled unitary can be decomposed as 
\begin{align}
U = \sum_{x \in \{0,1\}^m} U_{x} \otimes \dketbra{x} = \prod_{x \in \{0,1\}^m} \big[ U_{x} \otimes \dketbra{x} + I \otimes (I_{2^m} - \dketbra{x}) \big],
\end{align}
where $I$ is the identity operator of the same size as $U_{x}$.
Hence, if we can perform $m$-controlled unitaries C$^m$-$U_x$'s, that are controlled on $x = [1,1,\ldots,1]$, then each of the component gates in the above product can be obtained by ``sandwiching'' appropriate Pauli $X$ gates, similar to Ci-$U$ above where we had $m=1$.

\begin{table}

\begin{tabular}{lc|cr}
(a) \ 
\begin{tikzcd}
\lstick{$X$} & \gate{H} & \qw
\end{tikzcd}
=
\begin{tikzcd}
\qw & \gate{H} & \qw\rstick{$Z$}
\end{tikzcd}
         &  &  &
\begin{tikzcd}
\lstick{$Z$} & \gate{H} & \qw
\end{tikzcd}
=
\begin{tikzcd}
\qw & \gate{H} & \qw\rstick{$X$}
\end{tikzcd}
        \\ & & & \\         
(b) \ 
\begin{tikzcd}
\lstick{$X$} & \gate{P} & \qw
\end{tikzcd}
=
\begin{tikzcd}
\qw & \gate{P} & \qw\rstick{$Y$}
\end{tikzcd}
         &  &  &
\begin{tikzcd}
\lstick{$Z$} & \gate{P} & \qw
\end{tikzcd}
=
\begin{tikzcd}
\qw & \gate{P} & \qw\rstick{$Z$}
\end{tikzcd}
        \\ & & & \\         
(c) \ 
\begin{tikzcd}
\lstick{$X$} & \gate{T} & \qw
\end{tikzcd}
=
\begin{tikzcd}
\qw & \gate{T} & \qw\rstick{$e^{-\imath\pi/4} YP$}
\end{tikzcd}
         &  &  &
\begin{tikzcd}
\lstick{$Z$} & \gate{T} & \qw
\end{tikzcd}
=
\begin{tikzcd}
\qw & \gate{T} & \qw\rstick{$Z$}
\end{tikzcd}
        \\ &  &  &  \\
         \hline
           &  &  &  \\
CX :  \hspace*{-0.4cm}
\begin{tikzcd}
\lstick{~} & \ctrl{1} & \qw \\
\lstick{~} & \gate{X} & \qw
\end{tikzcd}
$\equiv$
\begin{tikzcd}
\qw & \ctrl{1} & \qw \\
\qw & \targ{} & \qw
\end{tikzcd}
         &  &  &
CZ :  \hspace*{-0.4cm}
\begin{tikzcd}
\lstick{~} & \ctrl{1} & \qw \\
\lstick{~} & \gate{Z} & \qw
\end{tikzcd}
$\equiv$
\begin{tikzcd}
\qw & \ctrl{1} & \qw \\
\qw & \control{} & \qw
\end{tikzcd} \hspace*{0.4cm}
        \\ & & & \\         
(d) \ 
\begin{tikzcd}
\lstick{$X$} & \ctrl{1} & \qw \\
\lstick{} & \targ{} & \qw
\end{tikzcd}
=
\begin{tikzcd}
\qw & \ctrl{1} & \qw\rstick{$X$} \\
\qw & \targ{} & \qw\rstick{$X$}
\end{tikzcd}
         &  &  &
\begin{tikzcd}
\lstick{$X$} & \ctrl{1} & \qw \\
\lstick{} & \control{} & \qw
\end{tikzcd}
=
\begin{tikzcd}
\qw & \ctrl{1} & \qw\rstick{$X$} \\
\qw & \control{} & \qw\rstick{$Z$}
\end{tikzcd}
        \\ & & & \\         
(e) \ 
\begin{tikzcd}
\lstick{} & \ctrl{1} & \qw \\
\lstick{$X$} & \targ{} & \qw
\end{tikzcd}
=
\begin{tikzcd}
\qw & \ctrl{1} & \qw\rstick{} \\
\qw & \targ{} & \qw\rstick{$X$}
\end{tikzcd}
         &  &  &
\begin{tikzcd}
\lstick{} & \ctrl{1} & \qw \\
\lstick{$X$} & \control{} & \qw
\end{tikzcd}
=
\begin{tikzcd}
\qw & \ctrl{1} & \qw\rstick{$Z$} \\
\qw & \control{} & \qw\rstick{$X$}
\end{tikzcd}
        \\ & & & \\         
(f) \ 
\begin{tikzcd}
\lstick{$Z$} & \ctrl{1} & \qw \\
\lstick{} & \targ{} & \qw
\end{tikzcd}
=
\begin{tikzcd}
\qw & \ctrl{1} & \qw\rstick{$Z$} \\
\qw & \targ{} & \qw\rstick{}
\end{tikzcd}
         &  &  &
\begin{tikzcd}
\lstick{$Z$} & \ctrl{1} & \qw \\
\lstick{} & \control{} & \qw
\end{tikzcd}
=
\begin{tikzcd}
\qw & \ctrl{1} & \qw\rstick{Z} \\
\qw & \control{} & \qw\rstick{}
\end{tikzcd}
        \\ & & & \\         
(g) \ 
\begin{tikzcd}
\lstick{} & \ctrl{1} & \qw \\
\lstick{$Z$} & \targ{} & \qw
\end{tikzcd}
=
\begin{tikzcd}
\qw & \ctrl{1} & \qw\rstick{$Z$} \\
\qw & \targ{} & \qw\rstick{$Z$}
\end{tikzcd}
         &  &  &
\begin{tikzcd}
\lstick{} & \ctrl{1} & \qw \\
\lstick{$Z$} & \control{} & \qw
\end{tikzcd}
=
\begin{tikzcd}
\qw & \ctrl{1} & \qw\rstick{} \\
\qw & \control{} & \qw\rstick{$Z$}
\end{tikzcd}
        \\ & & & \\         
(h) \ 
\begin{tikzcd}
\lstick{$X$} & \ctrl{1} & \qw \\
\lstick{$Z$} & \targ{} & \qw
\end{tikzcd}
=
\begin{tikzcd}
\qw & \ctrl{1} & \qw\rstick{$-Y$} \\
\qw & \targ{} & \qw\rstick{$Y$}
\end{tikzcd}
         &  &  &
\begin{tikzcd}
\lstick{$X$} & \ctrl{1} & \qw \\
\lstick{$X$} & \control{} & \qw
\end{tikzcd}
=
\begin{tikzcd}
\qw & \ctrl{1} & \qw\rstick{$Y$} \\
\qw & \control{} & \qw\rstick{$Y$}
\end{tikzcd}
        \\ & & & \\         
%
\end{tabular}
    
\caption[Commonly used identities for the standard gates. Circuits are drawn using the ``Quantikz'' package]{\label{tab:circuit_identities}Commonly used identities for the standard gates. Also note that $(I_2 \otimes H) \, \text{CX}\, (I_2 \otimes H) = \text{CZ}$. Circuits are drawn using the ``Quantikz'' package~\cite{Kay-arxiv18}.}
\end{table}

\section{Measurements}
\label{sec:measurement}

A \emph{generalized measurement} is described by a set of measurement operators $\{ M_i \}$ that satisfy the \emph{completeness relation} $\sum_i M_i^{\dagger} M_i = \mathbbm{I}$, where $\mathbbm{I}$ is the identity operator~\cite{Nielsen-2010}.
If a generalized measurement is performed on a system, and the result of the measurement is $i$, then the resulting bag of states is given by 
\begin{align}
\left\{ p_x, \ \dket{\psi_x} \right\} \longmapsto \left\{ p_{x|i}, \  \dket{\psi_x^i} \coloneqq \frac{M_i \dket{\psi_x}}{\sqrt{\dbra{\psi_x} M_i^{\dagger} M_i \dket{\psi_x}}} \right\}.
\end{align}
If the initial state was $\dket{\psi_x}$ then the probability of measuring output $i$ is given by $p_{i|x} \coloneqq \dbra{\psi_x} M_i^{\dagger} M_i \dket{\psi_x}$, and the post-measurement state is $\dket{\psi_x^i}$.
Hence, we can calculate
\begin{align}
p_{x|i} = \frac{p_x \cdot p_{i|x}}{\sum_x p_x \cdot p_{i|x}} = \frac{p_x \cdot \dbra{\psi_x} M_i^{\dagger} M_i \dket{\psi_x}}{\sum_x p_x \cdot \dbra{\psi_x} M_i^{\dagger} M_i \dket{\psi_x}} = \frac{p_x \cdot \dbra{\psi_x} M_i^{\dagger} M_i \dket{\psi_x}}{\text{Tr}\left[ M_i^{\dagger} M_i \rho \right]}.
\end{align}
Therefore, given measurement result $i$, the density matrix evolves as
\begin{align}
\rho_i = \sum_x p_{x|i} \dketbra{\psi_x^i} = \sum_x \frac{p_x \cdot \dbra{\psi_x} M_i^{\dagger} M_i \dket{\psi_x}}{\text{Tr}\left[ M_i^{\dagger} M_i \rho \right]} \frac{M_i \dketbra{\psi_x} M_i^{\dagger}}{\dbra{\psi_x} M_i^{\dagger} M_i \dket{\psi_x}} = \frac{M_i \rho M_i^{\dagger}}{\text{Tr}\left[ M_i^{\dagger} M_i \rho \right]},
\end{align}
and the total probability of obtaining measurement result $i$ is $p_i = \text{Tr}\left[ M_i^{\dagger} M_i \rho \right]$.

A \emph{projective measurement} is described by a set of projection operators $\{ \Pi_i \}$ such that $0 \leq \Pi_i \leq \mathbbm{I}, \Pi_i^2 = \Pi_i, \Pi_i \Pi_j = \delta_{ij} \Pi_i$ and $\sum_i \Pi_i = \mathbbm{I}$, where $\mathbbm{I}$ is the identity operator~\cite{Nielsen-2010}.
For projective measurements the post-measurement states and probabilities simplify as
\begin{align}
p_{i|x} = \dbra{\psi_x} \Pi_i \dket{\psi_x},\ \dket{\psi_x^i} = \frac{\Pi_i \dket{\psi_x}}{\sqrt{ \dbra{\psi_x} \Pi_i \dket{\psi_x} }}, \ p_i = \text{Tr}\left[ \Pi_i \rho \right], \ \rho_i = \frac{\Pi_i \rho \Pi_i}{\text{Tr}\left[ \Pi_i \rho \right]}.
\end{align}

Equivalently, such a measurement can be described by a single \emph{observable} $O$, which is a self-adjoint (Hermitian) operator.
In this case, the projectors in the measurement are taken to be the projectors onto each eigenspace of $O$ with a distinct eigenvalue.
For example, if we measure a single qubit in the computational basis, then the observable is $Z$ and the projectors are $\{ \dketbra{0}, \dketbra{1} \}$.
The measurement results in this case are usually taken to be $0$ and $1$.
Similarly, if we measure in the conjugate basis, then the observable is $X$ and the projectors are $\{ \dketbra{+}, \dketbra{-} \}$.
The measurement results in this case are usually taken to be $+$ and $-$.
In this dissertation we will mostly consider measurement in these two bases.

In circuits, $Z$ and $X$ measurements are denoted as shown below.
\begin{align}
\begin{tikzcd}
\lstick{$\dket{\psi}$} & \meter{$0/1$} & \cw 
\end{tikzcd}
\quad
\begin{tikzcd}
\lstick{$\dket{\psi}$} & \meter{$0/1$} & \qw 
\end{tikzcd}
\quad
\begin{tikzcd}
\lstick{$\dket{\psi}$} & \meter{$+/-$} & \cw 
\end{tikzcd}
\quad
\begin{tikzcd}
\lstick{$\dket{\psi}$} & \meter{$+/-$} & \qw 
\end{tikzcd}
\end{align}
The double line indicates that the result is just the classical measurement value that can then be used to classically control future operations.
In this case the post-measurement state, i.e., just the qubit being measured, is implicitly discarded.
Whenever we perform future operations on the post-measurement state, we use the second and fourth forms above (resp. for $Z$ and $X$ measurements) as single lines (wires) always indicate qubits.
Since the Hadamard gate swaps the $Z$ and $X$ bases, the following identity automatically holds.
\begin{align}
\begin{tikzcd}
\lstick{$\dket{\psi}$} & \meter{$+/-$} & \qw 
\end{tikzcd}
=
\begin{tikzcd}
\lstick{$\dket{\psi}$} & \gate{H} & \meter{$0/1$} & \gate{H} & \qw 
\end{tikzcd}
\end{align}

In many situations, we might only measure a system at the end of an experiment (circuit) so that we do not care about the post-measurement state.
However, we would still like to know the probability of different outcomes under the given measurement.
Consider once again measurement operators $\{ M_i \}$ on an input state $\rho$ so that the probability of outcome $i$ is given by $p_i = \text{Tr}\left[ M_i^{\dagger} M_i \rho \right]$.
If we define $\tilde{M}_i \coloneqq M_i^{\dagger} M_i$, then it can be verified that $\tilde{M}_i$ is positive (for all $i$) such that $\sum_i \tilde{M}_i = \mathbbm{I}$ and $p_i = \text{Tr}\left[ \tilde{M}_i \rho \right]$.
Hence, the set of operators $\{ \tilde{M}_i \}$ suffices to determine the probabilities of different outcomes. 
This set is called a \emph{POVM (Positive Operator-Valued Measure)} and its elements are called \emph{POVM elements}~\cite{Nielsen-2010}.
By this terminology, the POVM for measuring $Z$ is $\{ \dketbra{0}, \dketbra{1} \}$ and the POVM for measuring $X$ is $\{ \dketbra{+}, \dketbra{-} \}$.

%% file: ch3_bpqm.tex

\label{ch:ch3_bpqm}

\emph{Message-passing} algorithms are classical methods that are used to efficiently compute certain quantities related to problems defined on graphs\footnote{Part of this work has been accepted to the 2020 IEEE International Symposium on Information Theory~\cite{Rengaswamy-arxiv20*3}.}.
They work by passing messages between nodes of the graph. 
For example, these algorithms have been successfully used for statistical inference, optimization, constraint-satisfaction problems and the graph isomorphism problem among several other applications~\cite{Yedidia-03,Yedidia-it05,Globerson-nips07,Lu-aller08,Donoho-itw10,Bayati-it11,Yedidia-jsp11,Mansour-arxiv17}.
In particular, \emph{belief-propagation (BP)} is a message-passing algorithm for efficiently marginalizing joint probability density functions in statistical inference applications.
The algorithm derives its name from the fact that the messages used by BP are ``local'' probabilities or ``beliefs''.
An application of BP that is relevant to this chapter is the decoding of classical linear codes by calculating the posterior bitwise marginals given the outputs of the (classical) channel~\cite{RU-2008}.
It is well-known that BP exactly performs the task of optimal bitwise maximum-a-posteriori (bit-MAP) decoding in cases where the factor graph for the code is a tree.
However, since codes with tree factor graphs have poor minimum distance~\cite{RU-2008}, BP is often applied to codes whose factor graphs have cycles, e.g, low-density parity-check (LDPC) codes.
Although BP does not compute the exact marginals in such cases, it is computationally efficient and usually performs quite well.
In fact, it has been proven that BP achieves the optimal MAP performance for spatially-coupled LDPC codes on the binary erasure channel~\cite{Kudekar-it11} and binary memoryless symmetric channels~\cite{Kudekar-it13,Kumar-it14}.

Given the success of BP, it is natural to ask if it can be generalized to the quantum setting and employed to perform optimal and efficient inference.
Consider the problem of distinguishing the outputs of a quantum channel where it is well-known that the error probability is minimized by the joint Helstrom measurement on all the outputs of the channel~\cite{Helstrom-jsp69,Helstrom-ieee70}.
Such a collective measurement enhances the capability to distinguish transmitted codewords beyond the performance achievable by optimal single symbol measurements followed by optimal classical processing.
This in turn translates to a significant increase in the classical communication capacity over the channel.
Indeed, for a binary modulation alphabet, e.g., BPSK optical coherent states, the ratio of $C_{\infty}$ (Holevo capacity) to $C_1$ (capacity attained with symbol by symbol measurements) approaches infinity at low mean photon number per mode (i.e., overlap between the states close to $1$)~\cite{Guha-isit12}.
The optimal Helstrom measurement can be calculated mathematically using the Yuen-Kennedy-Lax (YKL) conditions~\cite{Yuen-it75,Krovi-pra15}, but this can still be computationally challenging.
Even in cases where it is possible to compute this measurement, it can be hard to translate the mathematical description into a physical receiver design.
Therefore, it is of significant practical interest to find an efficient and physically realizable detector whose performance is close to optimal.

Renes~\cite{Renes-njp17} recently proposed a quantum generalization of BP by considering the transmission of a classical code over a specific classical-quantum (CQ) channel called the \emph{pure-state channel}.
The inputs to this channel are classical bits and the outputs are (tensor products of) qubits that are modulated by the input bits.
The pure-state channel model captures the characteristics of the binary phase shift keying (BPSK) modulated pure-loss optical channel that manifests in deep-space optical communications~\cite{Guha-isit12}.
Renes' algorithm is well-defined on a tree factor graph and works by passing qubits and classical information between nodes of the graph.
In the past, there have been other works that discuss ``quantum belief-propagation'' algorithms~\cite{Hastings-prb07,Leifer-annphy08}. 
But, as emphasized in~\cite{Renes-njp17}, all of them employ classical BP to quantum problems where, for example, the goal might be to compute the marginals of quantum states.
Therefore, Renes' algorithm appears to be the first BP algorithm that works by passing \emph{quantum messages} over a factor graph.
Since ``quantum belief-propagation'' already refers to a different algorithm in the literature~\cite{Leifer-annphy08}, we will refer to Renes' algorithm as \emph{belief-propagation with quantum messages (BPQM)}.


In~\cite{Renes-njp17}, the first step in developing BPQM was to interpret the BP message combining operations at nodes as ``channel combining'' rules that execute a local inference procedure.
This perspective on the BP node updates has close connections to the channel combining operation defined by Arikan for polar codes~\cite{Arikan-it09}.
The second step was to appropriately generalize these channel combining rules by considering the messages to be qubit density matrices.
These rules give a description of the channel that gets induced at each node when (qubit) messages arrive at it.
Finally, the third step was to define appropriate unitary operations at the nodes, which process the output of the induced channels and provide the necessary messages to be passed on to the next nodes.
While this approach provides a natural way to extend classical BP to the quantum setting in a precise mathematical fashion, the goal of BPQM remains unclear.
This is because, unless we measure the output of the channel, we do not ``observe'' the received signal in the quantum case, unlike the classical setting.
Therefore, there does not seem to be a clear definition of a posterior distribution that can then be ``marginalized'' by BPQM, to retain the spirit of BP.

On the other hand, from a practical standpoint, BPQM distinguishes itself from algorithms that achieve only the $C_1$ capacity by passing quantum messages between nodes of the factor graph. 
This enables it to behave like a legitimate joint measurement, and potentially perform better than symbol-by-symbol measurements followed by optimal classical processing such as block-MAP decoding.
Since BPQM has explicitly defined messages and rules for combining them at nodes of the graph, it can potentially be transformed into a physically realizable (optical) quantum receiver circuit.
Leveraging the work on ``cat basis'' quantum logic, we can translate the single- and two-qubit gates of BPQM on qubits into operations in the span of coherent states $\dket{\alpha}$ and $\dket{-\alpha}$~\cite{Ralph-pra03,Gilchrist-joptb04}.
However, BPQM has not been transformed into a circuit composed of elementary single- and two-qubit gates, even at the level of qubits.
Also, there has not been any work that discusses the performance of BPQM and compares it to that of optimal quantum processing as well as symbol-by-symbol detection followed by optimal classical processing.

\begin{figure}
    \centering
    
    \includegraphics[scale=0.75,keepaspectratio]{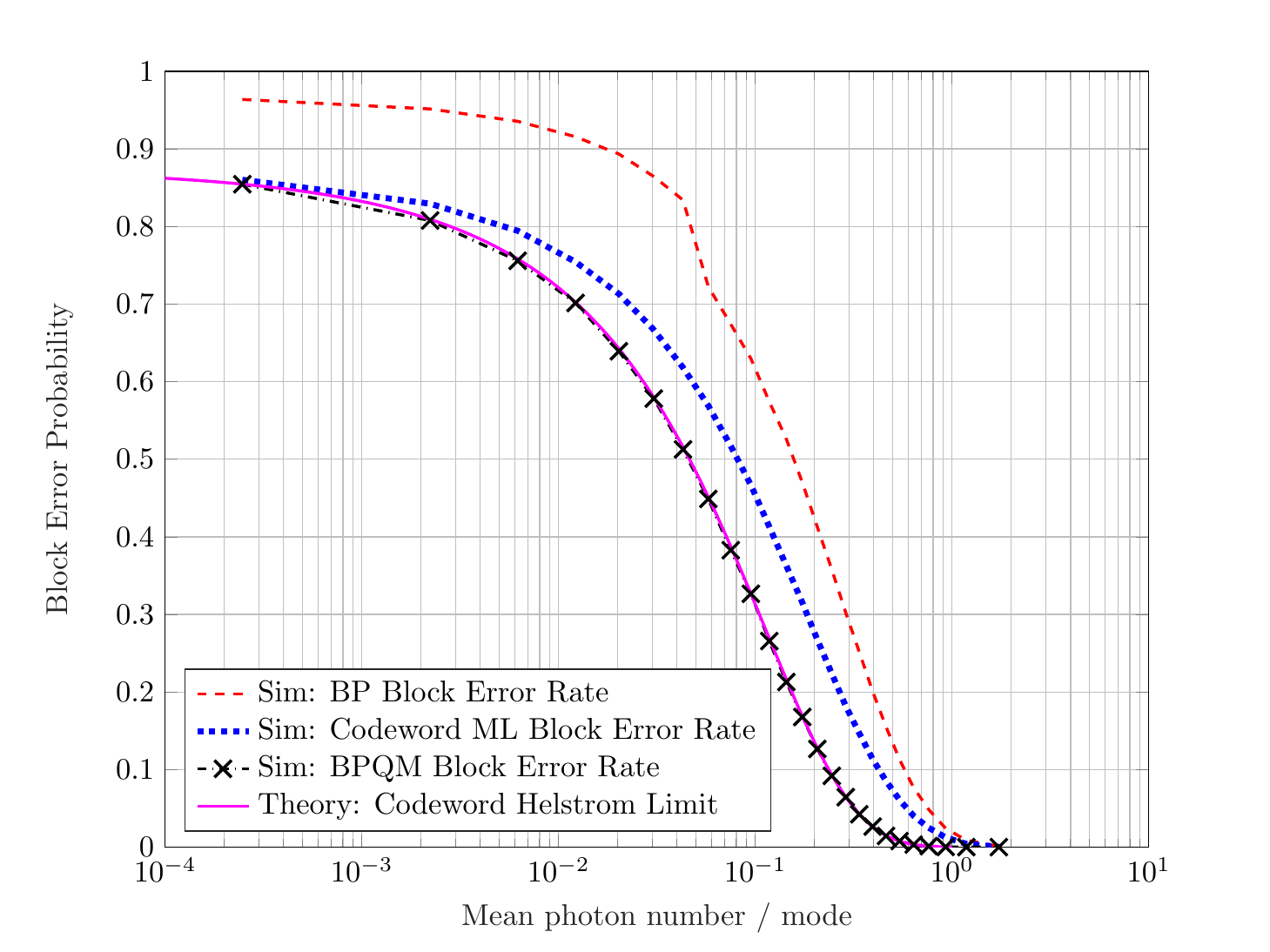}
    
    
    
    
    \includegraphics[scale=0.75,keepaspectratio]{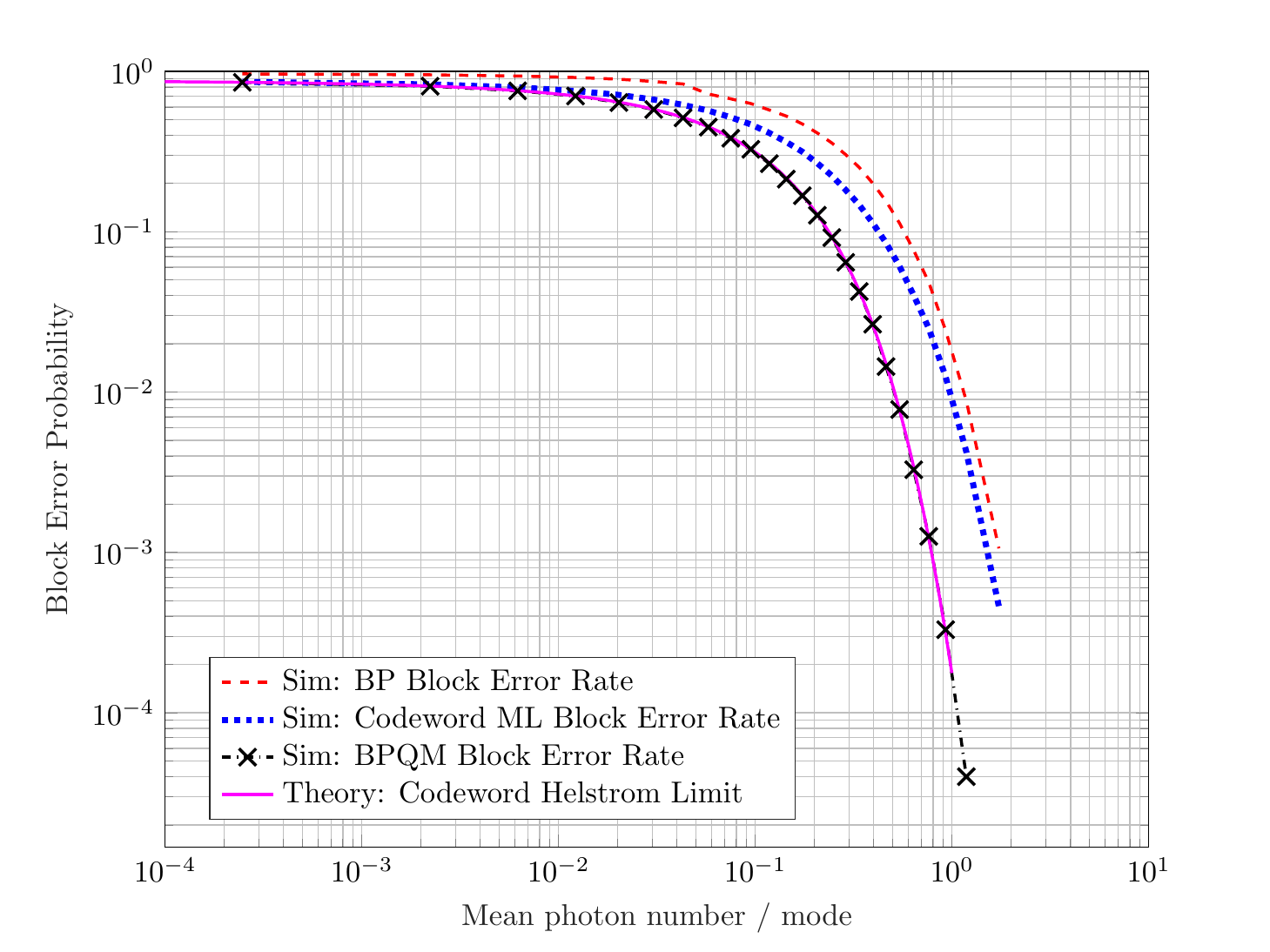}
    
    \caption[The overall block error rate of BPQM compared with the optimal one and strategies involving symbol-by-symbol Helstrom measurements.]{\label{fig:bpqm_perr_vs_nbar} The overall block error rate of BPQM along with those of quantum optimal joint Helstrom, symbol-by-symbol Helstrom followed by classical optimal block-MAP, and symbol-by-symbol Helstrom followed by classical BP.}
\end{figure}

The main purpose of this work is to provide a deeper look into BPQM, both into its sequence of operations and into its performance, and identify some of the important problems that need a detailed investigation.
We begin by explaining BP with the example of a $5$-bit linear code, and then develop the aforementioned perspective of BP that is particularly useful to understand the quantum generalization.
Subsequently, we discuss BPQM using the same example on the pure-state channel, and provide circuits for implementing BPQM.
In particular, through our analysis, we introduce a coherent rotation to be performed after decoding bit $1$, which is new and not part of Renes' original BPQM scheme.
This might be important in generalizing BPQM to more general codes than the specific example considered here.
We explicitly provide the quantum state density matrices that characterize the performance of BPQM for this example code.
For decoding bit $1$, we even derive an analytical expression for the BPQM success probability.
The benchmark for decoding each bit is the performance of the Helstrom measurement that optimally distinguishes the density matrices corresponding to the two values of the bit.
Here we show that BPQM is optimal for deciding the value of each of the $5$ bits.
We plot the performance curves for the following strategies in Fig.~\ref{fig:bpqm_perr_vs_nbar} to ascertain the ``global'' performance of BPQM for the 5-bit code in terms of block (codeword) error rates rather than the individual bit error rates. 
\begin{enumerate}
    \item[(a)] Collective (optimal) Helstrom measurement on all channel outputs corresponding to the transmitted codeword.
    
    \item[(b)] BPQM on all channel outputs corresponding to the transmitted codeword.
    
    \item[(c)] Symbol-by-symbol (optimal) Helstrom measurement followed by classical (optimal) block-MAP decoding.
    
    \item[(d)] Symbol-by-symbol (optimal) Helstrom measurement followed by classical BP.
\end{enumerate}
Note that for the last two schemes, classical processing is performed essentially on the binary symmetric channel (BSC) induced by measuring each qubit output by the pure-state channel.
Our implementation of the BPQM algorithm for the $5$-bit code is available at \url{https://github.com/nrenga/bpqm}.

As we expect, the block error probabilities are in increasing order from (a) through (d).
The plot shows that \emph{BPQM is strictly better than MAP and even as good as the optimal joint Helstrom measurement on the outputs of the channel}.
This is an important result because, it demonstrates that if we can construct a receiver for BPQM then it will outperform any known physically realizable receiver for this channel.
Therefore, BPQM provides a different kind of ``quantum advantage'' over classical processing than the commonly discussed applications in the literature, which usually involve speeding up a classical algorithm using quantum phenomena.
We provide more detailed observations at the end of Section~\ref{sec:BPQM_perf_all_bits}.

\section{Pure-State Classical-Quantum Channel}
\label{sec:pure_state_channel}

Classical-quantum (CQ) channels are defined by the density matrix of the observation given each classical input.
The output density matrix of the \emph{pure-state channel}, for classical inputs $x \equiv \dketbra{x}, x \in \{0,1\}$, is given by
\begin{align}
W(x) & \coloneqq \dketbra{\theta}{0} \cdot \dketbra{x} \cdot \dketbra{0}{\theta} + \dketbra{-\theta}{1} \cdot \dketbra{x} \cdot \dketbra{1}{-\theta} \\
  & = \dbraket{x}{0} \cdot \dketbra{\theta} + \dbraket{x}{1} \cdot \dketbra{-\theta} \\
  & = \dketbra{(-1)^x \theta}, \\
\dket{\pm \theta} & \coloneqq \cos\frac{\theta}{2} \dket{0} \pm \sin\frac{\theta}{2} \dket{1}.
\end{align}
Every quantum channel $\mathcal{E}$ can be expressed through its action on an input density matrix $\rho$ by $\mathcal{E}(\rho) = \sum_i A_i \rho A_i^{\dagger}$, where the complex matrices $A_i$ are called \emph{Kraus operators} and satisfy the completeness relation $\sum_i A_i^{\dagger} A_i = \mathbbm{I}$, where $\mathbbm{I}$ is the identity operator.
Hence, the Kraus operators for the pure-state channel can be taken to be $M_0 = \dketbra{\theta}{0}, M_1 = \dketbra{-\theta}{1}$.
If the input system to $W$ is denoted by $X$ and the output system by $B$, then the joint density matrix $\rho_{XB}$ that characterizes the entropic quantities for this channel is given by
\begin{align}
\rho_{XB} & \coloneqq q \cdot \dketbra{0}_X \otimes \dketbra{\theta}_B + (1-q) \cdot \dketbra{1}_X \otimes \dketbra{-\theta}_B,
\end{align}
where $q$ is the prior probability for input $x = 0$.
The joint entropy~\cite{Wilde-2013} for $XB$ is 
\begin{align}
H(XB)_{\rho} & = H(X) + \sum_{x \in \{0,1\}} p_X(x) H(\rho_B^{(x)}) \\
  & = h_2(q) + q \cdot H(\dketbra{\theta}_B) + (1 - q) H(\dketbra{-\theta}_B) \\
  & = h_2(q),
\end{align}
where $h_2(q) = -q \log_2(q) - (1-q) \log_2(1-q)$ (bits) is the binary entropy function and $H(\dketbra{\theta}_B) = H(\dketbra{-\theta}_B) = 0$ since the states are pure.
Therefore, the quantum mutual information for the pure-state channel is the symmetric Holevo information~\cite{Wilde-2013}
\begin{align}
I(X;B)_{\rho} & \coloneqq H(X)_{\rho} + H(B)_{\rho} - H(XB)_{\rho} \\
  & = h_2(q) + H(\rho_B) - h_2(q) \\
  & = H\left( q \cdot \dketbra{\theta}_B + (1-q) \cdot \dketbra{-\theta}_B \right).
\end{align}
The ultimate Holevo capacity for this channel in the asymptotic limit of a large number of channel uses is $C_{\infty}(W) \coloneqq \max_{q \in [0,1]}\ I(X;B)_{\rho} = \max_{q \in [0,1]}\ H(\rho_B)$ (per use of the channel).
Since this channel is equivalent to the BPSK (binary phase shift keying) modulated pure-loss optical channel~\cite{Guha-isit12}, it is known that the maximum occurs at $q = 1/2$. 
Hence
\begin{align}
C_{\infty}(W) = H\left( \frac{1}{2} \cdot \dketbra{\theta}_B + \frac{1}{2} \cdot \dketbra{-\theta}_B \right) = h_2\left( \cos^2\frac{\theta}{2} \right) = h_2\left( \frac{1 + \sqrt{F(W)}}{2} \right),
\end{align}
where the \emph{fidelity} of the channel is $F(W) \coloneqq |\dbraket{\theta}{-\theta}|^2 = \cos^2\theta, \cos\theta = 2\cos^2\frac{\theta}{2} - 1$, and the subscript ``$\infty$'' indicates that one might need collective (or joint) measurements on all channel outputs to achieve capacity~\cite{Guha-isit12}.

\section{Helstrom Measurement and its Difficulty}

If, instead, we performed the optimal measurement at the output of each use of the channel, i.e., for each code bit sent through $W$, then this would induce a BSC($P_{\min}$) with $P_{\min} = (1 - \sqrt{1 - F(W)})/2$, which is the minimum probability of error to distinguish between the states $\{ \dket{\theta}, \dket{-\theta} \}$.
The \emph{Helstrom measurement}~\cite{Helstrom-jsp69,Helstrom-ieee70} to optimally distinguish between two density matrices $\rho_0$ and $\rho_1$ is defined by the projectors $\{ \Pi_{\text{Hel}}, \mathbb{I} - \Pi_{\text{Hel}} \}$:
\begin{align}
\label{eq:Helstrom_msmt}
\Pi_{\text{Hel}} \coloneqq \sum_{i \colon \lambda_i \geq 0} \dketbra{\psi_i}, \ \ (\rho_0 - \rho_1) \dket{\psi_i} = \lambda_i \dket{\psi_i}.
\end{align}
For the pure state channel, it is easy to calculate that $\rho_0  - \rho_1 = \dketbra{\theta} - \dketbra{-\theta} = \sin\theta \cdot X$ so that the Helstrom measurement is projecting onto the Pauli $X$ basis, i.e., the projectors are $\{ \dketbra{+}, \dketbra{-} \}$.
In practice, the \emph{Dolinar receiver}~\cite{Dolinar-1973} for the BPSK modulated pure-loss optical channel induces this BSC($P_{\min}$)~\cite{Guha-isit12}.
If we implemented a classical optimal (block-MAP) decoder on this induced BSC, then the capacity that is attainable is $C_1(W) = 1 - h_2(P_{\min})$, where the subscript ``1'' indicates that we are performing symbol-by-symbol measurements and not a collective measurement.
It can be easily checked that $C_1(W) \ll C_{\infty}(W)$, and classical-quantum polar codes equipped with a quantum successive-cancellation decoder close this gap~\cite{Guha-isit12,Wilde-it13*2}.
However, the optimal decoder for CQ polar codes is hard to realize in the lab. 
Hence, an interesting open problem is to analyze how much of this gap is closed by BPQM because it can be mapped into a ``successive-cancellation-type'' decoder as discussed in~\cite{Renes-njp17}.
In fact, we will see that BPQM has a successive-cancellation flavor by definition.

\section{Classical Belief-Propagation (BP)}
\label{sec:classical_bp}

\subsection{Decoding Linear Codes Using BP}
\label{sec:decode_linear_codes}

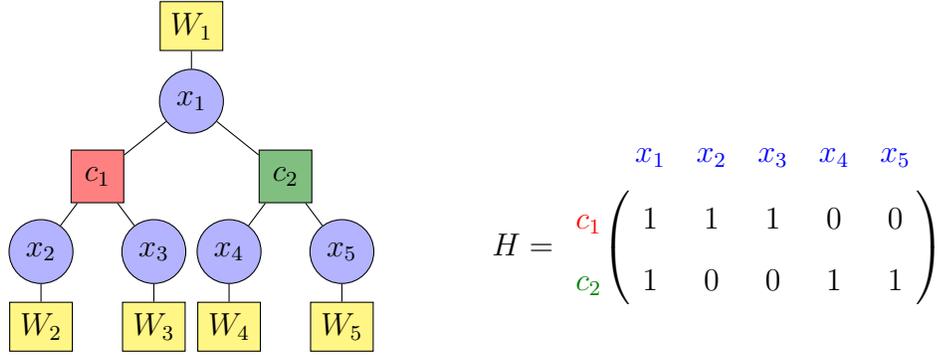
\begin{figure}
\begin{center}

\begin{tikzpicture}

\node[draw,rectangle,fill=yellow!60,minimum size=0.65cm] (W1) at (0,1) {$W_1$};

\node[draw,circle,fill=blue!30,minimum size=0.5cm] (x1) at (0,0) {$x_1$};
\node[draw,rectangle,fill=red!50,minimum size=0.7cm] (c1) at (-1.25,-1) {$c_1$};

\node[draw,circle,fill=blue!30,minimum size=0.5cm] (x2) at (-2,-2) {$x_2$};
\node[draw,rectangle,fill=yellow!60,minimum size=0.65cm] (W2) at (-2,-3) {$W_2$};
\node[draw,circle,fill=blue!30,minimum size=0.5cm] (x3) at (-0.5,-2) {$x_3$};
\node[draw,rectangle,fill=yellow!60,minimum size=0.65cm] (W3) at (-0.5,-3) {$W_3$};

\node[draw,rectangle,fill=darkgreen!50,minimum size=0.7cm] (c2) at (1.25,-1) {$c_2$};
\node[draw,circle,fill=blue!30,minimum size=0.5cm] (x4) at (0.5,-2) {$x_4$};
\node[draw,rectangle,fill=yellow!60,minimum size=0.65cm] (W4) at (0.5,-3) {$W_4$};
\node[draw,circle,fill=blue!30,minimum size=0.5cm] (x5) at (2,-2) {$x_5$};
\node[draw,rectangle,fill=yellow!60,minimum size=0.65cm] (W5) at (2,-3) {$W_5$};

\draw (x1) -- (c1);
\draw (x1) -- (c2);
\draw (c1) -- (x2);
\draw (c1) -- (x3);
\draw (c2) -- (x4);
\draw (c2) -- (x5);

\draw (W1) -- (x1);
\draw (W2) -- (x2);
\draw (W3) -- (x3);
\draw (W4) -- (x4);
\draw (W5) -- (x5);

\node[align=center] (H) at (7,-1.5) {$H = 
\bbordermatrix{
  & {\color{blue}x_1} & {\color{blue}x_2} & {\color{blue}x_3} & {\color{blue}x_4} & {\color{blue}x_5} \cr
{\color{red}c_1} & 1 & 1 & 1 & 0 & 0 \cr
{\color{darkgreen}c_2} & 1 & 0 & 0 & 1 & 1 \cr
}$};


\end{tikzpicture}

\caption{\label{fig:five_bit_code}Factor graph and parity-check matrix for the $5$-bit linear code $\mathcal{C}$.}

\end{center}
\end{figure}

An $[n,k,d]$ binary linear code $\MCC$ can be defined by a binary parity-check matrix $H$ as follows:
\begin{align}
\MCC \coloneqq \{ \vecnot{x} \in \{0,1\}^n \colon H \vecnot{x}^T = \vecnot{0}^T,\ H \in \{0,1\}^{(n-k) \times n} \}.
\end{align}
Such a code encodes $k$ message bits into $n$ code bits, and the minimum Hamming weight of any codeword $\vecnot{x} \in \MCC$ is $d$.
Consider the $[5,3,2]$ code, still denoted as $\MCC$, defined by
\begin{align}
H \coloneqq 
\begin{bmatrix}
1 & 1 & 1 & 0 & 0 \\
1 & 0 & 0 & 1 & 1
\end{bmatrix}.
\end{align}
A \emph{factor graph (FG)} for a linear code is a bipartite graph where the bits (variables) are represented by circle nodes and the checks (factors) are represented by square nodes.
The FG representation of (the above definition of) $\MCC$ is shown in Fig.~\ref{fig:five_bit_code}, where $c_1$ and $c_2$ represent the two parity checks on the $5$ bits of each codeword in $\MCC$.
Observe that every linear code has multiple associated parity-check matrices, each of which forms a generator matrix for its dual code. 
Hence, the FG representation of a code depends on the chosen parity-check matrix.

A discrete memoryless channel is represented as $W$ and is defined by the channel transition probability matrix $W(y|x) \coloneqq \mathbb{P}[Y = y | X = x]$, which represents the probability of observing $y \in \mathcal{Y}$ at the output of the channel when its input was $x \in \mathcal{X}$.
Here, $\mathcal{X}$ and $\mathcal{Y}$ represent the input and output alphabets of the channel, respectively.
A well-known example for such a channel is the \emph{binary symmetric channel} (BSC). 
For the BSC, $\mathcal{Y} = \{0,1\} = \mathcal{X}$ and the transition matrix is defined as
\begin{align}
W^{\text{BSC}} \coloneqq 
\begin{bmatrix}
1 - p & p \\
p & 1 - p
\end{bmatrix},
\end{align}
where the $(i,j)$-th entry is $W^{\text{BSC}}(y = j|x = i)$ and $0 \leq p \leq 0.5$.
The FG in Fig.~\ref{fig:five_bit_code} shows the channel $W_k$ associated with each bit $k$ as a separate factor node that provides the channel transition probability value for the observed output $y_k$ for input $x_k = 0$ (and $x_k = 1$).
These channels are not necessarily BSCs, but for simplicity we will assume that all bits go through the same channel, i.e., $W_k = W$ for all $k$.

Given the channel output vector, $\vecnot{y}$, the decoder tries to determine the codeword $\vecnot{x} \in \MCC$ that was actually sent at the input.
The \emph{block maximum-a-posteriori} (MAP) decoder calculates the posterior probability for each codeword in the code, given $\vecnot{y}$, and chooses the codeword with the maximum value.
This is the optimal decoder in terms of block error rate. 
For the example $5$-bit code $\MCC$, when all codewords are transmitted with equal probability, it calculates
\begin{align}
p(\vecnot{x} | \vecnot{y}) & = \frac{p(\vecnot{y} | \vecnot{x}) \cdot p(\vecnot{x})}{\sum_{\vecnot{x} \in \{0,1\}^5} p(\vecnot{y} | \vecnot{x}) \cdot p(\vecnot{x})} \\
  & = \frac{\prod_{k=1}^{5} W(y_k | x_k) \cdot \mathbb{P}[\vecnot{x} \in \MCC]}{p(\vecnot{y})} \\ 
\label{eq:map_propto}
  & \propto \prod_{k=1}^{5} W(y_k | x_k) \cdot \left[ \mathbb{I}(x_1 \oplus x_2 \oplus x_3 = 0) \mathbb{I}(x_1 \oplus x_4 \oplus x_5 = 0) \right] \\
  & = W(y_1 | x_1) \cdot \left[ \mathbb{I}(x_1 \oplus x_2 \oplus x_3 = 0) W(y_2 | x_2) W(y_3 | x_3) \right] \nonumber \\
  & \hspace*{3cm} \cdot \left[ \mathbb{I}(x_1 \oplus x_4 \oplus x_5 = 0) W(y_4 | x_4) W(y_5 | x_5) \right], \\
\hat{\vecnot{x}}^{\text{MAP}} & \coloneqq \underset{\vecnot{x} \in \{0,1\}^5}{\text{argmax}} \ p(\vecnot{x} | \vecnot{y}),
\end{align}
where the constant of proportionality in~\eqref{eq:map_propto} is independent of $\vecnot{x}$.
In general, the complexity of this scheme grows exponentially with the code dimension $k$ because the block-MAP decoder calculates the posterior probability for each codeword in the code.
A more efficient scheme is the \emph{bit-MAP} decoder which marginalizes the above joint posterior for each bit and makes a decision bit-wise.
Hence, to decode bit $1$, the bit-MAP decoder computes
\begin{align}
\hat{x_1}^{\text{MAP}} & \coloneqq \underset{x_1 \in \{0,1\}}{\text{argmax}} \ \sum_{x_2,x_3,x_4,x_5 \in \{0,1\}^4} p(\vecnot{x} | \vecnot{y}) \\
  & =\underset{x_1 \in \{0,1\}}{\text{argmax}} \ \Biggr\lbrace W(y_1 | x_1) \cdot \left[ \sum_{x_2,x_3 \in \{0,1\}^2} \mathbb{I}(x_1 \oplus x_2 \oplus x_3 = 0) W(y_2 | x_2) W(y_3 | x_3) \right] \nonumber \\
  & \hspace*{2.45cm} \cdot \left[ \sum_{x_4,x_5 \in \{0,1\}^2} \mathbb{I}(x_1 \oplus x_4 \oplus x_5 = 0) W(y_4 | x_4) W(y_5 | x_5) \right] \Biggr\rbrace.
\end{align}
Even though marginalizing a general joint probability distribution can have an exponentially scaling complexity in the number of involved variables, the idea of BP is that this can be done efficiently when the joint probability density factors into terms involving \emph{disjoint} sets of variables. For example, on a tree FG as in the above $5$-bit code, we can use the distributive property of addition over multiplication and compute the sums involved in the two square brackets \emph{simultaneously}.
Then the results can be pooled in one final step that takes their product and multiplies the result with $W(y_1|x_1)$.
This is exactly BP on this FG, since the two local sums can be interpreted as ``local beliefs'' about the variable $x_1$ that are propagated to be combined with the ``belief'' from the channel observation $y_1$.

Formally, the BP algorithm is initialized by setting the local channel factors $W_k = W(y_k|x_k)$ based on the observations $y_k$ for both $x_k = 0$ and $x_k = 1$.
Then the variables $x_k$ simply pass on the message $(W(y_k|x_k = 0), W(y_k|x_k = 1))$ to their associated factor node(s) (FNs).
In the first half-iteration of BP involving the FN update, the factors $c_a$ calculate the ``local belief'' $\sum_{x_i \in \{0,1\} \colon i \in \partial a \setminus \{1\}} \mathbb{I}(\oplus_{i} x_i = x_1) \prod_i W(y_i|x_i)$ for both $x_1 = 0$ and $x_1 = 1$, where $\partial a$ represents the indices of the set of variables attached to the factor $c_a, a \in \{1,2\}$.
In the next half-iteration of BP involving the variable node (VN) update, the VN $x_1$ combines all incoming local beliefs (including $W(y_1|x_1)$) by taking their product and renormalizing the result to make it the exact posterior marginal distribution for $x_1$.
Since this example has a tree FG, this completes BP for decoding the bit $x_1$.
A similar procedure can be executed for the other variables as well, and the whole scheme can be combined into a parallel BP schedule to compute marginals for all variables simultaneously.
If the FG were not a tree, then we usually run BP for multiple iterations and in many cases this procedure converges to a fixed point of the BP algorithm, but we will ignore these details for brevity.
Also, we note that it is common to take the BP messages to be log-likelihood ratios $\log \frac{W(y_k|x_k = 0)}{ W(y_k|x_k = 1)}$, but this is not very important for our purposes here.

\subsection{Induced Channels in BP}
\label{sec:induced_channels}

We will find it very convenient to represent the operations performed by BP at each FN and VN abstractly as ``local inference'' over a ``locally induced channel''.
For convenience, consider a VN attached to exactly two FNs since a degree-$d$ VN can always be analyzed sequentially with two FNs at a time.
Then we have the following representation of the information at the VN.

\begin{figure}[ht]
\begin{center}
\begin{tikzpicture}

\node[draw,circle] (x) at (0,0) {$x$};
\node[draw,rectangle] (c1) at (-1,-1) {$c_1$};
\node[draw,rectangle] (c2) at (1,-1) {$c_2$};
\node at (-1,-1.75) {$\cdots$};
\node at (1,-1.75) {$\cdots$};

\draw (x) -- (c1);
\draw (x) -- (c2);
\draw (c1) -- (-1.75,-2);
\draw (c1) -- (-0.25,-2);
\draw (c2) -- (0.25,-2);
\draw (c2) -- (1.75,-2);

\node at (2.25,-1) {$\equiv$};

\node[draw,circle] (x2) at (4.5,0) {$x$};
\node[draw,rectangle] (W) at (3.5,-1) {$W$};
\node[draw,rectangle] (Wp) at (5.5,-1) {$W'$};
\node[draw,circle] (y) at (3.5,-2) {$y$};
\node[draw,circle] (z) at (5.5,-2) {$z$};

\draw (x2) -- (W) -- (y);
\draw (x2) -- (Wp) -- (z);

\node at (6.5,-1) {$\equiv$};

\node[draw,circle] (x3) at (8,0) {$x$};
\node[draw,rectangle] (Wcircast) at (8,-1) {$W \circledast W'$};
\node[draw,circle] (yz) at (8,-2) {$w$};
\node (w) at (9.5,-2) {$w = (y,z)$};

\draw (x3) -- (Wcircast) -- (yz);

\end{tikzpicture}
\caption{\label{fig:induced_VN}Channel combining at a VN using the induced channels at the node.}
\end{center}
\end{figure}
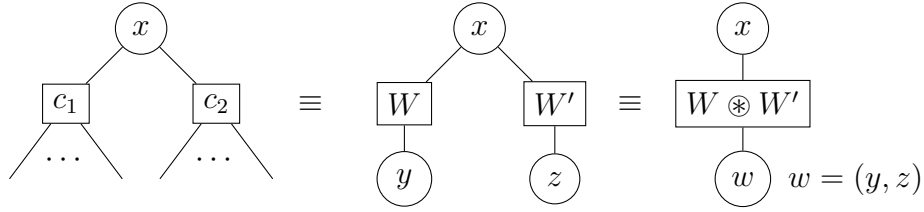

In Fig.~\ref{fig:induced_VN}, all nodes other than $x$ that are connected to $c_1$ and $c_2$ can be combined into a single value $y$ and $z$, respectively, which represent \emph{independent} observations of the variable $x$ through some induced channels $W$ and $W'$, respectively.
This is because the FNs $c_1$ and $c_2$ impose a simple even parity check that ensures all attached variables sum to $0$ (modulo $2$).
If the FNs represented a different check, then the induced channels have to be redefined accordingly.
Note that the independence is exact only when the full FG is a tree.
For the even parity check case, the two induced channels can be combined into a single channel $W \circledast W'$ whose outputs are the concatenation of $y$ and $z$:
\begin{align}
\label{eq:VN_conv}
[W \circledast W'](y,z|x) & = W(y|x) \cdot W'(z|x,y) 
                            = W(y|x) \cdot W'(z|x).
\end{align}
This is called as the \emph{variable node convolution} of two channels.
Hence, during the VN update of BP, the operation is simply performing local inference over the local channel $W \circledast W'$ i.e., calculating the local posterior for $x$ given $(y,z)$.

Similarly, at a (degree-$3$) FN we have a single input $x$ ``splitting'' into two outputs $u$ and $v$ (since they sum to $x$), whose independent observations through the underlying physical channel, as well as the remaining part of the FG, are obtained as $y$ and $z$.
Then we have only two possibilities, either $u = x$ and $v = 0$ or $u = x \oplus 1$ and $v = 1$, and both of them are equally likely assuming that the code does not have a trivial bit position where all codewords take the value $0$.
Hence, the \emph{factor node convolution} of two channels is
\begin{align}
[W \boxast W'](y,z | x) & = \frac{1}{2} W(y | u = x) \cdot W'(z | v = 0) + \frac{1}{2} W(y | u = x \oplus 1) \cdot W'(z | v = 1) \\
\label{eq:FN_conv}
  & = \frac{1}{2} W(y | x) \cdot W'(z | 0) + \frac{1}{2} W(y | x \oplus 1) \cdot W'(z | 1).
\end{align}

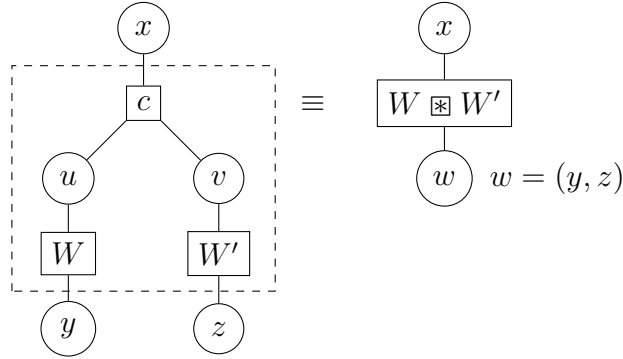
\begin{figure}[H]
\begin{center}
\begin{tikzpicture}

\node[draw,circle] (x) at (0,0) {$x$};
\node[draw,rectangle] (c) at (0,-1) {$c$};
\node[draw,circle] (u) at (-1,-2) {$u$};
\node[draw,circle] (v) at (1,-2) {$v$};
\node[draw,rectangle] (W) at (-1,-3) {$W$};
\node[draw,rectangle] (Wp) at (1,-3) {$W'$};
\node[draw,circle] (y) at (-1,-4) {$y$};
\node[draw,circle] (z) at (1,-4) {$z$};

\draw (x) -- (c);
\draw (c) -- (u) -- (W) -- (y);
\draw (c) -- (v) -- (Wp) -- (z);
\draw[dashed] (-1.75,-3.5) rectangle (1.75,-0.5);

\node at (2.25,-1) {$\equiv$};

\node[draw,circle] (x2) at (4,0) {$x$};
\node[draw,rectangle] (Wboxast) at (4,-1) {$W \boxast W'$};
\node[draw,circle] (yz) at (4,-2) {$w$};
\node (w) at (5.5,-2) {$w = (y,z)$};

\draw (x2) -- (Wboxast) -- (yz);

\end{tikzpicture}
\caption{\label{fig:induced_FN}Channel combining at a FN using the induced channels at the node.}
\end{center}
\end{figure}

We can perform a quick calculation using the FN update operation of BP to verify that BP is indeed performing local inference on this locally induced channel.
In Fig.~\ref{fig:five_bit_code}, consider the BP update at the FN $c_1$.
We observe that
\begin{align}
\sum_{x_2,x_3 \in \{0,1\}^2} & \mathbb{I}(x_1 \oplus x_2 \oplus x_3 = 0) W(y_2 | x_2) W(y_3 | x_3) \nonumber \\
  & = \sum_{x_2,x_3 \in \{0,1\}^2} \mathbb{I}(x_2 \oplus x_3 = x_1) W(y_2 | x_2) W(y_3 | x_3) \\
  & = W(y_2 | x_2 = x_1) \cdot W(y_3 | x_3 = 0) + W(y_2 | x_2 = x_1 \oplus 1) \cdot W(y_3 | x_3 = 1) \\
  & \propto [W \boxast W](y_2,y_3 | x_1),
\end{align}
where we need the factor $1/2$ to make sure that it is an exact marginal (which we had omitted at the beginning of BP, in the MAP formulation, for convenience), or equivalently to ensure that $W \boxast W'$ is indeed a channel.

\section{BP with Quantum Messages (BPQM)}
\label{sec:BPQM}

\subsection{Node Convolutions for Classical-Quantum Channels}

As discussed by Renes~\cite{Renes-njp17}, this ``induced channel'' perspective of BP crucially aids us in defining the quantum channel combining operations for a classical-quantum (CQ) channel~\cite{Wilde-2013} $W(x) \equiv W(\dketbra{x}), x \in \{0,1\}$, as follows:
\begin{align}
\label{eq:CQ_VN_conv}
[W \circledast W'](x) & \coloneqq W(x) \otimes W'(x), \\
\label{eq:CQ_FN_conv}
[W \boxast W'](x) & \coloneqq \frac{1}{2} W(x) \otimes W'(0) + \frac{1}{2} W(x \oplus 1) \otimes W'(1).
\end{align}
Here, we have adopted the same notation as in~\cite{Renes-njp17}, which suppresses the outputs ``$(y,z)$'' that were present in the classical channel convolutions~\eqref{eq:VN_conv} and~\eqref{eq:FN_conv}.
This is because we do not observe the output in the quantum case unless we measure it, and measuring each channel output is not always the optimal operation at the receiver.

\begin{remark}
\normalfont
Observe that for a general CQ channel $W(x)$ represents the output density matrix from the channel for the classical input $x \equiv \dketbra{x}$. 
Thus, even if the channel outputs are pure states, the induced channel at a FN still yields a mixed state.
Also note that these channels are ``automatically induced by the structure of the FG'', and do not require any operation to be performed on the received states.
Hence, the challenge is to identify the appropriate quantum ``local inference'' strategy, so that at the end of quantum BP we have performed appropriate statistical inference.
\end{remark}

The idea of Renes~\cite{Renes-njp17} is to generalize the classical BP algorithm, which is exact for marginalization over tree FGs, to the quantum scenario where the messages can be density matrices.
However, it is not immediately clear how the posterior distribution should be defined in the quantum case because, unlike the classical case, we do not get ``observations'' unless we make measurements.
If we indeed measure the output of each channel use, then we induce a binary-input binary-output channel whose transition probabilities depend on the measurement and the channel output density matrices $W(x),\ x\in\{0,1\}$.
But, it is well-known that this reduces the information capacity of the channel and that collective measurements on all channel outputs may be required to achieve capacity~\cite{Wilde-it13*2}.
Although this might be infeasible in practice, Renes shows that such a scheme can be simplified into a BP algorithm with quantum messages for the pure-state channel.

\subsection{Node Operations in BPQM}
\label{sec:BPQM_node_ops}

For the pure-state channel, the following operations are defined to be performed at variable nodes and factor nodes~\cite{Renes-njp17} (see Section~\ref{sec:node_convolutions} for detailed calculations).
At a VN, the convolution $W \circledast W'$ \emph{ideally} yields either $\dket{\theta} \otimes \dket{\theta'}$ or $\dket{-\theta} \otimes \dket{-\theta'}$.
Note that the local convolution is performed with respect to input $x = 0$ and $x = 1$ separately, respectively inducing signs $+$ and $-$.
We say ``ideally'' because we expect the signs of all incoming qubits at a VN to be the same, which means all independent local beliefs of the VN agree on the bit's value.
Since the pure-state channel does not introduce noise, and the only uncertainty arises from the non-orthogonality of $\dket{\theta}$ and $\dket{-\theta}$, the qubits always combine in this ideal fashion until the first bit is decoded.
But, whether this situation continues for the next bit depends upon whether the first bit was decoded to be a $0$ or $1$.
This is because, as mentioned earlier, the FN channel convolution in~\eqref{eq:CQ_FN_conv} is defined assuming that the FN imposes an even parity-check.
If, instead, it imposed an odd parity-check, as will happen when one of the bits is decoded to be $1$, then the FN convolution has to be modified appropriately.
Therefore, if the FN originally had degree $3$ and one of the bits have been estimated to be $1$, then we can remove the bit and update the FN to be an odd parity-check on two bits. 
This degree $2$ FN effectively induces a modified VN convolution with the signs of the two qubits in disagreement.

Given the (ideal) convolution outputs, the following unitary is applied to ``compress'' the information into one qubit and force the other system to be in state $\dket{0}$:
\begin{align}
U_{\circledast}(\theta,\theta') & \coloneqq 
\begin{bmatrix}
a_+ & 0 & 0 & a_{-} \\
a_{-} & 0 & 0 & - a_+ \\
0 & b_+ & b_{-} & 0 \\
0 & b_{-} & - b_+ & 0
\end{bmatrix}, \\
a_{\pm} \coloneqq \frac{1}{\sqrt{2}} \frac{\cos\left( \frac{\theta - \theta'}{2} \right) \pm \cos\left( \frac{\theta + \theta'}{2} \right)}{\sqrt{1 + \cos\theta \cos\theta'}}, & \quad  
b_{\pm} \coloneqq \frac{1}{\sqrt{2}} \frac{\sin\left( \frac{\theta + \theta'}{2} \right) \mp \sin\left( \frac{\theta - \theta'}{2} \right)}{\sqrt{1 - \cos\theta \cos\theta'}}.
\end{align}
Hence, we have $U_{\circledast}(\theta,\theta') \left( \dket{\pm \theta} \otimes \dket{\pm \theta'} \right) = \dket{\pm \theta^{\circledast}} \otimes \dket{0}$, where $\cos\theta^{\circledast} \coloneqq \cos\theta \cos\theta'$.
The VN update is then to pass the qubit from the first system and discard the second system.

At the FN, the induced mixed state $[W \boxast W'](x)$ can be transformed into the CQ state $\sum_{j \in \{0,1\}} p_j \dketbra{\pm \theta_j^{\boxast}} \otimes \dketbra{j}$ by performing $U_{\boxast} \coloneqq \cnot{W}{W'}$, the controlled-NOT gate with $W$ as control and $W'$ as target.
Hence, 
\begin{align}
U_{\boxast} \left( [W \boxast W'](x) \right) U_{\boxast}^{\dagger} & = \sum_{j \in \{0,1\}} p_j \dketbra{\pm \theta_j^{\boxast}} \otimes \dketbra{j}, \\
p_0 \coloneqq \frac{1}{2} (1 + \cos\theta \cos\theta'), \ p_1 \coloneqq 1 - p_0, & \quad
\cos\theta_0^{\boxast} \coloneqq \frac{\cos\theta + \cos\theta'}{1 + \cos\theta \cos\theta'}, \ 
\cos\theta_1^{\boxast} \coloneqq \frac{\cos\theta - \cos\theta'}{1 - \cos\theta \cos\theta'}.
\end{align}
Observe that for $j = 0$, the angle between the states has decreased, while for $j = 1$ the angle has increased.
The FN update is then to measure the second system and pass the resulting qubit from the first system as the message, along with the result of the classical measurement. 
This is because the VN update at the next stage needs to know the angle $\theta, \theta'$ of the incoming qubits.
When we have a degree $d$ node, these channel convolutions can be performed two at a time.
Equivalently, we can write down a circuit composed of CNOT operations and $Z$-basis measurements, and use the ``principle of deferred measurements''~\cite[Section 4.4]{Nielsen-2010} to delay the measurements until the end of the circuit.
For clarity, we will describe BPQM as a coherent operation that does not measure or discard qubits along the way at the nodes.

\begin{remark}
\normalfont
Note that in each half-iteration of BP, all VN or FN updates can be performed simultaneously since this involves only classical arithmetic and we can implicitly clone the values.
This is indeed a different update schedule when compared to slower sequential updates, but can always be implemented if desired.
However, in BPQM, we are forced to perform operations sequentially until one bit is decoded and then attempt to reverse the executed operations in a suitable manner.
Our subsequent analysis of BPQM will assume this sequential schedule.
Thus, BPQM appears closer to successive-cancellation than BP.
\end{remark}

\section{BPQM on the $5$-Bit Code}
\label{sec:BPQM_five_bit_code}

In this section we will use the above BPQM node operations to decode our running example.

\subsection{Decoding Bit $1$}
\label{sec:decode_bit_1}

\begin{figure}[ht]
\begin{center}

\begin{tikzcd}
\lstick{$1$} & \qw & \qw & \qw & \gate[wires=5]{V} \slice{(e)} & \gate{H} & \meter{$\{0,1\}$} & \cw \\
\lstick{$2$} \slice{(a)} & \ctrl{1} \slice{(b)} & \qw & \gate[wires=4]{U} \slice{(d)} &  & \qw & \qw & \qw \\
\lstick{$3$} & \targ{} & \swap{1} \slice{(c)} &  &  & \qw & \qw & \qw \\
\lstick{$4$} & \ctrl{1} & \targX{} &  &  & \qw & \qw & \qw \\
\lstick{$5$} & \targ{} & \qw &  &  & \qw & \qw & \qw
\end{tikzcd}

\caption{\label{fig:BPQM_circuit_bit1}The BPQM circuit to decode bit $1$ of the $5$-bit code in Fig.~\ref{fig:five_bit_code}.}

\vspace*{-0.4cm}

\end{center}
\end{figure}
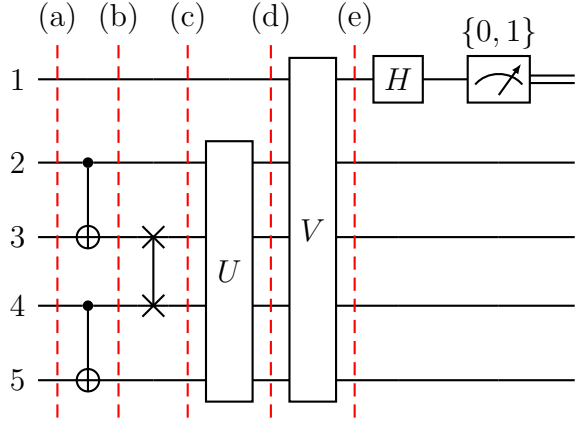

Let us begin by describing the procedure to decode bit $1$ of the $5$-bit code in our running example. 
Observe that the codewords belonging to the code are
\begin{align}
\MCC = \{ 00000, 00011, 01100, 01111, 10101, 10110, 11001, 11010 \}.
\end{align}
We assume that all the codewords are equally likely to be transmitted, just as in classical BP. Then the task of decoding the value of the first bit $x_1$ involves distinguishing between the density matrices $\rho_1^{(0)}$ and $\rho_1^{(1)}$, which are uniform mixtures of the states corresponding to the codewords that have $x_1 = 0$ and $x_1 = 1$, respectively, i.e.,
\begin{align}
\rho_1^{(0)} & = \dketbra{\theta}_1 \otimes \frac{1}{4} \bigg[ \dketbra{\theta}_2 \otimes \dketbra{\theta}_3 \otimes \dketbra{\theta}_4 \otimes \dketbra{\theta}_5 \nonumber \\
  & \hspace*{2cm} + \dketbra{\theta}_2 \otimes \dketbra{\theta}_3 \otimes \dketbra{-\theta}_4 \otimes \dketbra{-\theta}_5 \nonumber \\
  & \hspace*{2cm} + \dketbra{-\theta}_2 \otimes \dketbra{-\theta}_3 \otimes \dketbra{\theta}_4 \otimes \dketbra{\theta}_5 \nonumber \\
  & \hspace*{2cm} + \dketbra{-\theta}_2 \otimes \dketbra{-\theta}_3 \otimes \dketbra{-\theta}_4 \otimes \dketbra{-\theta}_5 \bigg], \\
\rho_1^{(1)} & = \dketbra{-\theta}_1 \otimes \frac{1}{4} \bigg[ \dketbra{\theta}_2 \otimes \dketbra{-\theta}_3 \otimes \dketbra{\theta}_4 \otimes \dketbra{-\theta}_5 \nonumber \\
  & \hspace*{3cm} + \dketbra{\theta}_2 \otimes \dketbra{-\theta}_3 \otimes \dketbra{-\theta}_4 \otimes \dketbra{\theta}_5 \nonumber \\
  & \hspace*{3cm} + \dketbra{-\theta}_2 \otimes \dketbra{\theta}_3 \otimes \dketbra{\theta}_4 \otimes \dketbra{-\theta}_5 \nonumber \\
  & \hspace*{3cm} + \dketbra{-\theta}_2 \otimes \dketbra{\theta}_3 \otimes \dketbra{-\theta}_4 \otimes \dketbra{\theta}_5 \bigg].
\end{align}
These density matrices can be written in terms of the FN channel convolution in (\ref{eq:CQ_FN_conv}) as $\rho_1^{(x_1)} =\rho_{\pm}= \dketbra{\pm \theta}_1 \otimes [W \boxast W](x_1)_{23} \otimes [W \boxast W](x_1)_{45}$, where we use the notation $\pm \equiv (-1)^{x_1}, \ x_1\in\{0,1\}$.
The BPQM circuit for decoding $x_1$ is shown in Fig.~\ref{fig:BPQM_circuit_bit1} along with the density matrix in each stage of the circuit denoted by (a) through (e).

\begin{itemize}

\item[(a)] $\rho_{\pm,a} = \dketbra{\pm \theta}_1 \otimes [W \boxast W](x_1)_{23} \otimes [W \boxast W](x_1)_{45}$

\item[(b)] $\rho_{\pm,b} = \dketbra{\pm \theta}_1 \otimes \sum_{j \in \{0,1\}} p_j \dketbra{\pm \theta_j^{\boxast}, j}_{23} \otimes \sum_{k \in \{0,1\}} p_k \dketbra{\pm \theta_k^{\boxast}, k}_{45}$

\item[(c)] $\rho_{\pm,c} = \dketbra{\pm \theta}_1 \otimes \sum_{j,k \in \{0,1\}^2} p_j p_k \dketbra{\pm \theta_j^{\boxast}}_2 \otimes \dketbra{\pm \theta_k^{\boxast}}_3 \otimes \dketbra{j}_4 \otimes \dketbra{k}_5$

\item[(d)] $\sigma_{\pm} = \sum_{j,k \in \{0,1\}^2} p_j p_k \dketbra{\pm \theta}_1 \otimes \dketbra{\pm \theta_{jk}^{\circledast}}_2 \otimes \dketbra{0}_3 \otimes \dketbra{jk}_{45}$ 

\item[(e)] $\Psi_{\pm} = \sum_{j,k \in \{0,1\}^2} p_j p_k \dketbra{\pm \varphi_{jk}^{\circledast}}_1 \otimes \dketbra{0}_2 \otimes \dketbra{0}_3 \otimes \dketbra{jk}_{45}$ 

\end{itemize}
We emphasize that at each stage, the density matrix is the \emph{expectation} over all pure states that correspond to transmitted codewords with the first bit taking value $x_1 \in \{ 0,1 \}$.

In the above circuit we have defined the following unitaries and angles:
\begin{align}
U & \coloneqq \sum_{j,k \in \{0,1\}^2} U_{\circledast}(\theta_j^{\boxast}, \theta_k^{\boxast})_{23} \otimes \dketbra{jk}_{45}, \\
\cos\theta_{jk}^{\circledast} & \coloneqq \cos\theta_j^{\boxast} \cos\theta_k^{\boxast}, \\
V & \coloneqq \sum_{j,k \in \{0,1\}^2} U_{\circledast}(\theta, \theta_{jk}^{\circledast})_{12} \otimes \dketbra{jk}_{45}, \\
\cos\varphi_{jk}^{\circledast} & \coloneqq \cos\theta \cos\theta_{jk}^{\circledast}.
\end{align}
Hence, the operations $U$ and $V$ are effectively two-qubit unitary operations, albeit controlled ones, and this phenomenon extends to any factor graph.
Evidently, BPQM compresses all the quantum information into system $1$ and the problem reduces to distinguishing between {$\Psi_{\pm}^{(1)} = \sum_{j,k \in \{0,1\}^2} p_j p_k \dketbra{\pm \varphi_{jk}^{\circledast}}_1$}, since the other systems are either trivial or completely classical and independent of $x_1$. 
Finally, system $1$ is measured by projecting onto the Pauli $X$ basis, which we know from the discussion in Section~\ref{sec:pure_state_channel} (after~\eqref{eq:Helstrom_msmt}) to be the Helstrom measurement to optimally distinguish between the states $\Psi_{\pm}^{(1)}$. 

\begin{figure}
\begin{center}

\includegraphics[scale=0.75,keepaspectratio]{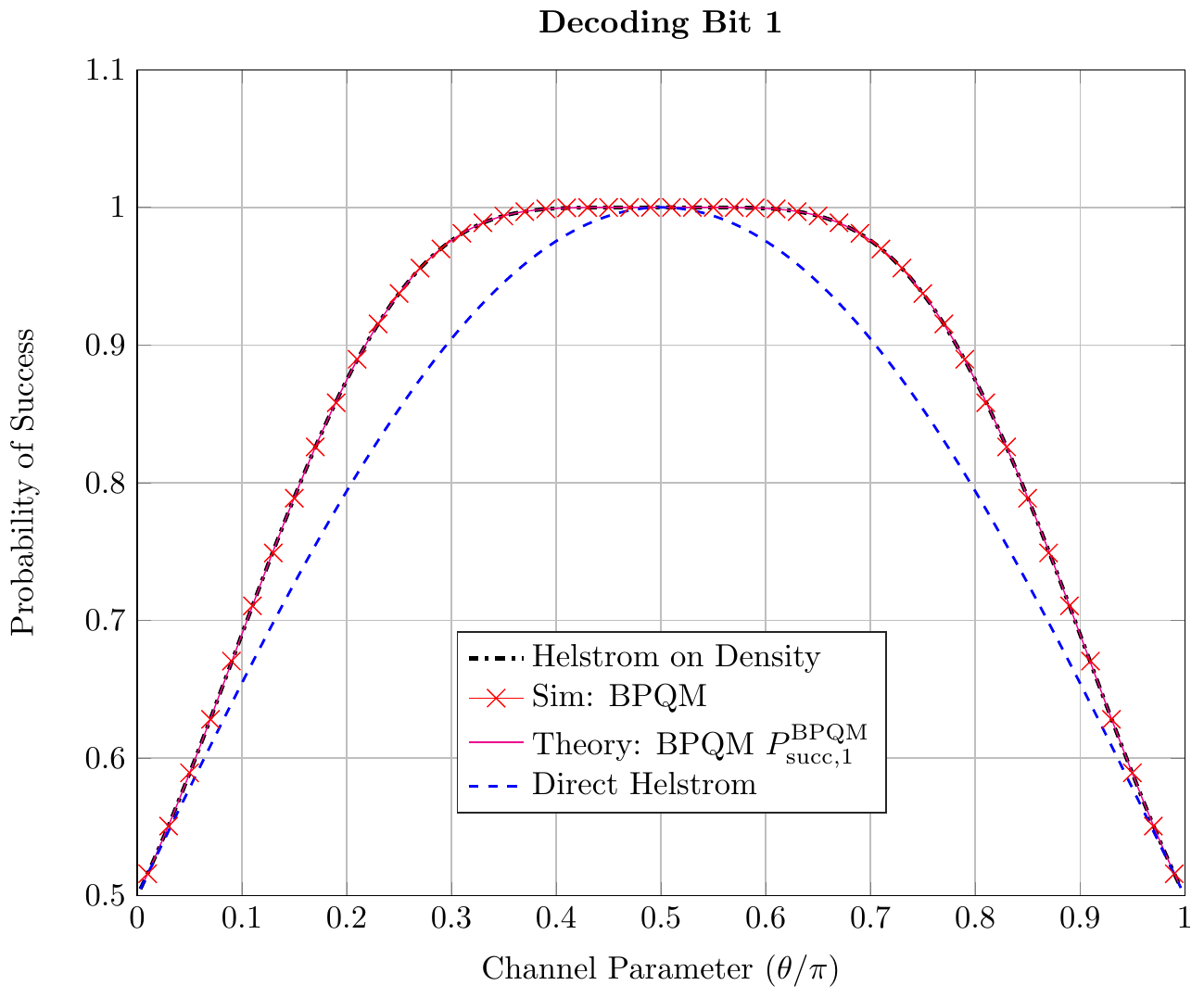}

\caption[The success probabilities for decoding the value of $x_1$ in the $5$-bit code.]{\label{fig:bit1_BPQM}The success probabilities for decoding the value of $x_1$ in the $5$-bit code. Here, ``Helstrom on Density'' represents $P_{\text{succ},1}^{\text{Hel}}$ and ``Direct'' represents the success probability when directly implementing the Helstrom measurement (i.e., measuring $X$) at the channel output on system $1$. The curve ``Sim: BPQM'' corresponds to a simulation that averaged each data point over $10^5$ codeword transmissions.}

\end{center}
\end{figure}

It is important to note that the optimal success probability of distinguishing between the density matrices $\rho_1^{(0)}$ and $\rho_1^{(1)}$ using a collective Helstrom measurement is~\cite{Helstrom-jsp69,Helstrom-ieee70}
\begin{align}
P_{\text{succ},1}^{\text{Hel}} = \frac{1}{2} + \frac{1}{4} \norm{\rho_1^{(0)} - \rho_1^{(1)}}_1, \ \ \norm{M}_1 \coloneqq \text{Tr}\left( \sqrt{M^{\dagger} M} \right).
\end{align}
The action of BPQM until the final measurement is unitary and the trace norm $\norm{\cdot}_1$ is invariant under unitaries. Thus, BPQM does not lose optimality until the final measurement. Since the final measurement is also optimal for distinguishing the two possible states at that stage (e), BPQM is indeed optimal in decoding the value of $x_1$. Thus, despite not performing a collective measurement, but rather only a \emph{single-qubit measurement} at the end of a sequence of unitaries motivated by the FG structure and induced channels in classical BP, BPQM is still optimal to determine whether $x_1 = 0$ or $1$. The performance curves plotted in Fig.~\ref{fig:bit1_BPQM} demonstrate this optimality.

\begin{remark}
\normalfont
Observe that in this quantum scenario, $\rho_1^{(x)}$ behave like a ``posterior'' for bit $x_1$.
However, these can be written down even before transmitting over the channel since they do not depend on the output of the channel.
Hence, it is unclear if there is a better notion of a true posterior which we can then show to be ``marginalized'' by BPQM.
\end{remark}



\subsection*{Performance Analysis}

For bit $x_1$, the probability to decode it as $\hat{x}_1 = 0$ is
\begin{align}
\mathbb{P}\left[ \hat{x}_1 = 0\, \vert \, \Psi_{\pm}^{(1)} \right] = \text{Tr}\left[ \Psi_{\pm}^{(1)} \dketbra{+} \right] = \sum_{j,k \in \{0,1\}^2} p_j p_k \left( \frac{1 \pm \sin\varphi_{jk}^{\circledast}}{2} \right).
\end{align}
Therefore, since there are $4$ codewords each that have $x_1 = 0$ and $x_1 = 1$, the prior for bit $x_1$ is $1/2$ and the probability of success for BPQM in decoding the bit $x_1$ is
\begin{align}
P_{\text{succ},1}^{\text{BPQM}} & = \mathbb{P}[x_1 = 0] \cdot \mathbb{P}[\hat{x}_1 = 0 \, | \, x_1 = 0] + \mathbb{P}[x_1 = 1] \cdot \mathbb{P}[\hat{x}_1 = 1 \, | \, x_1 = 1] \\
  & = \frac{1}{2} \left[ \sum_{j,k \in \{0,1\}^2} p_j p_k \left( \frac{1 + \sin\varphi_{jk}^{\circledast}}{2} \right) + \sum_{j,k \in \{0,1\}^2} p_j p_k \left( \frac{1 + \sin\varphi_{jk}^{\circledast}}{2} \right) \right] \\
  & = \frac{p_0^2}{2} \left( 1 + \sin\varphi_{00}^{\circledast} \right) + (1 - p_0^2) \\
  & = 1 - \frac{p_0^2}{2} (1 - \sin\varphi_{00}^{\circledast}),
\end{align}
where we have used the fact that since all channels are identically $W$, we have $\cos\theta_1^{\boxast} = 0$.
We can calculate
\begin{align}
\cos\varphi_{00}^{\circledast} & = \cos\theta \cos\theta_{00}^{\circledast} = \cos\theta \cos^2\theta_0^{\boxast} = \cos\theta \frac{4 \cos^2\theta}{(1 + \cos^2\theta)^2} \\
\Rightarrow \sin\varphi_{00}^{\circledast} & = \sqrt{1 - \frac{16 \cos^6\theta}{(1 + \cos^2\theta)^4}} = \sqrt{1 - \frac{(2p_0 - 1)^3}{p_0^4}} = \frac{\sqrt{p_0^4 - (2p_0 - 1)^3}}{p_0^2}.
\end{align}
Substituting back, we get the BPQM probability of success for bit $x_1$ to be
\begin{align}
P_{\text{succ},1}^{\text{BPQM}} = 1 - \frac{p_0^2 - \sqrt{p_0^4 - (2p_0 - 1)^3}}{2} = P_{\text{succ},1}^{\text{Hel}},
\end{align}
which is the curve plotted as ``Theory: BPQM $P_{\text{succ},1}^{\text{BPQM}}$'' in Fig.~\ref{fig:bit1_BPQM}.

Before measuring system $1$, the state of system $1$ is essentially $\Psi_{\pm}^{(1)} = p_0^2 \dketbra{\pm \varphi_{00}^{\circledast}} + (1 - p_0^2) \dketbra{\pm}$, since $\cos\varphi_{jk}^{\circledast} = 0$ whenever either $j$ or $k$ equals $1$ (or both) and hence $\dketbra{\pm \varphi_{jk}^{\circledast}} = \dketbra{\pm}$.
So, $p_0^2$ is the probability that the system ``confuses'' the decoder, and projection onto the $X$ basis essentially replaces the system with $\dketbra{m_1}$, where $m_1 = (-1)^{\hat{x}_1} \in \{+,-\}$.
The full post-measurement state (per Section~\ref{sec:measurement}) is given by
\begin{align}
\Phi_{m_1} = \sum_{j,k \in \{0,1\}^2} p_j p_k \frac{\left| \dbraket{m_1}{\pm \varphi_{jk}^{\circledast}} \right|^2}{\text{Tr}\left[ \Psi_{\pm}^{(1)} \dketbra{m_1} \right]} \dketbra{m_1}_1 \otimes \dketbra{00}_{23} \otimes \dketbra{jk}_{45}.
\end{align}
Note that in Fig.~\ref{fig:BPQM_circuit_bit1}, we need to apply a Hadamard after the $Z$-basis measurement in order to ensure that the effective projector is $H \dketbra{\hat{x}_1} H = \dketbra{m_1}$.

Let us denote the overall unitary operation performed in Fig.~\ref{fig:BPQM_circuit_bit1} until stage (e) as $B_1^{\text{BPQM}}$.
As mentioned earlier, the Helstrom measurement to optimally distinguish between $\rho_1^{(0)}$ and $\rho_1^{(1)}$ is given by the projectors $\{ \Pi_1^{\text{Hel}}, \mathbb{I} - \Pi_1^{\text{Hel}} \}$, where 
\begin{align}
\Pi_1^{\text{Hel}} \coloneqq \sum_{i \colon \lambda_i \geq 0} \dketbra{i}, \ \ (\rho_1^{(0)} - \rho_1^{(1)}) \dket{i} = \lambda_i \dket{i}.
\end{align}
So BPQM performs the final Helstrom measurement given by $\{ \tilde{\Pi}_1^{\text{Hel}}, \mathbb{I} - \tilde{\Pi}_1^{\text{Hel}} \}$, where 
\begin{align}
\tilde{\Pi}_1^{\text{Hel}} \coloneqq \sum_{j \colon \lambda_j \geq 0} \dketbra{j} = \dketbra{+}_1 \otimes (I_{16})_{2345}, \qquad \qquad  (\Psi_{+} - \Psi_{-}) \dket{j} & = \lambda_j \dket{j} \\
\Rightarrow \left[ B_1^{\text{BPQM}} \rho_1^{(0)} \left( B_1^{\text{BPQM}} \right)^{\dagger} - B_1^{\text{BPQM}} \rho_1^{(1)} \left( B_1^{\text{BPQM}} \right)^{\dagger} \right] \dket{j} & = \lambda_j \dket{j} \\
\Rightarrow (\rho_1^{(0)} - \rho_1^{(1)}) \left( B_1^{\text{BPQM}} \right)^{\dagger} \dket{j} & = \lambda_j \left( B_1^{\text{BPQM}} \right)^{\dagger} \dket{j}.
\end{align}
Thus, we can express the eigenvectors for $(\rho_1^{(0)} - \rho_1^{(1)})$ as $\dket{i} = \left( B_1^{\text{BPQM}} \right)^{\dagger} \dket{j}$.
This implies
\begin{align}
\Pi_1^{\text{Hel}} = \sum_{i \colon \lambda_i \geq 0} \dketbra{i} & = \left( B_1^{\text{BPQM}} \right)^{\dagger} \left[ \sum_{j \colon \lambda_j \geq 0} \dketbra{j} \right] B_1^{\text{BPQM}} \\
  & = \left( B_1^{\text{BPQM}} \right)^{\dagger} \left[ \dketbra{+}_1 \otimes (I_{16})_{2345} \right] B_1^{\text{BPQM}}.
\end{align}
Hence, in order to identically apply the Helstrom measurement $\Pi_1^{\text{Hel}}$, BPQM needs to first apply $B_1^{\text{BPQM}}$, then measure the first qubit in the $X$-basis, and finally invert $B_1^{\text{BPQM}}$ on the post-measurement state $\Phi_{m_1}$ above.
Although this is optimal for bit $1$, next we will see that it is beneficial to coherently rotate $\Phi_{m_1}$ before inverting $B_1^{\text{BPQM}}$, which sets up a better state discrimination problem for decoding bit $2$.

\subsection{Decoding Bits $2$ and $3$ (or $4$ and $5$)}
\label{sec:BPQM_bit2}

Next, in order to execute BPQM to decode bit $x_2$, we would ideally hope to change the state $\Phi_{m_1}$ back to the channel outputs.
However, this is impossible after having performed the measurement.
In the original BPQM algorithm~\cite{Renes-njp17}, the procedure to be performed at this stage is ambiguous, so we describe a strategy that treads closely along the path of performing the Helstrom measurement for bit $2$ as well, i.e., optimally distinguishing $\rho_2^{(0)}$ and $\rho_2^{(1)}$ evolved through $\tilde{A}_1^{\text{BPQM}} \coloneqq \left( B_1^{\text{BPQM}} \right)^{\dagger} \left[ \dketbra{m_1}_1 \otimes (I_{16})_{2345} \right] B_1^{\text{BPQM}}$.

In order to be able to run BPQM for bit $x_1$ in reverse to get ``as close'' to the channel outputs as possible, we need to make sure that the state $\Phi_{m_1}$ is modified to be compatible with the (angles used to define the) unitaries $V$ and $U$ in Fig.~\ref{fig:BPQM_circuit_bit1}.
Since we can keep track of the intermediate angles deterministically, we can conditionally rotate subsystem $1$ to be $\dketbra{m_1 \varphi_{00}^{\circledast}}_1$ for $\dketbra{jk}_{45} = \dketbra{00}_{45}$.
Note again that in $\Psi_{\pm}$, when either of $j$ or $k$ is $1$ (or both), $\varphi_{jk}^{\circledast} = \pi/2$ and hence $\dketbra{\pm \varphi_{jk}^{\circledast}} = \dketbra{\pm}$.
Therefore, if $\hat{x}_1$ is the wrong estimate for $x_1$, then $\dbraket{m_1}{\pm} = 0$ and the superposition in $\Phi_{m_1}$ collapses to a single term with $j = k = 0$.
More precisely, we can implement the unitary operation 
\begin{align}
\label{eq:cond_rotation}
M_{m_1} \coloneqq (K_{m_1})_1 \otimes \dketbra{00}_{45} + (I_2)_1 \otimes \left( \dketbra{01}_{45} + \dketbra{10}_{45} + \dketbra{11}_{45} \right),
\end{align}
where $K_+$ and $K_-$ are unitaries chosen to satisfy $K_+ \dket{+} = \dket{\varphi_{00}^{\circledast}}$ and $K_- \dket{-} = \dket{-\varphi_{00}^{\circledast}}$, respectively.
We can easily complete these partially defined unitaries with the conditions $K_+ \dket{-} = \sin\frac{\varphi_{00}^{\circledast}}{2} \dket{0} - \cos\frac{\varphi_{00}^{\circledast}}{2} \dket{1}$ and $K_- \dket{+} = \sin\frac{\varphi_{00}^{\circledast}}{2} \dket{0} + \cos\frac{\varphi_{00}^{\circledast}}{2} \dket{1}$.
Applying $M_{m_1}$ to $\Phi_{m_1}$ we get the desired state (compare to state $\Psi_{\pm}$ in stage (e) of Fig.~\ref{fig:BPQM_circuit_bit1})
\begin{align}
\tilde{\Psi}_{m_1} = \sum_{j,k \in \{0,1\}^2} p_j p_k \frac{\left| \dbraket{m_1}{\pm \varphi_{jk}^{\circledast}} \right|^2}{\text{Tr}\left[ \Psi_{\pm}^{(1)} \dketbra{m_1} \right]} \dketbra{m_1 \varphi_{jk}^{\circledast}}_1 \otimes \dketbra{00}_{23} \otimes \dketbra{jk}_{45}.
\end{align}
Now the BPQM circuit for bit $x_1$, shown in Fig.~\ref{fig:BPQM_circuit_bit1}, can be run in reverse from before the final measurement, i.e., from stage (e) back to stage (a).
Hence, the overall operation on the input state in Fig.~\ref{fig:BPQM_circuit_bit1} is
\begin{align}
\label{eq:A1_BPQM}
 A_1^{\text{BPQM}} \coloneqq \left( B_1^{\text{BPQM}} \right)^{\dagger} M_{m_1} \left[ \dketbra{m_1}_1 \otimes (I_{16})_{2345} \right] B_1^{\text{BPQM}}. 
\end{align}

Then we expect the state to be almost the same as the channel outputs, except that system $1$ will deterministically be in state $\dketbra{m_1 \theta}_1$.
However, a simple calculation shows that this is not completely true since the additional factor $\frac{\left| \dbraket{m_1}{\pm \varphi_{jk}^{\circledast}} \right|^2}{\text{Tr}\left[ \Psi_{\pm}^{(1)} \dketbra{m_1} \right]}$ prevents the density matrix to decompose into a tensor product of two $2$-qubit density matrices at stage (b) of Fig.~\ref{fig:BPQM_circuit_bit1}.
Specifically, when we take $\tilde{\Psi}_{m_1}$ at stage (e) back to stage (b) by inverting the BPQM operations, we arrive at the state
\begin{align}
\tilde{\rho}_{\pm, b}^{(m_1)} & = \dketbra{m_1 \theta}_1 \otimes \sum_{j,k \in \{0,1\}^2} p_j p_k \frac{\left| \dbraket{m_1}{\pm \varphi_{jk}^{\circledast}} \right|^2}{\text{Tr}\left[ \Psi_{\pm}^{(1)} \dketbra{m_1} \right]} \nonumber \\
  & \hspace{4.5cm} \dketbra{m_1 \theta_j^{\boxast}, j}_{23} \otimes \dketbra{m_1 \theta_k^{\boxast}, k}_{45} \\
  & = 
 \begin{cases}
  \dfrac{1}{P_{\text{succ}, 1}^{\text{BPQM}}} \dketbra{m_1 \theta}_1 \otimes \sum_{j,k \in \{0,1\}^2} p_j p_k \left| \dbraket{m_1}{\pm \varphi_{jk}^{\circledast}} \right|^2 &  \\
  \hspace{2.5cm} \dketbra{m_1 \theta_j^{\boxast}, j}_{23} \otimes \dketbra{m_1 \theta_k^{\boxast}, k}_{45}  & \text{if}\ \hat{x}_1 = x_1, \\
  \dketbra{m_1 \theta}_1 \otimes \dketbra{m_1 \theta_0^{\boxast}, 0}_{23} \otimes \dketbra{m_1 \theta_0^{\boxast}, 0}_{45}  & \text{if}\ \hat{x}_1 \neq x_1.
 \end{cases} 
\end{align}

\begin{lemma}
Let $C \coloneqq (I_2)_1 \, \cnot{2}{3} \, \cnot{4}{5},\ \dket{\Gamma_{\hat{x}_1}} \coloneqq \cos\frac{\theta_0^{\boxast}}{2} \dket{00} + (-1)^{\hat{x}_1} \sin\frac{\theta_0^{\boxast}}{2} \dket{11}$.
Then
\begin{align}
C \tilde{\rho}_{\pm, b}^{(m_1)} C^{\dagger} & = 
\begin{cases}
\dfrac{1}{P_{\text{succ}, 1}^{\text{BPQM}}} \dketbra{m_1 \theta}_1 \otimes [W \boxast W](\hat{x}_1)_{23} \otimes [W \boxast W](\hat{x}_1)_{45} & \\
 + \dfrac{p_0^2}{P_{\text{succ}, 1}^{\text{BPQM}}} \dfrac{\sin\varphi_{00}^{\circledast}}{2} \dketbra{m_1 \theta}_1 \otimes \dketbra{\Gamma_{\hat{x}_1}, \Gamma_{\hat{x}_1}}_{2345} & \text{if}\ \hat{x}_1 = x_1, \\
 \dketbra{m_1 \theta}_1 \otimes \dketbra{\Gamma_{\hat{x}_1}}_{23} \otimes \dketbra{\Gamma_{\hat{x}_1}}_{45} & \text{if}\ \hat{x}_1 \neq x_1.
\end{cases}
\end{align}
\begin{proof}
The definition of the factor node convolution operation of BPQM implies that
\begin{align}
& C \left( \dketbra{m_1 \theta}_1 \otimes [W \boxast W](\hat{x}_1)_{23} \otimes [W \boxast W](\hat{x}_1)_{45} \right) C^{\dagger} \nonumber \\
 & = \dketbra{m_1 \theta}_1 \otimes \sum_{j \in \{0,1\}} p_j \dketbra{m_1 \theta_j^{\boxast}, j}_{23} \otimes \sum_{k \in \{0,1\}} p_k \dketbra{m_1 \theta_k^{\boxast}, k}_{45} \\
 & = \rho_{m_1, b}.
\end{align}
This in turn implies that $C \rho_{m_1, b} C^{\dagger} = \dketbra{m_1 \theta}_1 \otimes [W \boxast W](\hat{x}_1)_{23} \otimes [W \boxast W](\hat{x}_1)_{45}$.
Ignoring the first qubit and the constant factor for simplicity, observe that
\begin{align}
\tilde{\rho}_{\pm, b}^{(m_1)} \bigg\vert_{\hat{x}_1 = x_1} & = \sum_{j,k \in \{0,1\}^2} p_j p_k \left( \left| \dbraket{m_1}{\pm \varphi_{jk}^{\circledast}} \right|^2 - 1 + 1 \right) \nonumber \\
  & \hspace{3.5cm} \dketbra{m_1 \theta_j^{\boxast}, j}_{23} \otimes \dketbra{m_1 \theta_k^{\boxast}, k}_{45}  \\
  & = \rho_{m_1, b} + p_0^2 \frac{\sin\varphi_{00}^{\circledast}}{2} \dketbra{m_1 \theta_0^{\boxast}, 0}_{23} \otimes \dketbra{m_1 \theta_0^{\boxast}, 0}_{45}.
\end{align}
We have used the fact that except when $j = k = 0$, assuming $\hat{x}_1 = x_1$, $\dbraket{m_1}{\pm \varphi_{jk}^{\circledast}} = \dbraket{m_1}{m_1 \varphi_{jk}^{\circledast}} = \dbraket{m_1}{m_1} = 1$.
Finally, using $\cnot{2}{3} \left( \dket{m_1 \theta_0^{\boxast}}_{2} \otimes \dket{0}_{3} \right) = \dket{\Gamma_{\hat{x}_1}}$, the result follows for both cases $\hat{x}_1 = x_1$ and $\hat{x}_1 \neq x_1$.
\end{proof}
\end{lemma}

Therefore, after reversing the operations of BPQM for bit $x_1$, the system is in the state
\begin{align}
\tilde{\rho}_{m_1,a} & = P_{\text{succ}, 1}^{\text{BPQM}} \cdot C \tilde{\rho}_{\pm, b}^{(m_1)} \bigg\vert_{\hat{x}_1 = x_1} C^{\dagger} + \left( 1 - P_{\text{succ}, 1}^{\text{BPQM}} \right) \cdot C \tilde{\rho}_{\pm, b}^{(m_1)} \bigg\vert_{\hat{x}_1 \neq x_1} C^{\dagger} \\
  & = \dketbra{m_1 \theta}_1 \otimes [W \boxast W](\hat{x}_1)_{23} \otimes [W \boxast W](\hat{x}_1)_{45} \nonumber \\
  & \qquad + p_0^2 \left[ 0.5 (1 + \sin\varphi_{00}^{\circledast}) - 1 \right] \dketbra{m_1 \theta}_1 \otimes \dketbra{\Gamma_{\hat{x}_1}}_{23} \otimes \dketbra{\Gamma_{\hat{x}_1}}_{45} \nonumber \\
  & \qquad + \left( 1 - P_{\text{succ}, 1}^{\text{BPQM}} \right) \cdot \dketbra{m_1 \theta}_1 \otimes \dketbra{\Gamma_{\hat{x}_1}}_{23} \otimes \dketbra{\Gamma_{\hat{x}_1}}_{45} \\
  & = \dketbra{m_1 \theta}_1 \otimes [W \boxast W](\hat{x}_1)_{23} \otimes [W \boxast W](\hat{x}_1)_{45},
\end{align}
since $P_{\text{succ}, 1}^{\text{BPQM}} = p_0^2 \cdot 0.5 (1 + \sin\varphi_{00}^{\circledast}) + (1 - p_0^2)$.

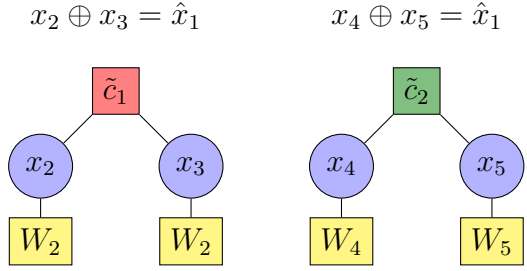
\begin{figure}
\begin{center}

\begin{tikzpicture}

\node at (0,1) {$x_2 \oplus x_3 = \hat{x}_1$};
\node[draw,rectangle,fill=red!50] (c1) at (0,0) {$\tilde{c}_1$};
\node[draw,circle,fill=blue!30] (x2) at (-1,-1) {$x_2$};
\node[draw,rectangle,fill=yellow!60] (W2) at (-1,-2) {$W_2$};
\node[draw,circle,fill=blue!30,fill=blue!30] (x3) at (1,-1) {$x_3$};
\node[draw,rectangle,fill=yellow!60] (W3) at (1,-2) {$W_2$};

\draw (c1) -- (x2);
\draw (c1) -- (x3);

\node at (4,1) {$x_4\oplus x_5 = \hat{x}_1$};
\node[draw,rectangle,fill=darkgreen!50] (c2) at (4,0) {$\tilde{c}_2$};
\node[draw,circle,fill=blue!30] (x4) at (3,-1) {$x_4$};
\node[draw,rectangle,fill=yellow!60] (W4) at (3,-2) {$W_4$};
\node[draw,circle,fill=blue!30] (x5) at (5,-1) {$x_5$};
\node[draw,rectangle,fill=yellow!60] (W5) at (5,-2) {$W_5$};

\draw (c2) -- (x4);
\draw (c2) -- (x5);

\draw (x2) -- (W2);
\draw (x3) -- (W3);
\draw (x4) -- (W4);
\draw (x5) -- (W5);

\end{tikzpicture}

\caption{\label{fig:fg_post_x1}The reduced factor graph after estimating bit $1$ to be $\hat{x}_1$.}

\end{center}
\end{figure}

At this point, we have decoded $\hat{x}_1 = 0$ if $m_1 = +$ and $\hat{x}_1 = 1$ if $m_1 = -$.
We can absorb the value of $\hat{x}_1$ in the FG by updating the parity checks $c_1$ and $c_2$ to impose $x_2 \oplus x_3 = \hat{x}_1$ and $x_4 \oplus x_5 = \hat{x}_1$, respectively.
Now we have two disjoint FGs as shown in Fig.~\ref{fig:fg_post_x1}.
It suffices to decode $x_2$ and $x_4$ since $\hat{x}_3 = \hat{x}_2 \oplus \hat{x}_1$ and $\hat{x}_5 = \hat{x}_4 \oplus \hat{x}_1$.
Also, due to symmetry, it suffices to analyze the success probability of decoding $x_2$ (resp. $x_4$) and $x_3$ (resp. $x_5$).
For this reduced FG, we need to split $\tilde{\rho}_{m_1, a}$ into two density matrices corresponding to the hypotheses $x_2 = 0$ and $x_2 = 1$.
If we revisit the density matrices $\rho_1^{(0)}$ and $\rho_1^{(1)}$, we observe that the $5$-qubit system at the channel output was exactly $\frac{1}{2} \rho_1^{(0)} + \frac{1}{2} \rho_1^{(1)}$.
Hence, for $x_2$, we accordingly split $[W \boxast W](\hat{x}_1)_{23}$ in $\tilde{\rho}_{m_1, a}$ and arrive at the two hypotheses states 
\begin{align}
\tilde{\Phi}_{x_2 = \hat{x}_1}(\hat{x}_1) & = \dketbra{m_1 \theta}_2 \otimes \dketbra{\theta}_3 \otimes [W \boxast W](\hat{x}_1)_{45}, \\
\tilde{\Phi}_{x_2 \neq \hat{x}_1}(\hat{x}_1) & = \dketbra{-m_1 \theta}_2 \otimes \dketbra{-\theta}_3 \otimes [W \boxast W](\hat{x}_1)_{45}.
\end{align}
We can deterministically apply $Z^{\hat{x}_1}$ to system $2$ in order to map this into the following state discrimination problem:
\begin{align}
\Phi_{\pm}(\hat{x}_1) = \dketbra{\pm \theta}_2 \otimes \dketbra{ \pm \theta}_3 \otimes [W \boxast W](\hat{x}_1)_{45},
\end{align}
where $\pm \equiv (-1)^{x_2 - \hat{x}_1}$.
Clearly, we can process systems $2$ and $3$ separately to decide $x_2$. 
Similarly, we can process systems $4$ and $5$ separately to decide $x_4$.
It is also clear that by performing the variable node operation $U_{\circledast}(\theta, \theta)$, we compress all the information into system $2$, i.e., produce $\dketbra{\pm \theta^{\circledast}}_2 \otimes \dketbra{ 0}_3$, which can then be optimally distinguished by measuring in the $X$-basis.
This agrees with the definition of node operations in BPQM as well because now the factor node $\tilde{c}_1$ in Fig.~\ref{fig:fg_post_x1} has degree $2$, and hence the optimal processing is to perform the variable node convolution between qubits $2$ and $3$.
We can incorporate the operation $Z^{\hat{x}_1}$ into BPQM by performing $U_{\circledast}(m_1 \theta, \theta)$ on systems $2$ and $3$ (and similarly on systems $4$ and $5$).
Although $U_{\circledast}(m_1 \theta, \theta) \neq U_{\circledast}(\theta, \theta) \cdot (Z^{\hat{x}_1} \otimes I_2)$, the two operations act identically on the states $\tilde{\Phi}_{x_2 = \hat{x}_1}(\hat{x}_1)$ and $\tilde{\Phi}_{x_2 \neq \hat{x}_1}(\hat{x}_1)$.

\begin{figure}
\begin{center}

\begin{tikzcd}
\lstick{$2$} & \gate[wires=2]{U_{\circledast}(m_1 \theta,\theta)} & \gate{H} & \meter{$\{0,1\}$} & \cw \\
\lstick{$3$} &  & \qw & \qw & \qw \\
\lstick{$4$} & \gate[wires=2]{U_{\circledast}(m_1 \theta,\theta)} & \gate{H} & \meter{$\{0,1\}$} & \cw \\
\lstick{$5$} &  & \qw & \qw & \qw
\end{tikzcd}

\caption{\label{fig:BPQM_circuit_post_x1}The circuit performed in BPQM after decoding bit $x_1$ and reversing the operations performed prior to measurement.}

\end{center}
\end{figure}
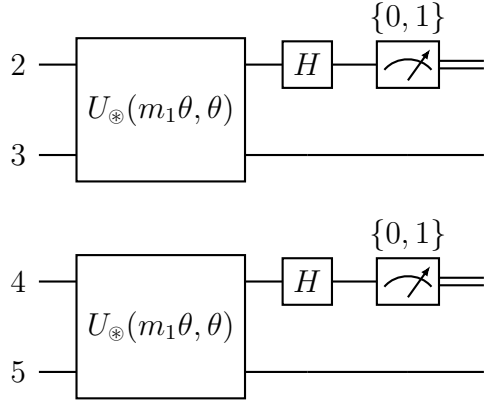

Hence, since the operation in Fig.~\ref{fig:BPQM_circuit_post_x1} is optimal (Helstrom) for distinguishing $\tilde{\Phi}_{x_2 = \hat{x}_1}(\hat{x}_1)$ and $\tilde{\Phi}_{x_2 \neq \hat{x}_1}(\hat{x}_1)$, we expect the BPQM success probability for decoding $\hat{x}_2 = x_2$ to be
\begin{align}
\mathbb{P}[\hat{x}_2 = x_2] & = \frac{1}{2} + \frac{1}{4} \norm{\tilde{\Phi}_{x_2 = \hat{x}_1}(\hat{x}_1) - \tilde{\Phi}_{x_2 \neq \hat{x}_1}(\hat{x}_1)}_1 \\
  & = \frac{1}{2} + \frac{1}{4} \norm{\Phi_{+}(\hat{x}_1) - \Phi_{-}(\hat{x}_1)}_1 \\
  & = \frac{1}{2} + \frac{1}{4} \norm{(\dketbra{\theta^{\circledast}} - \dketbra{-\theta^{\circledast}})_2 \otimes \dketbra{0}_3 \otimes [W \boxast W](\hat{x}_1)_{45}}_1 \\
  & = \frac{1}{2} + \frac{1}{4} \cdot \sin\theta^{\circledast} \norm{X}_1 \\
  & = \frac{1 + \sin\theta^{\circledast}}{2}
    = \frac{1 + \sqrt{1 - \cos^4\theta}}{2}.
\end{align}
Note that if we had defined the state $\tilde{\rho}_{m_1,a}$ to be conditioned on the cases $\hat{x}_1 = x_1$ and $\hat{x}_1 \neq x_1$ separately, then using a similar calculation as above we would still obtain 
\begin{align}
\mathbb{P}[\hat{x}_2 = x_2] & = \mathbb{P}[\hat{x}_1 = x_1] \cdot \mathbb{P}[\hat{x}_2 = x_2 \, | \, \hat{x}_1 = x_1] + \mathbb{P}[\hat{x}_1 \neq x_1] \cdot \mathbb{P}[\hat{x}_2 = x_2 \, | \, \hat{x}_1 \neq x_1] \\
  & = P_{\text{succ},1}^{\text{BPQM}} \cdot \frac{1}{2} \left(1 + \frac{\sin\theta^{\circledast}}{P_{\text{succ},1}^{\text{BPQM}}} \right) + \left(1 - P_{\text{succ},1}^{\text{BPQM}} \right) \cdot \frac{1}{2} 
  = \frac{1 + \sin\theta^{\circledast}}{2}.
\end{align}

\begin{figure}
\centering

\includegraphics[scale=0.75,keepaspectratio]{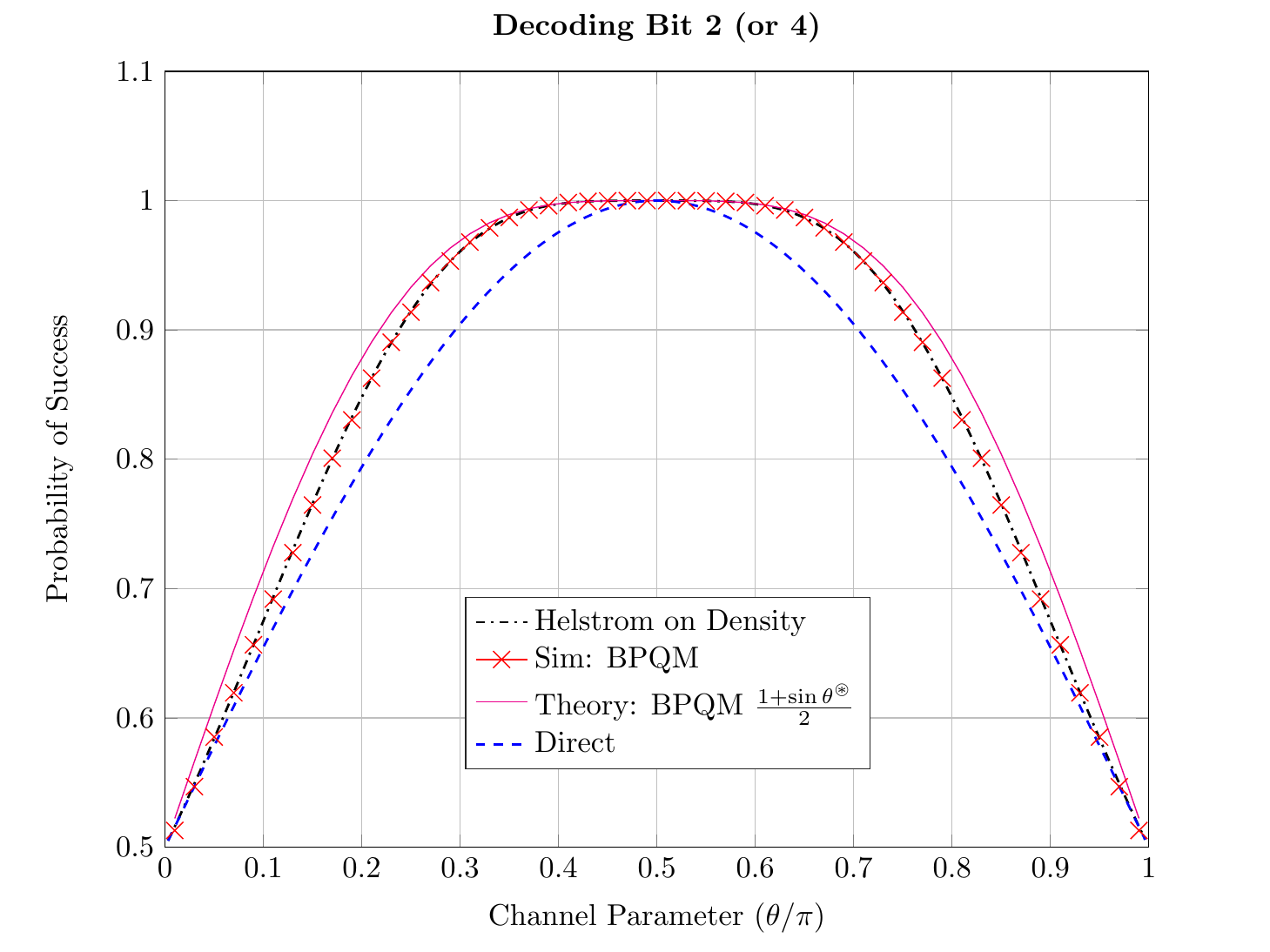}





\hspace{-0.65cm}
%
\includegraphics[scale=0.75,keepaspectratio]{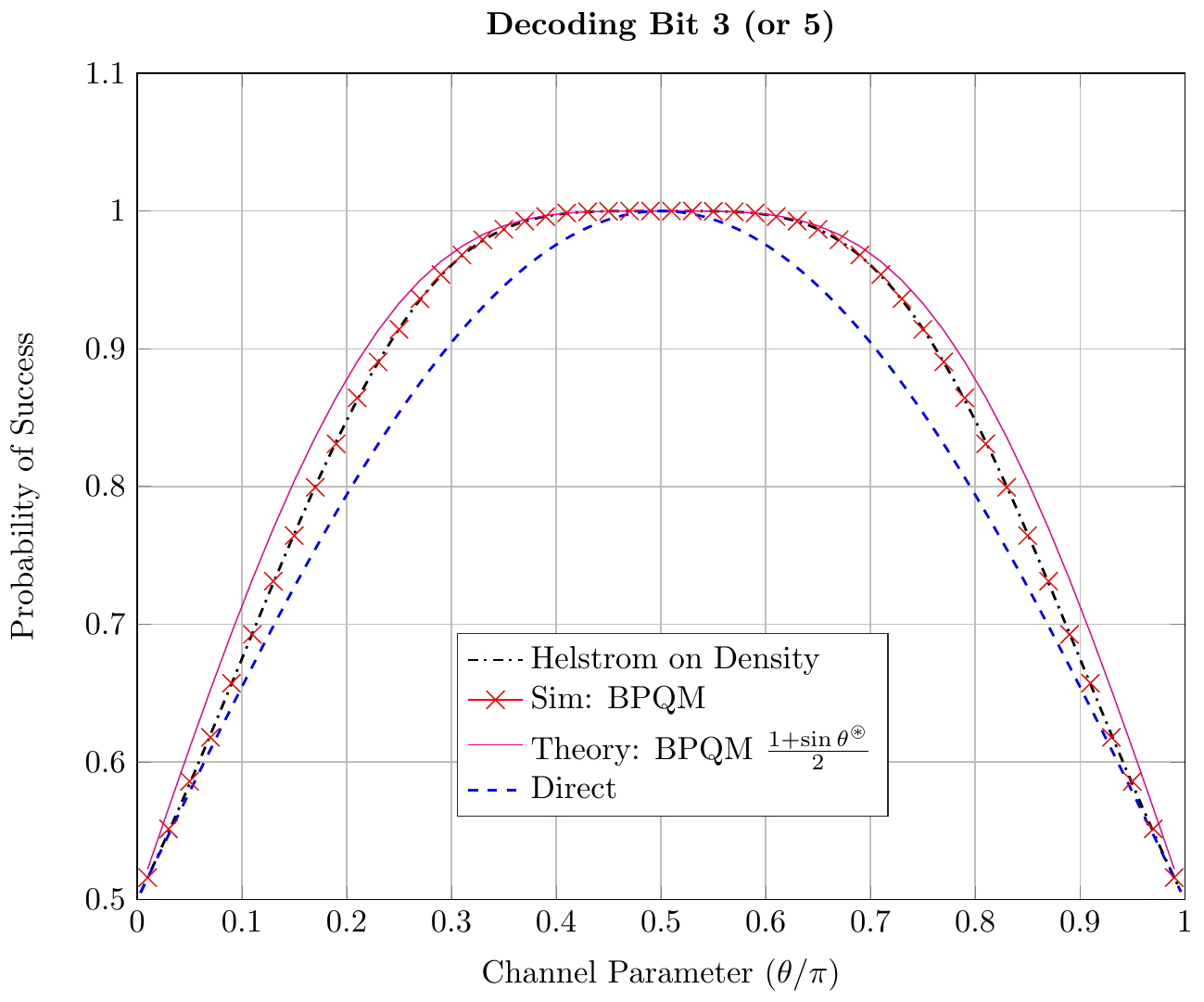}


\caption[The success probabilities for decoding the value of $x_2$ or $x_4$ (top) and $x_3$ or $x_5$ (bottom) in the $5$-bit code.]{\label{fig:bit23_BPQM}The success probabilities for decoding the value of $x_2$ or $x_4$ (top) and $x_3$ or $x_5$ (bottom) in the $5$-bit code. Here, ``Helstrom on Density'' represents $P_{\text{succ},k}^{\text{Hel}}$ and ``Direct'' represents the success probability when directly implementing the Helstrom measurement (i.e., measuring in the $X$ basis) at the channel output on system $k$, for $k \in \{2,3,4,5\}$. The curves ``Sim: BPQM'' correspond to a simulation that averaged each data point over $10^5$ uniformly random codeword transmissions.}

\end{figure}

Finally, bits $x_3$ and $x_5$ can be estimated simply as $\hat{x}_3 = \hat{x}_2 \oplus \hat{x}_1$ and $\hat{x}_5 = \hat{x}_4 \oplus \hat{x}_1$, without performing any quantum operations.
Due to symmetry, the success probabilities for both these bits are the same.
Bit $3$ is decoded correctly, i.e., $\hat{x}_3 = x_3$, if either $\hat{x}_1 = x_1$ and $\hat{x}_2 = x_2$ or $\hat{x}_1 \neq x_1$ and $\hat{x}_2 \neq x_2$.
Hence, we can write
\begin{align}
\mathbb{P}[\hat{x}_3 = x_3] & = \mathbb{P}[\hat{x}_1 = x_1] \cdot \mathbb{P}[\hat{x}_2 = x_2 \, | \, \hat{x}_1 = x_1] + \mathbb{P}[\hat{x}_1 \neq x_1] \cdot \mathbb{P}[\hat{x}_2 \neq x_2 \, | \, \hat{x}_1 \neq x_1] \\
  & = P_{\text{succ},1}^{\text{BPQM}} \cdot \frac{1}{2} \left(1 + \frac{\sin\theta^{\circledast}}{P_{\text{succ},1}^{\text{BPQM}}} \right) + \left(1 - P_{\text{succ},1}^{\text{BPQM}} \right) \cdot \frac{1}{2}
    = \frac{1 + \sin\theta^{\circledast}}{2}.
\end{align}
Again, $P_{\text{succ},5}^{\text{Hel}} = P_{\text{succ},3}^{\text{Hel}}$, and these probabilites are plotted in Fig.~\ref{fig:bit23_BPQM} (bottom).

We simulated the performance of BPQM by randomly generating $10^5$ codewords, passing each of them through $5$ independent copies of the pure state channel (for the $5$ bits), and employing the above sequence of unitaries and measurements for decoding all the bits.
We plot the empirical performance of BPQM for decoding bit $2$ in Fig.~\ref{fig:bit23_BPQM} (top), along with the predicted theoretical performance above, the performance for directly measuring the second qubit at the channel output, and also the Helstrom success probability $P_{\text{succ},2}^{\text{Hel}}$ for optimally distinguishing $\rho_2^{(0)}$ and $\rho_2^{(1)}$, where
\begin{align}
\rho_2^{(0)} & = \dketbra{\theta}_2 \otimes \bigg[ \frac{1}{2} \dketbra{\theta}_1 \otimes \dketbra{\theta}_3 \otimes [W \boxast W](0)_{45} \nonumber \\
  & \hspace{3cm} + \frac{1}{2} \dketbra{-\theta}_1 \otimes \dketbra{-\theta}_3 \otimes [W \boxast W](1)_{45} \bigg], \\
\rho_2^{(1)} & = \dketbra{-\theta}_2 \otimes \bigg[ \frac{1}{2} \dketbra{\theta}_1 \otimes \dketbra{-\theta}_3 \otimes [W \boxast W](0)_{45} \nonumber \\
  & \hspace{3cm} + \frac{1}{2} \dketbra{-\theta}_1 \otimes \dketbra{\theta}_3 \otimes [W \boxast W](1)_{45} \bigg].
\end{align}
We observe that BPQM performs significantly better than direct Helstrom measurement of the channel output, as we might expect.
Moreover, the performance of BPQM is indistinguishable (up to numerical precision) from that of the Helstrom measurement that optimally distinguishes between $\rho_2^{(0)}$ and $\rho_2^{(1)}$.
This suggests that BPQM might be provably optimal for decoding bit $2$ as well, even though the measurement for bit $1$ rendered the original channel outputs irrecoverable.

However, we also notice that the above prediction for the BPQM success probability for decoding bit $2$ is relatively higher than even the optimal Helstrom measurement.
This indicates that the state discrimination problem for bit $2$ discussed above is more ideal than the actual problem in hand.
Hence, next we analyze the true state discrimination problem for bit $2$ (or $4$) and clarify the observed performance in Fig.~\ref{fig:bit23_BPQM}.

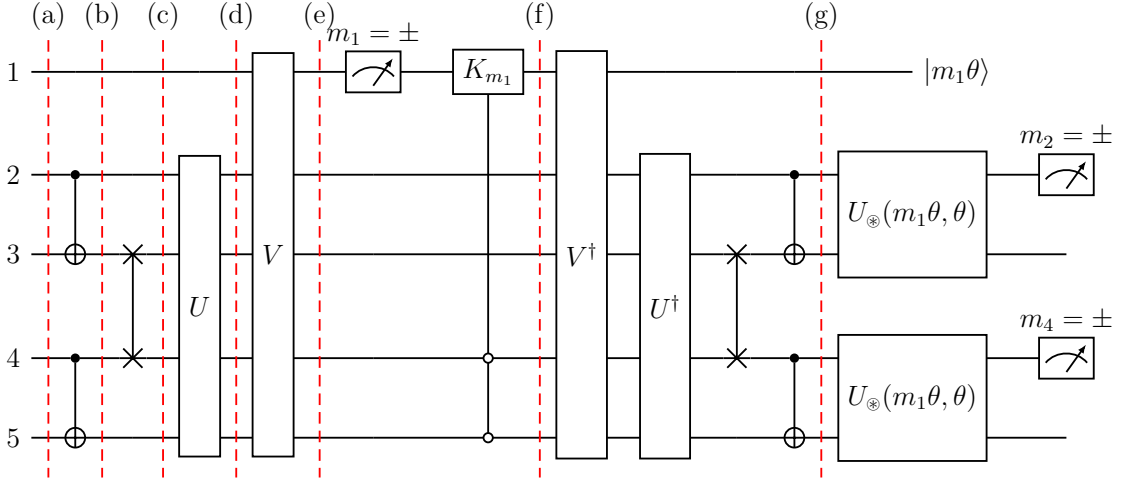
\begin{figure}
\centering

\scalebox{0.88}{%
\begin{tikzcd}
\lstick{$1$} \slice{(a)} & \qw & \qw & \qw & \gate[wires=5]{V} \slice{(e)} & \meter{$m_1 = \pm$} & \gate{K_{m_1}} \slice{(f)} & \gate[wires=5]{V^{\dagger}} & \qw & \qw & \qw & \qw\rstick{$\ket{m_1 \theta}$} \\
\lstick{$2$} & \ctrl{1} \slice{(b)} & \qw & \gate[wires=4]{U} \slice{(d)} &  & \qw & \qw &  & \gate[wires=4]{U^{\dagger}} & \qw & \ctrl{1} \slice{(g)} & \gate[wires=2]{U_{\circledast}(m_1 \theta, \theta)} & \meter{$m_2 = \pm$} \\
\lstick{$3$} & \targ{} & \swap{1} \slice{(c)} &  &  & \qw & \qw &  &  & \swap{1} & \targ{} &  & \qw \\
\lstick{$4$} & \ctrl{1} & \targX{} &  &  & \qw & \octrl{-3} &  &  & \targX{} & \ctrl{1} & \gate[wires=2]{U_{\circledast}(m_1 \theta, \theta)} & \meter{$m_4 = \pm$} \\
\lstick{$5$} & \targ{} & \qw &  &  & \qw & \octrl{-1} &  &  & \qw & \targ{} &  & \qw 
\end{tikzcd}
}

\caption[The full BPQM circuit to decode all bits of the $5$-bit code in Fig.~\ref{fig:five_bit_code}.]{\label{fig:BPQM_full_circuit}The full BPQM circuit to decode all bits of the $5$-bit code in Fig.~\ref{fig:five_bit_code}. The decoded values are related to the measurement results as $m_1 = (-1)^{\hat{x}_1}, m_2 = (-1)^{\hat{x}_2}, m_4 = (-1)^{\hat{x}_4}$, and $\hat{x}_3 = \hat{x}_1 \oplus \hat{x}_2, \hat{x}_5 = \hat{x}_1 \oplus \hat{x}_4$. The open-circled controls indicate that $K_{m_1}$ is coherently controlled by the last two qubits being in the state $\dket{00}_{45}$. The solid line before $K_{m_1}$ indicates that the controlled unitary is applied to the post-measurement state. See Fig.~\ref{fig:bpqm_decomposed} for the full decomposition.}

\end{figure}

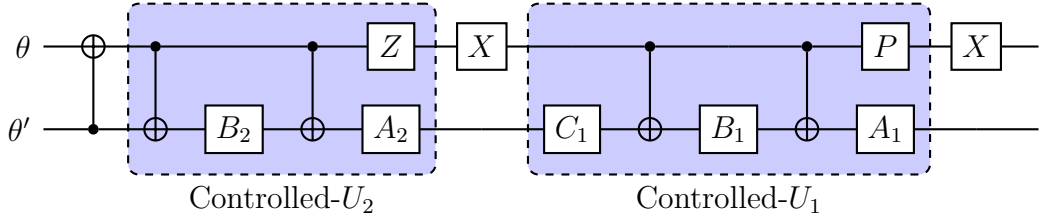
\begin{figure}

\centering

\begin{tikzcd}
\lstick{$\theta$} & \targ{} & \ctrl{1}\gategroup[2,steps=4,style={dashed, rounded corners,fill=blue!20, inner xsep=2pt}, background,label style={label position=below,anchor= north,yshift=-0.2cm}]{Controlled-$U_2$} & \qw & \ctrl{1} & \gate{Z} & \gate{X} & \qw\gategroup[2,steps=5,style={dashed, rounded corners,fill=blue!20, inner xsep=2pt}, background,label style={label position=below,anchor= north,yshift=-0.2cm}]{Controlled-$U_1$} & \ctrl{1} & \qw & \ctrl{1} & \gate{P} & \gate{X} & \qw \\
\lstick{$\theta'$} & \ctrl{-1} & \targ{} & \gate{B_2} & \targ{} & \gate{A_2} & \qw & \gate{C_1} & \targ{} & \gate{B_1} & \targ{} & \gate{A_1} & \qw & \qw
\end{tikzcd}

\caption[The circuit decomposition for the variable node unitary $U_{\circledast}(\theta,\theta')$.]{\label{fig:vn_unitary_circuit}The circuit decomposition for $U_{\circledast}(\theta,\theta')$, where we set  
$A_1 \coloneqq R_y(\gamma_1/2)$,  
$B_1 \coloneqq R_y(-\gamma_1/2) R_z(-\pi/2)$,   
$C_1 \coloneqq R_z(\pi/2)$, 
$A_2 \coloneqq R_z(\pi) R_y(\gamma_2/2)$,
$B_2 \coloneqq R_y(-\gamma_2/2) R_z(-\pi)$, and $P = \sqrt{Z}$ is the phase gate. See Section~\ref{sec:vn_unitary_decompose} for calculations and angles $\gamma_1, \gamma_2$. Note that, for example, $B_2 = R_y(-\gamma_2/2) R_z(-\pi)$ implies that $R_z(-\pi)$ must be applied first, then followed by $R_y(-\gamma_2/2)$.}

\end{figure}

\begin{figure}

\centering

\begin{tikzcd}
\lstick{$i$} & \gate{D} & \qw \\
\lstick{$4$} & \ctrl{-1} & \qw \\
\lstick{$5$} & \ctrl{-1} & \qw \\
\end{tikzcd}
\ \ =
\begin{tikzcd}
\lstick{$i$} & \qw & \gate{D} & \qw & \qw \\
\lstick{$\ket{0}$} & \targ{} & \ctrl{-1} & \targ{} & \qw \\
\lstick{$4$} & \ctrl{-1} & \qw & \ctrl{-1} & \qw \\
\lstick{$5$} & \ctrl{-1} & \qw & \ctrl{-1} & \qw \\
\end{tikzcd}

\begin{tikzcd}
\lstick{$i$} & \ctrl{1} & \qw \\
\lstick{$j$} & \targ{} & \qw \\
\lstick{$4$} & \ctrl{-1} & \qw \\
\lstick{$5$} & \ctrl{-1} & \qw \\
\end{tikzcd}
\ \ =
\begin{tikzcd}
\lstick{$i$} & \qw & \ctrl{1} & \qw & \qw \\
\lstick{$j$} & \qw & \targ{} & \qw & \qw \\
\lstick{$\ket{0}$} & \targ{} & \ctrl{-1} & \targ{} & \qw \\
\lstick{$4$} & \ctrl{-1} & \qw & \ctrl{-1} & \qw \\
\lstick{$5$} & \ctrl{-1} & \qw & \ctrl{-1} & \qw \\
\end{tikzcd}

\hspace{-5.15cm}
\begin{tikzcd}
\lstick{$1$} & \gate{K_{m_1}} & \qw \\
\lstick{$4$} & \octrl{-1} & \qw \\
\lstick{$5$} & \octrl{-1} & \qw 
\end{tikzcd}
\ \ =

\vspace{0.45cm}

\scalebox{0.85}{%
\begin{tikzcd}
\lstick{$1$} & \qw & \qw & \gate{X^{\hat{x}_1 \oplus 1}} \gategroup[2,steps=7,style={dashed, rounded corners,fill=blue!20, inner xsep=2pt}, background,label style={label position=below,anchor= north,yshift=-0.2cm}]{Controlled-$K_{m_1}$} & \gate{R_z(-\pi)} & \gate{R_y(\frac{-\gamma}{2})} & \targ{} & \gate{R_y(\frac{\gamma}{2})} & \gate{R_z(\pi)} & \gate{X^{\hat{x}_1}} & \qw & \qw & \qw \\
\lstick{$\ket{0}$} & \qw & \targ{} & \ctrl{-1} & \qw & \qw & \ctrl{-1} & \qw & \gate{Z} & \ctrl{-1} & \targ{} & \qw & \qw \\
\lstick{$4$} & \gate{X} & \ctrl{-1} & \qw & \qw & \qw & \qw & \qw & \qw & \qw & \ctrl{-1} & \gate{X} & \qw \\
\lstick{$5$} & \gate{X} & \ctrl{-1} & \qw & \qw & \qw & \qw & \qw & \qw & \qw & \ctrl{-1} & \gate{X} & \qw 
\end{tikzcd}
}

\caption[Decomposition of controlled gates using Toffoli (i.e., CCX) gates and an ancilla.]{\label{fig:controlled_unitary} Decomposition of controlled gates using Toffoli (i.e., CCX) gates and an ancilla~\cite[Fig. 4.10]{Nielsen-2010}, where $D \in \{A_1,B_1,C_1,A_2,B_2,P\}$ as appropriate ($P = \sqrt{Z}$) and $i,j \in \{1,2,3\}$. The top two identities can be used to implement each of the doubly-controlled $U_{\circledast}(\theta,\theta')$ operations appearing in Fig.~\ref{fig:bpqm_decomposed} by applying doubly-controlled versions of the components of $U_{\circledast}(\theta,\theta')$ in Fig.~\ref{fig:vn_unitary_circuit}. Note that $m_1 = (-1)^{\hat{x}_1}$ is the result of estimating $x_1$ to be $\hat{x}_1 \in \{0,1\}$. See Section~\ref{sec:vn_unitary_decompose} for the relevant calculations and the angle $\gamma$.}

\vspace{-0.5cm}
    
\end{figure}
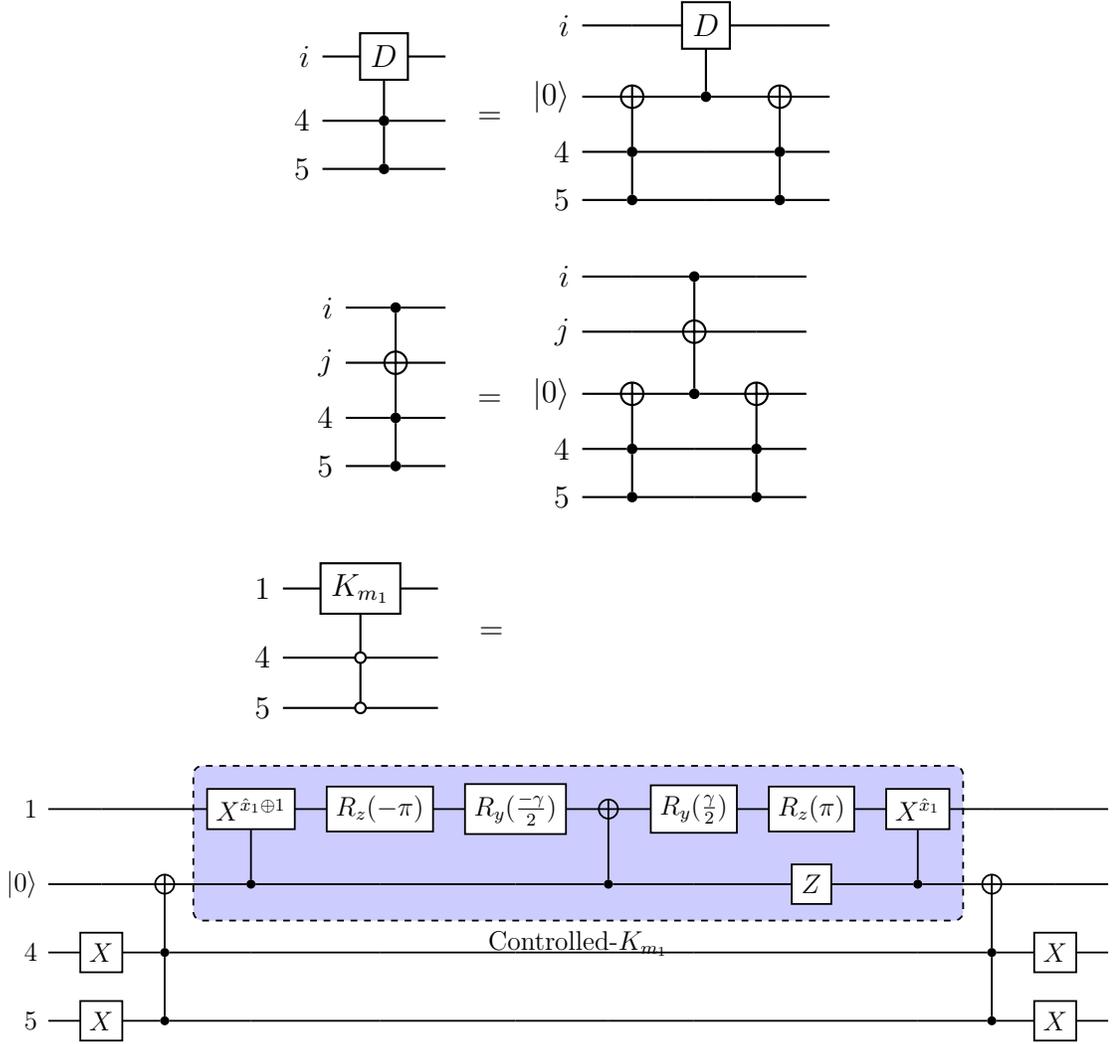

\begin{sidewaysfigure}

\centering

\scalebox{0.75}{
\begin{tikzcd}
\lstick{$1$}\slice{(a)} & \qw & \qw & \qw & \qw & \qw & \qw & \gate[wires=2]{U_{\circledast}(\theta, \theta_{00}^{\circledast})}\gategroup[5,steps=4,style={dashed, rounded corners,fill=red!20, inner xsep=2pt}, background,label style={label position=below,anchor= north,yshift=-0.2cm}]{$V$ in BPQM for bit $1$} & \gate[wires=2]{U_{\circledast}(\theta, \theta_{01}^{\circledast})} & \gate[wires=2]{U_{\circledast}(\theta, \theta_{11}^{\circledast})} & \gate[wires=2]{U_{\circledast}(\theta, \theta_{10}^{\circledast})} \slice{(e)} & \meter{$m_1 = \pm$} & \gate{K_{m_1}} & \qw \ \ldots \\
\lstick{$2$} & \ctrl{1}\slice{(b)} & \qw & \gate[wires=2]{U_{\circledast}(\theta_0^{\boxast}, \theta_0^{\boxast})}\gategroup[4,steps=4,style={dashed, rounded corners,fill=blue!20, inner xsep=2pt}, background,label style={label position=below,anchor= north,yshift=-0.2cm}]{$U$ in BPQM for bit $1$} & \gate[wires=2]{U_{\circledast}(\theta_0^{\boxast}, \theta_1^{\boxast})} & \gate[wires=2]{U_{\circledast}(\theta_1^{\boxast}, \theta_1^{\boxast})} & \gate[wires=2]{U_{\circledast}(\theta_1^{\boxast}, \theta_0^{\boxast})} \slice{(d)} &  &  &  &  & \qw & \qw & \qw \ \ldots \\
\lstick{$3$} & \targ{} & \swap{1}\slice{(c)} &  &  &  &  & \qw & \qw & \qw & \qw & \qw & \qw & \qw \ \ldots \\
\lstick{$4$} & \ctrl{1} & \targX{} & \octrl{-1} & \octrl{-1} & \ctrl{-1} & \ctrl{-1} & \octrl{-2} & \octrl{-2} & \ctrl{-2} & \ctrl{-2} & \qw & \octrl{-3} & \qw \ \ldots \\
\lstick{$5$} & \targ{} & \qw & \octrl{-1} & \ctrl{-1} & \ctrl{-1} & \octrl{-1} & \octrl{-1} & \ctrl{-1} & \ctrl{-1} & \octrl{-1} & \qw & \octrl{-1} & \qw \ \ldots 
\end{tikzcd}
}

\vspace{0.5cm}

\scalebox{0.75}{
\begin{tikzcd}
\slice{(f)} \ldots \ \qw & \gate[wires=2]{U_{\circledast}(\theta, \theta_{10}^{\circledast})^{\dagger}}\gategroup[5,steps=4,style={dashed, rounded corners,fill=red!20, inner xsep=2pt}, background,label style={label position=below,anchor= north,yshift=-0.2cm}]{$V^{\dagger}$ for $V$ in BPQM for bit $1$} & \gate[wires=2]{U_{\circledast}(\theta, \theta_{11}^{\circledast})^{\dagger}} & \gate[wires=2]{U_{\circledast}(\theta, \theta_{01}^{\circledast})^{\dagger}} & \gate[wires=2]{U_{\circledast}(\theta, \theta_{00}^{\circledast})^{\dagger}} & \qw & \qw & \qw & \qw & \qw & \qw & \qw\rstick{$\ket{m_1 \theta}$} \\
\ldots \ \qw &  &  &  &  & \gate[wires=2]{U_{\circledast}(\theta_1^{\boxast}, \theta_0^{\boxast})^{\dagger}}\gategroup[4,steps=4,style={dashed, rounded corners,fill=blue!20, inner xsep=2pt}, background,label style={label position=below,anchor= north,yshift=-0.2cm}]{$U^{\dagger}$ for $U$ in BPQM for bit $1$} & \gate[wires=2]{U_{\circledast}(\theta_1^{\boxast}, \theta_1^{\boxast})^{\dagger}} & \gate[wires=2]{U_{\circledast}(\theta_0^{\boxast}, \theta_1^{\boxast})^{\dagger}} & \gate[wires=2]{U_{\circledast}(\theta_0^{\boxast}, \theta_0^{\boxast})^{\dagger}} & \qw & \ctrl{1}\slice{(g)} & \gate[wires=2]{U_{\circledast}(m_1 \theta, \theta)} & \meter{$m_2 = \pm$} \\
\ldots \ \qw & \qw & \qw & \qw & \qw &  &  &  &  & \swap{1} & \targ{} &  & \qw \\
\ldots \ \qw & \ctrl{-2} & \ctrl{-2} & \octrl{-2} & \octrl{-2} & \ctrl{-1} & \ctrl{-1} & \octrl{-1} & \octrl{-1} & \targX{} & \ctrl{1} & \gate[wires=2]{U_{\circledast}(m_1 \theta, \theta)} & \meter{$m_4 = \pm$} \\
\ldots \ \qw & \octrl{-1} & \ctrl{-1} & \ctrl{-1} & \octrl{-1} & \octrl{-1} & \ctrl{-1} & \ctrl{-1} & \octrl{-1} & \qw & \targ{} &  & \qw
\end{tikzcd}
}

\caption[Full decomposition of the BPQM circuit in Fig.~\ref{fig:BPQM_full_circuit}.]{\label{fig:bpqm_decomposed} Full decomposition of the BPQM circuit in Fig.~\ref{fig:BPQM_full_circuit}. The variable node unitaries $U_{\circledast}(\theta,\theta')$ are decomposed in Fig.~\ref{fig:vn_unitary_circuit}. The two-qubit-controlled coherent versions of these unitaries as well as that of the single-qubit rotation $K_{m_1}$, which is a function of the measurement result $m_1$, are decomposed in Fig.~\ref{fig:controlled_unitary}.}

\end{sidewaysfigure}
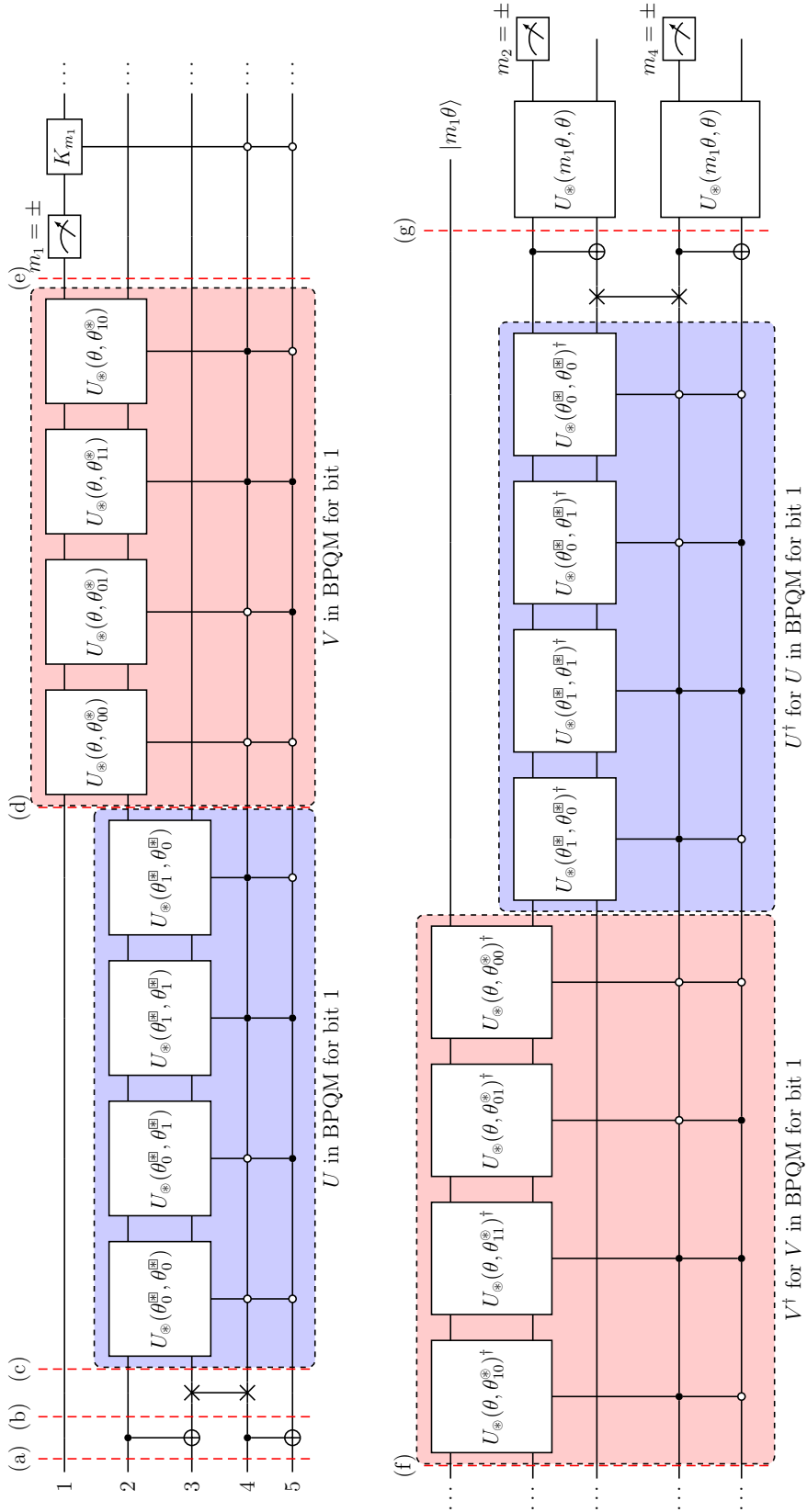

\newpage

\subsection{Analysis of BPQM Optimality for Decoding Bit $2$}
\label{sec:bpqm_bit2_optimal}

At the channel output, it is clear that the optimal strategy to decode bit $2$ is to perform the Helstrom measurement that distinguishes between $\rho_2^{(0)}$ and $\rho_2^{(1)}$.
However, since we performed BPQM operations to decode bit $1$ first, these two density matrices would have evolved through that process.
Therefore, the correct analysis is to derive the resulting states and then subject them to the BPQM strategy for decoding bit $2$ that was discussed above.
For simplicity, we only track $\rho_2^{(0)}$ through the different stages in Fig.~\ref{fig:BPQM_full_circuit} (or Fig.~\ref{fig:bpqm_decomposed}).
The corresponding states for $\rho_2^{(1)}$ can be ascertained from these.
Note that
\begin{align}
\frac{1}{2} \dketbra{\theta}_2 \otimes \dketbra{\theta}_3 & = [W \boxast W](0)_{23} - \frac{1}{2} \dketbra{-\theta}_2 \otimes \dketbra{-\theta}_3, \\
\frac{1}{2} \dketbra{\theta}_2 \otimes \dketbra{-\theta}_3 & = [W \boxast W](1)_{23} - \frac{1}{2} \dketbra{-\theta}_2 \otimes \dketbra{\theta}_3.
\end{align}
For brevity, we will use the notation $\text{CX}_{ij} \coloneqq \cnot{i}{j}$ and $\text{Swap}_{34} \coloneqq \text{CX}_{34}\, \text{CX}_{43}\, \text{CX}_{34}$.
Then
\begin{align}
\rho_{2,a}^{(0)} & = \frac{1}{2} \dketbra{\theta}_1 \otimes \dketbra{\theta}_2 \otimes \dketbra{\theta}_3 \otimes [W \boxast W](0)_{45} \nonumber \\
  & \hspace{2cm} + \frac{1}{2} \dketbra{-\theta}_1 \otimes \dketbra{\theta}_2 \otimes \dketbra{-\theta}_3 \otimes [W \boxast W](1)_{45}, \\
\rho_{2,b}^{(0)} & = \dketbra{\theta}_1 \otimes \left[ \sum_{j = 0,1} p_j \dketbra{\theta_j^{\boxast}}_2 \otimes \dketbra{j}_3 - \frac{1}{2} \text{CX}_{23}\, \dketbra{-\theta,-\theta}_{23}\, \text{CX}_{23} \right] \nonumber \\
  & \hspace{4cm} \otimes \left[ \sum_{k = 0,1} p_k \dketbra{\theta_k^{\boxast}}_4 \otimes \dketbra{k}_5 \right] \nonumber \\
  & \quad + \dketbra{-\theta}_1 \otimes \left[ \sum_{j = 0,1} p_j \dketbra{-\theta_j^{\boxast}}_2 \otimes \dketbra{j}_3 - \frac{1}{2} \text{CX}_{23}\, \dketbra{-\theta,\theta}_{23}\, \text{CX}_{23} \right] \nonumber \\
  & \hspace{4cm} \otimes \left[ \sum_{k = 0,1} p_k \dketbra{-\theta_k^{\boxast}}_4 \otimes \dketbra{k}_5 \right], \\
\rho_{2,c}^{(0)} & = \dketbra{\theta}_1 \otimes \sum_{j,k \in \{0,1\}^2} p_j p_k \dketbra{\theta_j^{\boxast}}_2 \otimes \dketbra{\theta_k^{\boxast}}_3 \otimes \dketbra{j}_4 \otimes \dketbra{k}_5 - \frac{1}{2} \dketbra{\theta}_1 \nonumber \\
  & \quad \hspace{1cm} \otimes \text{Swap}_{34}\, \left[ \text{CX}_{23} \dketbra{-\theta,-\theta}_{23} \text{CX}_{23} \otimes  \sum_{k = 0,1} p_k \dketbra{\theta_k^{\boxast}, k}_{45} \right] \text{Swap}_{34} \nonumber \\
  & \quad + \dketbra{-\theta}_1 \otimes \sum_{j,k \in \{0,1\}^2} p_j p_k \dketbra{-\theta_j^{\boxast}, -\theta_k^{\boxast}}_{23} \otimes \dketbra{j,k}_{45} - \frac{1}{2} \dketbra{-\theta}_1 \nonumber \\
  & \qquad  \otimes \text{Swap}_{34}\, \left[ \text{CX}_{23} \dketbra{-\theta,\theta}_{23} \text{CX}_{23} \otimes  \sum_{k = 0,1} p_k \dketbra{-\theta_k^{\boxast}, k}_{45} \right] \text{Swap}_{34}, \\
\rho_{2,e}^{(0)} & = \sum_{j,k \in \{0,1\}^2} p_j p_k \dketbra{ \varphi_{jk}^{\circledast}}_1 \otimes \dketbra{0}_2 \otimes \dketbra{0}_3 \otimes \dketbra{jk}_{45} \nonumber \\
  & \quad + \sum_{j,k \in \{0,1\}^2} p_j p_k \dketbra{- \varphi_{jk}^{\circledast}}_1 \otimes \dketbra{0}_2 \otimes \dketbra{0}_3 \otimes \dketbra{jk}_{45} \nonumber \\
  & \quad - \frac{1}{2} V U \Biggr\{ \dketbra{\theta}_1 \otimes \text{Swap}_{34}\, \Big[ \text{CX}_{23} \dketbra{-\theta,-\theta}_{23} \text{CX}_{23} \nonumber \\
  & \hspace{5cm} \otimes  \sum_{k = 0,1} p_k \dketbra{\theta_k^{\boxast}}_4 \otimes \dketbra{k}_5 \Big] \text{Swap}_{34} \nonumber \\
  & \quad + \dketbra{-\theta}_1 \otimes \text{Swap}_{34}\, \Big[ \text{CX}_{23} \dketbra{-\theta,\theta}_{23} \text{CX}_{23} \nonumber \\
  & \hspace{5cm} \otimes \sum_{k = 0,1} p_k \dketbra{-\theta_k^{\boxast}}_4 \otimes \dketbra{k}_5 \Big] \text{Swap}_{34} \Biggr\} U^{\dagger} V^{\dagger}.
\end{align}
Next we make an $X$-basis measurement on the first qubit, and for convenience we assume that the measurement result is $m_1 = +$.
The analysis for $m_1 = -$ is very similar and follows by symmetry.
We verified numerically that $\text{Tr}\left[ \dketbra{+}_1 \cdot \rho_{2,e}^{(0)} \right] = 0.5$, which we might expect since $\rho_{2,e}^{(0)}$ corresponds to $x_2 = 0$ and $x_2$ is independent from $x_1$.
Since $m_1 = +$, we follow the measurement with the conditional rotation $M_+$ in~\eqref{eq:cond_rotation} to obtain (Sp $\equiv$ Swap)
\begin{align}
\Phi_{2,m_1 = +}^{(0)} & = \frac{1}{0.5} \biggr[ \sum_{j,k \in \{0,1\}^2} p_j p_k \bigg\lvert \dbraket{+}{\varphi_{jk}^{\circledast}} \bigg\rvert^2 \dketbra{ \varphi_{jk}^{\circledast}}_1 \otimes \dketbra{00}_{23} \otimes \dketbra{jk}_{45} \nonumber \\
  & \qquad \qquad + \frac{p_0^2 (1 - \sin\varphi_{00}^{\circledast})}{2} \dketbra{\varphi_{00}^{\circledast}} \otimes \dketbra{00}_{23} \otimes \dketbra{00}_{45} \nonumber \\
  & \qquad \qquad - \frac{1}{2} M_+ \dketbra{+}_1 VU \Lambda_2^{(0)} U^{\dagger} V^{\dagger} \dketbra{+}_1 M_+^{\dagger} \biggr], \\
\Lambda_2^{(0)} \coloneqq \dketbra{\theta}_1 & \, \otimes \text{Sp}_{34}\, \left[ \text{CX}_{23} \dketbra{-\theta,-\theta}_{23} \text{CX}_{23} \otimes  \sum_{k = 0,1} p_k \dketbra{\theta_k^{\boxast}, k}_{45} \right] \text{Sp}_{34} \nonumber \\
    + \dketbra{-\theta}_1 & \, \otimes \text{Sp}_{34}\, \left[ \text{CX}_{23} \dketbra{-\theta,\theta}_{23} \text{CX}_{23} \otimes  \sum_{k = 0,1} p_k \dketbra{-\theta_k^{\boxast}, k}_{45} \right] \text{Sp}_{34}.
\end{align}
This is the state at stage (f) in Fig.~\ref{fig:BPQM_full_circuit}.
Hence, for $x_2 = 0$, the density matrix we have when $\hat{x}_1 = 0$ and we reverse the BPQM operations on $\Phi_{2,m_1 = +}^{(0)}$ is
\begin{align}
\tilde{\rho}_{2,m_1 = +}^{(0)} & = \frac{1}{0.5} \biggr[ \dketbra{\theta}_1 \otimes [W \boxast W](0)_{23} \otimes [W \boxast W](0)_{45} \nonumber \\
  & \qquad \qquad - \frac{1}{2} \text{CX}_{23}\, \text{CX}_{45}\, \text{Swap}_{34} U^{\dagger} V^{\dagger} M_+ \dketbra{+}_1 V U \Lambda_2^{(0)} \nonumber \\
  & \hspace{5cm} U^{\dagger} V^{\dagger} \dketbra{+}_1 M_+^{\dagger} V U \text{Swap}_{34}\, \text{CX}_{45}\, \text{CX}_{23} \biggr].
\end{align}
This is the state at stage (g) in Fig.~\ref{fig:BPQM_full_circuit}.
So this is the actual density matrix that BPQM encounters for $x_2 = 0$ after having estimated $\hat{x}_1 = 0$.
When compared with the earlier analysis, we observe numerically that this is close to $\tilde{\Phi}_{x_2 = \hat{x}_1}(\hat{x}_1)$ but is not exactly the same.
For example, when $\theta = 0.1\pi$, $\norm{\tilde{\rho}_{2,m_1 = +}^{(0)} - \tilde{\Phi}_{x_2 = 0}(0)}_{\text{Fro}} = 0.0542$, where ``Fro'' denotes the Frobenius norm, and only two of the distinct entries differ (slightly).
Similarly, 
\begin{align}
\tilde{\rho}_{2,m_1 = +}^{(1)} & = \frac{1}{0.5} \biggr[ \dketbra{\theta}_1 \otimes [W \boxast W](0)_{23} \otimes [W \boxast W](0)_{45} \nonumber \\
  & \qquad \qquad - \frac{1}{2} \text{CX}_{23}\, \text{CX}_{45}\, \text{Swap}_{34} U^{\dagger} V^{\dagger} M_+ \dketbra{+}_1 V U \Lambda_2^{(1)} \nonumber \\
  & \hspace{5cm} U^{\dagger} V^{\dagger} \dketbra{+}_1 M_+^{\dagger} V U \text{Swap}_{34}\, \text{CX}_{45}\, \text{CX}_{23} \biggr], \\
\Lambda_2^{(1)} \coloneqq \dketbra{\theta}_1 & \, \otimes \text{Sp}_{34}\, \left[ \text{CX}_{23} \dketbra{\theta,\theta}_{23} \text{CX}_{23} \otimes  \sum_{k = 0,1} p_k \dketbra{\theta_k^{\boxast}, k}_{45} \right] \text{Sp}_{34} \nonumber \\
  + \dketbra{-\theta}_1 & \, \otimes \text{Sp}_{34}\, \left[ \text{CX}_{23} \dketbra{\theta,-\theta}_{23} \text{CX}_{23} \otimes  \sum_{k = 0,1} p_k \dketbra{-\theta_k^{\boxast}, k}_{45} \right] \text{Sp}_{34}.
\end{align}
Therefore, the Helstrom measurement that optimally distinguishes between $\tilde{\rho}_{2,m_1 = +}^{(0)}$ and $\tilde{\rho}_{2,m_1 = +}^{(1)}$ only depends on
\begin{align}
\tilde{\rho}_{2,m_1 = +}^{(0)} - \tilde{\rho}_{2,m_1 = +}^{(1)} & = A \left[ \Lambda_2^{(1)} - \Lambda_2^{(0)} \right] A^{\dagger}, \ \ 
A \coloneqq \text{CX}_{23}\, \text{CX}_{45}\, \text{Swap}_{34} U^{\dagger} V^{\dagger} M_+ \dketbra{+}_1 V U.
\end{align}
By symmetry of $m_1 = +$ and $m_1 = -$, the optimal success probability to decide bit $2$ is
\begin{align}
P_{\text{succ},2}^{\text{Hel}} & = \frac{1}{2} + \frac{1}{4} \norm{\tilde{\rho}_{2,m_1 = +}^{(0)} - \tilde{\rho}_{2,m_1 = +}^{(1)}}_1 \\
  & = \frac{1}{2} + \frac{1}{4} \norm{A \left[ \Lambda_2^{(1)} - \Lambda_2^{(0)} \right] A^{\dagger}}_1 \\
  &= \frac{1}{2} + \frac{1}{4} \norm{L \left( \rho_{2}^{(0)} - \rho_{2}^{(1)} \right) L^{\dagger}}_1, \\
L & \coloneqq \text{CX}_{23}\, \text{CX}_{45}\, \text{Swap}_{34} U^{\dagger} V^{\dagger} M_+ \frac{\dketbra{+}_1}{\sqrt{0.5}} V U \text{Swap}_{34}\, \text{CX}_{45}\, \text{CX}_{23}.
\end{align}

Since $L$ is not unitary, we cannot directly apply the unitary invariance of the trace norm to conclude that there is no degradation in performance when compared to optimally distinguishing $\rho_{2}^{(0)}$ and $\rho_{2}^{(1)}$ at the channel output.
However, we observe numerically (even up to $12$ significant digits) that the operations in $L$ indeed ensure that $\norm{L \left( \rho_{2}^{(0)} - \rho_{2}^{(1)} \right) L^{\dagger}}_1 = \norm{\rho_{2}^{(0)} - \rho_{2}^{(1)}}_1$.
Moreover, we also observe that the BPQM operations for bit $2$ in Fig.~\ref{fig:BPQM_circuit_post_x1} achieve the same success probability, i.e., using $\pm \equiv (-1)^{x_2}$,
\begin{align}
P_{\text{succ}, 2}^{\text{BPQM}} & = \text{Tr}\left[ U_{\circledast}(m_1 \theta,\theta) \tilde{\rho}_{2,m_1}^{(x_2)} U_{\circledast}(m_1 \theta,\theta)^{\dagger} \cdot \dketbra{\pm}_2 \right] \\
  & = \frac{1}{2} + \frac{1}{4} \norm{L \left( \rho_{2}^{(0)} - \rho_{2}^{(1)} \right) L^{\dagger}}_1 \\
  & = \frac{1}{2} + \frac{1}{4} \norm{\rho_{2}^{(0)} - \rho_{2}^{(1)}}_1 \\
  & = P_{\text{succ},2}^{\text{Hel}}.
\end{align}
Finally, the simulation results in Fig.~\ref{fig:all_bits} clearly show that the overall block error rate of BPQM is almost indistinguishable from that of the quantum optimal joint Helstrom limit.
While we have verified the above equalities using the involved density matrices, it remains open to rigorously prove all of these observations.



\subsection{Overall Performance of BPQM}
\label{sec:BPQM_perf_all_bits}

\begin{figure}
\centering

\includegraphics[scale=0.7,keepaspectratio]{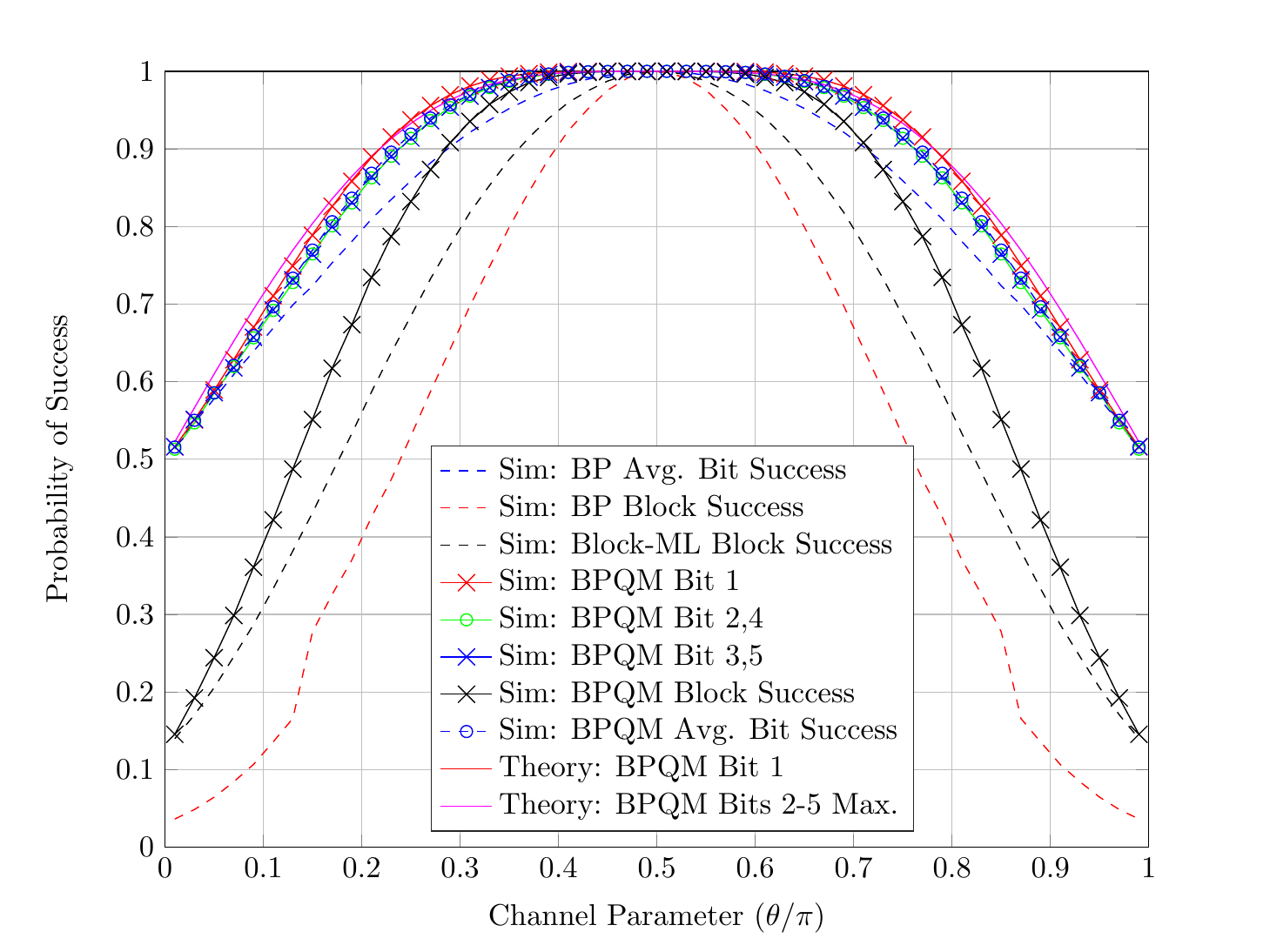}






\hspace{0.05cm}
%
\includegraphics[scale=0.7,keepaspectratio]{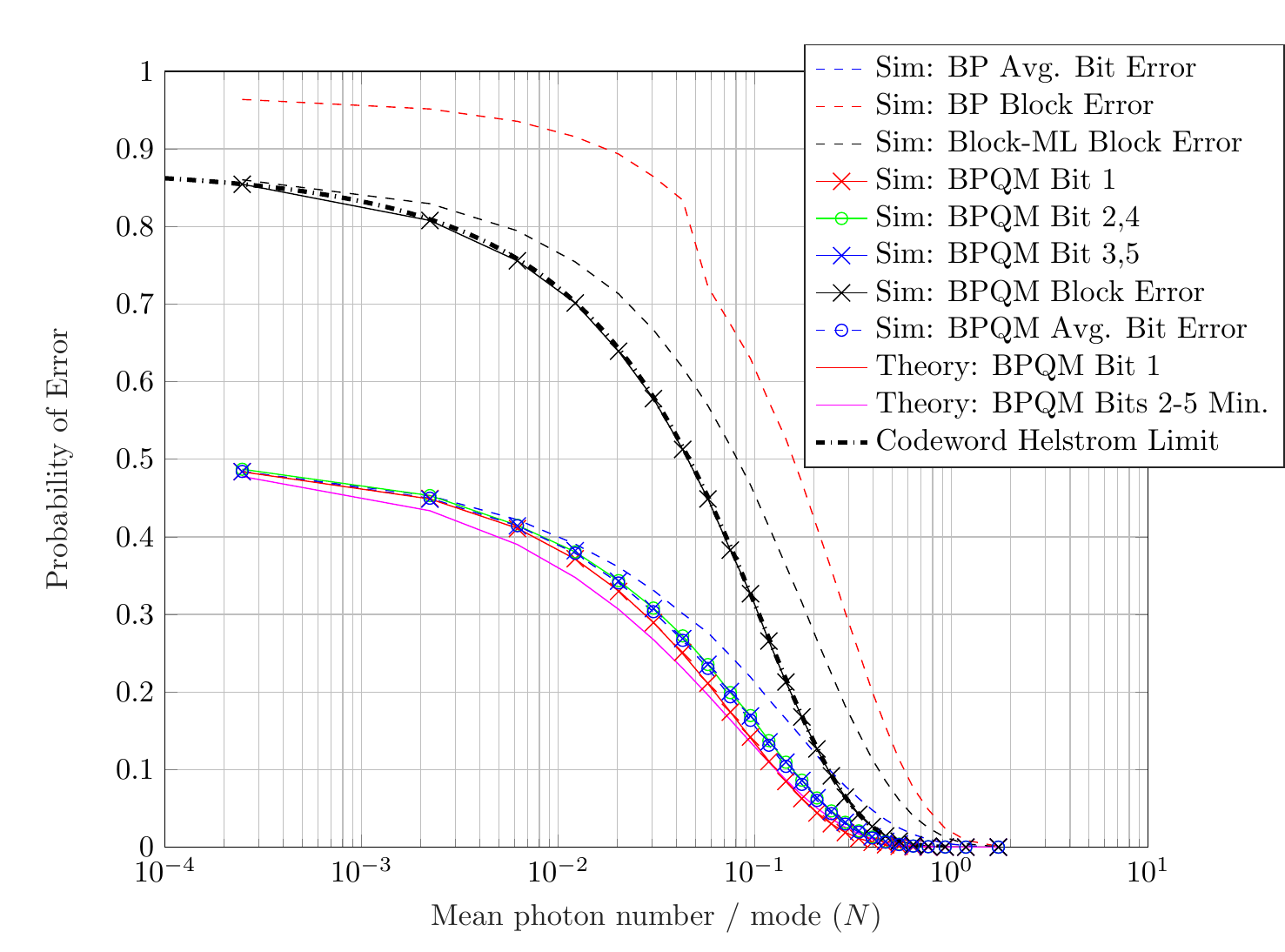}


\caption[The BPQM success probabilities for decoding each bit and its overall performance for the $5$-bit code.]{\label{fig:all_bits} (top) The BPQM success probabilities for decoding each bit and its overall performance for the $5$-bit code, the theoretical BPQM/Helstrom success rate for bit $1$, the initial theoretical prediction of $0.5(1 + \sin\theta^{\circledast})$ for BPQM for bits $2$-$5$, and the performance of BP and block-ML when applied to the directly measured channel outputs. (bottom) The same curves along with the joint Helstrom limit plotted against the mean photon number per mode $N$ ($\cos\theta = e^{-2N}$). Each simulation data point was obtained by averaging over $10^5$ uniformly random codeword transmissions.}

\end{figure}

We can calculate the probability that the full codeword $\vecnot{x}$ is decoded correctly as
\begin{align}
\mathbb{P}[\vecnot{\hat{x}} = \vecnot{x}] & = \mathbb{P}[\hat{x}_1 = x_1] \cdot \mathbb{P}[\hat{x}_2 = x_2 \, | \, \hat{x}_1 = x_1] \cdot \mathbb{P}[\hat{x}_4 = x_4 \, | \, \hat{x}_1 = x_1] \\
   & = P_{\text{succ},1}^{\text{Hel}} \left( \frac{1}{2} + \frac{1}{4 P_{\text{succ},1}^{\text{Hel}}} \norm{A \left[ \tilde{\Lambda}_2^{(1)} - \tilde{\Lambda}_2^{(0)} \right] A^{\dagger}}_1 \right)^2,
\end{align}
where $\tilde{\Lambda}_2^{(0)}$ and $\tilde{\Lambda}_2^{(1)}$ only contain the first (resp. second) term of $\Lambda_2^{(0)}$ and $\Lambda_2^{(1)}$ respectively, if $x_1 = 0$ (resp. $x_1 = 1$), since we have assumed that $\hat{x}_1 = x_1$.
The BPQM success probabilities for all bits as well as the empirical estimation of this overall block success probability are plotted in Fig.~\ref{fig:all_bits} (top). 
For convenience these performance curves are also plotted in a log-log scale along with the codeword Helstrom limit (bottom).
The plots show that even though BPQM is optimal for bits $x_2$ through $x_5$, it still performs slightly poorly when compared to the performance for $x_1$.
This might be attributed to the fact that in the chosen parity-check matrix, bit $x_1$ is involved in both checks whereas the other bits are involved in exactly one of the two checks.

\begin{remark}
\normalfont
The above analyses demonstrate that even though the measurement for each bit is irreversible, BPQM still decides each bit optimally in this $5$-bit example code.
In particular, it remains to be seen if the order in which the bits are decoded affects the overall performance.
This needs to be studied further and we need to analyze if BPQM always achieves the codeword Helstrom limit for all codes with tree factor graphs.
We emphasize that, while in classical BP there is no question of ordering and one makes hard decisions on all the bits simultaneously after several BP iterations, it appears that quantum BP always has a sequential nature due to the unitarity of operations and the no-cloning theorem.
Due to these facts, we expect that extending classical ideas for analyzing BP, such as density evolution~\cite{RU-2008}, will be challenging in the quantum setting.


In connection to this, Renes has recently developed a precise notion of duality between channels, and shown that classical channels need to be embedded in CQ channels in order to define their duals~\cite{Renes-arxiv17}.
An interesting fact that follows from this framework is that the dual of the pure-state channel is the classical binary symmetric channel (BSC).
Since we know that density evolution is a well-defined analysis technique for BP on BSCs, albeit sophisticated, it will be interesting to see if duality allows one to borrow from this literature and analyze BPQM on pure-state channels.
\end{remark}

Finally, as briefly discussed in the introduction, we compare the error rates of the following strategies:
\begin{enumerate}
    \item[(a)] Collective (optimal) Helstrom measurement on all channel outputs corresponding to the transmitted codeword.
    
    \item[(b)] BPQM on all channel outputs corresponding to the transmitted codeword.
    
    \item[(c)] Symbol-by-symbol (optimal) Helstrom measurement followed by classical (optimal) block-MAP decoding.
    
    \item[(d)] Symbol-by-symbol (optimal) Helstrom measurement followed by classical BP.
\end{enumerate}
For strategy (a), we calculated the performance of the collective Helstrom measurement using the Yuen-Kennedy-Lax (YKL) conditions~\cite{Yuen-it75} as discussed, for example, in~\cite{Krovi-pra15}.
Note that for the last two schemes, classical processing is performed essentially on the BSC induced by measuring each qubit output by the pure-state channel.
We had plotted the block error rates earlier in Fig.~\ref{fig:bpqm_perr_vs_nbar}, and we plot bit and block error rates in Fig.~\ref{fig:all_bits}.
The mean photon number per mode, $N$, relates to the pure-state channel parameter $\theta$ as $\cos\theta = e^{-2N}$ (e.g., see~\cite{Guha-isit12} for more information on this quantity).
Our implementation of the BPQM algorithm for the $5$-bit code is available at \url{https://github.com/nrenga/bpqm}.

We make the following observations from these performance curves.
\begin{enumerate}
    \item The block error rates are in increasing order from strategy (a) to (d), as we might expect. Even though classical BP is performed on a tree FG here, it only implements bit-MAP decoding and not block-MAP decoding. This is why it performs worse than block-ML (i.e., block-MAP with uniform prior on codewords) in this case.
    
    \item BPQM performs strictly better than symbol-by-symbol optimal detection followed by classical MAP decoding. This gives a clear demonstration that if one physically constructs a receiver for BPQM, then it will be the best known physically realizable receiver for the pure-state channel. For example, the \emph{Dolinar receiver}~\cite{Guha-isit12,Dolinar-1973} realizes only strategy (c). One can use our circuits to make such a physical realization.
    
    \item \emph{BPQM performs as well as the quantum optimal collective Helstrom measurement on the outputs of the channel.} This lends evidence to the conclusion that by passing \emph{quantum} messages, BPQM is able to behave like a collective measurement while still making only single-qubit Pauli measurements during the process. However, more careful analysis is required to characterize this in general for, say, the family of codes with tree FGs.
    
    
    \item As a first self-consistency check, observe that the block-ML curve asymptotes at roughly $0.875$ for low mean photon numbers per mode. This is because, in this regime, the BSC induced by the symbol-by-symbol measurement essentially has a bit-flip rate of $0.5$. Therefore, block-ML computes a posterior that is almost uniform on all codewords, and thus the block success probability is $1/|\MCC| = 1/8 = 0.125$.
    
    \item As another self-consistency check, note that the BP curve asymptotes at roughly $(1 - 1/32) = 0.9688$ for low mean photon numbers per mode. Since BP performs bit-MAP on this FG, and the induced BSC in this regime flips bits at a rate of almost $0.5$, BP essentially picks each bit uniformly at random, thereby returning a vector that is uniformly at random out of all the possible $2^5 = 32$ vectors of length $5$.
\end{enumerate}


\section{Conclusion}
\label{sec:conclusion}

This chapter began by reviewing classical belief-propagation, which is an algorithm to efficiently perform statistical inference by computing posterior marginal distributions of the involved variables.
Then, it discusses an induced channel perspective of understanding the action of classical BP, and generalizes it to the quantum case following Renes~\cite{Renes-njp17}.
Next, we use an example $5$-bit code to understand the recently introduced belief-propagation algorithm with quantum messages (BPQM) algorithm.
Since the action of BPQM after decoding the first bit is somewhat ambiguous in the original BPQM paper~\cite{Renes-njp17}, we introduce a perspective that provides a nice interpretation of BPQM.
We also provide a detailed analysis of the density matrices involved in BPQM, and show that BPQM is optimal for decoding all bits despite performing irreversible (single-qubit) measurements along the way.
Finally, we calculate the BPQM success probabilities for all bits and verify empirically that it performs as well as Helstrom's optimal joint measurement strategy for quantum systems.

This is the first practically feasible decoder that can be used to decode the classical-quantum polar codes on the pure state channel. 
Although CQ polar codes are known to achieve capacity on CQ channels when paired with a quantum successive cancellation decoder~\cite{Guha-isit12,Wilde-it13*2}, it is not clear whether BPQM retains this property.
This is because the quantum optimal nature of BPQM, at least for this $5$-bit code, does not immediately imply that it must also be a capacity-achieving decoder for CQ polar codes.
Beyond this question, it remains open as to how BPQM can be generalized to FGs with cycles and also for decoding over general CQ channels.
We will investigate these problems in future work. 

BPQM also has close connections with the recently introduced notion of \emph{channel} duality~\cite{Renes-arxiv17}. 
The resulting entropic relations could help characterize the performance of a code over a channel using the performance of its dual code over the dual channel.
Since the dual of the pure-state channel is the classical BSC, we believe it may be possible to extend classical techniques for analyzing BP (on BSC), such as density evolution, to analyze BPQM as well.

\section{Circuit Decomposition Calculations}
\label{sec:vn_unitary_decompose}

Let $\cnot{\theta'}{\theta} = I_2 \otimes \dketbra{0} + X \otimes \dketbra{1}$ be the controlled-NOT operation with the (second) qubit corresponding to angle $\theta'$ as the control qubit.
Then we observe that
\begin{align}
\tilde{U}_{\circledast}(\theta,\theta') \coloneqq U_{\circledast}(\theta,\theta')\, \cnot{\theta'}{\theta} & =
\begin{bmatrix}
a_+ & a_{-} & 0 & 0 \\
a_{-} & - a_+ & 0 & 0 \\
0 & 0 & b_{-} & b_+ \\
0 & 0 & - b_+ & b_{-}
\end{bmatrix}, \\
  & = \dketbra{0} \otimes 
\begin{bmatrix}
a_+ & a_{-} \\
a_{-} & -a_+
\end{bmatrix} + \dketbra{1} \otimes 
\begin{bmatrix}
b_{-} & b_+ \\
-b_+ & b_{-}
\end{bmatrix} \\
  & \eqqcolon \dketbra{0} \otimes U_1 + \dketbra{1} \otimes U_2 \\
  & = \left( \dketbra{0} \otimes U_1 + \dketbra{1} \otimes I_2 \right) \left( \dketbra{0} \otimes I_2 + \dketbra{1} \otimes U_2 \right).
\end{align}
Let $R_p(\theta) \coloneqq \exp\left( -\imath \frac{\theta}{2} p \right)$ denote Pauli rotations, where $p \in \{x,y,z\}$ and $\imath \coloneqq \sqrt{-1}$.
Then the $Z$-$Y$ decomposition for a single qubit~\cite[Theorem 4.1]{Nielsen-2010} implies that any unitary $U$ can be decomposed as
\begin{align}
U = e^{\imath \alpha} R_z(\beta) R_y(\gamma) R_z(\delta) = 
\begin{bmatrix}
e^{\imath (\alpha - \beta/2 - \delta/2)} \cos\frac{\gamma}{2} & -e^{\imath (\alpha - \beta/2 + \delta/2)} \sin\frac{\gamma}{2} \\
e^{\imath (\alpha + \beta/2 - \delta/2)} \sin\frac{\gamma}{2} & e^{\imath (\alpha + \beta/2 + \delta/2)} \cos\frac{\gamma}{2}
\end{bmatrix}.
\end{align}
Setting $\gamma_1 \coloneqq 2 \sin^{-1}(a_{-})$ and $\gamma_2 \coloneqq 2 \sin^{-1}(b_+)$, we observe that~\cite[Corollary 4.2]{Nielsen-2010}
\begin{align}
U_1 & = 
\begin{bmatrix}
\cos\frac{\gamma_1}{2} & \sin\frac{\gamma_1}{2} \\
\sin\frac{\gamma_1}{2} & -\cos\frac{\gamma_1}{2}
\end{bmatrix}
 = e^{\frac{\imath\pi}{2}} R_y(\gamma_1) R_z(\pi) \eqqcolon e^{\frac{\imath\pi}{2}} A_1 X B_1 X C_1, \\
A_1 & \coloneqq R_y\left( \frac{\gamma_1}{2} \right), \\ 
B_1 & \coloneqq R_y\left( \frac{-\gamma_1}{2} \right) R_z\left( \frac{-\pi}{2} \right), \\  
C_1 & \coloneqq R_z\left( \frac{\pi}{2} \right), \\
U_2 & =  
\begin{bmatrix}
\cos\frac{\gamma_2}{2} & \sin\frac{\gamma_2}{2} \\
-\sin\frac{\gamma_2}{2} & \cos\frac{\gamma_2}{2}
\end{bmatrix}
 = e^{\imath\pi} R_z(\pi) R_y(\gamma_2) R_z(\pi) \eqqcolon e^{\imath\pi} A_2 X B_2 X, \\
A_2 & \coloneqq R_z(\pi) R_y(\gamma_2/2), \\ 
B_2 & \coloneqq R_y(-\gamma_2/2) R_z(-\pi).
\end{align}
Then we can express the full circuit decomposition for $U_{\circledast}(\theta,\theta')$ as shown in Fig.~\ref{fig:vn_unitary_circuit}.

Similarly, the rotations $K_{+}$ and $K_{-}$ defined in~\eqref{eq:cond_rotation} can be expressed as
\begin{align}
K_{+} & = \frac{1}{\sqrt{2}}
\begin{bmatrix}
\cos\frac{\varphi_{00}^{\circledast}}{2} + \sin\frac{\varphi_{00}^{\circledast}}{2} & \cos\frac{\varphi_{00}^{\circledast}}{2} - \sin\frac{\varphi_{00}^{\circledast}}{2} \\
\sin\frac{\varphi_{00}^{\circledast}}{2} - \cos\frac{\varphi_{00}^{\circledast}}{2} & \sin\frac{\varphi_{00}^{\circledast}}{2} + \cos\frac{\varphi_{00}^{\circledast}}{2}
\end{bmatrix} \\
  & \eqqcolon 
\begin{bmatrix}
\cos\frac{\gamma}{2} & \sin\frac{\gamma}{2} \\
- \sin\frac{\gamma}{2} & \cos\frac{\gamma}{2}
\end{bmatrix} \\
  & = e^{\imath\pi} R_z(\pi) R_y(\gamma) R_z(\pi) \\
  & \eqqcolon e^{\imath\pi} A_{+} X B_{+} X, \\
K_{-} & = \frac{1}{\sqrt{2}}
\begin{bmatrix}
\sin\frac{\varphi_{00}^{\circledast}}{2} + \cos\frac{\varphi_{00}^{\circledast}}{2} & \sin\frac{\varphi_{00}^{\circledast}}{2} - \cos\frac{\varphi_{00}^{\circledast}}{2} \\
\cos\frac{\varphi_{00}^{\circledast}}{2} - \sin\frac{\varphi_{00}^{\circledast}}{2} & \cos\frac{\varphi_{00}^{\circledast}}{2} + \sin\frac{\varphi_{00}^{\circledast}}{2}
\end{bmatrix} \\
  & = K_{+}^{\dagger}, \\
\text{where} \ \ \gamma & \coloneqq 2 \sin^{-1}\left[ \frac{1}{\sqrt{2}} \left( \cos\frac{\varphi_{00}^{\circledast}}{2} - \sin\frac{\varphi_{00}^{\circledast}}{2} \right) \right], \\
A_{+} & \coloneqq R_z(\pi) R_y\left( \frac{\gamma}{2} \right), \\ 
B_{+} & \coloneqq R_y\left( \frac{-\gamma}{2} \right) R_z(-\pi).
\end{align}
The coherently controlled gate $M_{m_1}$ defined in~\eqref{eq:cond_rotation} is decomposed in Fig.~\ref{fig:controlled_unitary} using the above calculations.

\section{Calculations for BPQM Node Operations}
\label{sec:node_convolutions}

The variable and check node convolutions are given by
\begin{align}
\label{eq:vn_conv}
[W \circledast W'](x) &= W(x) \otimes W'(x) , \\
\label{eq:cn_conv}
[W \boxast W'](x) &= \frac{1}{2} ( W(x) \otimes W'(0) + W(x+1) \otimes W'(1) ).
\end{align}
The channel outputs for a pure state channel are $\dket{\pm \theta}$, where
\begin{equation}
\dket{\pm \theta} = \cos \frac{\theta}{2} \dket{0} \pm \sin \frac{\theta}{2} \dket{1} .
\end{equation}
The overlap between the two outputs is given by $\langle -\theta \vert \theta \rangle = \cos^2 \frac{\theta}{2} - \sin^2 \frac{\theta}{2} = \cos \theta$.
The Helstrom measurement projects onto $\dket{\pm \frac{\pi}{2}}$.

The Stinespring's representation for a classical-quantum (CQ) channel $W$ that maps density states from Hilbert space $\mathcal{H}_A$ to $\mathcal{H}_B$ is given by~\cite{Wilde-2013}
\begin{align}
W(x) \coloneqq {\rm Tr}_E \left[ V_{BE \mid A} \dketbra{x} V_{BE \mid A}^{\dagger} \right],
\end{align}
where $\mathcal{H}_E$ is an ancilla space and $V_{BE \mid A}$ is an isometry that maps $\mathcal{H}_A$ to $\mathcal{H}_B \otimes \mathcal{H}_E$.
Therefore, even for a pure state channel $W$, we must take its outputs to be density states $\dketbra{\pm \theta}$ and not $\dket{\pm \theta}$.
On the contrary, if we take it to be $\dket{\pm \theta}$ then we immediately notice that the output state of the check convolution in~\eqref{eq:cn_conv} above is not normalized and hence does not represent a physical operation.

\subsection{Variable Node Operation}

The convolution $W \circledast W'$ outputs (the density state for) either $\dket{\theta^{\circledast}} \coloneqq \dket{\theta} \otimes \dket{\theta'}$ or $\dket{-\theta^{\circledast}} \coloneqq \dket{-\theta} \otimes \dket{-\theta'}$, which are again two pure states with an overlap angle $\theta^{\circledast}$ given by 
\[ \cos \theta^{\circledast} \coloneqq \langle -\theta^{\circledast} \vert \theta^{\circledast} \rangle = (\dbra{-\theta} \otimes \dbra{-\theta'}) \cdot (\dket{\theta} \otimes \dket{\theta'}) = \langle -\theta \vert \theta \rangle \otimes \langle -\theta' \vert \theta' \rangle = \cos \theta \cos \theta' . \]
The following unitary transformation compresses the states to the first qubit, leaving the second in the state $\dket{0}$:
\begin{equation}
U_{\circledast}(\theta,\theta') \coloneqq 
\begin{bmatrix}
a_{+} & 0 & 0 & a_{-} \\
a_{-} & 0 & 0 & -a_{+} \\
0 & b_{+} & b_{-} & 0 \\
0 & b_{-} & -b_{+} & 0
\end{bmatrix}
,
\end{equation}
where 
\[ a_{\pm} = \frac{1}{\sqrt{2}} \frac{\cos \left( \frac{\theta-\theta'}{2} \right) \pm \cos \left( \frac{\theta+\theta'}{2} \right)}{\sqrt{1 + \cos \theta \cos \theta'}} , \quad b_{\pm} = \frac{1}{\sqrt{2}} \frac{\sin \left( \frac{\theta+\theta'}{2} \right) \mp \sin \left( \frac{\theta-\theta'}{2} \right)}{\sqrt{1 - \cos \theta \cos \theta'}} . \]
Let us verify by explicit calculation that 
\[ U_{\circledast}\left( \theta,\theta' \right) \left( \dketbra{\pm \theta} \otimes \dketbra{\pm \theta'} \right) U_{\circledast}^{\dagger}\left( \theta,\theta' \right) = \dketbra{\pm \theta^{\circledast}} \otimes \dketbra{0} , \] 
where $\dket{\pm \theta^{\circledast}} \coloneqq \sqrt{p_0} \dket{0} \pm \sqrt{p_1} \dket{1}$, $p_0 \coloneqq \frac{1}{2} (1+\cos\theta \cos\theta'), p_1 \coloneqq 1-p_0 = \frac{1}{2} (1-\cos\theta \cos\theta')$.
\begin{align}
\dket{\pm \theta^{\circledast}} \coloneqq 
\begin{bmatrix}
\sqrt{\frac{1 + \cos \theta \cos \theta'}{2}} \\
\pm \sqrt{\frac{1 - \cos \theta \cos \theta'}{2}}
\end{bmatrix}
=
\begin{bmatrix}
\sqrt{\frac{1 + \cos \theta^{\circledast}}{2}} \\
\pm \sqrt{\frac{1 - \cos \theta^{\circledast}}{2}}
\end{bmatrix}
=
\begin{bmatrix}
\cos \frac{\theta^{\circledast}}{2} \\
\pm \sin \frac{\theta^{\circledast}}{2}
\end{bmatrix}
=
\cos \frac{\theta^{\circledast}}{2} \dket{0} \pm \sin \frac{\theta^{\circledast}}{2} \dket{1}.
\end{align}

First, using the definitions for $\dket{\pm \theta}$ and $\dket{\pm \theta'}$ we have
\begin{align}
\dket{\pm \theta} \otimes \dket{\pm \theta'} &= \cos \frac{\theta}{2} \cos \frac{\theta'}{2} \dket{00} \pm \cos \frac{\theta}{2} \sin \frac{\theta'}{2} \dket{01} \pm \sin \frac{\theta}{2} \cos \frac{\theta'}{2} \dket{10} + \sin \frac{\theta}{2} \sin \frac{\theta'}{2} \dket{11} \\
 & = \left[ \cos \frac{\theta}{2} \cos \frac{\theta'}{2},\ \pm \cos \frac{\theta}{2} \sin \frac{\theta'}{2}, \ \pm \sin \frac{\theta}{2} \cos \frac{\theta'}{2}, \ \sin \frac{\theta}{2} \sin \frac{\theta'}{2} \right]^{\dagger} .
\end{align}
Hence we get
\begin{align}
\dket{\pm \psi} \coloneqq U_{\circledast}\left( \theta,\theta' \right) \left( \dket{\pm \theta} \otimes \dket{\pm \theta'} \right) & = 
\begin{bmatrix}[1.5]
a_{+} \cos \frac{\theta}{2} \cos \frac{\theta'}{2} + a_{-} \sin \frac{\theta}{2} \sin \frac{\theta'}{2} \\
a_{-} \cos \frac{\theta}{2} \cos \frac{\theta'}{2} - a_{+} \sin \frac{\theta}{2} \sin \frac{\theta'}{2} \\
\pm b_{+} \cos \frac{\theta}{2} \sin \frac{\theta'}{2} \pm b_{-} \sin \frac{\theta}{2} \cos \frac{\theta'}{2} \\
\pm b_{-} \cos \frac{\theta}{2} \sin \frac{\theta'}{2} \mp b_{+} \sin \frac{\theta}{2} \cos \frac{\theta'}{2}
\end{bmatrix}
\eqqcolon
\begin{bmatrix}[1.5]
\psi_{00} \\
\psi_{01} \\
\psi_{10} \\
\psi_{11}
\end{bmatrix}
.
\end{align}
For convenience let us make some definitions:
\begin{align}
\alpha \coloneqq \cos\frac{\theta-\theta'}{2} + \cos\frac{\theta+\theta'}{2} = \frac{1}{2} \cos \frac{\theta}{2} \cos \frac{\theta'}{2}, & \quad \beta \coloneqq \cos\frac{\theta-\theta'}{2} - \cos\frac{\theta+\theta'}{2} = \frac{1}{2} \sin \frac{\theta}{2} \sin \frac{\theta'}{2}, \nonumber \\
\gamma \coloneqq \sin\frac{\theta+\theta'}{2} - \sin\frac{\theta-\theta'}{2} = \frac{1}{2} \cos \frac{\theta}{2} \sin \frac{\theta'}{2}, & \quad \delta \coloneqq \sin\frac{\theta+\theta'}{2} + \sin\frac{\theta-\theta'}{2} = \frac{1}{2} \sin \frac{\theta}{2} \cos \frac{\theta'}{2}.
\end{align}
Then using the identities $\cos\theta = 2\cos^2 \frac{\theta}{2} - 1 = 1 - 2\sin^2 \frac{\theta}{2}$ we see that
\begin{align}
\psi_{00} & = \frac{1}{2} \left[ a_{+} \alpha + a_{-} \beta \right] = \frac{1}{2\sqrt{2}} \frac{2 \left[ \cos^2 \frac{\theta-\theta'}{2} + \cos^2 \frac{\theta+\theta'}{2} \right]}{\sqrt{1+\cos\theta \cos\theta'}} = \sqrt{\frac{1+\cos\theta \cos\theta'}{2}} , \\
\psi_{01} & = \frac{1}{2} \left[ a_{-} \alpha - a_{+} \beta \right] = \frac{1}{2\sqrt{2}} \frac{\alpha \beta - \beta \alpha}{\sqrt{1+\cos\theta \cos\theta'}} = 0, \\
\psi_{10} & = \frac{1}{2} \left[ \pm b_{+} \gamma \pm b_{-} \delta \right] = \frac{1}{2\sqrt{2}} \frac{2 \left[ \pm \sin^2 \frac{\theta+\theta'}{2} \pm \sin^2 \frac{\theta-\theta'}{2} \right]}{\sqrt{1-\cos\theta \cos\theta'}} = \pm \sqrt{\frac{1-\cos\theta \cos\theta'}{2}} , \\
\psi_{11} & = \frac{1}{2} \left[ \pm b_{-} \gamma \mp b_{+} \delta \right] = \frac{1}{2\sqrt{2}} \frac{\pm \gamma \delta \mp \delta \gamma}{\sqrt{1-\cos\theta \cos\theta'}} = 0 .
\end{align}
Therefore we find that $\dket{\pm \psi} = \dket{\pm \theta^{\circledast}} \otimes \dket{0}$ and hence
\[ U_{\circledast}\left( \theta,\theta' \right) \left( \dketbra{\pm \theta} \otimes \dketbra{\pm \theta'} \right) U_{\circledast}^{\dagger}\left( \theta,\theta' \right) = \dketbra{\pm \psi} = \dketbra{\pm \theta^{\circledast}} \otimes \dketbra{0} . \]

\subsection{Factor Node Operation}

The result of $W \boxast W$ is not pure and hence we would like to unitarily transform the output of the $\boxast$ convolution to a CQ state of the form
\begin{equation}
\label{eq:psi1}
\Psi_{\rm desired} \coloneqq \sum_{j \in \{0,1\}} p_j \dketbra{\pm \theta_{j}^{\boxast}} \otimes \dketbra{j} ,
\end{equation}
for some appropriate state $\dket{\pm \theta_{j}^{\boxast}}$ and probabilities $p_j$.
It turns out that the unitary operation $U_{\boxast} \coloneqq {\rm CNOT}_{1 \rightarrow 2}$ is the correct one.
Let us verify that by explicit calculation for inputs $x=0$ and $x=1$ simultaneously.
Using $\pm \equiv (-1)^x, \ \mp \equiv (-1)^{x \oplus 1}$ we have
\begin{align}
[W \boxast W'](x) &= \frac{1}{2} \left( W(x) \otimes W'(0) + W(x \oplus 1) \otimes W'(1) \right) \\
  & = \frac{1}{2} \biggr( \dketbra{\pm \theta} \otimes \dketbra{\theta'} + \dketbra{\mp \theta} \otimes \dketbra{-\theta'} \biggr) \\
  & \coloneqq \frac{1}{2} \left( \tilde{\varphi}_1 + \tilde{\varphi}_2 \right) . \\
\Rightarrow \Psi & \coloneqq U_{\boxast} \biggr( [W \boxast W'](x) \biggr) U_{\boxast}^{\dagger} \\
  & = \frac{1}{2} \left( U_{\boxast} \tilde{\varphi}_1 U_{\boxast}^{\dagger} + U_{\boxast} \tilde{\varphi}_2 U_{\boxast}^{\dagger} \right) \\
  & \coloneqq \frac{1}{2} \left( \varphi_1 + \varphi_2 \right) .
\end{align}
Now we calculate $\varphi_1$ and $\varphi_2$ separately.
We first have
\begin{align}
\tilde{\varphi}_1 & = \dketbra{\pm \theta} \otimes \dketbra{\theta'} \\
  & = \left( \dket{\pm \theta} \otimes \dket{\theta'} \right) \cdot \left( \dbra{\pm \theta} \otimes \dbra{\theta'} \right) \\
  & = \left[ \cos \frac{\theta}{2} \cos \frac{\theta'}{2} \dket{0}\dket{0} + \cos \frac{\theta}{2} \sin \frac{\theta'}{2} \dket{0}\dket{1} \pm \sin \frac{\theta}{2} \cos \frac{\theta'}{2} \dket{1}\dket{0} \pm \sin \frac{\theta}{2} \sin \frac{\theta'}{2} \dket{1}\dket{1} \right] \\
  & \quad \otimes \left[ \cos \frac{\theta}{2} \cos \frac{\theta'}{2} \dbra{0}\dbra{0} + \cos \frac{\theta}{2} \sin \frac{\theta'}{2} \dbra{0}\dbra{1} \pm \sin \frac{\theta}{2} \cos \frac{\theta'}{2} \dbra{1}\dbra{0} \pm \sin \frac{\theta}{2} \sin \frac{\theta'}{2} \dbra{1}\dbra{1} \right] \\
\Rightarrow \varphi_1 & = {\rm CNOT}_{1 \rightarrow 2} \left( \tilde{\varphi}_1 \right) {\rm CNOT}_{1 \rightarrow 2} \\
  & = \left[ \cos \frac{\theta}{2} \cos \frac{\theta'}{2} \dket{0}\dket{0} + \cos \frac{\theta}{2} \sin \frac{\theta'}{2} \dket{0}\dket{1} \pm \sin \frac{\theta}{2} \sin \frac{\theta'}{2} \dket{1}\dket{0} \pm \sin \frac{\theta}{2} \cos \frac{\theta'}{2} \dket{1}\dket{1} \right] \\
  & \quad \otimes \left[ \cos \frac{\theta}{2} \cos \frac{\theta'}{2} \dbra{0}\dbra{0} + \cos \frac{\theta}{2} \sin \frac{\theta'}{2} \dbra{0}\dbra{1} \pm \sin \frac{\theta}{2} \sin \frac{\theta'}{2} \dbra{1}\dbra{0} \pm \sin \frac{\theta}{2} \cos \frac{\theta'}{2} \dbra{1}\dbra{1} \right] \\
  & = \left[ \cos^2 \frac{\theta}{2} \cos^2 \frac{\theta'}{2} \dketbra{0} \pm \frac{1}{4} \sin\theta \sin\theta' (\dketbra{0}{1} + \dketbra{1}{0}) + \sin^2 \frac{\theta}{2} \sin^2 \frac{\theta'}{2} \dketbra{1} \right] \otimes \dketbra{0} \\
  & + \left[ \cos^2 \frac{\theta}{2} \sin^2 \frac{\theta'}{2} \dketbra{0} \pm \frac{1}{4} \sin\theta \sin\theta' (\dketbra{0}{1} + \dketbra{1}{0}) + \sin^2 \frac{\theta}{2} \cos^2 \frac{\theta'}{2} \dketbra{1} \right] \otimes \dketbra{1} \\
  & + \frac{1}{2} \left[ \cos^2 \frac{\theta}{2} \sin\theta' \dketbra{0} \pm \sin\theta \cos^2 \frac{\theta'}{2} \dketbra{0}{1} \pm \sin\theta \sin^2 \frac{\theta'}{2} \dketbra{1}{0} + \sin^2 \frac{\theta}{2} \sin\theta' \dketbra{1} \right] \nonumber \\
  & \hspace{9cm} \otimes \dketbra{0}{1} \\
  & + \frac{1}{2} \left[ \cos^2 \frac{\theta}{2} \sin\theta' \dketbra{0} \pm \sin\theta \sin^2 \frac{\theta'}{2} \dketbra{0}{1} \pm \sin\theta \cos^2 \frac{\theta'}{2} \dketbra{1}{0} + \sin^2 \frac{\theta}{2} \sin\theta' \dketbra{1} \right] \nonumber \\
  & \hspace{9cm} \otimes \dketbra{1}{0} .
\end{align}
Similarly we get
\begin{align}
\tilde{\varphi}_2 & = \dketbra{\mp \theta} \otimes \dketbra{-\theta'} \\
  & = \left( \dket{\mp \theta} \otimes \dket{-\theta'} \right) \cdot \left( \dbra{\mp \theta} \otimes \dbra{-\theta'} \right) \\
  & = \left[ \cos \frac{\theta}{2} \cos \frac{\theta'}{2} \dket{0}\dket{0} - \cos \frac{\theta}{2} \sin \frac{\theta'}{2} \dket{0}\dket{1} \mp \sin \frac{\theta}{2} \cos \frac{\theta'}{2} \dket{1}\dket{0} \pm \sin \frac{\theta}{2} \sin \frac{\theta'}{2} \dket{1}\dket{1} \right] \\
  & \quad \otimes \left[ \cos \frac{\theta}{2} \cos \frac{\theta'}{2} \dbra{0}\dbra{0} - \cos \frac{\theta}{2} \sin \frac{\theta'}{2} \dbra{0}\dbra{1} \mp \sin \frac{\theta}{2} \cos \frac{\theta'}{2} \dbra{1}\dbra{0} \pm \sin \frac{\theta}{2} \sin \frac{\theta'}{2} \dbra{1}\dbra{1} \right] \\
\Rightarrow \varphi_2 & = {\rm CNOT}_{1 \rightarrow 2} \left( \tilde{\varphi}_2 \right) {\rm CNOT}_{1 \rightarrow 2} \\
  & = \left[ \cos \frac{\theta}{2} \cos \frac{\theta'}{2} \dket{0}\dket{0} - \cos \frac{\theta}{2} \sin \frac{\theta'}{2} \dket{0}\dket{1} \pm \sin \frac{\theta}{2} \sin \frac{\theta'}{2} \dket{1}\dket{0} \mp \sin \frac{\theta}{2} \cos \frac{\theta'}{2} \dket{1}\dket{1} \right] \\
  & \quad \otimes \left[ \cos \frac{\theta}{2} \cos \frac{\theta'}{2} \dbra{0}\dbra{0} - \cos \frac{\theta}{2} \sin \frac{\theta'}{2} \dbra{0}\dbra{1} \pm \sin \frac{\theta}{2} \sin \frac{\theta'}{2} \dbra{1}\dbra{0} \mp \sin \frac{\theta}{2} \cos \frac{\theta'}{2} \dbra{1}\dbra{1} \right] \\
  & = \left[ \cos^2 \frac{\theta}{2} \cos^2 \frac{\theta'}{2} \dketbra{0} \pm \frac{1}{4} \sin\theta \sin\theta' (\dketbra{0}{1} + \dketbra{1}{0}) + \sin^2 \frac{\theta}{2} \sin^2 \frac{\theta'}{2} \dketbra{1} \right] \otimes \dketbra{0} \\
  & + \left[ \cos^2 \frac{\theta}{2} \sin^2 \frac{\theta'}{2} \dketbra{0} \pm \frac{1}{4} \sin\theta \sin\theta' (\dketbra{0}{1} + \dketbra{1}{0}) + \sin^2 \frac{\theta}{2} \cos^2 \frac{\theta'}{2} \dketbra{1} \right] \otimes \dketbra{1} \\
  & + \frac{1}{2} \bigg[ -\cos^2 \frac{\theta}{2} \sin\theta' \dketbra{0} \mp \sin\theta \cos^2 \frac{\theta'}{2} \dketbra{0}{1} \mp \sin\theta \sin^2 \frac{\theta'}{2} \dketbra{1}{0} \nonumber \\
  & \hspace{5.5cm} - \sin^2 \frac{\theta}{2} \sin\theta' \dketbra{1} \bigg] \otimes \dketbra{0}{1} \\
  & + \frac{1}{2} \bigg[ -\cos^2 \frac{\theta}{2} \sin\theta' \dketbra{0} \mp \sin\theta \sin^2 \frac{\theta'}{2} \dketbra{0}{1} \mp \sin\theta \cos^2 \frac{\theta'}{2} \dketbra{1}{0} \nonumber \\
  & \hspace{5.5cm} - \sin^2 \frac{\theta}{2} \sin\theta' \dketbra{1} \bigg] \otimes \dketbra{1}{0} .
\end{align}
Therefore we get
\begin{align}
\Psi & = \frac{1}{2} \left( \varphi_1 + \varphi_2 \right) \nonumber \\
  & = \left[ \cos^2 \frac{\theta}{2} \cos^2 \frac{\theta'}{2} \dketbra{0} \pm \frac{1}{4} \sin\theta \sin\theta' (\dketbra{0}{1} + \dketbra{1}{0}) + \sin^2 \frac{\theta}{2} \sin^2 \frac{\theta'}{2} \dketbra{1} \right] \otimes \dketbra{0} \nonumber \\
  & + \left[ \cos^2 \frac{\theta}{2} \sin^2 \frac{\theta'}{2} \dketbra{0} \pm \frac{1}{4} \sin\theta \sin\theta' (\dketbra{0}{1} + \dketbra{1}{0}) + \sin^2 \frac{\theta}{2} \cos^2 \frac{\theta'}{2} \dketbra{1} \right] \otimes \dketbra{1} \nonumber \\
  & = \frac{1}{4} \biggr[ (1+\cos \theta)(1+\cos \theta') \dketbra{0} \pm \sin \theta \sin \theta' (\dketbra{0}{1} + \dketbra{1}{0}) \nonumber \\
  & \hspace{5.5cm} + (1-\cos \theta)(1-\cos \theta') \dketbra{1} \biggr] \otimes \dketbra{0} \nonumber \\
\label{eq:psi2}
  & + \frac{1}{4} \biggr[ (1+\cos \theta)(1-\cos \theta') \dketbra{0} \pm \sin \theta \sin \theta' (\dketbra{0}{1} + \dketbra{1}{0}) \nonumber \\
  & \hspace{5.5cm} + (1-\cos \theta)(1+\cos \theta') \dketbra{1} \biggr] \otimes \dketbra{1} .
\end{align}
Now let us define two new angles $\theta_0^{\boxast},\theta_1^{\boxast}$ as
\begin{equation}
\cos \theta_0^{\boxast} \coloneqq \frac{\cos \theta + \cos \theta'}{1 + \cos \theta \cos \theta'} , \quad %
\cos \theta_1^{\boxast} \coloneqq \frac{\cos \theta - \cos \theta'}{1 - \cos \theta \cos \theta'} .
\end{equation}
Note that these are the two possible overlaps after applying the check node unitary.

This gives us the following identities:
\begin{align}
\cos \frac{\theta_0^{\boxast}}{2} = \sqrt{\frac{1}{2} (1 + \cos \theta_0^{\boxast})} &= \sqrt{\frac{1}{2} \frac{(1+\cos \theta) (1+\cos \theta')}{1 + \cos\theta \cos\theta'}} ; \ \cos \frac{\theta_1^{\boxast}}{2} = \sqrt{\frac{1}{2} \frac{(1+\cos \theta) (1-\cos \theta')}{1 - \cos\theta \cos\theta'}} \\
\sin \frac{\theta_0^{\boxast}}{2} = \sqrt{\frac{1}{2} (1 - \cos \theta_0^{\boxast})} &= \sqrt{\frac{1}{2} \frac{(1-\cos \theta) (1-\cos \theta')}{1 + \cos\theta \cos\theta'}} ; \ \sin \frac{\theta_1^{\boxast}}{2} = \sqrt{\frac{1}{2} \frac{(1-\cos \theta) (1+\cos \theta')}{1 - \cos\theta \cos\theta'}} \\
\cos \frac{\theta_0^{\boxast}}{2} \sin \frac{\theta_0^{\boxast}}{2} &= \frac{1}{2} \frac{\sin\theta \sin\theta'}{(1 + \cos\theta \cos\theta')} ; \quad \cos \frac{\theta_1^{\boxast}}{2} \sin \frac{\theta_1^{\boxast}}{2} = \frac{1}{2} \frac{\sin\theta \sin\theta'}{(1 - \cos\theta \cos\theta')}
\end{align}
Using these new angles and their identities in~\eqref{eq:psi2} we get
\begin{align}
\Psi &= \frac{1}{2} (1 + \cos\theta \cos\theta') \biggr[ \cos^2 \frac{\theta_0^{\boxast}}{2} \dketbra{0} \pm \cos \frac{\theta_0^{\boxast}}{2} \sin \frac{\theta_0^{\boxast}}{2} (\dketbra{0}{1} + \dketbra{1}{0} ) + \sin^2 \frac{\theta_0^{\boxast}}{2} \dketbra{1} \biggr] \nonumber \\
  & \hspace{8cm} \otimes \dketbra{0} \\
  & + \frac{1}{2} (1 - \cos\theta \cos\theta') \biggr[ \cos^2 \frac{\theta_1^{\boxast}}{2} \dketbra{0} \pm \cos \frac{\theta_1^{\boxast}}{2} \sin \frac{\theta_1^{\boxast}}{2} (\dketbra{0}{1} + \dketbra{1}{0} ) + \sin^2 \frac{\theta_1^{\boxast}}{2} \dketbra{1} \biggr] \nonumber \\
  & \hspace{8cm} \otimes \dketbra{1} \\
 & = p_0 \dketbra{\pm \theta_0^{\boxast}} \otimes \dketbra{0} + p_1 \dketbra{\pm \theta_1^{\boxast}} \otimes \dketbra{1} \\
\Rightarrow \Psi & = \sum_{j \in \{0,1\}} p_j \dketbra{\pm \theta_j^{\boxast}} \otimes \dketbra{j} ,
\end{align}
where $p_0 \coloneqq \frac{1}{2} (1+\cos\theta \cos\theta'), p_1 \coloneqq 1-p_0 = \frac{1}{2} (1-\cos\theta \cos\theta')$ and for $j=0,1$ we have
${\dket{\pm \theta_j^{\boxast}} \coloneqq \cos \frac{\theta_j^{\boxast}}{2} \dket{0} \pm \sin \frac{\theta_j^{\boxast}}{2} \dket{1}}$.
This is what we desired to get initially in~\eqref{eq:psi1}.

%% file: ch4_groups.tex

\label{ch:ch4_groups}

In this chapter we discuss the mathematical framework for quantum error correction introduced in~\cite{Calderbank-it98*2,Calderbank-physreva96,Gottesman-phd97} and described in detail in~\cite{Gottesman-arxiv97}.
We will use row vectors for binary and integer vectors and column vectors for quantum states.

\section{The Heisenberg-Weyl Group}
\label{sec:heisenberg_weyl}

In Chapter~\ref{ch:ch2_background} we defined qubits and Pauli matrices.
We will now formalize the representation of the $n$-qubit Pauli group which will prove very useful for the rest of this dissertation.

\begin{definition}
Given $a = [a_1,\ldots,a_n],\ b = [b_1,\ldots,b_n] \in \mathbb{F}_2^n$ define the operator
\begin{align}
\label{eq:d_ab}
D(a,b) \coloneqq X^{a_1} Z^{b_1} \otimes \cdots \otimes X^{a_n} Z^{b_n} .
\end{align}
For $N = 2^n$, the Heisenberg-Weyl group $HW_N$ of order $4N^2$ is defined as $HW_N \coloneqq \{ \imath^{\kappa} D(a,b) \mid a,b \in \mathbb{F}_2^n, \kappa \in \{0,1,2,3\} \}$.
This is also called the Pauli group on $n$ qubits.
\end{definition}

Note that the elements of $HW_N$ can be interpreted either as errors on the $n$ qubits or, in general, as $n$-qubit operators.
Since $X$ and $Z$ anti-commute, i.e., $XZ = -ZX$, multiplication in $HW_N$ satisfies the following identity:
\begin{align}
D(a, b) D(a', b') & = \left( \bigotimes_{j=1}^{n} X^{a_j} Z^{b_j} \right) \left( \bigotimes_{j=1}^{n} X^{a_j'} Z^{b_j'} \right) \nonumber \\
  & = \bigotimes_{j=1}^{n} X^{a_j} \left( Z^{b_j} X^{a_j'} \right) Z^{b_j'} \nonumber \\
  & = \bigotimes_{j=1}^{n} (-1)^{b_j a_j'} X^{a_j'} \left( X^{a_j} Z^{b_j'} \right) Z^{b_j}\ \left( \because Z^{b_j} X^{a_j'} = (-1)^{b_j a_j'} X^{a_j'} Z^{b_j} \right) \nonumber \\
  & = \prod_{j=1}^{n} (-1)^{a_j b_j'} (-1)^{b_j a_j'} \bigotimes_{j=1}^{n} X^{a_j'} Z^{b_j'} X^{a_j} Z^{b_j}\ \left(\because X^{a_j} Z^{b_j'} = (-1)^{a_j b_j'} Z^{b_j'} X^{a_j} \right) \nonumber \\
  & = (-1)^{a' b^T + b' a^T} \left( \bigotimes_{j=1}^{n} X^{a_j'} Z^{b_j'} \right) \left( \bigotimes_{j=1}^{n} X^{a_j} Z^{b_j} \right) \nonumber \\
\label{eq:hw_commute}
 & = (-1)^{a' b^T + b' a^T} D(a', b') D(a, b) .
\end{align}
Here we have used the property of the Kronecker product that $(A \otimes B)(C \otimes D) = AC \otimes BD$.
Similarly, it can be shown that elements of $HW_N$ also satisfy 
\begin{align}
D(a,b)^T = (-1)^{ab^T} D(a,b)\ \ \text{and} \ \ D(a,b) D(a',b') = (-1)^{a'b^T} D(a+a',b+b').
\end{align}

\begin{definition}
The standard symplectic inner product in $\mathbb{F}_2^{2n}$ is defined as 
\begin{align}
\label{eq:symp_inner_pdt}
\syminn{[a,b]}{[a',b']} \coloneqq a' b^T + b' a^T = [a,b]\ \Omega \ [a',b']^T \ (\bmod\ 2) ,
\end{align}
where the symplectic form in $\mathbb{F}_2^{2n}$ is $\Omega = 
\begin{bmatrix}
0 & I_n \\ 
I_n & 0
\end{bmatrix}$ (see~\cite{Calderbank-it98*2}).
\end{definition}

We observe that two operators $D(a,b)$ and $D(a',b')$ commute if and only if $\syminn{[a,b]}{[a',b']} = 0$.
Hence, the mapping $\gamma \colon HW_N/\langle \imath^{\kappa} I_N \rangle \rightarrow \mathbb{F}_2^{2n}$ defined by 
\begin{align}
\label{eq:gamma}
\gamma(D(a,b)) \coloneqq [a,b]
\end{align}
is an isomorphism that enables the representation of elements of $HW_N$ (up to multiplication by scalars) as binary vectors.
This also initiates the symplectic geometry connection.

\begin{remark}
\normalfont
Formally, a \emph{symplectic geometry} is a pair $(V,\beta)$ where $V$ is a finite-dimensional vector space over a field $K$ and $\beta \colon V \times V \rightarrow K$ is a non-degenerate alternating bilinear form (see~\cite[Chap. 1]{Gosson-2006}).
A vector space $V$ equipped with a non-degenerate alternating bilinear form is called a \emph{symplectic vector space}.
The function $\beta$ is \emph{bilinear} if for any $u,v,w \in V$ and any $k \in K$ it satisfies
\[ \beta(u + k v, w) = \beta(u, w) + k \beta(v, w) \ \text{and}\ \beta(w, u + k v) = \beta(w, u) + k \beta(w, v) . \]
It is \emph{alternating} if for any $v \in V$ it satisfies $\beta(v, v) = 0$.
Notice that expanding $\beta(v + w, v + w) = 0 \Rightarrow \beta(v, w) = - \beta(w, v)$ for any $v, w \in V$.
Finally, $\beta$ is \emph{non-degenerate} if $\beta(v, w) = 0$ for all $w \in V$ implies that $v = 0$.
For our purposes, we set $V = \mathbb{F}_2^{2n}, K = \mathbb{F}_2$ and $\beta = \syminn{\, \cdot \,}{\, \cdot \, }$ defined in~\eqref{eq:symp_inner_pdt}.
\end{remark}

\subsection{Hermitian Pauli Matrices}
\label{sec:hermitian_paulis}

We will often find it convenient to focus on Hermitian Pauli matrices,
\begin{align}
E(a,b) \coloneqq \imath^{ab^T \bmod 4} D(a,b) = \imath^{ab^T \bmod 4} D(a,0) D(0,b).
\end{align}
Since these are also unitary, these satisfy $E(a,b)^2 = I_N$, the $N \times N$ identity matrix.
Note that $D(a,0) = E(a,0)$ are permutation matrices that map $\ket{v} \mapsto \ket{v \oplus a}$, and $D(0,b) = E(0,b)$ are diagonal matrices that act like $D(0,b) \ket{v} = (-1)^{vb^T} \ket{v}$.
Hence we can write
\begin{align}
\label{eq:Eab_expansion}
E(a,0) = \sum_{v \in \mathbb{F}_2^n} \dketbra{v \oplus a}{v}, \quad E(0,b) = \sum_{v \in \mathbb{F}_2^n} (-1)^{vb^T} \dketbra{v}.
\end{align}
Computationally, substituting these definitions in the definition of $E(a,b)$ allows us to construct these matrices for fairly large $n$ using sparse matrix tools.

\begin{remark}
\label{rem:general_Eab}
\normalfont
It will be convenient to generalize the above definitions to vectors $a,b \in \MZ^n$.
We express these vectors as $a = a_0 + 2a_1 + 4a_2 + \ldots, b = b_0 + 2b_1 + 4b_2 + \ldots$, where $a_i, b_i \in \mathbb{Z}_2^n$.
Note that this does not distort these definitions since $X^2 = Z^2 = I_2$ implies $D(a,b)$ remains unchanged, while the exponent of $\imath$ for $E(a,b)$ will change to $(a_0+2a_1) (b_0+2b_1)^T = a_0 b_0^T + 2(a_0 b_1^T + a_1 b_0^T)\ (\bmod\ 4)$ which only ever introduces an additional $(-1)$ factor thereby ensuring that $E(a,b)$ is still Hermitian and $E(a,b)^2 = I_N$.
Henceforth, all inner (dot) products are performed over $\MZ$, unless mentioned otherwise, and if they occur in the exponent of a $2^{\ell}$-th root of unity then the result is automatically reduced modulo $2^{\ell}$.
\end{remark}

Hence, we can consolidate some useful identities for Hermitian Pauli matrices.
\begin{align}
\label{eq:Eab_multiply}
E(a,b) E(c,d) & = \imath^{bc^T - ad^T} E(a+c, b+d)
                = (-1)^{\syminn{[a,b]}{[c,d]}} E(c,d) E(a,b), \\
\label{eq:Eab_2x}
E(a, b + 2x) & = \imath^{a(b+2x)^T} D(a,b + 2x) = (-1)^{ax^T} \imath^{ab^T} D(a,b) = (-1)^{ax^T} E(a,b), \\
\label{eq:Eab_x}
E(a, b + x) & = E(a, b \oplus x + 2 (b \ast x)) = (-1)^{a(b \ast x)^T} E(a, b \oplus x),
\end{align}
where $(b \ast x) \in \mathbb{Z}^n$ represents the point-wise product of the two vectors.
We will continue to use $\oplus$ to denote addition modulo $2$ and $+$ to denote usual addition over the integers.
Identity~\eqref{eq:Eab_multiply} implies that if $E(a,b)$ and $E(a',b')$ commute then $E(a,b) E(a',b') = \pm E(a+a',b+b')$, and if not then $E(a,b) E(a',b') = \pm \imath E(a+a',b+b')$.

\begin{theorem}
\label{thm:Pauli_basis}
The normalized Hermitian Pauli matrices $\{ \frac{1}{\sqrt{N}} E(a,b) \colon a,b \in \mathbb{Z}_2^n \}$ form an orthonormal basis over $\mathbb{C}$ for all $N \times N$ matrices under the Frobenius inner product.
\begin{proof}
First it is easy to check that we need $N^2$ elements in the basis since the standard basis for matrices consists of matrices each containing a single entry $1$ and remaining entries $0$.
This exactly coincides with the number of vectors $a,b \in \mathbb{Z}_2^n$, and their corresponding Pauli matrices $E(a,b)$ are linearly independent.
Recollect that the Frobenius inner product between two matrices is given by $\langle A,B \rangle \coloneqq \text{Tr}(A^{\dagger} B)$.
For $n=1$, it is clear that $\text{Tr}(\frac{1}{\sqrt{2}} A \cdot \frac{1}{\sqrt{2}} A) = 1$ for $A \in \{ I_2, X, Z, Y \}$.
A quick calculation also shows that $\text{Tr}(\frac{1}{\sqrt{2}} A \cdot \frac{1}{\sqrt{2}} B) = 0$ whenever $A \neq B$ and $A,B \in \{ I_2, X, Z, Y \}$.
Therefore this proves that $\frac{1}{\sqrt{2}} \{ I_2, X, Z, Y \}$ forms an orthonormal basis for $n=1$.
For $n>1$, we recollect that $\text{Tr}(C \otimes D) = \text{Tr}(C) \text{Tr}(D)$.
Thus it still holds that $\text{Tr}(\frac{1}{\sqrt{N}} E(a,b) \cdot \frac{1}{\sqrt{N}} E(a,b)) = 1$ and $\text{Tr}(\frac{1}{\sqrt{N}} E(a,b) \cdot \frac{1}{\sqrt{N}} E(a',b')) = 0$ for $[a,b] \neq [a',b']$.
This completes the proof. 
\end{proof}
\end{theorem}

This is a very useful (and well-known) result that makes it convenient to analyze unitary operations that act non-trivially on the code subspace of quantum error-correcting codes.
As we will see in Chapters~\ref{ch:ch6_qfd_gates} and~\ref{ch:ch7_stabilizer_codes_qfd}, this essentially translates to understanding the action of these unitaries on all Pauli matrices.
Theorem~\ref{thm:Pauli_basis} also enables one to show that if a code corrects certain Pauli errors then it can also correct any linear combination of those Pauli errors~\cite{LB-2013}.
Thus, this is an important result that leads to the discretization of errors in quantum computers, thereby distinguishing them from classical analog computers.

\section{The Clifford Group}
\label{sec:clifford_gp}

\begin{definition}
Let $\mathbb{U}_N$ be the group of unitary operators on vectors in $\mathbb{C}^N$.
The Clifford group $\text{Cliff}_N \subset \mathbb{U}_N$ consists of all unitary matrices $g \in \mathbb{U}_N$ that permute the elements of $HW_N$ under conjugation.
Formally,
\begin{align}
\text{Cliff}_N \coloneqq \left\{ g \in \mathbb{U}_N \mid g D(a,b) g^{\dagger} \in HW_N \ \text{for\ all} \ D(a,b) \in HW_N \right\} ,
\end{align}
where $g^{\dagger}$ is the adjoint (or Hermitian transpose) of $g$~\cite{Gottesman-arxiv09}.
Note that $HW_N \subset \text{Cliff}_N$.
\end{definition}

\begin{definition}
Let $G$ and $H$ be subgroups of a group.
The set of elements $f \in G$ such that $f H f^{-1} = H$ is defined to be the normalizer of $H$ in $G$, denoted as $\mathcal{N}_G(H)$.
The condition $f H f^{-1} = H$ can be restated as requiring that the left coset $fH$ be equal to the right coset $Hf$.
If $H$ is a subgroup of $G$, then $\mathcal{N}_G(H)$ is also a subgroup containing $H$.
In this case, $H$ is a normal subgroup of $\mathcal{N}_G(H)$.
\end{definition}

Hence, the Clifford group is the \emph{normalizer} of $HW_N$ in the unitary group $\mathbb{U}_N$, i.e., $\text{Cliff}_N = \mathcal{N}_{\mathbb{U}_N}(HW_N)$.
We regard operators in $\text{Cliff}_N$ as physical operators acting on quantum states in $\mathbb{C}^N$, to be implemented by quantum circuits.

\begin{definition}
\label{def:automorphism}
Let $A$ be a collection of objects.
The automorphism group Aut($A$) of $A$ is the group of functions $f \colon A \rightarrow A$ (with the composition operation) that preserve the structure of $A$.
If $A$ is a group then every $f \in \text{Aut}(A)$ preserves the multiplication table of the group.
The inner automorphism group Inn($A$) is a subgroup of Aut($A$) defined as $\text{Inn}(A) \coloneqq \{ f_a \mid a \in A \}$, where $f_a \in \text{Aut}(G)$ is defined by $f_a(x) = a x a^{-1}$, i.e., these are automorphisms of $A$ induced by conjugation with elements of $A$.
\end{definition}

As a corollary of Theorem~\ref{thm:symp_action}, it holds that $\text{Aut}(HW_N) = \text{Cliff}_N$, i.e. the automorphisms induced by conjugation form the full automorphism group of $HW_N$ in $\mathbb{U}_N$.
We sometimes refer to elements of $\text{Cliff}_N$ as \emph{unitary automorphisms} of $HW_N$.

\begin{fact}
\label{fact:XnZn}
The action of any unitary automorphism of $HW_N$ is defined by its action on the following two maximal commutative subgroups of $HW_N$ that generate $HW_N$:
\begin{align}
\label{eq:XnZn}
X_N \coloneqq \left\{ D(a,0) \in HW_N \mid a \in \mathbb{F}_2^n \right\} , \ \ Z_N \coloneqq \left\{ D(0,b) \in HW_N \mid b \in \mathbb{F}_2^n \right\} .
\end{align}
\end{fact}

\begin{definition}
Given a group $G$ and an element $h \in G$, a conjugate of $h$ in $G$ is an element $ghg^{-1}$, where $g \in G$.
Conjugacy defines an equivalence relation on $G$ and the equivalence classes are called conjugacy classes of $G$.
\end{definition}

\begin{lemma}
The set of all conjugacy classes of $HW_N$ is given by $\text{Class}(HW_N) = $
\begin{align*}
\left( \bigcup_{\substack{D(a,b) \in HW_N\\(a,b) \neq (0,0)}} \left\{ \{ D(a,b), -D(a,b) \}, \{ \imath D(a,b), -\imath D(a,b) \} \right\} \right) \cup \left( \bigcup_{\kappa=0}^{3} \left\{ \left\{ \imath^\kappa I_N \right\} \right\} \right).
\end{align*}
\begin{proof}
For an element $D(a,b) \in HW_N$, its conjugacy class is given by
\[ \{ D(c,d) D(a,b) D(c,d)^{-1} \mid D(c,d) \in HW_N \} . \]
We know that $D(c,d) D(a,b) = (-1)^{cb^T + da^T} D(a,b) D(c,d)$ which implies
\[ D(c,d) D(a,b) D(c,d)^{-1} = (-1)^{cb^T + da^T} D(a,b) D(c,d) D(c,d)^{-1} = (-1)^{cb^T + da^T} D(a,b) . \]
Therefore $D(a,b)$ is mapped either to itself, if $cb^T + da^T = 0$, or to $- D(a,b)$, if $cb^T + da^T = 1$.
This does not change if $D(c,d)$ has a leading $\pm \imath$ because the inverse will cancel it.
So the conjugacy class of $\pm D(a,b)$ is $\{ D(a,b), -D(a,b) \}$ and that of $\pm \imath D(a,b)$ is $\{ \imath D(a,b), -\imath D(a,b) \}$.
For the special case of $D(a,b) = \imath^\kappa I_N$ with $\kappa \in \{ 0,1,2,3 \}$, the corresponding conjugacy class is a singleton $\{ \imath^\kappa I_N \}$ since $D(c,d) D(a,b) = \imath^\kappa D(c,d)$.
\end{proof}
\end{lemma}

\begin{corollary}
\label{cor:conjugacy}
The elements of $HW_N$ have order either $1, 2$ or $4$.
Any automorphism of $HW_N$ must preserve the order of all elements since by Definition~\ref{def:automorphism} it must preserve the multiplication table of the group.
Also, the inner automorphisms defined by conjugation with matrices $\imath^\kappa D(a,b) \in HW_N$ preserve every conjugacy class of $HW_N$, because~\eqref{eq:hw_commute} implies that elements in $HW_N$ either commute or anti-commute.
\end{corollary}

\subsection{The Map to Symplectic Matrices}

\begin{definition}
\label{def:symp_matrix}
An invertible matrix $F \in \mathbb{F}_2^{2n \times 2n}$ is said to be a symplectic matrix if it preserves the symplectic inner product between vectors in $\mathbb{F}_2^{2n}$ (see \cite{Can-2017a,Gottesman-arxiv09}).
In other words, a symplectic matrix $F$ satisfies $\syminn{[a,b]F}{[a',b']F} = \syminn{[a,b]}{[a',b']}$ for all $[a,b], [a',b'] \in \mathbb{F}_2^{2n}$.
This implies that $[a,b]\ F\Omega F^T \ [a',b']^T = [a,b]\ \Omega \ [a',b']^T$ and hence an equivalent condition for a symplectic matrix is that it must satisfy $F \Omega F^T = \Omega$ (mod $2$).
\end{definition}

If we represent a symplectic matrix as $F = 
\left[ \begin{array}{c|c}
A & B \\
\hline
C & D
\end{array} \right]$ then the condition $F \Omega F^T = \Omega$ can be explicitly written as 
\begin{align}
A B^T = B A^T, \ C D^T = D C^T, \ A D^T + B C^T = I_n\ (\bmod\ 2).
\end{align}

\begin{definition}
The group of symplectic $2n \times 2n$ binary matrices is called the symplectic group over $\mathbb{F}_2^{2n}$ and is denoted by $\text{Sp}(2n,\mathbb{F}_2)$.
It can be interpreted as a generalization of the orthogonal group to the symplectic inner product.
The size of the symplectic group is well-known to be $2^{n^2} \prod_{j=1}^{n} (4^j - 1)$ (e.g., see~\cite{Calderbank-it98*2}).
\end{definition}

\begin{definition}
\label{def:symp_basis}
A symplectic basis for $\mathbb{F}_2^{2n}$ is a set of pairs $\{ (v_1,w_1), (v_2,w_2), \ldots, (v_n,w_n) \}$ such that $\syminn{v_i}{w_j} = \delta_{ij}$ and $\syminn{v_i}{v_j} = \syminn{w_i}{w_j} = 0$, where $i,j \in \{1,\ldots,n\}$, and $\delta_{ij} = 1$ if $i=j$ and $0$ if $i \neq j$.
\end{definition}

Note that the rows of any matrix in $\text{Sp}(2n,\mathbb{F}_2)$ form a symplectic basis for $\mathbb{F}_2^{2n}$.
Also, the top and bottom halves of a symplectic matrix satisfy 
\[ [ A \mid B ] \ \Omega \ [ A \mid B ]^T = [ C \mid D ] \ \Omega \ [ C \mid D ]^T = 0. \]
There also exists a symplectic Gram-Schmidt orthogonalization procedure that produces a symplectic basis starting from the standard basis and an additional vector $v \in \mathbb{F}_2^{2n}$ (see~\cite{Koenig-jmp14}).

Next we state a classical result which forms the foundation for our algorithm in Chapter~\ref{ch:ch5_lcs_algorithm} to synthesize logical Clifford operators of stabilizer codes.
For completeness we also provide a proof here (adapted from~\cite{Can-2017a}).

\begin{theorem}
\label{thm:symp_action}
The automorphism induced by $g \in \text{Cliff}_N$ satisfies
\begin{align}
\label{eq:symp_action}
g E(a,b) g^{\dagger} = \pm E\left( [a,b] F_g \right) \ , \ \text{where} \ \ F_g = 
\begin{bmatrix}
A_g & B_g \\
C_g & D_g
\end{bmatrix} \in \text{Sp}(2n,\mathbb{F}_2).
\end{align} 
\begin{proof}
First we show that there exists a well-defined $2n \times 2n$ binary transformation $F$ such that for all $[a,b] \in \mathbb{F}_2^{2n}$ we have $g E(a,b) g^{\dagger} = \pm E\left( [a,b] F \right)$.
Let the standard basis vectors of $\mathbb{F}_2^{n}$ be denoted as $e_i$, which have an entry $1$ in the $i$-th position and entries $0$ elsewhere.
Then $\{E(e_1,0),\ldots,E(e_n,0)\}$ and $\{E(0,e_1),\ldots,E(0,e_n)\}$ form a basis for $X_N$ and $Z_N$ defined in~\eqref{eq:XnZn}, respectively.
Since $g \in \text{Cliff}_N$ is an automorphism of $HW_N$, by Corollary~\ref{cor:conjugacy} it preserves the order of every element of $HW_N$. 
Hence it maps $E(e_i,0) \mapsto \pm E(a_i',b_i'), E(0,e_j) \mapsto \pm E(c_j',d_j')$ for some $[a_i',b_i'], [c_j',d_j'] \in \mathbb{F}_2^{2n}$, where $i,j \in \{1,\ldots,n\}$.
Therefore we can express the action of $g$ as $g E(e_i,0) g^{\dagger} = \pm E(a_i',b_i'), g E(0,e_j) g^{\dagger} = \pm E(c_j',d_j')$.
Using the same $i,j$ define a matrix $F$ with the $i$-th row being $[a_i',b_i']$ and the $(n+j)$-th row being $[c_j',d_j']$.
This matrix satisfies 
\begin{align*}
g E(e_i,0) g^{\dagger} = \pm E\left( [e_i,0] F \right), \ \ g E(0,e_j) g^{\dagger} = \pm E\left( [0,e_j] F \right) .
\end{align*}
Using the fact that any $[a,b] \in \mathbb{F}_2^{2n}$ can be written as a linear combination of $[e_i,0]$ and $[0,e_j]$, the identity~\eqref{eq:Eab_multiply}, and Corollary~\ref{cor:conjugacy}, we have $g E(a,b) g^{\dagger} = \pm E\left( [a,b] F \right)$.
As the rows of $F$ are a result of the action of $g$ on $X_N$ and $Z_N$, we explicitly denote this dependence by $F_g$.
All that is left is to show that $F_g$ is symplectic.

Conjugating both sides of the equation~\eqref{eq:Eab_multiply} by $g$ we get
\begin{align*}
\left( g E(a,b) g^{\dagger} \right) \left( g E(a',b') g^{\dagger} \right) &= (-1)^{a'b^T} \imath^{\syminn{[a,b]}{[a',b']}} \left( g E(a+a',b+b') g^{\dagger} \right) \\
\Rightarrow E\left( [a,b] F_g \right) E\left( [a',b'] F_g \right) &= \pm (-1)^{a'b^T} \imath^{\syminn{[a,b]}{[a',b']}} E\left( [a+a',b+b'] F_g \right) .
\end{align*}
Now apply~\eqref{eq:Eab_multiply} to $E\left( [a,b] F_g \right), E\left( [a',b'] F_g \right)$, and define $[c,d] \coloneqq [a,b] F_g, [c',d'] \coloneqq [a',b'] F_g$:
\begin{align*}
E\left( [a,b] F_g \right) E\left( [a',b'] F_g \right) = (-1)^{c'd^T} \imath^{\syminn{[a,b] F_g}{[a',b'] F_g}} E\left( [a+a',b+b'] F_g \right) .
\end{align*}
Equating the two expressions on the right hand side for all $[a,b],[a',b']$ we get $F_g \Omega F_g^T = \Omega$, or that $F_g$ is symplectic.
\end{proof}
\end{theorem}

\begin{table}
{
\begin{tabular}{ccccc}
~ & ~ & ~ & ~ & ~ \\
Symplectic Matrix $F_g$ & \hspace*{5mm} & Clifford Operator $g$ & \hspace*{5mm} & Circuit Element \\
~ & ~ & ~ & ~ & ~ \\
\hline
~ & ~ & ~ & ~ & ~ \\
$\Omega = \begin{bmatrix} 0 & I_n \\ I_n & 0 \end{bmatrix}$ & \hspace*{5mm} & $H_N = H_2^{\otimes n}$ & \hspace*{5mm} & Full Hadamard \\
~ & ~ & ~ & ~ & ~ \\
\makecell{$L_Q = \begin{bmatrix} Q & 0 \\ 0 & Q^{-T} \end{bmatrix}$\\ $(Q^{-T} = (Q^{-1})^T = (Q^T)^{-1})$} & \hspace*{5mm} & $\ell_Q = \sum_{v \in \mathbb{F}_2^n} \dketbra{v Q}{v}$ & \hspace*{5mm} & \makecell{Controlled-X (CX),\\ Permutations} \\
~ & ~ & ~ & ~ & ~ \\
\makecell{$T_R = \begin{bmatrix} I_n & R \\ 0 & I_n \end{bmatrix}$\\ $(R = R^T)$} & \hspace*{5mm} & $\begin{aligned} t_R & = \text{diag}\left( \imath^{v R v^T \bmod 4} \right) \\ & = \sum_{v \in \mathbb{F}_2^n} \imath^{v R v^T} \dketbra{v} \end{aligned}$ & \hspace*{5mm} &  \makecell{Controlled-$Z$ (CZ),\\ Phase ($P$) gates}\\
~ & ~ & ~ & ~ & ~ \\
$G_t = \begin{bmatrix} L_{n-t} & U_t \\ U_t & L_{n-t} \end{bmatrix}$ & \hspace*{5mm} & $g_t = H_{2^t} \otimes I_{2^{n-t}}$ & \hspace*{5mm} & Partial Hadamard \\
~ & ~ & ~ & ~ & ~ \\
\hhline{=====}
\end{tabular}

\caption[A generating set of binary symplectic matrices and their corresponding unitary operators.]{\label{tab:std_symp}\small A generating set of binary symplectic matrices and their corresponding unitary operators. The number of $1$s in $Q$ and $R$ directly relates to number of gates involved in the circuit realizing the respective unitary operators (see Section~\ref{sec:elem_symp}). The $N$ coordinates are indexed by binary vectors $v \in \mathbb{F}_2^n$.
Here $H_{2^t}$ denotes the Walsh-Hadamard matrix of size $2^t$, $U_t = \text{diag}\left( I_t, 0_{n-t} \right)$ and $L_{n-t} = \text{diag}\left( 0_t, I_{n-t} \right)$, where $I_t$ is the $t \times t$ identity matrix and $0_t$ is the $t \times t$ matrix with all zero entries. }
%
}
\end{table}

Since the action of any $g \in \text{Aut}(HW_N)$ on $HW_N$ can be expressed as mappings $[e_i,0] \mapsto [a_i',b_i'], [0,e_j] \mapsto [c_j',d_j']$ in $\mathbb{F}_2^{2n}$, that can be induced by a $g \in \text{Cliff}_N$ (under conjugation), we have $\text{Aut}(HW_N) = \text{Cliff}_N$.
As discussed after Theorem~\ref{thm:Pauli_basis}, since Hermitian Paulis form an orthonormal basis, the above result implies that a binary symplectic matrix $F_g$ contains most of the information (except signs) about the corresponding Clifford operator $g$.
The fact that $F_g$ is symplectic expresses the property that the automorphism induced by $g$ must respect commutativity in $HW_N$.
By Corollary~\ref{cor:conjugacy}, any inner automorphism induced by conjugation with $g \in HW_N$ has $F_g = I_{2n}$.
So the map $\phi \colon \text{Cliff}_N \rightarrow \text{Sp}(2n,\mathbb{F}_2)$ defined by 
\begin{align}
\label{eq:phi}
\phi(g) \coloneqq F_g
\end{align}
is a homomorphism with kernel $HW_N$, and every Clifford operator maps down to a matrix $F_g$.
Hence, $HW_N$ is a normal subgroup of $\text{Cliff}_N$ and $\text{Cliff}_N/HW_N \cong \text{Sp}(2n,\mathbb{F}_2)$.

In summary, every unitary automorphism $g \in \text{Cliff}_N$ of $HW_N$ induces a symplectic transformation $F_g \in \text{Sp}(2n,\mathbb{F}_2)$ via the map $\phi$, and two automorphisms that induce the same symplectic transformation can differ only by an inner automorphism that is also an element of $HW_N$.
The symplectic group $\text{Sp}(2n,\mathbb{F}_2)$ is generated by the set of elementary symplectic transformations given in the first column of Table~\ref{tab:std_symp} (also see~\cite{Dehaene-physreva03}).
The second column lists their corresponding unitary automorphisms, under the relation proven in Theorem~\ref{thm:symp_action}, i.e., $g E(a,b) g^{\dagger} = \pm E\left( [a,b] F_{g} \right)$.
The third column relates these elementary forms with the elementary quantum gates that they can represent.
A discussion of these transformations and their circuits is provided in the next subsection.

\begin{definition}
\label{def:depth}
In a circuit, any set of gates acting sequentially on disjoint subsets of qubits can be applied concurrently.
These gates constitute one stage of the circuit, and the number of such stages is defined to be the depth of the circuit.
\end{definition}

For example, consider $n=4$ and the quantum circuit $H_1$ --- $\cz{2}{3}$ --- $P_4$, where the subscripts indicate the qubit(s) on which the gate is applied.
More precisely, we can explicitly write $H_1 = H \otimes I_2 \otimes I_2 \otimes I_2$, where the subscript $2$ indicates the $2 \times 2$ identity matrix.
This is a circuit of depth $1$.
However the circuit $H_2$ --- $\cz{2}{3}$ --- $P_4$ has depth $2$.
Note that the unitary operator is computed as the matrix product $U = P_4 \cz{2}{3} H_2 = \cz{2}{3} P_4 H_2 = \cz{2}{3} H_2 P_4$, since $U$ acts on a quantum state $\dket{\psi}$ as $U \dket{\psi}$, i.e., on the right.
However, in the circuit the state goes through from left to right and hence the order is reversed.

\subsection{Circuits for Elementary Symplectic Matrices}
\label{sec:elem_symp}

In this section we verify that the physical operators listed in Table~\ref{tab:std_symp} are associated with the corresponding symplectic transformation.
Furthermore, we also provide circuits that realize these physical operators (also see~\cite{Dehaene-physreva03}).


\begin{enumerate}

\item $H_N = H^{\otimes n} \colon$ The Hadamard operator $H \coloneqq \frac{1}{\sqrt{2}} \begin{bmatrix}
1 & 1 \\
1 & -1
\end{bmatrix}$ satisfies $H X H^{\dagger} = Z$, $H Z H^{\dagger} = X$. 
Hence the action of $H_N$ on a $HW_N$ element $D(a,b)$ is given by
\begin{align*}
H_N D(a,b) H_N^{\dagger} = H_N D(a,0) D(0,b) H_N^{\dagger} & = (H_N D(a,0) H_N^{\dagger}) (H_N D(0,b) H_N^{\dagger}) \\
  & = D(0,a) D(b,0) = (-1)^{ab^T} D(b,a) \\
\Rightarrow  H_N D(a,b) H_N^{\dagger} &= (-1)^{ab^T} D \left( [a,b] \Omega \right).
\end{align*}
The circuit for $H_N$ is just $H$ applied to each of the $n$ qubits.

\item $GL(n,\mathbb{F}_2) \colon$ Each non-singular $n \times n$ binary matrix $Q$ is associated with a symplectic transformation $L_Q$ given by 
\begin{align*}
L_Q = 
\begin{bmatrix} 
Q & 0 \\ 
0 & Q^{-T} 
\end{bmatrix} ,
\end{align*}
where $Q^{-T} = (Q^T)^{-1} = (Q^{-1})^T$.
The matrix $Q$ is also associated with the unitary operator $\ell_Q$ which realizes the mapping $\dket{v} \mapsto \dket{vQ}$.
We verify the correspondence $\ell_Q \rightarrow L_Q$ as follows.
Note that $D(c,0) \dket{v} = \dket{v \oplus c}$ and $D(0,d) \dket{v} = (-1)^{vd^T} \dket{v}$.
\begin{align}
(\ell_Q D(c,d) \ell_Q^{\dagger}) \dket{v} & = \ell_Q D(c,0) D(0,d) \dket{vQ^{-1}} \\
                               & = \ell_Q (-1)^{cd^T} D(0,d) D(c,0) \dket{vQ^{-1}} \\
                               & = (-1)^{cd^T} \ell_Q (-1)^{(vQ^{-1}+c) d^T} \dket{vQ^{-1} \oplus c} \\
                               & = (-1)^{cd^T} (-1)^{(v+cQ) Q^{-1}d^T} \dket{v \oplus cQ} \\
                               & = (-1)^{cd^T} D(0,d(Q^{-1})^T) D(cQ,0) \dket{v} \\
                               & = D(cQ,dQ^{-T}) \dket{v} \\
                               & = D \left( [c,d] L_Q \right) \dket{v}.
\end{align}
Since the operator $\ell_Q$ realizes the map $\dket{v} \mapsto \dket{vQ}$, the circuit for the operator is equivalent to the binary circuit that realizes $v \mapsto vQ$.
Evidently, this elementary transformation encompasses CNOT operations and qubit permutations.
For the latter, $Q$ will be a permutation matrix.
Note that if $\ell_Q$ preserves the code space of a CSS code then the respective permutation must be in the automorphism group of the constituent classical code.
This is the special case that is discussed in detail by Grassl and Roetteler in~\cite{Grassl-isit13}.

For a general $Q$, one can use the LU decomposition over $\mathbb{F}_2$ to obtain $P_{\pi} Q = LU$, where $P_{\pi}$ is a permutation matrix, $L$ is lower triangular and $U$ is upper triangular.
Note that $L_{ii} = U_{ii} = 1 \ \forall \ i \in \{1,\ldots,n\}$.
Then the circuit for $Q$ first involves the permutation $P_{\pi}^T$ (or $\pi^{-1}$), then CNOTs for $L$ with control qubits in the order $1,2,\ldots,n$ and then CNOTs for $U$ with control qubits in reverse order $n,n-1,\ldots,1$.
The order is important because an entry $L_{ji} = 1$ implies a CNOT gate with qubit $j$ controlling qubit $i$ (with $j > i$), i.e, $\cnot{j}{i}$, and similarly $L_{kj} = 1$ implies the gate $\cnot{k}{j}$ (with $k > j$).
Since the gate $\cnot{j}{i}$ requires the value of qubit $j$ \emph{before} it is altered by $\cnot{k}{j}$, it needs to be implemented first.
A similar reasoning applies to the reverse order of control qubits for $U$.

\item $t_R = \text{diag}\left( \imath^{vRv^T} \right) \colon$ Each symmetric matrix $R \in \mathbb{F}_2^{n \times n}$ is associated with a symplectic transformation $T_R$ given by
\begin{align*}
T_R = 
\begin{bmatrix} 
I_n & R \\ 
0 & I_n
\end{bmatrix} ,
\end{align*}
and with a unitary operator $t_R$ that realizes the map $\dket{v} \mapsto \imath^{vRv^T} \dket{v}$.
We now verify that conjugation by $t_R$ induces the symplectic transformation $T_R$.
\begin{align}
(t_R D(a,b) t_R^{\dagger}) \dket{v} & = \imath^{-vRv^T} t_R (-1)^{ab^T} D(0,b) D(a,0) \dket{v} \\
                               & = \imath^{-vRv^T} (-1)^{ab^T} t_R (-1)^{(v \oplus a)b^T} \dket{v \oplus a} \\
                               & = (-1)^{ab^T} \imath^{-vRv^T} (-1)^{(v + a)b^T} \imath^{(v \oplus a)R(v \oplus a)^T} \dket{v \oplus a} \\
                               & = (-1)^{ab^T} \imath^{-vRv^T} (-1)^{(v+a)b^T} \imath^{(v+a)R(v+a)^T} \dket{v \oplus a} \\
                               & = (-1)^{ab^T} \imath^{aRa^T} (-1)^{vRa^T + (v+a)b^T} \dket{v \oplus a} \\
                               & = (-1)^{ab^T} \imath^{-aRa^T} (-1)^{(v+a)(b+aR)^T} \dket{v \oplus a} \\
                               & = (-1)^{ab^T} \imath^{-aRa^T}  D(0,b+aR) D(a,0) \dket{v} \\
                               & = (-1)^{ab^T} \imath^{-aRa^T} (-1)^{a(b+aR)^T} D(a,b+aR) \dket{v} \\
                               & = \imath^{aRa^T} D \left( [a,b] T_R \right) \dket{v} .
\end{align}
Note that for the fourth equality we have used $v \oplus a = v + a - 2(v \ast a)$.
Hence, for $E(a,b) \coloneqq \imath^{ab^T} D(a,b)$, we have $t_R E(a,b) t_R^{\dagger} = \imath^{ab^T} \imath^{aRa^T} D(a,b+aR) = E \left( [a,b] T_R \right)$ as required, where $[a,b] T_R$ is over $\mathbb{Z}$.
We derive the circuit for this unitary operator by observing the action of $T_R$ on the standard basis vectors $[\vecnot{e}_1,\vecnot{0}],\ldots,[\vecnot{e}_n,\vecnot{0}]$, $[\vecnot{0},\vecnot{e}_1],\ldots,[\vecnot{0},\vecnot{e}_n]$ of $\mathbb{F}_2^{2n}$, where $i \in \{1,\ldots,n\}$, which captures the effect of $t_R$ on the (basis) elements $X_1,\ldots,X_n$, $Z_1,\ldots,Z_n$ of $HW_N$, respectively, under conjugation.

Assume as the first special case that $R$ has non-zero entries only in its (main) diagonal.
If $R_{ii} = 1$ then we have $[\vecnot{e}_i,\vecnot{0}] T_R = [\vecnot{e}_i, \vecnot{e}_i]$.
This indicates that $t_R$ maps $X_i = E(\vecnot{e}_i,0) \mapsto E(\vecnot{e}_i, \vecnot{e}_i) = Y_i$.
Since we know that the phase gate $P_i$ on the $i$-th qubit performs exactly this map under conjugation, we conclude that the circuit for $t_R$ involves $P_i$.
We proceed similarly for every $i \in \{1,\ldots,n\}$ such that $R_{ii} = 1$.

Now consider the case where $R_{ij} = R_{ji} = 1$ (since $R$ is symmetric).
Then we have ${[\vecnot{e}_i,\vecnot{0}] T_R = [\vecnot{e}_i,\vecnot{e}_j], \ [\vecnot{e}_j,\vecnot{0}] T_R = [\vecnot{e}_j,\vecnot{e}_i]}$.
This indicates that $t_R$ maps $X_i \mapsto X_i Z_j$ and $X_j \mapsto Z_i X_j$.
Since we know that the controlled-$Z$ gate $\text{CZ}_{ij}$ on qubits $(i,j)$ performs exactly this map under conjugation, we conclude that the circuit for $t_R$ involves $\text{CZ}_{ij}$.
We proceed similarly for every pair $(i,j)$ such that $R_{ij} = R_{ji} = 1$.

Finally, we note that the symplectic transformation associated with the operator $H_N t_R H_N$ is 
$\Omega\, T_R\, \Omega = \begin{bmatrix} 
I_n & 0 \\ 
R & I_n 
\end{bmatrix}$.

\item $g_t = H_{2^t} \otimes I_{2^{n-t}} \colon$ Since $H_{2^t}$ is the $t$-fold Kronecker product of $H$ we have
\begin{align}
g_t D(a,b) g_t^{\dagger} & = \left( Z^{a_1} X^{b_1} \otimes \cdots \otimes Z^{a_t} X^{b_t} \right) \otimes \left( X^{a_{t+1}} Z^{b_{t+1}} \otimes \cdots \otimes X^{a_n} Z^{b_n} \right) \\
  & = \left( (-1)^{a_1 b_1} X^{b_1} Z^{a_1} \otimes \cdots \otimes (-1)^{a_t b_t} X^{b_t} Z^{a_t} \right) \nonumber \\
  & \hspace{4cm} \otimes \left( X^{a_{t+1}} Z^{b_{t+1}} \otimes \cdots \otimes X^{a_n} Z^{b_n} \right) .
\end{align}
We write $(a,b) = (\hat{a} \bar{a}, \hat{b} \bar{b})$ where $\hat{a} \coloneqq [a_1, \cdots, a_t], \bar{a} \coloneqq [a_{t+1}, \cdots, a_n], \ \hat{b} \coloneqq [b_1, \cdots, b_t]$, $\bar{b} \coloneqq [b_{t+1}, \cdots, b_n]$.
Then we have 
\begin{align}
g_t D(\hat{a} \bar{a}, \hat{b} \bar{b}) g_t^{\dagger} & = (-1)^{\hat{a} \hat{b}^T} D(\hat{b} \bar{a}, \hat{a} \bar{b}) = (-1)^{\hat{a} \hat{b}^T} D \left( [\hat{a} \bar{a}, \hat{b} \bar{b}] G_t \right) , \\
\text{where} \ 
G_t & = 
\begin{bmatrix}
0 & 0 & I_t & 0 \\
0 & I_{n-t} & 0 & 0 \\
I_t & 0 & 0 & 0 \\
0 & 0 & 0 & I_{n-t}
\end{bmatrix} .
\end{align}
Defining $U_t \coloneqq \begin{bmatrix} I_t & 0 \\ 0 & 0 \end{bmatrix}, L_{n-t} \coloneqq \begin{bmatrix} 0 & 0 \\ 0 & I_{n-t} \end{bmatrix}$, we then write $G_t = 
\begin{bmatrix} 
L_{n-t} & U_t \\ 
U_t & L_{n-t} 
\end{bmatrix}$.
Similar to part 1 above, the circuit for $g_t$ is simply $H$ applied to each of the first $t$ qubits.
Although this is a special case where the Hadamard operator was applied to consecutive qubits, we note that the symplectic transformation for Hadamards applied to arbitrary non-consecutive qubits can be derived in a similar fashion.

\end{enumerate}

Hence we have demonstrated the elementary symplectic transformations in $\text{Sp}(2n,\mathbb{F}_2)$ that are associated with arbitrary Hadamard, Phase, Controlled-$Z$ and Controlled-NOT gates.
Since we know that these gates (which include $HW_N$) generate the full Clifford group~\cite{Gottesman-arxiv09}, these elementary symplectic transformations form a universal set corresponding to physical operators in the Clifford group.

\subsection{Decomposition of Clifford Operators}
\label{sec:cliff_decompose}

We can synthesize a physical operator $g \in \text{Cliff}_N$ by factoring $F_{g}$ into elementary symplectic matrices of the types listed in Table~\ref{tab:std_symp}.
Three algorithms for this purpose are given in~\cite{Dehaene-physreva03},\cite{Maslov-arxiv17} and~\cite{Can-2017a}.
We restate the relevant result given by Can in~\cite{Can-2017a} and include the proof for completeness.
The decomposition in~\cite{Dehaene-physreva03} and~\cite{Maslov-arxiv17} are similar.

\begin{theorem}
\label{thm:Trung}
Any binary symplectic transformation $F$ can be expressed as
\begin{align*}
F = L_{Q_1} \Omega\, T_{R_1} G_k T_{R_2} L_{Q_2} ,
\end{align*}
as per the notation used in Table~\ref{tab:std_symp}, where the integer $k$, invertible matrices $Q_1, Q_2$ and symmetric matrices $R_1,R_2$ are chosen appropriately.
\begin{proof}
The idea is to perform row and column operations on the matrix $F$ via left and right multiplication by elementary symplectic transformations from Table~\ref{tab:std_symp}, and bring the matrix $F$ to the standard form $\Omega\, T_{R_1} \Omega$.

Let $F = \begin{bmatrix}
A & B \\
C & D
\end{bmatrix}$ so that $[ A \mid B ] \ \Omega \ [ A \mid B ]^T = 0$ and $[ C \mid D ] \ \Omega \ [ C \mid D ]^T = 0$.
If rank$(A) = k$ then there exists a row transformation $Q_{11}^{-1}$ and a column transformation $Q_2^{-1}$ such that
\begin{align}
Q_{11}^{-1} A Q_2^{-1} = 
\begin{bmatrix}
I_k & 0 \\
0 & 0
\end{bmatrix} .
\end{align}
Using the notation for elementary symplectic transformations discussed above, we apply $Q_{11}^{-1}$ and $L_{Q_2^{-1}}$ to $[ A \mid B ]$ and obtain
\begin{align}
\begin{bmatrix} Q_{11}^{-1} A & Q_{11}^{-1} B \end{bmatrix} \begin{bmatrix} Q_2^{-1} & 0 \\ 0 & Q_2^T \end{bmatrix} =
\left[ \begin{array}{cc|cc}
I_k & 0 & R_k & E' \\
0 & 0 & E & B_{n-k}
\end{array} \right] \coloneqq \begin{bmatrix} A' & B' \end{bmatrix} ,
\end{align}
where $B_{n-k}$ is an $(n-k) \times (n-k)$ matrix.
Since the above result is again the top half of a symplectic matrix, we have $[ A' \mid B' ] \ \Omega \ [ A' \mid B' ]^T = 0$ which implies $R_k$ is symmetric, $E = 0$ and hence rank$(B_{n-k}) = n-k$.
Therefore we determine an invertible matrix $Q_{n-k}$ which transforms $B_{n-k}$ to $I_{n-k}$ under row operations.
Then we apply $Q_{12}^{-1} \coloneqq \begin{bmatrix}
I_k & 0 \\
0 & Q_{n-k}
\end{bmatrix}$ on the left of the matrix $[ A \mid B ]$ to obtain
\begin{align}
\begin{bmatrix} Q_{12}^{-1} Q_{11}^{-1} A & Q_{12}^{-1} Q_{11}^{-1} B \end{bmatrix} \begin{bmatrix} Q_2^{-1} & 0 \\ 0 & Q_2^T \end{bmatrix} =
\left[ \begin{array}{cc|cc}
I_k & 0 & R_k & E' \\
0 & 0 & 0 & I_{n-k}
\end{array} \right] .
\end{align}
Now we observe that we can apply row operations to this matrix and transform $E'$ to $0$.
We left multiply by $Q_{13}^{-1} \coloneqq \begin{bmatrix}
I_k & E' \\
0 & I_{n-k}
\end{bmatrix}$ to obtain
\begin{align}
\begin{bmatrix} Q_{13}^{-1} Q_{12}^{-1} Q_{11}^{-1} A & Q_{13}^{-1} Q_{12}^{-1} Q_{11}^{-1} B \end{bmatrix} \begin{bmatrix} Q_2^{-1} & 0 \\ 0 & Q_2^T \end{bmatrix} =
\left[ \begin{array}{cc|cc}
I_k & 0 & R_k & 0 \\
0 & 0 & 0 & I_{n-k}
\end{array} \right] .
\end{align}
Since the matrix $R_2 \coloneqq \begin{bmatrix}
R_k & 0 \\
0 & 0
\end{bmatrix}$ is symmetric, we apply the elementary transformation $T_{R_2}$ from the right to obtain
\begin{align}
\left[ \begin{array}{cc|cc}
I_k & 0 & R_k & 0 \\
0 & 0 & 0 & I_{n-k}
\end{array} \right] 
\left[ \begin{array}{cc|cc}
I_k & 0 & R_k & 0 \\
0 & I_{n-k} & 0 & 0 \\
\hline
0 & 0 & I_k & 0 \\
0 & 0 & 0 & I_{n-k}
\end{array} \right] 
=
\left[ \begin{array}{cc|cc}
I_k & 0 & 0 & 0 \\
0 & 0 & 0 & I_{n-k}
\end{array} \right] .
\end{align}
Finally we apply the elementary transformation $G_k \Omega = \begin{bmatrix} 
U_k & L_{n-k} \\
L_{n-k} & U_k  
\end{bmatrix}$ to obtain
\begin{align}
\left[ \begin{array}{cc|cc}
I_k & 0 & 0 & 0 \\
0 & 0 & 0 & I_{n-k}
\end{array} \right]
\left[ \begin{array}{cc|cc}
I_k & 0 & 0 & 0 \\
0 & 0 & 0 & I_{n-k} \\
\hline
0 & 0 & I_k & 0 \\
0 & I_{n-k} & 0 & 0
\end{array} \right] 
=
\left[ \begin{array}{cc|cc}
I_k & 0 & 0 & 0 \\
0 & I_{n-k} & 0 & 0
\end{array} \right]
=
\begin{bmatrix}
I_n & 0
\end{bmatrix} .
\end{align}
Hence we have transformed the matrix $F$ to the form $\Omega \, T_{R_1} \Omega = \begin{bmatrix}
I_n & 0 \\
R_1 & I_n
\end{bmatrix}$, i.e., if we define $Q_1^{-1} \coloneqq Q_{13}^{-1} Q_{12}^{-1} Q_{11}^{-1}$ then we have
\begin{align}
L_{Q_{1}^{-1}} F L_{Q_2^{-1}} T_{R_2} G_k \Omega = \Omega \, T_{R_1} \Omega .
\end{align} 
Rearranging terms and noting that $L_{Q}^{-1} = L_{Q^{-1}},  \Omega^{-1} = \Omega, G_k^{-1} = G_k, T_{R_2}^{-1} = T_{R_2}$ we obtain
\begin{IEEEeqnarray*}{rCl+x*}
F & = & L_{Q_1} \Omega \, T_{R_1} \Omega^2 G_k T_{R_2} L_{Q_2} = L_{Q_1} \Omega \, T_{R_1} G_k T_{R_2} L_{Q_2} . & \qedhere
\end{IEEEeqnarray*} 
\end{proof}
\end{theorem}

\section{Stabilizer Codes}
\label{sec:stabilizer_codes}

Stabilizer codes were introduced by Gottesman~\cite{Gottesman-phd97} and independently by Calderbank, Rains, Shor and Sloane~\cite{Calderbank-it98*2} in the framework of \emph{additive codes over GF(4)}, the Galois field with $4$ elements.
These form the quantum analogues of classical linear codes.
Most works in the literature primarily consider either stabilizer codes or \emph{subsystem codes}~\cite{Bacon-pra06}.
The latter is a generalization of stabilizer codes that provides a more general partition of the underlying Hilbert space $\mathcal{H}$ as $\mathcal{H} = \mathcal{H}_A \otimes \mathcal{H}_B \oplus \mathcal{H}_C$.
If $\mathcal{H}_B$ is trivial then subsystem codes become stabilizer codes, where $\mathcal{H}_C = \mathcal{H}_A^{\perp}$.
Since stabilizer codes protect the information by storing it in the subspace $\mathcal{H}_A$, they are also referred to as \emph{subspace codes}.
Subsystem codes provide an additional ``gauge'' degree of freedom via the space $\mathcal{H}_B$ where information is not stored.
Thus they are defined by a \emph{gauge group} $\mathcal{G} \subset HW_N$ that is not commutative unless it is a stabilizer code, and the stabilizer group $S \subset \mathcal{G}$.
In this dissertation we will primarily consider stabilizer codes. 
Note that extending our results (and methods) to subsystem codes can be non-trivial and is not merely a technical exercise.

\begin{definition}
A stabilizer group $S$ is a commutative subgroup of the Pauli group $HW_N$ with Hermitian elements, and does not contain $-I_N$.
\end{definition}

If $S$ has dimension $r$, then it can be generated as $S = \langle \nu_i E(c_i,d_i) ; i = 1,\ldots,r \rangle$, where $\nu_i \in \{ \pm 1 \}$ and $E(c_i,d_i), E(c_j,d_j)$ commute for all $i \neq j, i,j \in \{1,\ldots,r\}$, i.e., $\syminn{[c_i,d_i]}{[c_j,d_j]} = c_i d_j^T + d_i c_j^T = 0$ (mod $2$).
Recollect that commuting $N \times N$ Hermitian matrices can be simultaneously diagonalized, and hence they have a common basis of eigenvectors that span $\mathbb{C}^N$.

\begin{definition}
Given a stabilizer group $S$, the corresponding stabilizer code~\cite{Nielsen-2010} is the subspace $V(S)$ spanned by all eigenvectors in the common eigenbasis of $S$ that have eigenvalue $+1$ with all elements in $S$, i.e., $V(S) \coloneqq \{ \ket{\psi} \in \mathbb{C}^N \colon g \ket{\psi} = \ket{\psi}\ \text{for\ all}\ g \in S \}$.
The subspace $V(S)$ is called an $\llbr n,k,d \rrbr$ stabilizer code, that encodes $k \coloneqq n-r$ logical qubits into $n$ physical qubits, and $d$ is the minimum weight of any operator in $\mathcal{N}_{HW_N}(S) \setminus S$.
\end{definition}

Here $\mathcal{N}_{HW_N}(S)$ denotes the normalizer of $S$ inside $HW_N$, i.e., 
\begin{align}
\label{eq:normalizer_S}
\mathcal{N}_{HW_N}(S) \coloneqq \{ \imath^{\kappa} E(a,b) \in HW_N \colon E(a,b) E(c,d) E(a,b) = E(c,d)\ \forall\ \nu E(c,d) \in S,\ \kappa \in \mathbb{Z}_4 \}.
\end{align}

\begin{definition}
Weight of a Pauli operator refers to the number of qubits on which it acts non-trivially, i.e., as $X, Z$ or $Y$.
More precisely, weight of $E(a,b)$ is $w_H(a\, \vee\, b)$, where $w_H(\cdot)$ denotes the Hamming weight and $\vee$ denotes the logical OR operation.
\end{definition}

If a stabilizer is generated as $S = \langle \nu_i E(c_i,d_i) ; i = 1,\ldots,r \rangle$, then a binary \emph{stabilizer generator} matrix or \emph{parity-check} matrix for the resulting stabilizer code is given by
\begin{align}
G_S \coloneqq \left[
\begin{array}{ccc|ccc}
\llongdash & c_1 & \rlongdash & \llongdash & d_1 & \rlongdash \\ \llongdash & c_2 & \rlongdash & \llongdash & d_2 & \rlongdash \\ & \vdots &  &  & \vdots & \\ \llongdash & c_r & \rlongdash & \llongdash & d_r & \rlongdash
\end{array} \right] \in \mathbb{F}_2^{r \times 2n}, \quad G_S\, \Omega \, G_S^T = 0.
\end{align}
The above condition holds for $G_S$ because all the $r$ rows have to commute pairwise, and this means their symplectic inner products must be $0$.
Note that $G_S$ ignores signs $\nu_i$.

Given a Hermitian Pauli matrix $E(c,d)$, it is easy to show that $\frac{I_N + \nu E(c,d)}{2}$ is the projector on to the $\nu$-eigenspace of $E(c,d)$, where $\nu \in \{ \pm 1 \}$.
Therefore, the projector on to the code subspace $V(S)$ of the stabilizer code defined by $S$ is given by
\begin{align}
\Pi_S \coloneqq \prod_{i=1}^{r} \frac{\left( I_N + \nu_i E(c_i,d_i) \right)}{2} = \frac{1}{2^r} \sum_{j = 1}^{2^r} \epsilon_j E(a_j, b_j).
\end{align}
Here $\epsilon_j \in \{ \pm 1 \}$ in the last equality is a character of the group $S$, and hence is determined by the product of signs of the generators of $S$ that produce $E(a_j,b_j)$, i.e., $\epsilon_j E(a_j,b_j) = \prod_{t \in J \subseteq \{1,\ldots,r\}} \nu_t E(c_t,d_t)$ for a unique subset $J$.
Hence, a feasible method to initialize the $n$ physical qubits in a code state is to start with an arbitrary state and then apply the above projector by measuring each stabilizer generator $\nu_i E(c_i,d_i)$.
If we also measure the \emph{logical} $Z$ (resp. $X$) generators, each of which corresponds to one of the $k$ logical qubits, then we can establish the logical $Z$ (resp. $X$) basis state in which the code has been initialized.
These measurements can be performed using the tools introduced in Chapter~\ref{ch:ch2_background} (see~\cite{Nielsen-2010}).

\subsection{Decoding a Stabilizer Code}

Using the result in Theorem~\ref{thm:Pauli_basis}, we can restrict to Pauli errors since all other errors are linear combinations of these.
Note that, even for a general quantum channel that does not just introduce a unitary error, the Kraus operators can be expressed using Pauli matrices; hence, this is indeed a statement on a general error channel~\cite[Section 2.6]{LB-2013}.
Assume that we were initially in a code state $\dket{\psi}$ and then a Pauli error $E(e,f)$ occurred.
Then using the identity~\eqref{eq:Eab_multiply} we observe that
\begin{align}
E(e,f) \dket{\psi} = E(e,f) \cdot \nu_i E(c_i,d_i) \dket{\psi} = (-1)^{\syminn{[e,f]}{[c_i,d_i]}} \nu_i E(c_i,d_i) \left( E(e,f) \dket{\psi} \right).
\end{align}
Hence the stabilizer generators for the state $E(e,f) \dket{\psi}$ are $\left\{ (-1)^{\syminn{[e,f]}{[c_i,d_i]}} \nu_i E(c_i,d_i) \right\}$.
In order to detect this change, we measure each stabilizer generator $\nu_i E(c_i,d_i)$ and observe the measurement result $(-1)^{\syminn{[e,f]}{[c_i,d_i]}}$.
These are the \emph{syndromes} for the error $E(e,f)$.
Note that these can also be computed via the product $G_S\, [f,e]^T = G_S\, \Omega\, [e,f]^T$.
Then a standard procedure, as in classical coding theory, is to identify the smallest weight error $E(e',f')$ such that $(-1)^{\syminn{[e',f']}{[c_i,d_i]}} = (-1)^{\syminn{[e,f]}{[c_i,d_i]}}$ for all $i$.
We apply the \emph{correction operator} $E(e',f')$ to return back to the code subspace.
If $E(e',f') E(e,f) \in S$ then we have corrected the error successfully. 
Otherwise, we have returned to an incorrect state since $E(e',f') E(e,f) \in \mathcal{N}_{HW_N}(S) \setminus S$.
This is because, by definition, the only operators that commute with all stabilizers, but are not themselves stabilizers, belong in $\mathcal{N}_{HW_N}(S) \setminus S$.
In other words, these are \emph{undetectable errors} and they define the distance of the stabilizer code.
We will not discuss decoding much more since this dissertation is primarily about synthesizing logical operators for stabilizer codes.
For more details, see~\cite{Nielsen-2010,LB-2013}.

\subsection{Logical Qubits and their Operators}
\label{sec:logical_qubits}

We introduced a qubit in Chapter~\ref{ch:ch2_background} as a normalized vector in $\mathbb{C}^2$.
Although this arises directly from the postulates of quantum mechanics, it is useful to think of a qubit abstractly as a pair of bases $(X,Z)$.
More precisely, note that a qubit $\dket{\psi}$ can be expressed as
\begin{align}
\dket{\psi} = \alpha \dket{0} + \beta \dket{1} = (\alpha_I I_2 + \alpha_X X + \imath \alpha_Z Z + \alpha_Y Y) \dket{0},
\end{align}
where $\alpha, \beta \in \mathbb{C}$ but $\alpha_I, \alpha_X, \alpha_Z, \alpha_Y \in \mathbb{R}$ satisfying constraints such that the above linear combination is unitary.
If any operation $U$ is to be performed on $\dket{\psi}$, then its effect can be calculated by understanding its action on just $X$ and $Z$, i.e.,
\begin{align}
U \dket{\psi} = U (\alpha_I I_2 + \alpha_X X + \imath \alpha_Z Z + \alpha_Y Y) \dket{0} = \left( U (\alpha_I I_2 + \alpha_X X + \imath \alpha_Z Z + \alpha_Y Y) U^{\dagger} \right) U \dket{0}.
\end{align}
Here $U \dket{0}$ is a \emph{deterministic} state, given $U$, and all the information is gathered by understanding the conjugations $UXU^{\dagger}$ and $UZU^{\dagger}$ (since $Y = \imath X Z$).
The only requirements for the abstract bases $X$ and $Z$ are that $XZ = -ZX$ on the same qubit, and that $X$ and $Z$ on different qubits commute.
This abstraction is reminiscent of the difference between the \emph{Schr{\"o}dinger} perspective, which tracks the evolution of states, and the \emph{Heisenberg} perspective, which tracks the evolution of operators (i.e., bases here).
In other words, after $U$ is applied, the new pair of bases is $(UXU^{\dagger}, UZU^{\dagger})$, and this defines the evolved qubit.
If we only considered one of the two bases, say $X$, then the system is \emph{classical}.

Hence, once we define a $\llbr n,k,d \rrbr$ stabilizer code, we need to define the $k$ logical qubits by identifying $k$ pairs of bases $(\bar{X}_i,\bar{Z}_i), i = 1,\ldots,k$, that satisfy $\bar{X}_i \bar{Z}_j = (-1)^{\delta_{ij}} \bar{Z}_j \bar{X}_i$.
These operators also need to commute with all stabilizers.
Here, $\bar{X}_i$ and $\bar{Z}_i$ are $n$-qubit operators that form physical representatives of the ($k$-qubit) logical $X_i^L$ and $Z_i^L$, where $X_i^L$ and $Z_i^L$ are simply the Pauli $X$ and $Z$ gates on the $i$-th logical qubit.
For brevity, $\bar{X}_i$ and $\bar{Z}_i$ are themselves called the \emph{logical Pauli} operators.
These can be determined by using either Gottesman's~\cite{Gottesman-phd97} or Wilde's algorithm~\cite{Wilde-physreva09}.
Therefore, a particular logical computational basis state $\dket{x_1 x_2 \cdots x_k}_L, x_i \in \{0,1\}$, is defined by the $n$-dimensional (maximal) commutative group $\langle \nu_1 E(c_1,d_1), \ldots, \nu_r E(c_r,d_r), (-1)^{x_1} \bar{Z}_1, (-1)^{x_2} \bar{Z}_2, \ldots, (-1)^{x_k} \bar{Z}_k \rangle$.

With this perspective, let us revisit the procedure to initialize a system into a code state.
First we start in an arbitrary $n$-qubit state and then measure all stabilizer generators.
This projects the state onto the code subspace, up to an overall Pauli error inferred from the syndromes, but we do not know the equivalent logical $k$-qubit state.
We can apply a Pauli correction based on the syndromes to return exactly into the code subspace.
Finally, we measure the logical $Z$ operators $\bar{Z}_i$ for all $i$ and note down the measurement results $(-1)^{x_i}$.
These values dictate the state $\ket{x_1 x_2 \cdots x_k}_L$ in which we have initialized the system.

Beyond logical Pauli operators, we need to be able to perform arbitrary quantum computation on the $k$ logical qubits.
A necessary condition for a physical (i.e., $n$-qubit) operator $U$ to induce a logical operation is to preserve the code subspace.
This translates to the condition that $U$ commute with the code projector, i.e., 
\begin{align}
U \Pi_S U^{\dagger} = \Pi_S \Rightarrow \frac{1}{2^r} \prod_{i=1}^{r} \left( I_N + \nu_i U E(c_i,d_i) U^{\dagger} \right) = \frac{1}{2^r} \sum_{j=1}^{2^r} \epsilon_j E(a_j,b_j).
\end{align}
If $U \in HW_N$ then this condition effectively implies that $U \in \mathcal{N}_{HW_N}(S)$, since Paulis commute or anti-commute.
If $U \in \text{Cliff}_N$ then this condition implies that $U$ must normalize the stabilizer, i.e., $U \in \mathcal{N}_{\text{Cliff}_N}(S)$.
We will use this condition in Chapter~\ref{ch:ch5_lcs_algorithm} for synthesizing logical Clifford gates.
For non-Clifford operations $U$, one needs to understand their action on Pauli matrices and use that to solve the above equality.
This is the approach we will take in Chapter~\ref{ch:ch7_stabilizer_codes_qfd} for the case when $U$ is composed of $T$ and $T^{\dagger}$ gates.
Finally, the induced logical operation can be determined from $U \bar{X}_i U^{\dagger}, U \bar{Z}_i U^{\dagger}$ (for all $i$).

\subsection{Calderbank-Shor-Steane (CSS) Codes}

A \emph{CSS code} is a special type of stabilizer code defined by a stabilizer $S$ whose generators split into strictly $X$-type and strictly $Z$-type operators.
Consider two classical binary codes $C_1,C_2$ such that $C_2 \subset C_1$, and let $C_1^{\perp}, C_2^{\perp}$ represent their respective dual codes ($C_1^{\perp} \subset C_2^{\perp}$).
Define the stabilizer $S \coloneqq \langle \nu_c E(c,0), \nu_d E(0,d), c \in C_2, d \in C_1^{\perp} \rangle$ for some suitable $\nu_c, \nu_d \in \{ \pm 1 \}$.
Let $C_1$ be an $[n,k_1]$ code and $C_2$ be an $[n,k_2]$ code such that $C_1$ and $C_2^{\perp}$ can correct up to $t$ errors.
Then $S$ defines an $\llbr n, k_1-k_2, \geq 2t+1 \rrbr$ CSS code that we will represent as CSS($X,C_2 ; Z, C_1^{\perp}$).
We say that the distance is \emph{at least} $2t+1$ because the distance of the code is the minimum weight of any vector in $(C_1 \setminus C_2) \cup (C_2^{\perp} \setminus C_1^{\perp})$, and not just $C_1 \cup C_2^{\perp}$.
These spaces $(C_1 \setminus C_2)$ and $(C_2^{\perp} \setminus C_1^{\perp})$ exactly form the pure $X$-type and pure $Z$-type parts of $\mathcal{N}_{HW_N}(S) \setminus S$, respectively, for the CSS code; these and their combinations that form mixed-type Paulis constitute all of $\mathcal{N}_{HW_N}(S) \setminus S$.
If $G_2$ and $G_1^{\perp}$ represent generator matrices for the codes $C_2$ and $C_1^{\perp}$, respectively, then a binary parity-check matrix for CSS($X,C_2 ; Z, C_1^{\perp}$) can be written as
\begin{align}
\setlength\aboverulesep{0pt}\setlength\belowrulesep{0pt}
    \setlength\cmidrulewidth{0.5pt}
G_S \coloneqq 
\begin{blockarray}{c c c}
 n & n &   \\
\begin{block}{[c | c] c}
\hspace*{1cm} & G_1^{\perp} & n - k_1 \\
\cmidrule(lr){1-2}
G_2 & \hspace*{1cm} & k_2 \\
\end{block}
\end{blockarray}.
\end{align}

For CSS codes it is also possible to write down a mapping from the logical computational basis states $\{ \dket{x_1 x_2 \cdots x_k}_L \colon x_i \in \{0,1\} \ \text{for all}\ i = 1,\ldots,k \}$ into the physical code states, i.e., the definition of an encoder for the code.
In order to do this, it will be convenient to split the generator matrix for $C_1$ into the generator matrix for $C_2 \subset C_1$ and the generator matrix for the coset representatives of the quotient group $C_1/C_2$, i.e.,
\begin{align}
G_1 = 
\begin{bmatrix}
G_{C_1/C_2} \\ G_2
\end{bmatrix}.
\end{align}
Now we can define an encoding map as follows, where $x = [x_1,\ldots,x_k] \in \mathbb{F}_2^k$.
\begin{align}
\dket{x_1 x_2 \cdots x_k}_L \mapsto \dket{\psi_x} \equiv
\dket{x \cdot G_{C_1/C_2} \oplus C_2} & \coloneqq \frac{1}{\sqrt{|C_2|}} \sum_{c \in C_2} \dket{x \cdot G_{C_1/C_2} \oplus c} \\
  & = \frac{1}{\sqrt{|C_2|}} \sum_{y \in \mathbb{F}_2^{k_2}} \dket{x \cdot G_{C_1/C_2} \oplus y \cdot G_2}.
\end{align}
Since the cosets of $C_2$ in $C_1$ are distinct for distinct $x$, it can be easily verified that $\dbraket{\psi_x}{\psi_{x'}} = 0$ for all $x \neq x'$.
Therefore, any code state of the code CSS($X, C_2 ; Z, C_1^{\perp}$) is a (complex) linear combination of the states $\{ \dket{\psi_x} \colon x \in \mathbb{F}_2^k \}$.

For decoding a CSS code, we need to measure the syndrome for an error $E(e,f)$ as $G_S\, \Omega\, [e,f]^T$. 
Hence, we can use the classical parity-check matrix of $C_2^{\perp}$ to correct $Z$ errors and the classical parity-check matrix of $C_1$ to correct $X$ errors.
This procedure ignores correlations in $X$ and $Z$ errors that can cause $Y$ errors, and several works in the literature have studied such general Pauli error models.

For a CSS code, the logical Pauli operators can also be constructed from the component classical codes.
We can define the rows of the coset generator matrix $G_{C_1/C_2}$ to be the $k$ logical $X$ generators, i.e., $\bar{X}_i \coloneqq E(x_i,0)$ where $x_i$ is the $i$-th row of $G_{C_1/C_2}$.
Then we can determine a coset generator matrix $G_{C_2^{\perp}/C_1^{\perp}}$ and define its rows to be the $k$ logical $Z$ generators, i.e., $\bar{Z}_i \coloneqq E(0,z_i)$ where $z_i$ is the $i$-th row of $G_{C_2^{\perp}/C_1^{\perp}}^T$. 
This matrix needs to be such that $G_{C_1/C_2} \cdot G_{C_2^{\perp}/C_1^{\perp}}^T = I_k$, which is required because the logical Paulis must satisfy 
\begin{align}
\bar{X}_i \bar{Z}_j = (-1)^{\delta_{ij}} \bar{Z}_j \bar{X}_i \Rightarrow \syminn{[x_i,0]}{[0,z_j]} = x_i z_j^T = \delta_{ij}.
\end{align}
For a more detailed discussion on logical Paulis for CSS codes, see~\cite{Rengaswamy-arxiv18*2}.
Although this is not the first work on this subject, in~\cite{Rengaswamy-arxiv18*2} we have elaborated on the above perspective in a rigorous fashion.

The CSS formalism is well-established to translate classical codes into stabilizer codes. 
In Chapter~\ref{ch:ch7_stabilizer_codes_qfd}, we will refine the construction to introduce \emph{CSS-T} codes where a transversal $T$ gate induces fault-tolerant logical (non-Clifford) operations.

\subsection{Fault-Tolerance and Transversality}

The goal of coded quantum computation is to be able to perform arbitrary quantum computations on the $k$ encoded qubits of a code by implementing the relevant $n$-qubit physical operations in an error-resilient manner.
The notion of \emph{fault-tolerance} formalizes this idea and we need to perform logical operations and error-correction fault-tolerantly in a universal fault-tolerant quantum computer.

\begin{definition}
Given a $\llbr n,k,d \rrbr$ quantum error-correcting code, a fault-tolerant physical circuit for a logical operation or error-correction must ensure that any $t \leq \lfloor \frac{d-1}{2} \rfloor$ Pauli errors in the circuit remain correctable after the procedure.
\end{definition}

The above definition implicitly uses the identities in Table~\ref{tab:circuit_identities} to track Pauli errors through standard gates in a quantum circuit.
If the physical circuit is Clifford, as is the case for error-correction or logical Clifford gates for stabilizer codes, then we can track Pauli errors in the middle of the circuit using the above identities to arrive at an effective Pauli error at the end of the circuit.
Fault-tolerance implies that this effective Pauli error should have weight at most $t$, or at least remain correctable if errors in the circuit have a specific pattern as in the case of \emph{flag fault-tolerant computation}~\cite{Chao-arxiv17a,Chao-arxiv17b,Tansuwannont-arxiv18,Chao-arxiv19}.
The simplest fault-tolerant physical circuit is a \emph{transversal} operation.

\begin{definition}
A transversal operation is one that uses separate operations on the $n$ physical qubits.
If $m > 1$ code blocks are involved, then a transversal operation acts separately on each $m$-tuple of corresponding qubits across the $m$ blocks.
\end{definition}

For example, if we had two code blocks and performed $n$ CNOT operations between the corresponding qubits on the two blocks, then a single error in any code block remains at most a single error in each code block.
In an ideal world we will be able to perform universal quantum computation just using transversal physical operations.
However, this is forbidden by the Eastin-Knill theorem~\cite{Eastin-prl09}. 
Thus, other fault-tolerant methods are needed to complement transversality.
For example, CSS codes with $C_2 = C_1^{\perp}$ can often implement all logical Clifford gates transversally but not the logical $T$ gate.

%% file: ch5_lcs_algorithm.tex

\label{ch:ch5_lcs_algorithm}

\section{Motivation}

The most common approach to building a universal fault-tolerant quantum computer (UFTQC) involves choosing a stabilizer (or subsystem) code to encode the information and then performing coded quantum computation in a fault-tolerant manner\footnote{Part of this work was presented at the 2018 IEEE International Symposium on Information Theory~\cite{Rengaswamy-isit18}.}.
In this dissertation we only consider stabilizer codes.
Once a particular $\llbr n,k,d \rrbr$ stabilizer code is chosen, one needs to find fault-tolerant $n$-qubit physical operators that realize a generating set for universal quantum computation on the $k$ logical qubits.
While fault-tolerance is very important, it is also important to have code parameters $(n,k,d)$ that minimize (qubit) overhead and to have efficient decoders that correct the most likely errors.
In this regard, it is often useful to think beyond individual codes and instead think of \emph{families of codes}. 
This enables us to analyze the scaling of code parameters $(\frac{k}{n}, \frac{d}{n})$ and decoder complexity for diverging $n$.
Hence, when we consider such code families, we need a systematic approach to translate logical operators to physical operators that works beyond operator optimization for individual codes.
In this chapter, we propose such an approach for synthesizing logical \emph{Clifford} operators for stabilizer codes.

For the task of synthesizing the logical Pauli operators for stabilizer codes, the first algorithm was introduced by Gottesman~\cite[Sec.~4]{Gottesman-phd97} and subsequently, another algorithm based on symplectic geometry was proposed by Wilde~\cite{Wilde-physreva09}. 
The latter is closely related to earlier work by Brun et al.~\cite{Brun-science06,Brun-it14}.
Since the logical Paulis are inputs to our algorithm that synthesizes logical Clifford operators for stabilizer codes, we will consider the above two procedures to be ``preprocessors'' for our algorithm.

Given the logical Pauli operators for an $\llbr n,k \rrbr$ stabilizer code, as we discussed in Chapter~\ref{ch:ch4_groups}, physical Clifford realizations of Clifford operators on the logical qubits can be represented by $2n \times 2n$ binary symplectic matrices, thereby reducing the complexity \emph{dramatically} from $2^{2n}$ complex variables to $4n^2$ binary variables (see~\cite{Calderbank-physrevlett97,Gottesman-arxiv09}).
We exploit this fact to propose an algorithm that efficiently assembles \emph{all} $2^{r(r+1)/2}$, where $r = n-k$, symplectic matrices representing physical Clifford operators (circuits) that realize a given logical Clifford operator on the protected qubits.
We will refer to this procedure as the \emph{Logical Clifford Synthesis (LCS) algorithm}.
Here, each symplectic solution represents an equivalence class of Clifford circuits, all of which ``propagate'' input Pauli operators through them in an identical fashion.
As we will discuss later in the context of the algorithm, the degrees of freedom not exploited by our algorithm are those provided by stabilizers (see Remark~\ref{rem:stab_freedom}). 
But, at the cost of some increased computational complexity, the algorithm can easily be modified to account for these stabilizer degrees of freedom.
Hence, our work makes it possible to optimize the choice of circuit with respect to a suitable metric, that might be a function of the quantum hardware.

We note that there are several works that focus on exactly decomposing, or approximating, an arbitrary unitary operator as a sequence of operators from a fixed \emph{instruction set}, such as Clifford + $T$~\cite{Kliuchnikov-prl13,Amy-tcad13,Maslov-arxiv17,Fagan-eptcs19,Iten-arxiv19,Duncan-arxiv19}.
However, these works do not consider the problem of circuit synthesis or optimization over different realizations of unitary operators on the \emph{encoded} space.
We also note that there exists several works in the literature that study this problem for specific codes and operations, e.g., see~\cite{Gottesman-phd97,Bacon-pra06,Fowler-arxiv12,Grassl-isit13,Kubica-pra15,Yoder-arxiv17,Chao-arxiv17b}. 
However, we believe our work is the first to propose a systematic framework to address this problem for general stabilizer codes, and hence enable automated circuit synthesis for encoded Clifford operators.
This procedure is more systematic in considering all degrees of freedom than conjugating the desired logical operator by the encoding circuit for the QECC.

Recently, we have used the LCS algorithm to translate the unitary $2$-design we constructed from classical Kerdock codes into a \emph{logical} unitary $2$-design~\cite{Can-arxiv19}, and in general any design consisting of only Clifford elements can be transformed into a logical design using our algorithm. 
An implementation of the design is available at: \url{https://github.com/nrenga/symplectic-arxiv18a}.
This finds direct application in the \emph{logical randomized benchmarking} protocol proposed by Combes et al.~\cite{Combes-arxiv17}.
This protocol is a more robust procedure to estimate logical gate fidelities than extrapolating results from randomized benchmarking performed on physical gates~\cite{Magesan-physreva12}.
Now we discuss some more motivations and potential applications for the LCS algorithm.


\subsection{Noise Variation in Quantum Systems}

\begin{figure}
\begin{center}

\scalebox{1}{%
\begin{tikzcd}
\lstick{$1$} & \gate{P} & \ctrl{1} & \ctrl{2} & \ctrl{3} & \qw & \qw \\
\lstick{$2$} & \gate{P} & \control{} & \qw & \qw & \qw & \qw \\
\lstick{$3$} & \gate{P} & \qw & \control{} & \qw & \qw & \qw \\
\lstick{$4$} & \gate{P} & \qw & \qw & \control{} & \gate{Z} & \qw
\end{tikzcd}
$\equiv$
\begin{tikzcd}
\lstick{$1$} & \qw & \qw & \qw & \ghost{Z}\qw & \qw \\
\lstick{$2$} & \ctrl{1} & \ctrl{2} & \qw & \ghost{Z}\qw & \qw \\
\lstick{$3$} & \control{} & \qw & \ctrl{1} & \ghost{Z}\qw & \qw \\
\lstick{$4$} & \qw & \control{} & \control{} & \gate{Z} & \qw 
\end{tikzcd}
}

\caption{\label{fig:cz_412}Two physical circuits that realize the CZ gate on the two logical qubits of the $\llbr 4,2,2 \rrbr$ code.}
\vspace*{-0.35cm}
\end{center}
\end{figure}
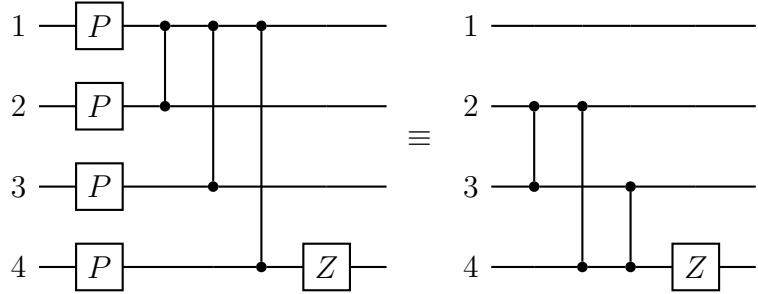

Although depth or the number of two-qubit gates might appear to be natural metrics for optimization, near-term quantum computers can also benefit from more nuanced metrics depending upon the physical system.
For example, it is now established that the noise in the \emph{IBM Q Experience} computers varies widely among qubits and also with time, and that circuit optimizations might have to be done in regular time intervals in order to exploit the current noise characteristics of the hardware~\cite{Murali-arxiv19}.
In such a scenario, if we need to implement a specific logical operator at the current time, and if it is the case that some specific qubits or qubit-links in the system are particularly unreliable, then it might be better to sacrifice depth and identify an equivalent logical operator that avoids those qubits or qubit-links (if possible).
As an example, for the well-known $\llbr 4,2,2 \rrbr$ code~\cite{Gottesman-phd97,Chao-arxiv17b}, whose stabilizer group is generated as $S = \langle X_1 X_2 X_3 X_4, Z_1 Z_2 Z_3 Z_4 \rangle$, two implementations of the logical controlled-$Z$ ($\lcz{1}{2}$) operation on the two logical qubits are shown in Fig.~\ref{fig:cz_412}.
The logical Pauli operators in this case are $\lX_1 = X_1 X_2, \lX_2 = X_1 X_3, \lZ_1 = Z_2 Z_4, \lZ_2 = Z_3 Z_4$.

Assuming that single-qubit gates do not contribute to complexity (or difficulty of implementation), we observe that both choices have the same number of two-qubit gates and depth.
More interestingly, we see that the second choice completely avoids the first physical qubit while realizing the same logical CZ operation.
Therefore, if either the first qubit itself has poor fidelity or coupling to it does, then clearly the second choice is more appropriate.
Preliminary experiments on the IBM system confirm this advantage when qubits are mapped appropriately.
Note that even if we use a QECC that protects a single qubit but has a transversal CZ implementation, i.e., the logical CZ is a CZ between corresponding physical qubits in two \emph{separate} code blocks, this incurs a larger overhead than the above scheme. 
We identified this example by using our open-source implementation of our LCS algorithm, that is available at: \url{https://github.com/nrenga/symplectic-arxiv18a}.
In order to identify (or construct) more interesting codes that exhibit a ``rich'' set of choices for each logical operator, one needs a better understanding of the geometry of the space of symplectic solutions. 
We believe this is an important open problem arising from our work.



For near-term \emph{NISQ} (Noisy Intermediate-Scale Quantum~\cite{Preskill-nisq18}) era of quantum computers, a lot of current research is focused on equipping compilers with routines that optimize circuits for depth and two-qubit gates, and the mapping of qubits from the algorithm to the hardware, while all taking into account the specific characteristics and noise in the hardware~\cite{Paler-arxiv18,Shi-arch19,Murali-micro19,Nash-arxiv19,Murali-arxiv19}.
Although employing QECCs is considered to be beyond the NISQ regime, exploiting simple codes such as the $\llbr 4,2,2 \rrbr$ code and using post-selection provides increased reliability than uncoded computation (as Harper and Flammia have demonstrated~\cite{Harper-prl19,Linke-sciadv17}). 
Therefore, our efficient LCS algorithm might find an application in such quantum compilers, where the utility is to determine the best physical realization of a logical operator with respect to current system characteristics.
Specifically, this allows \emph{dynamic compilation} (i.e., during program execution) that could provide significant reliability gains in practice.

In light of such applications, our software currently allows one to determine only one physical realization in cases where the number of solutions is prohibitively large, specifically for QECCs with large-dimension stabilizers $(r = n-k \gg 1)$. 
However, this single solution does not come with any explicit guarantees regarding depth or number of two-qubit gates or avoiding certain physical qubits.
Therefore, even developing heuristics to directly optimize for a ``good enough'' solution, instead of assembling all solutions and searching over them, will have a significant impact on the efficiency of compilers.

\subsection{QECCs for Universal Quantum Computation}

Physical single-qubit rotation gates on trapped-ion qubits are natural, reliable and have a long history~\cite{Ozeri-cphy11}.
Recently, it has also been observed that small-angle M{\o}lmer-S{\o}rensen gates, i.e., $XX_{ij}(\theta) = \cos\frac{\theta}{2} \cdot I_4 - \imath \sin\frac{\theta}{2} \cdot X_i X_j$ for small $\theta$, are more reliable than the maximally-entangling $XX_{ij}(\frac{\pi}{2})$ gate~\cite{Nam-arxiv19}.
Since these are the primitive operations in trapped-ion systems~\cite{Linke-nas17}, codes that support a transversal $T = \text{diag}(1,\exp(\frac{\imath\pi}{4}))$ gate, such as tri-orthogonal codes~\cite{Bravyi-pra12}, could be directly used for computation rather than being dedicated for expensive magic state distillation~\cite{Bravyi-pra12,Haah-quantum17,Haah-quantum17b,Gidney-arxiv18}.
However, it is well-known that there exists no single QECC that supports a universal set of gates where all of them have a transversal implementation at the logical level~\cite{Zeng-it07,Eastin-prl09,Newman-arxiv17}.
Therefore, there is a natural tradeoff between exploiting transversality for logical non-Clifford operations versus Clifford operations.

Indeed, this will be a realistic alternative only if the logical Clifford operations on these codes are ``error-resilient'', by which we mean that for at least constant-depth circuits, the most likely errors remain correctable and do not propagate catastrophically through the Clifford sections of these logical circuits.
For this purpose, our LCS algorithm can be a supportive tool to investigate properties of stabilizer QECCs that guarantee error-resilience of their logical Clifford operators.
Note that constant-depth circuits have been shown to provide a quantum advantage over classical computation~\cite{Bravyi-science18}.
In fact, it has been shown that the advantage persists even if those circuits are noisy~\cite{Bravyi-arxiv19}, and the proof involves a QECC which admits constant-depth logical Clifford gates.

Next we begin by demonstrating our LCS algorithm on a simple $\llbr 6,4,2 \rrbr$ code.

\section{LCS for a $\llbr 6,4,2 \rrbr$ CSS Code}
\label{sec:css642}

As stated in Section~\ref{sec:clifford_gp}, the logical Clifford group $\text{Cliff}_{2^4}$ is generated by the operations $g^L \in \{ HW_{2^4}, P^L, H^L, \text{CZ}^L, \text{CNOT}^L \}$.
We will first discuss the construction of the stabilizer $S$ and the physical implementations $\bar{h} \in HW_{2^6}$ of the logical Pauli operators $h^L \in HW_{2^4}$ for this code. 
These are synthesized using the generator matrices of the classical codes from which the $\llbr 6,4,2 \rrbr$ code is constructed.
Then we will demonstrate our algorithm by determining the symplectic matrices corresponding to the physical equivalents  $\bar{g} \in \{ \lP_1, \lH_1, \lcz{1}{2}, \lcnot{2}{1} \}$ of the above generating set, where the subscripts indicate the logical qubit(s) involved in the logical operations realized by these physical operators.
The Clifford gates on other logical qubits can be synthesized via a similar procedure.

\subsection{Stabilizer of the Code}
\label{sec:css642_stabilizer}

The $[6,5,2]$ classical binary single parity-check code $\MCC$ is generated by $G_{\MCC}$, where
\begin{equation}
\label{eq:Cgen}
G_{\MCC} =
\begin{bmatrix}
H_{\MCC} \\
G_{\MCC/\MCCd}
\end{bmatrix}
\ ; \
G_{\MCC/\MCCd} \coloneqq
\begin{bmatrix}
1 & 1 & 0 & 0 & 0 & 0 \\
1 & 0 & 1 & 0 & 0 & 0 \\
1 & 0 & 0 & 1 & 0 & 0 \\
1 & 0 & 0 & 0 & 1 & 0
\end{bmatrix} 
 ,
\end{equation}
and $H_{\MCC} = [1\ 1\ 1\ 1\ 1\ 1]$ is the parity-check matrix for $\MCC$.
Hence, the dual (repetition) code $\MCCd = \{ 000000, 111111 \}$ of $\MCC$ is generated by the matrix $H_{\MCC}$.
The rows $h_i$ of $G_{\MCC/\MCCd}$, for $i=1,2,3,4$, generate all coset representatives for $\MCCd$ in $\MCC$, which determine the physical states of the code.
The CSS construction~\cite{Calderbank-physreva96,Steane-physreva96,Nielsen-2010} provides an $\llbr n,k \rrbr = \llbr 6,4 \rrbr$ stabilizer code $\MCQ$ spanned by the set of basis vectors $\{ \ket{\psi_x} \mid x \in \mathbb{F}_2^4 \}$, where $x \coloneqq [x_1,\ x_2,\ x_3,\ x_4]$ and
\begin{equation}
\label{eq:css_state}
\ket{\psi_x} \coloneqq
\frac{1}{\sqrt{2}} \ket{(000000) + \sum_{j=1}^{4} x_j h_j} + \frac{1}{\sqrt{2}} \ket{(111111) + \sum_{j=1}^{4} x_j h_j} . 
\end{equation}
Let $X_t$ and $Z_t$ denote the $X$ and $Z$ operators, respectively, acting on the $t$-th physical qubit.
Then the physical operators defined by the row of $H_{\MCC}$ are
\begin{align} 
\label{eq:642stabilizer}
\bg^X = E(111111,000000) = X_1 X_2 \cdots X_6\ , \  \bg^Z = E(000000,111111) = Z_1 Z_2 \cdots Z_6 .
\end{align}
These generate the stabilizer group $S$ that determines $\MCQ$.
The notation $X_1 X_2 \cdots X_6$ is commonly used in the literature to represent $X \otimes X \otimes \cdots \otimes X$.
If subscript $i \in \{1,\ldots,n\}$ is omitted, then it implies that the operator $I_2$ acts on the $i$-th qubit.


For the generating set $g^L \in \{ X^L, Z^L, P^L, H^L, \text{CZ}^L, \text{CNOT}^L \}$ of logical Clifford operators $\text{Cliff}_{2^4}$, we now synthesize their corresponding physical operators $\bar{g}$ that realize the action of $g^L$ on the protected qubits.
Since the operator $\bar{g}$ must also preserve $\MCQ$, conjugation by $\bar{g}$ must preserve both the stabilizer $S$ and hence its normalizer $S^{\perp}$ in $HW_N$~\cite{Calderbank-it98*2}.
We note that $\bar{g}$ need not commute with every element of the stabilizer $S$, i.e., \emph{centralize} $S$, although this can be enforced if necessary (see Theorem~\ref{thm:normalize_centralize}).

\subsection{Logical Paulis}
\label{sec:css_pauli}

Let $\ket{x}_L,\ x = [x_1,x_2,x_3,x_4] \in \mathbb{F}_2^4,$ be the logical state protected by the physical state $\ket{\psi_x}$ defined in~\eqref{eq:css_state}. 
Then the generating set $\{ X_j^L, Z_j^L \in HW_{2^4} \mid j=1,2,3,4 \}$ is defined by
\begin{align}
X_j^L \ket{x}_L = \ket{x'}_L ,\ & \text{where}\ x_i' = \begin{cases} x_j \oplus 1 &,\  \text{if}\ i=j \\ x_i &,\ \text{if}\ i \neq j \end{cases} \ \ 
\text{and}\ \  Z_j^L \ket{x}_L = (-1)^{x_j} \ket{x}_L .
\end{align}
We denote their corresponding physical operators as $\lX_j$ and $\lZ_j$, respectively.
The rows of
\begin{align}
\label{eq:GCZ}
G_{\MCC/\MCCd}^X \coloneqq G_{\MCC/\MCCd} =
\begin{bmatrix}
1 & 1 & 0 & 0 & 0 & 0 \\
1 & 0 & 1 & 0 & 0 & 0 \\
1 & 0 & 0 & 1 & 0 & 0 \\
1 & 0 & 0 & 0 & 1 & 0
\end{bmatrix} \ \text{and} \ \ 
G_{\MCC/\MCCd}^Z \coloneqq
\begin{bmatrix}
0 & 1 & 0 & 0 & 0 & 1 \\
0 & 0 & 1 & 0 & 0 & 1 \\
0 & 0 & 0 & 1 & 0 & 1 \\
0 & 0 & 0 & 0 & 1 & 1
\end{bmatrix} 
\end{align}
are used to define these physical implementations $\lX_j, \lZ_j, j=1,2,3,4$ as follows.
\begin{equation}
\label{eq:css_pauli}
\begin{array}{cc|cc}
\lX_1 \coloneqq E(110000,000000) = X_1 X_2 & \quad & \quad & \lZ_1 \coloneqq E(000000,010001) = Z_2 Z_6 \\
\lX_2 \coloneqq E(101000,000000) = X_1 X_3 & \quad & \quad & \lZ_2 \coloneqq E(000000,001001) = Z_3 Z_6 \\
\lX_3 \coloneqq E(100100,000000) = X_1 X_4 & \quad & \quad & \lZ_3 \coloneqq E(000000,000101) = Z_4 Z_6 \\
\lX_4 \coloneqq E(100010,000000) = X_1 X_5 & \quad & \quad & \lZ_4 \coloneqq E(000000,000011) = Z_5 Z_6 
\end{array} . 
\end{equation}
Although these are the physical realizations of logical Pauli operators, it is standard practice in the literature to refer to $\lX_j, \lZ_j$ itself as the logical Pauli operators.
These operators commute with every element of the stabilizer $S$ and satisfy, as required, $\lX_i \lZ_j = (-1)^{\delta_{ij}} \lZ_j \lX_i$.
Note that this is a translation of the commutation relations between $X_j^L$ and $Z_j^L$ as discussed in Section~\ref{sec:logical_qubits}.
In general, to define valid logical Pauli operators, we need 
\begin{equation}
G_{\MCC/\MCCd}^X \left( G_{\MCC/\MCCd}^Z \right)^T = I_{k} ,\ \ \text{and}\ \ 
G_{\MCC}^Z =
\begin{bmatrix}
H_{\MCC} \\
G_{\MCC/\MCCd}^Z
\end{bmatrix}
\end{equation} 
must form another generator matrix for the (classical) code $\MCC$.
It can be verified that the above matrices satisfy these conditions and hence the operators in~\eqref{eq:css_pauli} indeed form a generating set for all logical Pauli operators.
Note that $S^{\perp}$ is generated by $S, \lX_i, \lZ_i$, i.e., $S^{\perp} = \langle S, \lX_i, \lZ_i ; i=1,2,3,4 \rangle$ (see~\cite{Gottesman-arxiv09}).
This completes the synthesis of logical Paulis.

Now we discuss the synthesis of physical operators $\bar{g} \in \{ \lP_1, \lH_1, \lcz{1}{2}, \lcnot{2}{1} \}$ corresponding to the generating set $\{ HW_{2^4}, P^L, H^L, \text{CZ}^L, \text{CNOT}^L \}$ for $\text{Cliff}_{2^4}$.

\subsection{Logical Phase Gate}
\label{sec:css_phase}

The phase gate $\bar{g} = \lP_1$ on the first logical qubit is defined by the actions (Section~\ref{ch:ch2_background})
\begin{align}
\label{eq:phase}
\lP_1 \lX_j \lP_1^{\dagger} & = 
\begin{cases}
\lY_j & \text{if}\ j=1, \\
\lX_j & \text{if}\ j\neq 1,
\end{cases} \\ 
\lP_1 \lZ_j \lP_1^{\dagger} & = \lZ_j \ \forall \ j=1,2,3,4 .
\end{align}
Again, this is a translation of the relations $P_1^L X_j^L (P_1^L)^{\dagger}$ to the physical space.
One can express $\lP_1$ in terms of its action on the physical Paulis $X_t, Z_t$ as follows.
The condition $\lP_1 \lX_1 \lP_1^{\dagger} = \lY_1$ implies $\lP_1$ must transform $\lX_1 = X_1 X_2$ into $\lY_1 \coloneqq \imath \lX_1 \lZ_1 = \imath X_1 X_2 Z_2 Z_6 = X_1 (\imath X_2 Z_2) Z_6 = X_1 Y_2 Z_6$.
Similarly, the other conditions imply that all other $\lX_j$s and all $\lZ_j$s must remain unchanged.
Hence we can explicitly write the mappings as below.
\begin{equation}
\label{eq:phase_maps}
\hspace*{-0.25cm}
\begin{array}{lc|cl}
\lX_1 = X_1 X_2 \overset{\lP_1}{\longmapsto} \lX_1' = X_1 Y_2 Z_6 & & & \lZ_1 = Z_2 Z_6 \overset{\lP_1}{\longmapsto} \lZ_1' = Z_2 Z_6 \\
\lX_2 = X_1 X_3 \overset{\lP_1}{\longmapsto} \lX_2' = X_1 X_3 & & & \lZ_2 = Z_3 Z_6 \overset{\lP_1}{\longmapsto} \lZ_2' = Z_3 Z_6 \\
\lX_3 = X_1 X_4 \overset{\lP_1}{\longmapsto} \lX_3' = X_1 X_4 & & & \lZ_3 = Z_4 Z_6 \overset{\lP_1}{\longmapsto} \lZ_3' = Z_4 Z_6 \\
\lX_4 = X_1 X_5 \overset{\lP_1}{\longmapsto} \lX_4' = X_1 X_5 & & & \lZ_4 = Z_5 Z_6 \overset{\lP_1}{\longmapsto} \lZ_4' = Z_5 Z_6 
\end{array} . 
\end{equation}

Direct inspection of these conditions yields the circuit given below.
First we find an operator which transforms $X_2$ to $Y_2$ and leaves other Paulis unchanged; this is $P_2$, the phase gate on the second physical qubit.
Then we find an operator that transforms $Y_2$ into $Y_2 Z_6$, which is $\text{CZ}_{26}$ as $X_2 \text{CZ}_{26} X_2^{\dagger} = X_2 Z_6$ and $Z_i \text{CZ}_{26} Z_i^{\dagger} = Z_i, i=1,2,\ldots,6$.
Here $\text{CZ}_{26}$ is the controlled-$Z$ gate on physical qubits $2$ and $6$.
But this also transforms $X_6$ into $Z_2 X_6$ and hence the circuit $\text{CZ}_{26} P_2$ does not fix the stabilizer $\bg^X$.
Therefore, we include $P_6$ so that the full circuit $\lP_1 = P_6 \text{CZ}_{26} P_2$ fixes $\bg^X$, fixes $\bg^Z$, and realizes $P_1^L$.
%
\begin{center}







\begin{tikzcd}
\lstick{$2$} & \gate{P} & \ctrl{1} & \qw & \qw \\
\lstick{$6$} & \qw & \control{} & \gate{P} & \qw \\
\end{tikzcd}
\ $\equiv$
\begin{tikzcd}
\lstick{$\dket{x_1}_L$} & \gate{P} & \qw \\
\end{tikzcd}
\end{center}
See Table~\ref{tab:circuit_identities} for the circuit identities used above. 
We now describe how the same circuit can be synthesized via symplectic geometry.
Let $F = \begin{bmatrix} A & B \\ C & D \end{bmatrix}$ be the symplectic matrix corresponding to $\lP_1$.
Using~\eqref{eq:symp_action}, the conditions imposed in~\eqref{eq:phase} on $\lX_j, j=1,2,3,4$, give
\begin{align}
[110000,000000] F = [110000,010001] &
\Rightarrow [110000] A = [110000], [110000] B = [010001] , \nonumber \\
[101000,000000] F = [101000,000000] &
 \Rightarrow [101000] A = [101000], [101000] B = [000000] , \nonumber \\
[100100,000000] F = [100100,000000] &
 \Rightarrow [100100] A = [100100], [100100] B = [000000] , \nonumber \\
[100010,000000] F = [100010,000000] &
 \Rightarrow [100010] A = [100010], [100010] B = [000000] .
\end{align}
Let $\vecnot{e}_i \in \mathbb{F}_2^6$ be the standard basis vector with entry $1$ in the $i$-th location and zeros elsewhere, for $i=1,\ldots,6$.
Then the above conditions can be rewritten compactly as
\begin{align}
(\vecnot{e}_1 + \vecnot{e}_2) A & = \vecnot{e}_1 + \vecnot{e}_2 ,\ (\vecnot{e}_1 + \vecnot{e}_2) B = \vecnot{e}_2 + \vecnot{e}_6 , \nonumber \\ 
(\vecnot{e}_1 + \vecnot{e}_i) A & = \vecnot{e}_1 + \vecnot{e}_i,\ (\vecnot{e}_1 + \vecnot{e}_i) B = \vecnot{0},\ i=3,4,5.
\end{align}
Similarly, the conditions imposed on $\lZ_j, j=1,2,3,4$, give
\begin{align}
[000000,010001] F = [000000,010001] &
 \Rightarrow [010001] C = [000000], [010001] D = [010001] , \nonumber \\
[000000,001001] F = [000000,001001] &
 \Rightarrow [001001] C = [000000], [001001] D = [001001] , \nonumber \\
[000000,000101] F = [000000,000101] &
 \Rightarrow [000101] C = [000000], [000101] D = [000101] , \nonumber \\
[000000,000011] F = [000000,000011] &
 \Rightarrow [000011] C = [000000], [000011] D = [000011] . 
\end{align}
Again, these can be rewritten compactly as
\begin{align}
(\vecnot{e}_i + \vecnot{e}_6) C = 0,\ (\vecnot{e}_i + \vecnot{e}_6) D = \vecnot{e}_i + \vecnot{e}_6,\ i=2,3,4,5.
\end{align}
Although it is sufficient for $\lP_1$ to only normalize $S$, we can always require that $\lP_1$ commute with every stabilizer element (see Theorem~\ref{thm:normalize_centralize}).
This gives the centralizing conditions
\begin{align}
[111111,000000] F = [111111,000000] &
 \Rightarrow [111111] A = [111111], [111111] B = [000000] , \nonumber \\
[000000,111111] F = [000000,111111] &
 \Rightarrow [111111] C = [000000], [111111] D = [111111] .
\end{align}
Again, these can be rewritten compactly as
\begin{align}
(\vecnot{e}_1 + \ldots + \vecnot{e}_6) A & = \vecnot{e}_1 + \ldots + \vecnot{e}_6 = (\vecnot{e}_1 + \ldots + \vecnot{e}_6) D , \nonumber \\
(\vecnot{e}_1 + \ldots + \vecnot{e}_6) B & = \vecnot{0} = (\vecnot{e}_1 + \ldots + \vecnot{e}_6) C .
\end{align}
Note that, in addition to these linear constraints, $F$ also needs to satisfy the symplectic constraint $F \Omega F^T = \Omega$.
We obtain one solution using Algorithm~\ref{alg:transvec} as $F = T_{B}$ (see Table~\ref{tab:std_symp}),
\begin{equation}
\label{eq:css_phase_B}
B \coloneqq B_{P} = 
\begin{bmatrix}
0 & 0 & 0 & 0 & 0 & 0 \\
0 & 1 & 0 & 0 & 0 & 1 \\
0 & 0 & 0 & 0 & 0 & 0 \\
0 & 0 & 0 & 0 & 0 & 0 \\
0 & 0 & 0 & 0 & 0 & 0 \\
0 & 1 & 0 & 0 & 0 & 1 
\end{bmatrix} \Rightarrow F = \begin{bmatrix} I_6 & B_P \\ 0 & I_6 \end{bmatrix}  .
\end{equation} 
The resulting physical operator $\lP_1 = \text{diag}\left( \imath^{vB_{P}v^T} \right)$ satisfies $\lP_1 = P_6 \text{CZ}_{26} P_2$ and hence coincides with the above circuit (see the discussion in Sections~\ref{sec:elem_symp} and~\ref{sec:cliff_decompose} for this circuit decomposition).
Note that there can be multiple symplectic solutions to the set of linear constraints derived from~\eqref{eq:phase_maps}, and each symplectic solution could correspond to multiple circuits depending on its decomposition into elementary symplectic forms from Table~\ref{tab:std_symp}.
The set of all symplectic solutions for $\lP_1$ were obtained using the result of Theorem~\ref{thm:symp_lineq_all} in Section~\ref{sec:generic_algorithm} below, and these are listed in Appendix~\ref{sec:css_phase_all}.
The above solution is the least complex in this set in terms of the depth of the circuit (see Def.~\ref{def:depth}).

The above example also provides the basic idea for the LCS algorithm.
We will see that each logical gate below reveals a different scenario for the LCS algorithm.
Henceforth, for any logical operator in $\text{Cliff}_{2^4}$, we refer to its physical implementation $\bar{g}$ itself as the logical operator, since this is common terminology in the literature.

\subsection{Logical Controlled-$Z$ (CZ)}
\label{sec:css_cz}

The logical operator $\bar{g} = \lcz{1}{2}$ is defined by its action on the logical Paulis as 
\begin{align}
\label{eq:cz}
\lcz{1}{2} \lX_j \lcz{1}{2}^{\dagger} & = 
\begin{cases}
\lX_1 \lZ_2 & \text{if}\ j=1, \\
\lZ_1 \lX_2 & \text{if}\ j=2,\\
\lX_j & \text{if}\ j\neq 1,2,
\end{cases} \\ 
\lcz{1}{2} \lZ_j \lcz{1}{2}^{\dagger} & = \lZ_j \ \forall \ j=1,2,3,4 .
\end{align}
We first express the logical operator $\lcz{1}{2}$, on the first two logical qubits, in terms of its action on the physical Pauli operators $X_t, Z_t$.
\begin{equation}
\label{eq:cz_maps}
\begin{array}{lc|cl}
\lX_1 = X_1 X_2 \overset{\lcz{1}{2}}{\longmapsto} X_1 X_2 Z_3 Z_6 & & & \lZ_1 = Z_2 Z_6 \overset{\lcz{1}{2}}{\longmapsto} Z_2 Z_6 \\
\lX_2 = X_1 X_3 \overset{\lcz{1}{2}}{\longmapsto} X_1 X_3 Z_2 Z_6 & & &  \lZ_2 = Z_3 Z_6 \overset{\lcz{1}{2}}{\longmapsto} Z_3 Z_6 \\
\lX_3 = X_1 X_4 \overset{\lcz{1}{2}}{\longmapsto} X_1 X_4 & & &  \lZ_3 = Z_4 Z_6 \overset{\lcz{1}{2}}{\longmapsto} Z_4 Z_6 \\
\lX_4 = X_1 X_5 \overset{\lcz{1}{2}}{\longmapsto} X_1 X_5 & & &  \lZ_4 = Z_5 Z_6 \overset{\lcz{1}{2}}{\longmapsto} Z_5 Z_6 
\end{array} . 
\end{equation}
As with the phase gate, we translate these conditions into linear equations involving the constituents of the corresponding symplectic matrix $F$.
The conditions from $\lX_j$s are
\begin{align}
(\vecnot{e}_1 + \vecnot{e}_i) A & = \vecnot{e}_1 + \vecnot{e}_i, \ i=2,3,4,5, \ (\vecnot{e}_1 + \vecnot{e}_2) B = \vecnot{e}_3 + \vecnot{e}_6, \nonumber \\ 
(\vecnot{e}_1 + \vecnot{e}_3) B & = \vecnot{e}_2 + \vecnot{e}_6, \ (\vecnot{e}_1 + \vecnot{e}_i) B = \vecnot{0}, \ i=4,5 .
\end{align}
The conditions imposed by the $\lZ_j$s are
\begin{align}
(\vecnot{e}_i + \vecnot{e}_6) C = \vecnot{0}, \ (\vecnot{e}_i + \vecnot{e}_6) D = \vecnot{e}_i + \vecnot{e}_6, \ i=2,3,4,5 .
\end{align}
As for the phase gate, we require that $\lcz{1}{2}$ centralize the stabilizer, i.e.,
\begin{align}
(\vecnot{e}_1 + \ldots + \vecnot{e}_6) A & = \vecnot{e}_1 + \ldots + \vecnot{e}_6 = (\vecnot{e}_1 + \ldots + \vecnot{e}_6) D , \nonumber \\ 
(\vecnot{e}_1 + \ldots + \vecnot{e}_6) B & = \vecnot{0} = (\vecnot{e}_1 + \ldots + \vecnot{e}_6) C .
\end{align}
We again obtain one solution using Algorithm~\ref{alg:transvec} as $F = T_{B}$, where
\begin{equation}
B \coloneqq B_{\text{CZ}} = 
\begin{bmatrix}
0 & 0 & 0 & 0 & 0 & 0 \\
0 & 0 & 1 & 0 & 0 & 1 \\
0 & 1 & 0 & 0 & 0 & 1 \\
0 & 0 & 0 & 0 & 0 & 0 \\
0 & 0 & 0 & 0 & 0 & 0 \\
0 & 1 & 1 & 0 & 0 & 0 
\end{bmatrix} .
\end{equation} 
We find that the physical operator $\lcz{1}{2} = \text{diag}\left( \imath^{vB_{\text{CZ}}v^T} \right)$ commutes with the stabilizer $\bg^Z$ but not with $\bg^X$; it takes $X^{\otimes 6}$ to $-X^{\otimes 6}$.
This is remedied through post multiplication by $Z_6$ to obtain $\lcz{1}{2} = \text{diag}\left( \imath^{vB_{\text{CZ}}v^T} \right) Z_6$, which does not modify the symplectic matrix $F$ as $Z_6 \in HW_N$ and $HW_N$ is the kernel of the map $\phi$ defined in~\eqref{eq:phi}.
The resulting physical operator $\lcz{1}{2}$ corresponds to the same circuit obtained by Chao and Reichardt in~\cite{Chao-arxiv17b}. 
\begin{center}







\begin{tikzcd}
\lstick{$2$} & \ctrl{1} & \ctrl{2} & \qw & \qw \\
\lstick{$3$} & \control{} & \qw & \ctrl{1} & \qw \\
\lstick{$6$} & \gate{Z} & \control{} & \control{} & \qw \\
\end{tikzcd}
\ $\equiv$
\begin{tikzcd}
\lstick{$\dket{x_1}_L$} & \ctrl{1} & \qw \\
\lstick{$\dket{x_2}_L$} & \control{} & \qw \\
\end{tikzcd}
\end{center}
The set of all symplectic solutions for $\lcz{1}{2}$ were obtained using the result of Theorem~\ref{thm:symp_lineq_all} in Section~\ref{sec:generic_algorithm} below, and these are listed in Appendix~\ref{sec:css_cz_all}.
As for $\lP_1$, the above solution has the smallest depth.
Hence, this example shows that the LCS algorithm might have to correct signs using Pauli matrices, since the homomorphism $\phi$ in~\eqref{eq:phi} is agnostic to signs.

\subsection{Logical Controlled-NOT (CNOT)}
\label{sec:css_cnot}

The logical operator $\bar{g} = \lcnot{2}{1}$, where logical qubit $2$ controls $1$, is defined by
\begin{align}
\label{eq:lcnot21}
\lcnot{2}{1} \lX_j \lcnot{2}{1}^{\dagger} & = 
\begin{cases}
\lX_1 \lX_2 & \text{if}\ j=2,\\
\lX_j & \text{if}\ j\neq 2,
\end{cases}  \\  
\lcnot{2}{1} \lZ_j \lcnot{2}{1}^{\dagger} & = 
\begin{cases}
\lZ_1 \lZ_2 & \text{if}\ j=1,\\
\lZ_j & \text{if}\ j\neq 1 .
\end{cases}
\end{align}
We approach synthesis via symplectic geometry, and express the operator $\lcnot{2}{1}$ in terms of its action on the physical operators $X_t, Z_t$ as shown below.
\begin{equation}
\label{eq:lcnot21_maps}
\begin{array}{lc|cl}
\lX_1 = X_1 X_2 \overset{2 \rightarrow 1}{\longmapsto} X_1 X_2 & \ & \ & \lZ_1 = Z_2 Z_6 \overset{2 \rightarrow 1}{\longmapsto} Z_2 Z_3 \\
\lX_2 = X_1 X_3 \overset{2 \rightarrow 1}{\longmapsto} X_2 X_3 & \ & \ &  \lZ_2 = Z_3 Z_6 \overset{2 \rightarrow 1}{\longmapsto} Z_3 Z_6 \\
\lX_3 = X_1 X_4 \overset{2 \rightarrow 1}{\longmapsto} X_1 X_4 & \ & \ &  \lZ_3 = Z_4 Z_6 \overset{2 \rightarrow 1}{\longmapsto} Z_4 Z_6 \\
\lX_4 = X_1 X_5 \overset{2 \rightarrow 1}{\longmapsto} X_1 X_5 & \ & \ &  \lZ_4 = Z_5 Z_6 \overset{2 \rightarrow 1}{\longmapsto} Z_5 Z_6 
\end{array} . 
\end{equation}
Note that only $\lX_2$ and $\lZ_1$ are modified by $\lcnot{2}{1}$.
As before, we translate these conditions into linear equations involving the constituents of the corresponding symplectic transformation $F$.
The conditions imposed by $\lX_j$s are
\begin{align}
(\vecnot{e}_1 + \vecnot{e}_3) A = \vecnot{e}_2 + \vecnot{e}_3, \ (\vecnot{e}_1 + \vecnot{e}_i) A & = \vecnot{e}_1 + \vecnot{e}_i, \ i=2,4,5, \nonumber \\
(\vecnot{e}_1 + \vecnot{e}_i) B & = \vecnot{0}, \ i=2,3,4,5 .
\end{align}
The conditions imposed by $\lZ_j$s are
\begin{align}
(\vecnot{e}_i + \vecnot{e}_6) C = \vecnot{0}, \ i=2,3,4,5, \ (\vecnot{e}_2 + \vecnot{e}_6) D & = \vecnot{e}_2 + \vecnot{e}_3, \nonumber \\
(\vecnot{e}_i + \vecnot{e}_6) D & = \vecnot{e}_i + \vecnot{e}_6, \ i=3,4,5 .
\end{align}
Again we require that $\lcnot{2}{1}$ commute with every stabilizer element, i.e.,
\begin{align}
(\vecnot{e}_1 + \ldots + \vecnot{e}_6) A & = \vecnot{e}_1 + \ldots + \vecnot{e}_6 = (\vecnot{e}_1 + \ldots + \vecnot{e}_6) D , \nonumber \\ 
(\vecnot{e}_1 + \ldots + \vecnot{e}_6) B & = \vecnot{0} = (\vecnot{e}_1 + \ldots + \vecnot{e}_6) C .
\end{align}
We again obtain one solution using Algorithm~\ref{alg:transvec} as $F = \begin{bmatrix} A & 0 \\ 0 & A^{-T} \end{bmatrix}$, where
\begin{equation}
A = 
\begin{bmatrix}
1 & 0 & 0 & 0 & 0 & 0 \\
0 & 1 & 0 & 0 & 0 & 0 \\
1 & 1 & 1 & 0 & 0 & 0 \\
0 & 0 & 0 & 1 & 0 & 0 \\
0 & 0 & 0 & 0 & 1 & 0 \\
1 & 1 & 0 & 0 & 0 & 1 
\end{bmatrix} ,\ A^{-T} = 
\begin{bmatrix}
1 & 0 & 1 & 0 & 0 & 1 \\
0 & 1 & 1 & 0 & 0 & 1 \\
0 & 0 & 1 & 0 & 0 & 0 \\
0 & 0 & 0 & 1 & 0 & 0 \\
0 & 0 & 0 & 0 & 1 & 0 \\
0 & 0 & 0 & 0 & 0 & 1 
\end{bmatrix} . 
\end{equation} 
The action of $\lcnot{2}{1}$ on logical qubits is related to the action on physical qubits through the generator matrix $G_{\MCC/\MCCd}$.
The map $v \mapsto v A$ fixes the code $\MCC$ (i.e., $\ket{v} \mapsto \ket{v A}$ fixes $\MCQ$ and hence its stabilizers $\bg^X$ and $\bg^Z$) and induces a linear transformation on the coset space $\MCC/\MCCd$ (which defines the CSS state).
The action $K$ on logical qubits is related to the action $A$ on physical qubits by ${K \cdot G_{C/C^{\perp}}^X = G_{C/C^{\perp}}^X \cdot A}$, and we obtain
\begin{align}
K = 
\begin{bmatrix}
1 & 0 & 0 & 0 \\
1 & 1 & 0 & 0 \\
0 & 0 & 1 & 0 \\
0 & 0 & 0 & 1
\end{bmatrix}
\end{align}
as desired.
The circuit on the left below implements the operator $\dket{v} \mapsto \dket{vA}$, i.e., $\lcnot{2}{1}$.
The circuit on the right implements $\dket{x} \mapsto \dket{xK}$, i.e., $\cnot{2}{1}^L$, where $x \in \mathbb{F}_2^4$.
\begin{center}







\begin{tikzcd}
\lstick{$1$} & \qw & \targ{} & \qw & \targ{} & \qw \\
\lstick{$2$} & \targ{} & \qw & \targ{} & \qw & \qw \\
\lstick{$3$} & \ctrl{-1} & \ctrl{-2} & \qw & \qw & \qw \\
\lstick{$6$} & \qw & \qw & \ctrl{-2} & \ctrl{-3} & \qw \\
\end{tikzcd}
\ $\equiv$
\begin{tikzcd}
\lstick{$\dket{x_1}_L$} & \targ{} & \qw \\
\lstick{$\dket{x_2}_L$} & \ctrl{-1} & \qw \\
\end{tikzcd}
\end{center}
We note that~\cite{Grassl-isit13} discusses codes and operators where $A$ is a permutation matrix corresponding to an automorphism of $\MCC$.
The set of all symplectic solutions for $\lcnot{2}{1}$ were obtained using the result of Theorem~\ref{thm:symp_lineq_all} in Section~\ref{sec:generic_algorithm} below, and these are listed in Appendix~\ref{sec:css_cnot_all}.
As for $\lP_1$ and $\lcz{1}{2}$, the above solution has the smallest circuit depth.

\begin{remark}
\normalfont
To implement $\cnot{2}{1}^L$ we can also use the circuit identity (see Table~\ref{tab:circuit_identities})
\begin{center}







\begin{tikzcd}
\lstick{$\dket{x_1}_L$} & \targ{} & \qw \\
\lstick{$\dket{x_2}_L$} & \ctrl{-1} & \qw \\
\end{tikzcd}
\ $=$
\begin{tikzcd}
\lstick{$\dket{x_1}_L$} & \gate{H} & \control{} & \gate{H} & \qw \\
\lstick{$\dket{x_2}_L$} & \qw & \ctrl{-1} & \qw & \qw \\
\end{tikzcd}
\end{center}
where $H_1^L$ is synthesized below.
However, this construction might require more gates.
\end{remark}

\subsection{Logical Targeted Hadamard}
\label{sec:css_had}

The Hadamard gate $\bar{g} = \lH_1$ on the first logical qubit is defined by the actions
\begin{equation}
\label{eq:hadamard}
\lH_1 \lX_j \lH_1^{\dagger} = 
\begin{cases}
\lZ_j & \text{if}\ j=1, \\
\lX_j & \text{if}\ j\neq 1,
\end{cases}  \ \ 
\lH_1 \lZ_j \lH_1^{\dagger} = 
\begin{cases}
\lX_j & \text{if}\ j=1, \\
\lZ_j & \text{if}\ j\neq 1.
\end{cases} 
\end{equation}
As for the other gates, we express $\lH_1$ in terms of its action on the physical Paulis $X_t, Z_t$:
\begin{equation}
\label{eq:hadamard_maps}
\begin{array}{lc|cl}
\lX_1 = X_1 X_2 \overset{\lH_1}{\longmapsto} Z_2 Z_6 & \ & \ &  \lZ_1 = Z_2 Z_6 \overset{\lH_1}{\longmapsto} X_1 X_2 \\
\lX_2 = X_1 X_3 \overset{\lH_1}{\longmapsto} X_1 X_3 & \ & \ &  \lZ_2 = Z_3 Z_6 \overset{\lH_1}{\longmapsto} Z_3 Z_6 \\
\lX_3 = X_1 X_4 \overset{\lH_1}{\longmapsto} X_1 X_4 & \ & \ &  \lZ_3 = Z_4 Z_6 \overset{\lH_1}{\longmapsto} Z_4 Z_6 \\
\lX_4 = X_1 X_5 \overset{\lH_1}{\longmapsto} X_1 X_5 & \ & \ &  \lZ_4 = Z_5 Z_6 \overset{\lH_1}{\longmapsto} Z_5 Z_6 
\end{array} . 
\end{equation}
As before, we translate these conditions into linear equations involving the constituents of the corresponding symplectic transformation $F$.
The conditions imposed by $\lX_j$s are
\begin{align}
(\vecnot{e}_1 + \vecnot{e}_2) A & = \vecnot{0}, \ (\vecnot{e}_1 + \vecnot{e}_i) A = \vecnot{e}_1 + \vecnot{e}_i, \ i=3,4,5, \nonumber \\ 
(\vecnot{e}_1 + \vecnot{e}_2) B & = \vecnot{e}_2 + \vecnot{e}_6, \ (\vecnot{e}_1 + \vecnot{e}_i) B = \vecnot{0}, \ i=3,4,5 .
\end{align}
The conditions imposed by $\lZ_j$s are
\begin{align}
(\vecnot{e}_2 + \vecnot{e}_6) C & = \vecnot{e}_1 + \vecnot{e}_2, \ (\vecnot{e}_i + \vecnot{e}_6) C = \vecnot{0}, \ i=3,4,5, \nonumber \\ 
(\vecnot{e}_2 + \vecnot{e}_6) D & = \vecnot{0}, \ (\vecnot{e}_i + \vecnot{e}_6) D = \vecnot{e}_i + \vecnot{e}_6, \ i=3,4,5 .
\end{align}
Again we require $\lH_1$ to centralize the stabilizer, which yields the conditions
\begin{align}
(\vecnot{e}_1 + \ldots + \vecnot{e}_6) A & = \vecnot{e}_1 + \ldots + \vecnot{e}_6 = (\vecnot{e}_1 + \ldots + \vecnot{e}_6) D , \nonumber \\
(\vecnot{e}_1 + \ldots + \vecnot{e}_6) B & = \vecnot{0} = (\vecnot{e}_1 + \ldots + \vecnot{e}_6) C .
\end{align}
We again obtain one solution using Algorithm~\ref{alg:transvec} as 
\begin{align}
\label{eq:css_H1}
A = 
\begin{bmatrix}
1 & 0 & 0 & 0 & 0 & 0 \\
1 & 0 & 0 & 0 & 0 & 0 \\
0 & 0 & 1 & 0 & 0 & 0 \\
0 & 0 & 0 & 1 & 0 & 0 \\
0 & 0 & 0 & 0 & 1 & 0 \\
1 & 1 & 0 & 0 & 0 & 1 
\end{bmatrix} , &\  
B = 
\begin{bmatrix}
0 & 0 & 0 & 0 & 0 & 0 \\
0 & 1 & 0 & 0 & 0 & 1 \\
0 & 0 & 0 & 0 & 0 & 0 \\
0 & 0 & 0 & 0 & 0 & 0 \\
0 & 0 & 0 & 0 & 0 & 0 \\
0 & 1 & 0 & 0 & 0 & 1  
\end{bmatrix} , \nonumber \\ 
C = 
\begin{bmatrix}
1 & 1 & 0 & 0 & 0 & 0 \\
1 & 1 & 0 & 0 & 0 & 0 \\
0 & 0 & 0 & 0 & 0 & 0 \\
0 & 0 & 0 & 0 & 0 & 0 \\
0 & 0 & 0 & 0 & 0 & 0 \\
0 & 0 & 0 & 0 & 0 & 0  
\end{bmatrix} , & \ 
D = 
\begin{bmatrix}
1 & 1 & 0 & 0 & 0 & 1 \\
0 & 0 & 0 & 0 & 0 & 1 \\
0 & 0 & 1 & 0 & 0 & 0 \\
0 & 0 & 0 & 1 & 0 & 0 \\
0 & 0 & 0 & 0 & 1 & 0 \\
0 & 0 & 0 & 0 & 0 & 1 
\end{bmatrix} .  
\end{align}
The unitary operation corresponding to this solution commutes with each stabilizer element.
Another solution for $\lH_1$ which fixes $Z^{\otimes 6}$ but takes $X^{\otimes 6} \leftrightarrow (111111,000000)$ to $Y^{\otimes 6} \leftrightarrow (111111,111111)$ is given by just changing $B$ above to
\begin{align}
B =
\begin{bmatrix}
0 & 0 & 0 & 0 & 0 & 1 \\
0 & 1 & 0 & 0 & 0 & 0 \\
0 & 0 & 0 & 0 & 0 & 1 \\
0 & 0 & 0 & 0 & 0 & 1 \\
0 & 0 & 0 & 0 & 0 & 1 \\
1 & 0 & 1 & 1 & 1 & 1  
\end{bmatrix} .  
\end{align}
However, for both these solutions the resulting symplectic transformation does not correspond to any of the elementary forms in Table~\ref{tab:std_symp}.
Hence the unitary needs to be determined by expressing $F$ as a sequence of elementary transformations and then multiplying the corresponding unitaries.
An algorithm for this is given by Can~\cite{Can-2017a} which we restated in Theorem~\ref{thm:Trung} earlier.
For the solution~\eqref{eq:css_H1}, we verified that the symplectic matrix corresponds to the following circuit on the top given by Chao and Reichardt~\cite{Chao-arxiv17b}.
On the bottom we produce the circuit obtained by using Theorem~\ref{thm:Trung} for decomposition.
%
%
%
%
%
%
%
%
%
\begin{center}
\begin{tikzcd}
\lstick{$1$} & \gate{H} & \ctrl{1} & \qw & \ctrl{1} & \gate{H} & \gate{X} & \qw \\
\lstick{$2$} & \ctrl{1} & \targ{} & \gate{H} & \targ{} & \ctrl{1} & \qw & \qw \\
\lstick{$6$} & \control{} & \qw & \qw & \qw & \control{} & \gate{Z} & \qw \\
\end{tikzcd}
\ \ $\equiv$
\begin{tikzcd}
\lstick{$1$} & \qw & \targ{} & \targ{} & \qw & \gate{H} & \ctrl{2} & \qw & \gate{H} & \qw & \qw \\
\lstick{$2$} & \gate[swap]{} & \ctrl{-1} & \qw & \ctrl{1} & \gate{H} & \qw & \ctrl{1} & \gate{H} & \ctrl{1} & \qw \\
\lstick{$6$} &   & \qw & \ctrl{-2} & \targ{} & \gate{H} & \control{} & \control{} & \qw & \targ{} & \qw \\
\end{tikzcd}
\end{center}
Although our decomposition provides a circuit with more operations in this case, it is a systematic procedure that can be applied to any symplectic matrix.
We think that further research into the structure of symplectic matrices might aid in taking advantage of possible symmetries in the code and the operation.
Note that these are only two of all possible circuits, even given the specific symplectic matrix~\eqref{eq:css_H1}.
The set of all symplectic solutions for $\lH_1$ were obtained using the result of Theorem~\ref{thm:symp_lineq_all} in Section~\ref{sec:generic_algorithm} below, and these are listed in Appendix~\ref{sec:css_had_all}.

\subsection{Logical Transversal Hadamard}

As noted in~\cite{Chao-arxiv17b}, for this code, the logical transversal Hadamard operator $\lH^{\otimes 4}$, applied to all logical qubits simultaneously, is easy to construct.
This operator must satisfy the conditions $\lH_j \lX_j \lH_j = \lZ_j, \lH_j \lZ_j \lH_j = \lX_j$ for $j=1,2,3,4$.
If we apply the physical Hadamard operator $H$ transversally, i.e. $H_1 H_2 \cdots H_6$, we get the mappings
\[ X_1 X_{i+1} \mapsto Z_1 Z_{i+1} \quad , \quad Z_{i+1} Z_6 \mapsto X_{i+1} X_6 . \]
To complete the logical transversal Hadamard we now have to just swap physical qubits $1$ and $6$.
We note from Table~\ref{tab:std_symp} that the symplectic matrices associated with physical transversal Hadamard and swapping qubits $1$ and $6$ are $\Omega$ and $\begin{bmatrix} A & 0 \\ 0 & A \end{bmatrix}$, respectively, where 
\begin{align}
A = 
\begin{bmatrix}
0 & 0 & 0 & 0 & 0 & 1 \\
0 & 1 & 0 & 0 & 0 & 0 \\
0 & 0 & 1 & 0 & 0 & 0 \\
0 & 0 & 0 & 1 & 0 & 0 \\
0 & 0 & 0 & 0 & 1 & 0 \\
1 & 0 & 0 & 0 & 0 & 0 
\end{bmatrix} .
\end{align}
Hence the symplectic matrix associated with the logical transversal Hadamard is 
\begin{align}
{F = \begin{bmatrix} 0 & I_6 \\ I_6 & 0 \end{bmatrix} \begin{bmatrix} A & 0 \\ 0 & A \end{bmatrix} = \begin{bmatrix} 0 & A \\ A & 0 \end{bmatrix}} .
\end{align}
Note that this solution swaps $X^{\otimes 6}$ and $Z^{\otimes 6}$ and hence only normalizes the stabilizer.
Therefore, in general, the simplest circuit to realize a logical operator might not always fix the stabilizer element-wise, i.e., it might not centralize the stabilizer.

The important tool in translating the above procedure into a systematic algorithm is symplectic transvections, which we describe next.

\section{Symplectic Transvections}
\label{sec:symp_transvec}

\begin{definition}
Given a vector $h \in \mathbb{F}_2^{2n}$, a \emph{symplectic transvection}~\cite{Koenig-jmp14} is a map $Z_h \colon \mathbb{F}_2^{2n} \rightarrow \mathbb{F}_2^{2n}$ defined by
\begin{align}
\label{eq:symp_transvec}
Z_h(x) \coloneqq x + \syminn{x}{h} h \ \Leftrightarrow \ F_h \coloneqq I_{2n} + \Omega h^T h ,
\end{align}
where $F_h$ is its associated symplectic matrix.
A transvection does not correspond to a single elementary Clifford operator.
\end{definition}

\begin{fact}[{\hspace{1sp}\cite[Theorem 2.10]{Salam-laa08}}]
The symplectic group $\text{Sp}(2n,\mathbb{F}_2)$ is generated by the family of symplectic transvections.
\end{fact}

An important result that is involved in the proof of this fact is the following theorem from~\cite{Salam-laa08,Koenig-jmp14}, which we restate here for $\mathbb{F}_2^{2n}$ since we will build on this result to state and prove Theorem~\ref{thm:symp_lineq}.

\begin{theorem}
\label{thm:symp_transvec}
Let $x,y \in \mathbb{F}_2^{2n}$ be two non-zero vectors. Then $x$ can be mapped to $y$ by a product of at most two symplectic transvections.
\begin{proof}
There are two possible cases: $\syminn{x}{y} = 1$ or $0$.
For $\syminn{x}{y} = 1$, $h \coloneqq x + y$, observe
\begin{align}
x F_h = Z_h(x) = x + \syminn{x}{x+y} (x+y) & = x + \left( \syminn{x}{x} + \syminn{x}{y} \right) (x+y) \\
  & = x + (0+1) (x+y) = y .
\end{align}
Next assume $\syminn{x}{y} = 0$.
Define $h_1 \coloneqq w + y, h_2 \coloneqq x + w$, where $w \in \mathbb{F}_2^{2n}$ is chosen such that $\syminn{x}{w} = \syminn{y}{w} = 1$.
Then we have
\begin{align}
x F_{h_1} F_{h_2} & = Z_{h_2} \left( Z_{h_1} (x) \right) \\
  & = Z_{h_2} \left(x + \syminn{x}{w+y} (w+y) \right) \\
  & = (x + w+ y) + \syminn{(x+w)+y}{x+w} (x+w) \\
  & = y.  \qedhere    
\end{align}
\end{proof}
\end{theorem}

In Section~\ref{sec:generic_algorithm} we will use the above result to propose an algorithm (Alg.~\ref{alg:transvec}) which determines a symplectic matrix $F$ that satisfies $x_i F = y_i,\ i=1,2,\ldots,t \leq 2n$, where $x_i$ are linearly independent and satisfy $\syminn{x_i}{x_j} = \syminn{y_i}{y_j}$ for all $i,j \in \{1,\ldots,t\}$.

\subsection{M{\o}lmer-S{\o}rensen Gates as Transvections}

In trapped-ion quantum computing, the native two-qubit gate is the M{\o}lmer-S{\o}rensen gate
\begin{align}
\text{XX}_{ij}(\theta) \coloneqq \exp\left( -\imath \frac{\theta}{2} X_i X_j \right) = \cos\frac{\theta}{2} I_4 - \imath \sin\frac{\theta}{2} E(e_i + e_j, 0), 
\end{align}
where $e_i, e_j \in \{0,1\}^n$ represent the standard basis vectors with a single $1$ in the $i$-th and $j$-th position, respectively.
It is easy to check that $\text{XX}_{ij}(\theta)$ is Clifford only when $\theta = \frac{\pi}{2}$.
Now consider the action of $W \coloneqq \frac{1}{\sqrt{2}} (I_N - \imath E(a,b))$ on an arbitrary Pauli matrix $E(x_1,x_2)$.
If $\syminn{[a,b]}{[x_1,x_2]} = a x_2^T + b x_1^T = 0$ then $E(a,b)$ and $E(x_1,x_2)$ commute so that
\begin{align}
W E(x_1,x_2) W^{\dagger} = E(x_1,x_2) W W^{\dagger} = E(x_1,x_2).
\end{align}
If $\syminn{[a,b]}{[x_1,x_2]} = a x_2^T + b x_1^T = 1$ then $E(a,b)$ and $E(x_1,x_2)$ anti-commute so that
\begin{align}
W E(x_1,x_2) W^{\dagger} & = E(x_1,x_2) \, \frac{1}{\sqrt{2}} (I_N + \imath E(a,b)) \cdot \frac{1}{\sqrt{2}} (I_N + \imath E(a,b)) \\
  & = E(x_1,x_2) \, \frac{1}{2} \left[ I_N + 2\imath E(a,b) - I_N \right] \\
  & = E(x_1,x_2) \, \imath E(a,b) \\
  & = \imath^{1 + a x_2^T - b x_1^T} E(a+x_1, b+x_2) \\
  & = \pm E(a \oplus x_1, b \oplus x_2),
\end{align}
since $a x_2^T - b x_1^T \in \{1,3\}$ (mod $4$) and the result of the conjugation must be Hermitian, i.e., there cannot be an overall factor $\imath$.
Thus, we have $W E(x_1,x_2) W^{\dagger} = \pm E([x_1,x_2] F_h)$, where $h \coloneqq [a,b]$. 
Hence, the Clifford M{\o}lmer-S{\o}rensen gates map to symplectic transvections in the binary symplectic group.
Note that the definition of $W$ generalizes beyond two qubits.

\section{The Generic Algorithm}
\label{sec:generic_algorithm}

The synthesis of logical Paulis by Gottesman~\cite{Gottesman-phd97} and by Wilde~\cite{Wilde-physreva09} exploits symplectic geometry over the binary field. 
Building on their work we have demonstrated, using the $\llbr 6,4,2 \rrbr$ code as an example, that symplectic geometry provides a systematic framework for synthesizing physical implementations of any logical operator in the logical Clifford group $\text{Cliff}_{2^k}$ for stabilizer codes. 
In other words, symplectic geometry provides a \emph{control plane} where effects of Clifford operators can be analyzed efficiently.

\subsection{A Naive Approach}

For each logical Clifford operator, one can obtain all symplectic solutions using the algorithm below.

\begin{enumerate}

\item Collect all the linear constraints on $F$, obtained from the conjugation relations of the desired Clifford operator with the stabilizer generators and logical Paulis, to obtain a system of equations $UF = V$.

\item Then vectorize both sides to get $\left( I_{2n} \otimes U \right) \text{vec}(F) = \text{vec}(V)$.

\item Perform Gaussian elimination on the augmented matrix $\left[ \left( I_{2n} \otimes U \right),\ \text{vec}(V) \right]$.
If $\ell$ is the number of non-pivot variables in the row-reduced echelon form, then there are $2^{\ell}$ solutions to the linear system.

\item For each such solution, check if it satisfies $F \Omega F^T = \Omega$.
If it does, then it is a feasible symplectic solution for $\bar{g}$.

\end{enumerate}

Clearly, this algorithm is not very efficient since $\ell$ could be very large.
Specifically, for codes that do not encode many logical qubits this number will be very large as the system $UF = V$ will be very under-constrained.
We now prove two theorems that enable us to determine all symplectic solutions for each logical Clifford operator much more efficiently.

\subsection{Solving a Symplectic Linear System of Equations}

\begin{theorem}
\label{thm:symp_lineq}
Let $x_i, y_i \in \mathbb{F}_2^{2n}, i=1,2,\ldots,t \leq 2n$ be a collection of (row) vectors such that $\syminn{x_i}{x_j} = \syminn{y_i}{y_j}$.
Assume that the $x_i$ are linearly independent.
Then a solution $F \in \text{Sp}(2n,\mathbb{F}_2)$ to the system of equations $x_i F = y_i$ can be obtained as the product of a sequence of at most $2t$ symplectic transvections $F_{h} \coloneqq I_{2n} + \Omega h^T h$, where $h \in \mathbb{F}_2^{2n}$.
\begin{proof}
We will prove this result by induction.
For $i=1$ we can simply use Theorem~\ref{thm:symp_transvec} to find $F_1 \in \text{Sp}(2n,\mathbb{F}_2)$ as follows. 
If $\syminn{x_1}{y_1} = 1$ then $F_1 \coloneqq F_{h_1}$ with $h_1 \coloneqq x_1 + y_1$, or if $\syminn{x_1}{y_1} = 0$ then $F_1 \coloneqq F_{h_{11}} F_{h_{12}}$ with $h_{11} \coloneqq w_1 + y_1, h_{12} \coloneqq x_1 + w_1$, where $w_1$ is chosen such that $\syminn{x_1}{w_1} = \syminn{y_1}{w_1} = 1$.
In any case $F_1$ satisfies $x_1 F_1 = y_1$.
Next consider $i = 2$.
Let $\tilde{x}_2 \coloneqq x_2 F_1$ so that $\syminn{x_1}{x_2} = \syminn{y_1}{y_2} = \syminn{y_1}{\tilde{x}_2}$, since $F_1$ is symplectic and hence preserves symplectic inner products.
Similar to Theorem~\ref{thm:symp_transvec} we have two cases: $\syminn{\tilde{x}_2}{y_2} = 1$ or $0$.
For the former, we set $h_2 \coloneqq \tilde{x}_2 + y_2$ so that we clearly have $\tilde{x}_2 F_{h_2} = Z_{h_2}(\tilde{x}_2) = y_2$ (see Section~\ref{sec:symp_transvec} for the definition of $Z_h(\cdot)$).
We also observe that
\begin{align*}
y_1 F_{h_2} = Z_{h_2}(y_1) = y_1 + \syminn{y_1}{\tilde{x}_2 + y_2} (\tilde{x}_2 + y_2) = y_1 + (\syminn{y_1}{y_2} + \syminn{y_1}{y_2}) (\tilde{x}_2 + y_2) = y_1 .
\end{align*}
Hence in this case $F_2 \coloneqq F_1 F_{h_2}$ satisfies $x_1 F_2 = y_1, x_2 F_2 = y_2$.
For the case $\syminn{\tilde{x}_2}{y_2} = 0$ we again find a $w_2$ that satisfies $\syminn{\tilde{x}_2}{w_2} = \syminn{y_2}{w_2} = 1$ and set $h_{21} \coloneqq w_2 + y_2, h_{22} \coloneqq \tilde{x}_2 + w_2$.
Then by Theorem~\ref{thm:symp_transvec} we clearly have $\tilde{x}_2 F_{h_{21}} F_{h_{22}} = y_2$.
For $y_1$ we observe that
\begin{align}
y_1 F_{h_{21}} F_{h_{22}} & = Z_{h_{22}} \left( Z_{h_{21}}(y_1) \right) \\
  & = Z_{h_{22}} \left( y_1 + \syminn{y_1}{w_2 + y_2} (w_2 + y_2) \right) \\
  & = y_1 + \syminn{y_1}{w_2 + y_2} (w_2 + y_2) + \big( \syminn{y_1}{\tilde{x}_2 + w_2} + \nonumber \\
  & \hspace{2cm} \syminn{y_1}{w_2 + y_2} \syminn{w_2 + y_2}{\tilde{x}_2 + w_2} \big) (\tilde{x}_2 + w_2) \\
  & = y_1 + \syminn{y_1}{w_2 + y_2} (\tilde{x}_2 + y_2) \\
  & = y_1 \ \text{if and only if}\ \syminn{y_1}{w_2} = \syminn{y_1}{y_2} .
\end{align}
The penultimate equality holds because $\syminn{y_1}{\tilde{x}_2} = \syminn{y_1}{y_2}, \syminn{w_2 + y_2}{\tilde{x}_2 + w_2} = 1+0+0+1 = 0$.
Hence, we pick a $w_2$ such that $\syminn{\tilde{x}_2}{w_2} = \syminn{y_2}{w_2} = 1$ and $\syminn{y_1}{w_2} = \syminn{y_1}{y_2}$, and then set $F_2 \coloneqq F_1 F_{h_{21}} F_{h_{22}}$.
Again, for this case $F_2$ satisfies $x_1 F_2 = y_1, x_2 F_2 = y_2$.

By induction, assume $F_{i-1}$ satisfies $x_j F_{i-1} = y_j$ for all $j=1,\ldots,i-1$, where $i \geq 3$.
Using the same idea as for $i=2$ above, let $x_i F_{i-1} = \tilde{x}_i$. 
If $\syminn{\tilde{x}_i}{y_i} = 1$, we simply set $F_i \coloneqq F_{i-1} F_{h_i}$, where $h_i \coloneqq \tilde{x}_i + y_i$. 
If $\syminn{\tilde{x}_i}{y_i} = 0$, we find a $w_i$ that satisfies $\syminn{\tilde{x}_i}{w_i} = \syminn{y_i}{w_i} = 1$ and $\syminn{y_j}{w_i} = \syminn{y_j}{y_i} \ \forall \ j < i$. 
Then we define $h_{i1} \coloneqq w_i + y_i, h_{i2} \coloneqq \tilde{x}_i + w_i$ and observe that for $j < i$ we have
\begin{align}
y_j F_{h_{i1}} F_{h_{i2}} = Z_{h_{i2}} \left( Z_{h_{i1}}(y_j) \right) = y_j + \syminn{y_j}{w_i + y_i} (\tilde{x}_i + y_i) = y_j .
\end{align}
Again, by Theorem~\ref{thm:symp_transvec}, we clearly have $\tilde{x}_i F_{h_{i1}} F_{h_{i2}} = y_i$.
Hence we set $F_i \coloneqq F_{i-1} F_{h_{i1}} F_{h_{i2}}$ in this case. 
In both cases $F_i$ satisfies $x_j F_i = y_j \ \forall \ j=1,\ldots,i$. 
Setting $F \coloneqq F_t$ completes the inductive proof and it is clear that $F$ is the product of at most $2t$ transvections.
\end{proof}
\end{theorem}

\renewcommand{\algorithmicrequire}{\textbf{Input:}}
\renewcommand{\algorithmicensure}{\textbf{Output:}}

The algorithm defined implicitly by the above proof is stated explicitly in Algorithm~\ref{alg:transvec}.
\begin{algorithm}
{\singlespacing
\caption{Algorithm to find $F \in \text{Sp}(2n,\mathbb{F}_2)$ satisfying a linear system of equations, using Theorem~\ref{thm:symp_lineq}}
}
\label{alg:transvec}
  \begin{algorithmic}[1]	
\REQUIRE $x_i ,y_i \in \mathbb{F}_2^{2n}$ s.t. $\syminn{x_i}{x_j} = \syminn{y_i}{y_j} \ \forall \ i,j \in \{1,\ldots,t\}$. 
\ENSURE $F \in \text{Sp}(2n,\mathbb{F}_2)$ satisfying $x_i F = y_i \ \forall \ i \in \{1,\ldots,t\}$
\IF{$\syminn{x_1}{y_1} = 1$}
	\STATE set $h_1 \coloneqq x_1 + y_1$ and $F_1 \coloneqq F_{h_1}$.		
\ELSE
	\STATE $h_{11} \coloneqq w_1 + y_1, h_{12} \coloneqq x_1 + w_1$ and $F_1 \coloneqq F_{h_{11}} F_{h_{12}}$.
\ENDIF
		
\FOR{$i = 2,\ldots,t$}
	\STATE Calculate $\tilde{x}_i \coloneqq x_i F_{i-1}$ and $\syminn{\tilde{x}_i}{y_i}$.
	\IF{$\tilde{x}_i = y_i$}
		\STATE Set $F_i \coloneqq F_{i-1}$. \textbf{Continue}.
	\ENDIF
	\IF{$\syminn{\tilde{x}_i}{y_i} = 1$}
		\STATE Set $h_i \coloneqq \tilde{x}_i + y_i, F_i \coloneqq F_{i-1} F_{h_i}$.
	\ELSE
		\STATE Find a $w_i$ s.t. $\syminn{\tilde{x}_i}{w_i} = \syminn{y_i}{w_i} = 1$ and $\syminn{y_j}{w_i} = \syminn{y_j}{y_i} \ \forall \ j < i$. 
		\STATE Set $h_{i1} \coloneqq w_i + y_i, h_{i2} \coloneqq \tilde{x}_i + w_i, F_i \coloneqq F_{i-1} F_{h_{i1}} F_{h_{i2}}$.
	\ENDIF
\ENDFOR
\RETURN $F \coloneqq F_t$.
\end{algorithmic}
\end{algorithm}

Now we state our main theorem, which enables one to determine all symplectic solutions for a system of linear equations.

\begin{theorem}
\label{thm:symp_lineq_all}
Let $\{(u_a, v_a),\ a \in \{1,\ldots,n\}\}$ be a collection of pairs of (row) vectors that form a symplectic basis for $\mathbb{F}_2^{2n}$, where $u_a, v_a \in \mathbb{F}_2^{2n}$. 
Consider the system of linear equations $u_i F = u_i', v_j F = v_j'$, where $i \in \mathcal{I} \subseteq \{1,\ldots,n\}, {j \in \mathcal{J} \subseteq \{1,\ldots,n\}}$ and $F \in \text{Sp}(2n,\mathbb{F}_2)$.
Assume that the given vectors satisfy $\syminn{u_{i_1}}{u_{i_2}} = \syminn{u_{i_1}'}{u_{i_2}'} = 0, \syminn{v_{j_1}}{v_{j_2}} = \syminn{v_{j_1}'}{v_{j_2}'} = 0, \syminn{u_{i}}{v_{j}} = \syminn{u_{i}'}{v_{j}'} = \delta_{ij}$, where $i_1,i_2 \in \mathcal{I}, \ j_1,j_2 \in \mathcal{J}$, since symplectic transformations $F$ must preserve symplectic inner products.
Let $\alpha \coloneqq |\bar{\mathcal{I}}| + |\bar{\mathcal{J}}|$, where $\bar{\mathcal{I}}, \bar{\mathcal{J}}$ denote the set complements of $\mathcal{I},\mathcal{J}$ in $\{1,\ldots,n\}$, respectively.
Then there are $2^{\alpha(\alpha+1)/2}$ solutions $F$ to the given linear system.
\begin{proof}
By the definition of a symplectic basis (Definition~\ref{def:symp_basis}), we have $\syminn{u_a}{v_b} = \delta_{ab}$ and $\syminn{u_a}{u_b} = \syminn{v_a}{v_b} = 0$, where $a,b \in \{1,\ldots,n\}$.
The same definition extends to any (symplectic) subspace of $\mathbb{F}_2^{2n}$.
The linear system under consideration imposes constraints only on $u_i, i \in \mathcal{I}$ and $v_j, j \in \mathcal{J}$.
Let $W$ be the subspace of $\mathbb{F}_2^{2n}$ spanned by the symplectic pairs $(u_c,v_c)$ where $c \in \mathcal{I} \cap \mathcal{J}$ and $W^{\perp}$ be its orthogonal complement under the symplectic inner product, i.e., $W \coloneqq \langle \{ (u_c,v_c),\ c \in \mathcal{I} \cap \mathcal{J} \} \rangle$ and $W^{\perp} \coloneqq \langle \{ (u_d, v_d),\ d \in \bar{\mathcal{I}} \cup \bar{\mathcal{J}} \} \rangle$, where $\bar{\mathcal{I}}, \bar{\mathcal{J}}$ denote the set complements of $\mathcal{I},\mathcal{J}$ in $\{1,\ldots,n\}$, respectively.

Using the result of Theorem~\ref{thm:symp_lineq}, we first compute one solution $F_0$ for the given system of equations.
In the subspace $W$, $F_0$ maps $(u_c,v_c) \mapsto (u_c',v_c')$ for all $c \in \mathcal{I} \cap \mathcal{J}$ and hence we now have $W = \langle \{ (u_c',v_c'),\ c \in \mathcal{I} \cap \mathcal{J} \} \rangle$ spanned by its new basis pairs $(u_c',v_c')$.
However in $W^{\perp}$, $F_0$ maps $(u_d,v_d) \mapsto (u_d',\tilde{v}_d')$ or $(u_d,v_d) \mapsto (\tilde{u}_d',v_d')$ or $(u_d,v_d) \mapsto (\tilde{u}_d',\tilde{v}_d')$ depending on whether $d \in \mathcal{I} \cap \bar{\mathcal{J}}$ or $d \in \bar{\mathcal{I}} \cap \mathcal{J}$ or $d \in \bar{\mathcal{I}} \cap \bar{\mathcal{J}}$, respectively ($d \notin \mathcal{I} \cap \mathcal{J}$  by definition of $W^{\perp}$).
Note however that the subspace $W^{\perp}$ itself is fixed.
We observe that such $\tilde{u}_d'$ and $\tilde{v}_d'$ are not specified by the given linear system and hence form only a particular choice for the new symplectic basis of $W^{\perp}$. 
These can be mapped to arbitrary choices $\tilde{u}_d$ and $\tilde{v}_d$, while fixing other $u_d'$ and $v_d'$, as long as the new choices still complete a symplectic basis for $W^{\perp}$.
Hence, these form the degrees of freedom for the solution set of the given system of linear equations.
The number of such ``free'' vectors is exactly $|\bar{\mathcal{I}}| + |\bar{\mathcal{J}}| = \alpha$.
This can be verified by observing that the number of basis vectors for $W^{\perp}$ is $2 |\bar{\mathcal{I}} \cup \bar{\mathcal{J}}|$ and by calculating
\begin{align}
\text{Number\ of\ constrained} &\ \text{vectors\ in\ the\ new\ basis\ for}\ W^{\perp} \\
  &= |\mathcal{I} \setminus \mathcal{J}| + |\mathcal{J} \setminus \mathcal{I}| \\
  &= |\mathcal{I}| - |\mathcal{I} \cap \mathcal{J}| + |J| - |\mathcal{I} \cap \mathcal{J}| \\
  &= (m - |\bar{\mathcal{I}}|) + (m - |\bar{\mathcal{J}}|) - 2 (m - |\bar{\mathcal{I}} \cup \bar{\mathcal{J}}|) \\
  &= 2 |\bar{\mathcal{I}} \cup \bar{\mathcal{J}}| - (|\bar{\mathcal{I}}| + |\bar{\mathcal{J}}|) \\
  &= 2 |\bar{\mathcal{I}} \cup \bar{\mathcal{J}}| - \alpha .
\end{align}

Let  $d, d_1, d_2 \in \bar{\mathcal{I}} \cup \bar{\mathcal{J}}$ be indices of some symplectic basis vectors for $W^{\perp}$.
Then, the constraints on free vectors $\tilde{u}_d$ and $\tilde{v}_d$ are that $\syminn{\tilde{u}_{d_1}}{v_{d_2}'} = \syminn{u_{d_1}'}{\tilde{v}_{d_2}} = \syminn{\tilde{u}_{d_1}}{\tilde{v}_{d_2}} = \delta_{d_1 d_2}$ and all other pairs of vectors in the new basis set for $W^{\perp}$ be orthogonal to each other.
In the $d$-th symplectic pair --- $(\tilde{u}_d,v_d')$ or $(u_d',\tilde{v}_d)$ or $(\tilde{u}_d,\tilde{v}_d)$ --- of its new symplectic basis there is at least one free vector --- $\tilde{u}_d$ or $\tilde{v}_d$ or both, respectively.
For the first of the $\alpha$ free vectors, there are $2 |\bar{\mathcal{I}} \cup \bar{\mathcal{J}}| - \alpha$ symplectic inner product constraints (which are linear constraints) imposed by the $2 |\bar{\mathcal{I}} \cup \bar{\mathcal{J}}| - \alpha$ constrained vectors $u_d',v_d'$. 
Since $W^{\perp}$ has (binary) vector space dimension $2 |\bar{\mathcal{I}} \cup \bar{\mathcal{J}}|$ and each linearly independent constraint decreases the dimension by $1$, this leads to $2^{\alpha}$ possible choices for the first free vector.
For the second free vector, there are $\alpha-1$ degrees of freedom as it has an additional inner product constraint from the first free vector.
This leads to $2^{\alpha-1}$ possible choices for the second free vector, and so on.
Therefore, the given linear system has at least $\prod_{\ell=1}^{\alpha} 2^{\ell} = 2^{\alpha(\alpha+1)/2}$ symplectic solutions.

We will now argue that there cannot be more solutions.
The given system of equations can be represented compactly as $UF = V$, where $U, V \in \mathbb{F}_2^{(2n - \alpha) \times 2n}$ and $F$ is symplectic.
Observe that for each valid choice of $V$ the set of symplectic solutions is disjoint, and hence they form a partition of the binary symplectic group $\text{Sp}(2n,\mathbb{F}_2)$.
Therefore, it is enough to show that the product of the number of such valid matrices $V$ and $2^{\alpha(\alpha + 1)/2}$ is equal to the size of $\text{Sp}(2n,\mathbb{F}_2)$.
By defining $k \coloneqq n - \alpha$, the number of such valid matrices $V$ is given by
\begin{align}
& \left[ (2^{2n} - 1) \cdot (2^{2n-1} - 2^1) \cdot (2^{2n-2} - 2^2) \cdots (2^{2n-(n-1)} - 2^{n-1}) \right] \nonumber \\
  & \hspace{5cm} \times \left[ 2^{2n-n} \cdot 2^{2n-(n+1)} \cdots 2^{2n - (n+k-1)} \right] \\
  & = \left( 2^0 (2^{2n} - 1) \cdot 2^1(2^{2n-2} - 1) \cdot 2^2(2^{2n-4} - 1) \cdots 2^{n-1}(2^{(n+1) - (n-1)} - 1) \cdot 2^n \right) \nonumber \\
  & \hspace{7cm} \cdot 2^{n-1} \cdots 2^{n-(k-1)} \\
  & = 2^{\frac{n(n+1)}{2} + (k-1)n - \frac{k(k-1)}{2}} \prod_{j=1}^n (4^j - 1).
\end{align}
The counting in the first line is as follows.
First, we assume without loss of generality that the pairs of rows $i$ and $(n + i)$ of $V$ form a symplectic pair, for $i = 1,\ldots,k$, and the rows $k+1,\ldots,n$ are orthogonal to all rows of $V$ under the symplectic inner product.
More precisely, the inner products between pairs of rows of $V$ must be the same as those between corresponding pairs of rows of $U$.
But, we assume that we can perform a symplectic Gram-Schmidt process on $U$ so that the above assumption is valid.
For the first row of $V$, we can choose any non-zero vector and there are $(2^{2n} - 1)$ of them.
For the second row, we need to restrict to vectors that are orthogonal to the first row, and we need to eliminate the subspace generated by the first row.
Similarly, for the third row until the $n$-th row, we keep restricting to the subspace of vectors orthogonal to all previous rows and eliminate the subspace generated by all previous rows.
For the $(n+1)$-th row, it needs to be orthogonal to all rows starting from the second to the $n$-th, but it needs to have symplectic inner product $1$ with the first row.
Hence the dimension decreases by $n$ from $2n$, but notice that the subspace generated by the first $n$ rows cannot have any vector that has symplectic inner product $1$ with the first row.
Therefore, we need not subtract this subspace and this gives the count $2^{2n - n}$ for the $(n+1)$-th row.
A similar argument can be made for all remaining rows and this completes the argument for counting.
(It is easy to verify that by substituting $k = n$ above we obtain the size of $\text{Sp}(2n,\mathbb{F}_2)$ exactly.)
Now we expand the exponent of $2$ above to obtain $\frac{1}{2}(n^2 + 2nk - k^2 - n + k)$.

Recollect that the size of the symplectic group is $2^{n^2} \prod_{j=1}^n (4^j - 1)$, and we need to check that the number obtained by dividing this by $2^{\alpha(\alpha + 1)/2}$ is equal to the above number, for $\alpha = n - k$.
Since the product $\prod_{j=1}^n (4^j - 1)$ matches with the expression in the count above, we only have to check that the exponents of $2$ match.
Here, the exponent of $2$ is given by 
\begin{align}
n^2 - \frac{(n-k)(n-k+1)}{2} & = \frac{1}{2} \left( 2n^2 - (n^2 - 2nk + k^2 + n - k) \right) \\
  & = \frac{1}{2} \left(n^2 + 2nk - k^2 - n + k \right),
\end{align}
which equals the exponent calculated above.
This completes the proof that the given system has exactly $2^{\alpha(\alpha + 1)/2}$ solutions.

Finally, we show how to get each symplectic solution $F$ for the given linear system. 
First form the matrix $A$ whose rows are the new symplectic basis vectors for $\mathbb{F}_2^{2n}$ obtained under the action of $F_0$, i.e., the first $n$ rows are $u_c',u_d',\tilde{u}_d'$ and the last $n$ rows are $v_c',v_d',\tilde{v}_d'$. 
Observe that this matrix is symplectic and invertible.
Then form a matrix $B = A$ and replace the rows corresponding to free vectors with a particular choice of free vectors, chosen to satisfy the conditions mentioned above.
Note that $B$ and $A$ differ in exactly $\alpha$ rows, and that $B$ is also symplectic and invertible.
Determine the symplectic matrix $F' = A^{-1} B$ which fixes all new basis vectors obtained for $W$ and $W^{\perp}$ under $F_0$ except the free vectors in the basis for $W^{\perp}$.
Then this yields a new solution $F = F_0 F'$ for the given system of linear equations.
Note that if $\tilde{u}_d = \tilde{u}_d'$ and $\tilde{v}_d = \tilde{v}_d'$ for all free vectors, where $\tilde{u}_d', \tilde{v}_d'$ were obtained under the action of $F_0$ on $W^{\perp}$, then $F' = I_{2n}$.
Repeating this process for all $2^{\alpha(\alpha+1)/2}$ choices of free vectors enumerates all the solutions for the linear system under consideration.
\end{proof}
\end{theorem}

\begin{remark}
\normalfont
For any system of symplectic linear equations $x_i F = y_i,\ i=1,\ldots,t$ where the $x_i$ do not form a symplectic basis for $\mathbb{F}_2^{2n}$, we first calculate a symplectic basis $(u_j,v_j), \ j=1,\ldots,m$ using the symplectic Gram-Schmidt orthogonalization procedure discussed in~\cite{Koenig-jmp14}.
Then we transform the given system into an equivalent system of constraints on these basis vectors $u_j,v_j$ and apply Theorem~\ref{thm:symp_lineq_all} to obtain all symplectic solutions.
\end{remark}

The algorithm defined implicitly by the above proof is stated explicitly in Algorithm~\ref{alg:symp_lineq_all}.

\begin{algorithm}
{\singlespacing
\caption{Algorithm to determine all $F \in \text{Sp}(2n,\mathbb{F}_2)$ satisfying a linear system of equations, using Theorem~\ref{thm:symp_lineq_all}}
}
\label{alg:symp_lineq_all}
\begin{algorithmic}[1]

\REQUIRE $u_a,v_b \in \mathbb{F}_2^{2n}$ s.t. $\syminn{u_a}{v_b} = \delta_{ab}$ and $\syminn{u_a}{u_b} = \syminn{v_a}{v_b} = 0$, where $a,b \in \{1,\ldots,n\}$.

$u_i', v_j' \in \mathbb{F}_2^{2n}$ s.t. $\syminn{u_{i_1}'}{u_{i_2}'} = 0, \syminn{v_{j_1}'}{v_{j_2}'} = 0, \syminn{u_{i}'}{v_{j}'} = \delta_{ij}$, where $i,i_1,i_2 \in \mathcal{I}, \ j,j_1,j_2 \in \mathcal{J},\ \mathcal{I}, \mathcal{J} \subseteq \{1,\ldots,n\}$.

\ENSURE $\mathcal{F} \subset \text{Sp}(2n,\mathbb{F}_2)$ such that each $F \in \mathcal{F}$ satisfies $u_i F = u_i' \ \forall\ i \in \mathcal{I}$, and $v_j F = v_j' \ \forall \ j \in \mathcal{J}$.

\STATE Determine a particular symplectic solution $F_0$ for the linear system using Algorithm~\ref{alg:transvec}.

\STATE Form the matrix $A$ whose $a$-th row is $u_a F_0$ and $(n+b)$-th row is $v_b F_0$, where $a,b \in \{1,\ldots,n\}$. 

\STATE Compute the inverse of this matrix, $A^{-1}$, in $\mathbb{F}_2$.

\STATE Set $\mathcal{F} = \phi$ and $\alpha \coloneqq |\bar{\mathcal{I}}| + |\bar{\mathcal{J}}|$, where $\bar{\mathcal{I}}, \bar{\mathcal{J}}$ denote the set complements of $\mathcal{I},\mathcal{J}$ in $\{1,\ldots,n\}$, respectively.

\FOR{$\ell = 1,\ldots,2^{\alpha(\alpha+1)/2}$}
	
	\STATE Form a matrix $B_{\ell} = A$.

	\STATE For $i \notin \mathcal{I}$ and $j \notin \mathcal{J}$ replace the $i$-th and $(n+j)$-th rows of $B_{\ell}$ with arbitrary vectors such that $B_{\ell} \Omega B_{\ell}^T = \Omega$ and $B_{\ell} \neq B_{\ell'}$ for $1 \leq \ell' < \ell$.  
	
	\text{$\boldsymbol{/\ast}$ See proof of Theorem~\ref{thm:symp_lineq_all} for details or Appendix~\ref{sec:alg2_matlab} for example}   
	
	\text{\texttt{MATLAB\textsuperscript{\textregistered}} code $\boldsymbol{\ast /}$}
	
	\STATE Compute $F' = A^{-1} B$.
	
	\STATE Add $F_{\ell} \coloneqq F_0 F'$ to $\mathcal{F}$.
	
\ENDFOR

\RETURN $\mathcal{F}$

\end{algorithmic}
\end{algorithm}

For a given system of linear (independent) equations, if $\alpha = 0$ then the symplectic matrix $F$ is fully constrained and there is a unique solution.
Otherwise, the system is partially constrained and we refer to a solution $F$ as a \emph{partial} symplectic matrix.

\emph{Example}:
As an application of this theorem, we discuss the procedure to determine all symplectic solutions for the logical Phase gate $\lP_1$ discussed in Section~\ref{sec:css_phase}.
First we define a symplectic basis for $\mathbb{F}_2^{2n}$ using the binary vector representation of the logical Pauli operators and stabilizer generators of the $\llbr 6,4,2 \rrbr$ code.
\begin{align}
u_1 \coloneqq [110000,000000] \quad &, \quad v_1 \coloneqq [000000,010001] , \nonumber \\ 
u_2 \coloneqq [101000,000000] \quad &, \quad v_2 \coloneqq [000000,001001] , \nonumber \\ 
u_3 \coloneqq [100100,000000] \quad &, \quad v_3 \coloneqq [000000,000101] , \nonumber \\ 
u_4 \coloneqq [100010,000000] \quad &, \quad v_4 \coloneqq [000000,000011] , \nonumber \\ 
u_5 \coloneqq [111111,000000] \quad &, \quad v_5 \coloneqq [000000,000001] , \nonumber \\ 
u_6 \coloneqq [100000,000000] \quad &, \quad v_6 \coloneqq [000000,111111] .
\end{align}
Note that $v_5$ and $u_6$ do not correspond to either a logical Pauli operator or a stabilizer element but were added to complete a symplectic basis.
Hence we have $\mathcal{I} = \{1,2,3,4,5\}, \mathcal{J} = \{1,2,3,4,6\}$ and $\alpha = 1 + 1 = 2$.
As discussed in Section~\ref{sec:css_phase}, we impose constraints on all $u_i,v_j$ except for $i=6$ and $j=5$.
Therefore, as per the notation in the above proof, we have $W \coloneqq \langle \{(u_1,v_1), \ldots, (u_4,v_4)\} \rangle$ and $W^{\perp} \coloneqq \langle \{(u_5,v_5), (u_6,v_6)\} \rangle$.
Using Algorithm~\ref{alg:transvec} we obtain a particular solution $F_0 = T_B$ where $B$ is given in~\eqref{eq:css_phase_B}.
Then we compute the action of $F_0$ on the bases for $W$ and $W^{\perp}$ to get
\begin{align}
u_i F_0 \coloneqq u_i', & \ v_j F_0 \coloneqq v_j',\ i \in \mathcal{I},\ j \in \mathcal{J}, \ \ \text{and} \nonumber \\ 
u_6 F_0 = [100000,000000] \coloneqq \tilde{u}_6', & \ v_5 F_0 = [000000,000001] \coloneqq \tilde{v}_5' ,
\end{align}
where $u_i', v_j'$ are the vectors obtained in Section~\ref{sec:css_phase}.
Then we identify $\tilde{v}_5$ and $\tilde{u}_6$ to be the free vectors and one particular solution is $\tilde{v}_5 = \tilde{v}_5', \tilde{u}_6 = \tilde{u}_6'$.
In this case we have $2^{\alpha} = 2^2 = 4$ choices to pick $\tilde{v}_5$ (since we need $\syminn{u_5}{\tilde{v}_5} = 1, \syminn{v_6}{\tilde{v}_5} = 0$) and for each such choice we have $2^{\alpha-1} = 2$ choices for $\tilde{u}_6$.
Next we form the matrix $A$ whose $i$-th row is $u_i'$ and $(6+j)$-th row is $v_j'$, where $i \in \mathcal{I},j \in \mathcal{J}$.
We set the $6$th row to be $\tilde{u}_6'$ and the $11$th row to be $\tilde{v}_5'$.
Then we form a matrix $B = A$ and replace rows $6$ and $11$ by one of the $8$ possible pair of choices for $\tilde{u}_6$ and $\tilde{v}_5$, respectively.
This yields the matrix $F' = A^{-1} B$ and the symplectic solution $F = F_0 F'$.
Looping through all the $8$ choices we obtain the solutions listed in Appendix~\ref{sec:css_phase_all}.

\begin{theorem}
\label{thm:stabilizer_solutions}
For an $\llbr n,k \rrbr$ stabilizer code, the number of solutions for each logical Clifford operator is $2^{r(r+1)/2}$, where $r = n-k$.
\begin{proof}
Let $u_i, v_i \in \mathbb{F}_2^{2n}$ represent the logical Pauli operators $\lX_i, \lZ_i$, for $i=1,\ldots,k$, respectively, i.e., $\gamma(\lX_i) = u_i, \gamma(\lZ_i) = v_i$, where $\gamma$ is the map defined in~\eqref{eq:gamma}.
Since $\lX_i \lZ_i = -\lZ_i \lX_i$ and $\lX_i \lZ_j = \lZ_j \lX_i$ for all $j \neq i$, it is clear that $\syminn{u_i}{v_j} = \delta_{ij}$ for $i,j \in \{1,\ldots,k\}$ and hence they form a partial symplectic basis for $\mathbb{F}_2^{2n}$.
Let $u_{k+1},\ldots,u_n$ represent the stabilizer generators, i.e., $\gamma(S_j) = u_{k+j}$ where the stabilizer group is $S = \langle S_1,\ldots,S_r \rangle$.
Since by definition $\lX_i, \lZ_i$ commute with all stabilizer elements, it is clear that $\syminn{u_i}{u_j} = \syminn{v_i}{u_j} = 0$ for $i \in \{1,\ldots,k\}, j \in \{k+1,\ldots,n\}$.
To complete the symplectic basis we find vectors $v_{k+1},\ldots,v_n$ s.t. $\syminn{u_i}{v_j} = \delta_{ij} \ \forall \ i,j \in \{1,\ldots,n\}$.
Now we note that for any logical Clifford operator, the conjugation relations with logical Paulis yield $2k$ constraints, on $u_i,v_i$ for $i \in \{1,\ldots,k\}$, and the normalization condition on the stabilizer yields $r$ constraints, on $u_{k+1},\ldots,u_n$.
Hence we have $\bar{\mathcal{I}} = \phi, \bar{\mathcal{J}} = \{k+1,\ldots,n\}$, as per the notation in Theorem~\ref{thm:symp_lineq_all}, and thus $\alpha = |\bar{\mathcal{I}}| + |\bar{\mathcal{J}}| = n - k = r$.
\end{proof}
\end{theorem}

Note that for each symplectic solution there are multiple decompositions into elementary forms (from Table~\ref{tab:std_symp}) possible, and one possibility is given in Theorem~\ref{thm:Trung}.
Although each decomposition yields a different circuit, all of them will act identically on $X_N$ and $Z_N$, defined in~\eqref{eq:XnZn}, under conjugation.
Once a logical Clifford operator is defined by its conjugation with the logical Pauli operators, a physical realization of the operator could either normalize the stabilizer or centralize it, i.e., fix each element of the stabilizer group under conjugation.
We show that any obtained normalizing solution can be converted into a centralizing solution.

\begin{theorem}
\label{thm:normalize_centralize}
For an $\llbr n,k \rrbr$ stabilizer code with stabilizer $S$, each physical realization of a given logical Clifford operator that normalizes $S$ can be converted into a circuit that centralizes $S$ while realizing the same logical operation.
\begin{proof}
Let the symplectic solution for a specific logical Clifford operator $\bar{g} \in \text{Cliff}_N$ that normalizes the stabilizer $S$ be denoted by $F_n$. 
Define the logical Pauli groups $\lX \coloneqq \langle \lX_1,\ldots,\lX_{k} \rangle$ and $\lZ \coloneqq \langle \lZ_1,\ldots,\lZ_{k} \rangle$. 
Let $\gamma(\lX)$ and $\gamma(\lZ)$ denote the matrices whose rows are $\gamma(\lX_i)$ and $\gamma(\lZ_i)$, respectively, for $i=1,\ldots,k$, where $\gamma$ is the map defined in~\eqref{eq:gamma}.
Similarly, let $\gamma(S)$ denote the matrix whose rows are the images of the stabilizer generators under the map $\gamma$.
Then, by stacking these matrices as in the proof of Theorem~\ref{thm:stabilizer_solutions}, we observe that $F_n$ is a solution of the linear system
\begin{align*}
\begin{bmatrix}
\gamma(\lX) \\
\gamma(S) \\
\gamma(\lZ)
\end{bmatrix} F_n = 
\begin{bmatrix}
\gamma(\lX') \\
\gamma(S') \\
\gamma(\lZ')
\end{bmatrix} ,
\end{align*}
where $\lX', \lZ'$ are defined by the conjugation relations of $\bar{g}$ with the logical Paulis, i.e., $\bar{g} \lX_i \bar{g}^{\dagger} = \lX_i', \bar{g} \lZ_i \bar{g}^{\dagger} = \lZ_i'$, and $S'$ denotes the stabilizer group of the code generated by a different set of generators than that of $S$.
Note, however, that as a group $S' = S$.
The goal is to find a different solution $F_c$ that centralizes $S$, i.e., we replace $\gamma(S')$ with $\gamma(S)$ above.

We first find a matrix $K \in \text{GL}(r, \mathbb{F}_2)$ such that $K \gamma(S') = \gamma(S)$, which always exists since generators of $S'$ span $S$ as well.
Then we determine a symplectic solution $H$ for
\begin{align*}
\begin{bmatrix}
\gamma(\lX) \\
\gamma(S) \\
\gamma(\lZ)
\end{bmatrix} H = 
\begin{bmatrix}
\gamma(\lX) \\
K \gamma(S) \\
\gamma(\lZ)
\end{bmatrix} ,
\end{align*}
so that $H$ satisfies $K \gamma(S) = \gamma(S) H$ while fixing $\gamma(\lX)$ and $\gamma(\lZ)$.
Then, since $K$ is invertible,
\begin{align*}
\begin{bmatrix}
I_{k} &   &         \\
        & K &         \\
        &   & I_{k}
\end{bmatrix}
\begin{bmatrix}
\gamma(\lX) \\
\gamma(S) \\
\gamma(\lZ)
\end{bmatrix} F_n = 
\begin{bmatrix}
I_{k} &   &         \\
        & K &         \\
        &   & I_{k}
\end{bmatrix}
\begin{bmatrix}
\gamma(\lX') \\
\gamma(S') \\
\gamma(\lZ')
\end{bmatrix}
\Rightarrow 
\begin{bmatrix}
\gamma(\lX) \\
\gamma(S) \\
\gamma(\lZ)
\end{bmatrix} H F_n = 
\begin{bmatrix}
\gamma(\lX') \\
\gamma(S) \\
\gamma(\lZ')
\end{bmatrix} .
\end{align*}
Hence $F_c \coloneqq H F_n$ is a centralizing solution for $\bar{g}$.
Note that there are $2^{r(r+1)/2}$ solutions for $H$, as per the result of Theorem~\ref{thm:stabilizer_solutions}, with the operator being the identity operator on the logical qubits, and these produce all centralizing solutions for $\bar{g}$.
\end{proof}
\end{theorem}

The above result demonstrates the relationship between the two solutions for the targeted Hadamard operator discussed in Section~\ref{sec:css_had}.
As noted in that section, after the logical transversal Hadamard operator, although any normalizing solution can be converted into a centralizing solution, the optimal solution with respect to a suitable metric need not always centralize the stabilizer.
Anyhow, we can always setup the problem of identifying a symplectic matrix, representing the physical circuit, by constraining it to centralize the stabilizer.

\subsection{The LCS Algorithm}

The general procedure to determine all symplectic solutions, and their circuits, for a logical Clifford operator for a stabilizer code is summarized in Algorithm~\ref{alg:log_ops}.
For the $\llbr 6,4,2 \rrbr$ CSS code, we employed Algorithm~\ref{alg:log_ops} to determine the solutions listed in Appendix~\ref{sec:css642_all_phy_ops} for each of the operators discussed before.

\begin{algorithm}
\caption{Algorithm to determine logical Clifford operators for a stabilizer code}
\label{alg:log_ops}

\begin{algorithmic}[1]

\STATE Determine the target logical operator $\bar{g}$ by specifying its action on logical Paulis $\lX_i, \lZ_i$~\cite{Gottesman-arxiv09}: $\bar{g} \lX_i \bar{g}^{\dagger} = \lX_i', \bar{g} \lZ_i \bar{g}^{\dagger} = \lZ_i'$ .

\STATE Transform the above relations into linear equations on $F \in \text{Sp}(2n,\mathbb{F}_2)$ using the map $\gamma$ in~\eqref{eq:gamma} and the result of Theorem~\ref{thm:symp_action}, i.e., $\gamma(\lX_i) F = \gamma(\lX_i'), \gamma(\lZ_i) F = \gamma(\lZ_i')$. 
Add the conditions for normalizing the stabilizer $S$, i.e., $\gamma(S) F = \gamma(S')$.

\STATE Calculate the feasible symplectic solution set $\mathcal{F}$ using Algorithm~\ref{alg:symp_lineq_all} by mapping $\lX_i, S, \lZ_i$ to $u_i, v_i$ as in Theorem~\ref{thm:stabilizer_solutions}.

\STATE Factor each $F \in \mathcal{F}$ into a product of elementary symplectic transformations listed in Table~\ref{tab:std_symp}, possibly using the algorithm given in~\cite{Can-2017a} (which is restated in Theorem~\ref{thm:Trung} here), and compute the physical Clifford operator $\bar{g}$.

\STATE Check for conjugation of $\bar{g}$ with the stabilizer generators and for the conditions derived in step 1.
If some signs are incorrect, post-multiply by an element from $HW_N$ as necessary to satisfy all these conditions (apply~\cite[Proposition 10.4]{Nielsen-2010} for $S^{\perp} = \langle S, \lX_i, \lZ_i \rangle$, using~\eqref{eq:gamma}). 
Since $HW_N$ is the kernel of the map $\phi$ in~\eqref{eq:phi}, post-multiplication does not change $F$.

\STATE Express $\bar{g}$ as a sequence of physical Clifford gates corresponding to the elementary symplectic matrices obtained from the factorization in step 4 (see Section~\ref{sec:elem_symp} for the circuits for these matrices).

\end{algorithmic}
\end{algorithm}

The \texttt{MATLAB\textsuperscript{\textregistered}} programs for all algorithms in this work are available at \url{https://github.com/nrenga/symplectic-arxiv18a}.
We executed our programs on a laptop running the Windows 10 operating system (64-bit) with an Intel\textsuperscript{\textregistered} Core\textsuperscript{\texttrademark} i7-5500U @ 2.40GHz processor and 8GB RAM.
For the $\llbr 6,4,2 \rrbr$ CSS code, it takes about 0.5 seconds to generate all 8 symplectic solutions and their circuits for one logical Clifford operator.
For the $\llbr 5,1,3 \rrbr$ perfect code, it takes about 20 seconds to generate all 1024 solutions and their circuits.
Note that for step 5 in Algorithm~\ref{alg:log_ops}, we use 1-qubit and 2-qubit unitary matrices (from $\text{Cliff}_{2^2}$) to calculate conjugations for the Pauli operator on each qubit, at each circuit element at each depth (see Def.~\ref{def:depth}), and then combine the results to compute the conjugation of $\bar{g}$ with a stabilizer generator or logical Pauli operator.
Owing to our naive implementation, we observe that most of the time is consumed in computing Kronecker products and not in calculating the symplectic solutions.

\begin{remark}
\label{rem:stab_freedom}
\normalfont
Observe that, in our LCS algorithm, we are not taking into account the degrees of freedom provided by stabilizers.
That is, if the logical operator $\bar{g}$ is required to map $\lX_i \mapsto \lX_i'$, then an equivalent condition is to map $\lX_i \mapsto \lX_i' \cdot \boldsymbol{s}$, where $\boldsymbol{s} \in S$ is any stabilizer element for the given code.
A similar statement is true for $\lZ_i \mapsto \lZ_i'$.
An explicit example for this scenario is the $\lcnot{1}{2}$ for the $\llbr 4,2,2 \rrbr$ code with the logical Paulis defined instead as $\lX_1 = X_1 X_2, \lX_2 = X_2 X_4, \lZ_1 = Z_1 Z_3, \lZ_2 = Z_3 Z_4$.
The operation $\lcnot{1}{2}$ can simply be defined as swapping qubits $2$ and $4$, but this maps $\lZ_2 \mapsto Z_2 Z_3 = \lZ_1 \lZ_2 \cdot \bg^Z$, where $\bg^Z = Z_1 Z_2 Z_3 Z_4$, instead of just $\lZ_2 \mapsto \lZ_1 \lZ_2$ as the above algorithm would require.

In principle, the LCS algorithm can be easily modified to consider these possibilities, but this significantly increases the computational complexity of the algorithm.
A better understanding of the structure of logical Clifford operators for a given general stabilizer code, or even heuristics developed to identify which degrees of freedom are worth considering for a given code, would greatly improve the quality of solutions produced by the algorithm.
\end{remark}

\section{Some Thoughts for Future Work}

In this chapter we have developed a systematic algorithm that translates logical Clifford operators into their physical realizations on stabilizer codes.
We used binary symplectic matrices to make this translation efficient by working with only $4n^2$ binary variables rather than $2^{2n}$ complex variables for the Clifford unitary on $n$ qubits.
While several works in the literature have performed this translation for specific codes and operators, the LCS algorithm is applicable to generic stabilizer codes.
However, the circuits produced by the algorithm are not necessarily fault-tolerant.
In this regard, a better understanding of symplectic matrices and their relation to circuit structure might lead to a systematic approach for fault-tolerant logical Clifford operators.
A specific direction of research could be an attempt to incorporate flag fault-tolerant schemes~\cite{Chao-arxiv17a,Chao-arxiv17b,Chamberland-quantum18,Tansuwannont-arxiv18,Chao-arxiv19} in the symplectic framework.
Since ensuring fault-tolerance in such schemes involves combinatorial arguments on Clifford circuits, this incorporation might not be straightforward.
But a formalism that takes advantage of the symplectic framework and flag-based schemes would go a long way in producing fault-tolerant logical Clifford gates for arbitrary stabilizer codes.
This will provide us access to a larger variety of code parameters rather than restricting ourselves to well-known codes and code families.

A common approach to perform fault-tolerant logical Clifford gates on topological codes or hypergraph product codes is to exploit carefully constructed measurements~\cite{Fowler-arxiv12,Krishna-arxiv19a,Krishna-arxiv19b}.
Since these are not unitary operations, they are outside the purview of the symplectic framework.
Hence, even if the LCS algorithm is refined to produce fault-tolerant circuits, the schemes cannot be compared to measurement-based schemes such as braiding.
So, it is worth investigating if there is a theoretical (algebraic) framework that incorporates both Clifford operations and (Pauli) measurements simultaneously.
Such a framework would enable resource comparisons between a larger variety of schemes.

%% file: ch6_qfd_gates.tex

\label{ch:ch6_qfd_gates}

\section{Challenges in Extending LCS Algorithm}

In Chapter~\ref{ch:ch5_lcs_algorithm} we discussed a systematic approach to synthesize logical Clifford gates for arbitrary stabilizer codes.
We know that the Clifford group does not provide universal quantum computation, and that we need to implement at least one non-Clifford logical gate.
However, we encounter at least two issues if we attempt to directly generalize the LCS algorithm.
Firstly, for realizing a logical non-Clifford gate on a stabilizer code we need to implement a physical non-Clifford gate.
But such a physical operation might not map all stabilizers to stabilizers since non-Clifford gates can map a Pauli operator outside of the Heisenberg-Weyl group (under conjugation).
Hence, we need to find an alternative to our condition in the LCS algorithm that the physical operator normalize (or centralize) the stabilizer.
Secondly, the key tool used in the LCS algorithm is the binary symplectic framework for Clifford gates.
However, there is no clear symplectic connection for non-Clifford gates and this makes the extension of the LCS algorithm challenging.
We will address this latter problem in this chapter and discuss the former in the next chapter.
We will begin by describing some additional motivation for this work.

\section{Motivation: The Clifford Hierarchy}

Gottesman and Chuang showed~\cite{Gottesman-nature99} that universal quantum computation can be achieved via quantum teleportation if one has access to Bell-state preparation, Bell-basis measurements, and arbitrary single-qubit operations on \emph{known} ancilla states.
Their protocol involved construction of the \emph{Clifford hierarchy}.
By definition of the hierarchy, when elements in the $\ell$-th level act by conjugation on Pauli matrices, they produce a result in the $(\ell-1)$-th level.
The first level is the Heisenberg-Weyl group of Pauli matrices and the second level is the Clifford group, i.e., $\MCC^{(1)} \coloneqq HW_N$ and the higher levels $\ell > 1$ are defined recursively by
\begin{align}
\MCC^{(\ell)} \coloneqq \{ U \in \mathbb{U}_N \colon U E(a,b) U^{\dagger} \in \MCC^{(\ell-1)} \ \forall \ E(a,b) \in \MCC^{(1)} \},
\end{align}
where $\mathbb{U}_N$ denotes the group of all $N \times N$ unitary matrices~\cite{Gottesman-nature99}.

It is known that for $\ell \geq 3$ the unitaries at a level do not form a group~\cite{Zeng-physreva08}. 
The \emph{Gottesman-Knill} theorem~\cite{Gottesman-arxiv98} established that the Clifford group can be efficiently simulated classically and hence does not provide a significant quantum advantage over classical computation (also see~\cite{Aaronson-pra04} for a classical simulator of such circuits).
But the Clifford group combined with \emph{any} unitary outside the group enables arbitrarily good approximation of any other unitary, thus enabling universal quantum computation given the ability to execute a \emph{finite} set of gates~\cite{Boykin-arxiv99}.
The standard choice outside the group is the ``$\pi/8$''- or $T$-gate which belongs to the third level of the Clifford hierarchy.
However, unitaries decomposed with this fixed set of gates could result in circuits with large depth that are especially hard to implement reliably in near-term quantum computers.
It is now established that constant-depth circuits indeed provide a quantum advantage over classical computation~\cite{Bravyi-science18}.
Hence, it is imperative to understand the structure of this hierarchy in order to leverage higher level unitaries and obtain smaller depth circuits.
Moreover, \emph{native} operations in quantum technologies might not belong to the Clifford+$T$ set of gates but to higher levels of the hierarchy, e.g., $X$- and $Z$-rotations of arbitrary angles in trapped-ion systems~\cite{Linke-nas17}. 
Since any circuit must eventually be translated to such native operations by a compiler, this provides us an opportunity to directly consider such operations in circuit decompositions.

There have been several attempts at understanding the structure of the hierarchy~\cite{Zeng-physreva08,Bengtsson-jphysa14,Cui-physreva17}, but the complete structure still remains elusive.
Since the Clifford group is the normalizer of the Pauli group in the unitary group, it permutes maximal commutative subgroups of the Pauli group under conjugation.
Zeng et al.~\cite{Zeng-physreva08} considered a class of unitaries called the \emph{semi-Clifford} operations, which are defined as those unitaries that map \emph{at least one} maximal commutative subgroup of the Pauli group to another maximal commutative subgroup of the Pauli group.
While Gottesman and Chuang~\cite{Gottesman-nature99} used the standard two-ancilla quantum teleportation circuit to demonstrate universal computation, Zhou et al.~\cite{Zhou-pra00} showed that these semi-Clifford operations can be applied via teleportation with \emph{one less} ancilla qubit.
Zeng et al. showed that for $n=1,2$, the unitaries at any level $\ell$ of the hierarchy are semi-Clifford, and that for $n=3$ all the unitaries in level $\ell=3$ are semi-Clifford.
For $n > 2$ and $\ell = 3$, they conjectured that all unitaries are semi-Clifford operations as well and we believe this question remains open.
Furthermore, they also defined \emph{generalized} semi-Clifford operations to be those unitaries that map the span of at least one maximal commutative subgroup of the Pauli group to the span of another maximal commutative subgroup of the Pauli group, where span refers to the group algebra over the complex field.
For $n > 2$ and $\ell > 3$ they conjectured that all unitaries are generalized semi-Clifford operations but, to the best of our knowledge, this question also remains open.

\emph{Stabilizer states} are the unit vectors that belong to the orbit of the computational basis state $\dket{0}^{\otimes n}$ under Clifford operations~\cite{Aaronson-pra04,Dehaene-physreva03}.
Equivalently, they are the common eigenvectors of the commuting Hermitian matrices forming maximal commutative subgroups of the Pauli group.
It is well-known that certain stabilizer states can be grouped and arranged to form mutually unbiased bases (MUBs), which means pairs of vectors within a group are orthogonal and pairs formed from different groups have a small inner product~\cite{Calderbank-seta10,Tirkkonen-isit17}.
The images of stabilizer states under the action of a third level unitary from the Clifford hierarchy are known to produce the states in Alltop's construction of MUBs~\cite{Bengtsson-jphysa14}.
These MUBs are exactly a type of ``magic states'' that provide an alternative path to universal quantum computation~\cite{Bravyi-pra05}.
Bengtsson et al.~\cite{Bengtsson-jphysa14} studied the role of order-$3$ Clifford operators, their relation to Alltop MUBs, and a deep connection between Alltop MUBs and symmetric informationally complete (SIC) measurements. 

\subsection{Our Contributions}

The starting point for our contributions is~\cite{Cui-physreva17}, where Cui et al. revealed the structure of the \emph{diagonal} gates in each level of the Clifford hierarchy.
For a single qudit with prime dimension $p$, they constructed a new hierarchy from unitaries of the form $U_{\ell,a} \coloneqq \sum_{j \in \mathbb{Z}_p} \exp\left( \frac{2\pi \imath}{p^{\ell}} j^a \right) \dketbra{j}$, where $\mathbb{Z}_p \coloneqq \{0,1,\ldots,p-1\}, \imath \coloneqq \sqrt{-1}$, and $a$ is an integer such that $1 \leq a \leq p-1$.
They showed that such unitaries determine all diagonal unitaries in the level $(p-1)(\ell-1)+a$ of the Clifford hierarchy, and they also extended the result to multiple qudits.
In this chapter, we provide a simpler description of certain diagonal unitaries (for qubits, i.e., $p=2$) and reveal their structure more explicitly by making a connection to symmetric matrices $R$ over the ring $\MZ_{2^{\ell}}$ of integers modulo $2^{\ell}$.
We define diagonal unitaries of the form $\tau_R^{(\ell)} \coloneqq \text{diag}\left( \xi^{v R v^T \bmod 2^{\ell}} \right) = \sum_{v \in \MZ_2^n} \xi^{v R v^T \bmod 2^{\ell}} \dketbra{v}$, where $\xi \coloneqq e^{2\pi\imath/2^{\ell}}$ and $v$ is a binary (row) vector indexing the rows of the matrix, and prove that all two-local and certain higher locality diagonal unitaries in the $\ell$-th level can be described in this form (see Theorem~\ref{thm:diagonals} and Remark~\ref{rem:u_ccz}).
We derive precise formulas for their action on Pauli matrices, and show that the result naturally involves a unitary of the form $\tau_{\tilde{R}}^{(\ell-1)}$, thereby yielding a recursion, where $\tilde{R}$ is a symmetric matrix in $\MZ_{2^{\ell-1}}$ that is a function of $R$ and the Pauli matrix (see Corollary~\ref{cor:tauR_conjugate}).
Hence the matrix $R$ contains \emph{all} the information about the diagonal unitary $\tau_R^{(\ell)}$.
Finally, we formally prove that these diagonal unitaries form a subgroup of all diagonal gates in the $\ell$-th level, and that the map from these diagonal unitaries to symmetric matrices is an isomorphism.

During this process, we obtain a function $q^{(\ell-1)}(v; R, a, b)$ (that fully characterizes $\tau_{\tilde{R}}^{(\ell-1)}$), where $(a,b)$ represents a Pauli matrix, and we demonstrate some of its properties.
We also provide examples of matrices $R$ for some standard gates, and for the non-Clifford ``$\pi/8$''-gate we clarify the connection between our formula and the well-known action of this gate on the Pauli $X$ matrix.
These symmetric matrices identify symplectic matrices over $\MZ_{2^{\ell}}$, and this approach \emph{unifies} these diagonal elements of the Clifford hierarchy with the Clifford group that can be mapped to binary symplectic matrices~\cite{Dehaene-physreva03,Gottesman-arxiv09,Rengaswamy-isit18}. 
We believe this is the first work that provides such a unification, and our results indicate that some non-diagonal unitaries in the Clifford hierarchy might be explored by extending other binary symplectic matrices to rings $\MZ_{2^{\ell}}$.

\subsection{Other Potential Applications}

Zeng et al. showed that a semi-Clifford operator $g$ is of the form $g = C_1 D C_2$, where $C_1, C_2$ are Cliffords and $D$ is a diagonal unitary~\cite{Zeng-physreva08}. 
Hence, using calculations similar to those in Section~\ref{sec:discuss} it might be possible to explore the above conjectures by Zeng et al. on semi-Cliffords.
Furthermore, binary symplectic matrices have been used to efficiently decompose Clifford unitaries into circuits composed of standard gates~\cite{Dehaene-physreva03,Can-sen18,Rengaswamy-isit18}.
Using our unification, a better understanding of the interaction between binary and integer symplectic matrices might produce efficient algorithms to decompose unitaries into Cliffords and diagonal gates, thereby also reducing circuit depth.

As another application, classical simulation of quantum circuits is currently an important research topic since it serves at least two purposes: (i) it provides a method to check the integrity of the results produced by near-term quantum computers, and (ii) it refines our understanding of the kind of quantum circuits that indeed provide a computational advantage over classical computation.
Bravyi et al.~\cite{Bravyi-arxiv18} have developed a comprehensive mathematical framework of the notion of \emph{stabilizer rank}, which measures the number of stabilizer states required to express the output state of a given unitary operator, acting on $\dket{0}^{\otimes n}$ without loss of generality.
(Recollect that since Clifford operations can be efficiently simulated classically, each stabilizer state can be easily handled by the \texttt{CHP} simulator of Aaronson and Gottesman~\cite{Aaronson-pra04}, the package on which Bravyi et al. build.)
Using this notion, they have developed a powerful simulator of quantum circuits that can currently handle about $40$-$50$ qubits and over $60$ non-Clifford gates without resorting to high-performance computers.
As they highlight, a key feature of their simulator and a reason for its efficiency is the decomposition of unitaries into Cliffords and arbitrary diagonal gates, such as arbitrary angle $Z$-rotations and controlled-controlled-$Z$ (CCZ) gates, instead of just Cliffords and $T$-gates.
Hence, it is natural to investigate whether our symplectic representation of certain diagonal unitaries can be used to extend their simulator.

\section{QFD Gates}
\label{sec:diag_cliff}

Let $\xi \coloneqq \exp\left( \frac{2\pi \imath}{2^{\ell}} \right)$ and $R$ be an $n \times m$ \emph{symmetric} matrix over $\MZ_{2^{\ell}}$. 
Consider the diagonal unitary matrix
\begin{align}
\label{eq:qfd_definition}
\tau_R^{(\ell)} \coloneqq \text{diag}\left( \xi^{v R v^T \bmod 2^{\ell}} \right) = \sum_{v \in \MZ_2^n} \xi^{v R v^T} \dketbra{v},
\end{align}
where $v \in \MZ_2^n$ indexes the rows of $\tau_R^{(\ell)}$.
We will derive the action of $\tau_R^{(\ell)}$ on $E(a,b)$ under conjugation, prove that $\tau_R^{(\ell)} \in \MCC_d^{(\ell)}$, and argue that all two-local and certain higher locality diagonal gates can be represented in this form. 
Finally, we will show that the map $\gamma \colon \MCC_{d,\text{sym}}^{(\ell)} \rightarrow \MZ_{2^{\ell},\text{sym}}^{n \times n}$ defined by $\gamma(\tau_R^{(\ell)}) \coloneqq R$ is an isomorphism, where the subscript ``sym'' denotes symmetric matrices whose diagonal entries are in $\MZ_{2^{\ell}}$ and off-diagonal entries are in $\MZ_{2^{\ell-1}}$, and $\MCC_{d,\text{sym}}^{(\ell)} \subset \MCC_d^{(\ell)}$ is the subgroup of all unitaries of the form $\tau_R^{(\ell)}$. 
Owing to the exponent involved in~\eqref{eq:qfd_definition}, we refer to these as \emph{Quadratic Form Diagonal (QFD)} gates.

Given two vectors $v, w \in \MZ_2^n$, their binary sum can be expressed over $\MZ_{2^{\ell}}$ as
\begin{align}
v \oplus w & = v + w - 2 (v \ast w)\ (\bmod\ 2^{\ell}),
\end{align}
where $v \ast w$ represents the element-wise product of $v$ and $w$, i.e., $v \ast w = [v_1 w_1, v_2 w_2, \cdots, v_n w_n]$.


\begin{lemma}
\label{lem:binary_quad}
For any $v, w \in \MZ_2^n$, symmetric $R \in \MZ_{2^{\ell}}^{n \times n}$, and $\ell \in \mathbb{N}$, the following holds:
\begin{align}
(v \oplus w) R (v \oplus w)^T & \equiv (v + w) R (v + w)^T - 4 \eta(v ; R,w) \ (\bmod\ 2^{\ell}), \\
\text{where}\ \ \eta(v; R,w) & \coloneqq [(v + w) - (v \ast w)] R (v \ast w)^T.
\end{align}
\begin{proof}
We observe that
\begin{align}
(v \oplus w) R (v \oplus w)^T & = [(v + w) - 2(v \ast w)] R [(v + w) - 2(v \ast w)]^T  \\
  & = (v + w) R (v + w)^T - 4 (v + w) R (v \ast w)^T + 4 (v \ast w) R (v \ast w)^T  \\
  & = (v + w) R (v + w)^T - 4 [(v + w) - (v \ast w)] R (v \ast w)^T  \\
  & = (v + w) R (v + w)^T - 4 (v\ \text{OR}\ w) R (v\ \text{AND}\ w)^T  \\
  & = (v + w) R (v + w)^T - 4 \eta(v ; R,w) \ (\bmod\ 2^{\ell}).  \tag*{\qedhere}
\end{align}
\end{proof}
\end{lemma}

For a given binary vector $x$, let $D_x \coloneqq \text{diag}(x)$ denote the diagonal matrix with the diagonal set to $x$. 
Then $D_w$ projects onto $w$ so that $D_w v^T = (v \ast w)^T$.
Similarly, $D_{\bar{w}}$ projects onto $\bar{w} = w \oplus \vecnot{1} = \vecnot{1} - w$ so that $v D_{\bar{w}} = v \ast (\vecnot{1} - w) = v - (v \ast w)$, where $\vecnot{1}$ denotes the vector with all entries $1$.
Also, by observing that $v_i^2 = v_i$ for all $i \in \{1,\ldots,n\}$, the inner product $uv^T$ can be expressed as the quadratic form $v D_u v^T$, where $u \in \MZ_2^n$.
Thus, for any $v, w \in \MZ_2^n$, we can write $w R (v \ast w)^T = w R D_w v^T = v D_{wRD_w} v^T$.
It follows that
\begin{align}
\eta(v; R,w) & \coloneqq [(v + w) - (v \ast w)] R (v \ast w)^T \nonumber \\
             & = v \, [D_{\bar{w}} R D_w + D_{wRD_w}] \, v^T \nonumber \\
             & = v \, [D_{w} R D_{\bar{w}} + D_{wRD_w}] \, v^T.
\end{align}
%
Next we determine the action of $\tau_R^{(\ell)}$ on $E(a,b)$ under conjugation (see Section~\ref{sec:elem_symp} to compare with the calculation for $t_R \in \text{Cliff}_N$ listed in Table~\ref{tab:std_symp}). 


\begin{lemma}
\label{lem:tauR_conjugate}
Let $\ell \geq 2, v \in \MZ_2^n, a = a_0 + 2a_1 +4a_2 + \ldots, b = b_0 + 2b_1 + 4b_2 + \ldots,$ and $a_i, b_i \in \MZ_2^n$.
Then,
\begin{align}
\biggr(\tau_R^{(\ell)} E(a,b) (\tau_R^{(\ell)})^{\dagger} \biggr) \dket{v} & = \xi^{q^{(\ell-1)}(v; R,a,b)} E([a_0,b_0] \Gamma_R) \dket{v} \nonumber \\
                                                                & = \xi^{q^{(\ell-1)}(v; R,a,b)} E(a_0,b_0 + a_0 R) \dket{v},
\end{align}
where 
$\Gamma_R \coloneqq 
\begin{bmatrix}
I_n & R \\
0 & I_n
\end{bmatrix} \in \MZ_{2^{\ell}}^{2n \times 2n}$ and 
\begin{align}
q^{(\ell-1)}(v; R,a,b) & \coloneqq (1 - 2^{\ell-2}) a_0 R a_0^T + 2^{\ell-1} (a_0 b_1^T + b_0 a_1^T) \nonumber \\
                    & \qquad \qquad + (2 + 2^{\ell-1}) vRa_0^T - 4 \eta(v; R,a_0).
\end{align}
\begin{proof}
We observe $D(a,0) \dket{v} = \dket{v \oplus a_0}, D(0,b) \dket{v} = (-1)^{vb_0^T} \dket{v}, \xi^{2^{\ell-2}} = \imath, \xi^{2^{\ell-1}} = -1$:
\begin{align}
& \biggr(\tau_R^{(\ell)} E(a,b) (\tau_R^{(\ell)})^{\dagger} \biggr) \dket{v} \nonumber \\
  & \overset{\text{(i)}}{=} \imath^{ab^T} \xi^{-vRv^T} \tau_R^{(\ell)} (-1)^{ab^T} D(0,b) D(a,0) \dket{v}  \\
  & = \imath^{ab^T} \xi^{-vRv^T} (-1)^{a_0 b_0^T} \tau_R^{(\ell)} (-1)^{(v \oplus a_0)b_0^T} \dket{v \oplus a_0}  \\
  & = \imath^{ab^T} \xi^{-vRv^T} (-1)^{a_0 b_0^T} (-1)^{(v + a_0)b_0^T} \xi^{(v \oplus a_0) R (v \oplus a_0)^T} \dket{v \oplus a_0}  \\
  & \overset{\text{(ii)}}{=} \xi^{-4 \eta(v; R,a_0)} \imath^{ab^T} (-1)^{a_0 b_0^T} (-1)^{(v + a_0)b_0^T} \xi^{2vRa_0^T + a_0 R a_0^T} \dket{v \oplus a_0}  \\
  & \overset{\text{(iii)}}{=} \xi^{a_0 R a_0^T - 4 \eta(v; R,a_0)} \imath^{ab^T} (-1)^{a_0 b_0^T} (-1)^{(v + a_0) (b_0 + a_0 R)^T} (-1)^{a_0 R a_0^T} \xi^{(2 + 2^{\ell-1}) vRa_0^T} \dket{v \oplus a_0}  \\
  & \overset{\text{(iv)}}{=} \xi^{a_0 R a_0^T + (2 + 2^{\ell-1}) vRa_0^T - 4 \eta(v; R,a_0)} \imath^{ab^T} (-1)^{a_0 (b_0 + a_0 R)^T} D(0, b_0 + a_0 R) D(a_0,0) \dket{v}  \\
  & = \xi^{a_0 R a_0^T + (2 + 2^{\ell-1}) vRa_0^T - 4 \eta(v; R,a_0)} \imath^{a_0 b_0^T + 2(a_0 b_1^T + b_0 a_1^T)} D(a_0, b_0 + a_0 R) \dket{v}  \\
  & \overset{\text{(v)}}{=} \xi^{(1 - 2^{\ell-2}) a_0 R a_0^T + 2^{\ell-1} (a_0 b_1^T + b_0 a_1^T) + (2 + 2^{\ell-1}) vRa_0^T - 4 \eta(v; R,a_0)} \imath^{a_0 (b_0 + a_0 R)^T} D(a_0, b_0 + a_0R) \dket{v}  \\
  & = \xi^{q^{(\ell-1)}(v; R,a,b)} E(a_0,b_0 + a_0 R) \dket{v}.
\end{align}
In (i), we have applied $(\tau_R^{(\ell)})^{\dagger}$ to $\dket{v}$ to get the phase $\xi^{-vRv^T}$ and used the fact that $D(a,b) = (-1)^{ab^T} D(0,b) D(0,a)$.
In (ii), we have used Lemma~\ref{lem:binary_quad} to express $(v \oplus a_0) R (v \oplus a_0)^T$ and canceled the factor $\xi^{vRv^T}$ that results with the existing $\xi^{-vRv^T}$.
In (iii), we have rewritten $(v + a_0)b_0^T$ as $(v + a_0)(b_0 + a_0 R)^T - v R a_0^T - a_0 R a_0^T$ and rewritten $(-1)$ as $\xi^{2^{\ell-1}}$ for the exponent $v R a_0^T$.
In (iv), we have collected all the exponents of $\xi$ and $(-1)$, and then used the fact that $D(0, b_0 + a_0 R) D(a_0,0) \dket{v} = (-1)^{(v + a_0) (b_0 + a_0 R)^T} \dket{v \oplus a_0}$.
In (v), we have added and subtracted $a_0 R a_0^T$ in the exponent of $\imath$ and again used the fact that $\xi^{2^{\ell-2}} = \imath$.
Finally, we have applied the (generalized) definition of $E(a,b)$ (i.e., Remark~\ref{rem:general_Eab}).
\end{proof}
\end{lemma}

\begin{remark}
\normalfont
Consider $\ell = 2$ so that $\tau_R^{(2)} \in \text{Cliff}_N$ (by Theorem~\ref{thm:diagonals}), and let $a,b \in \MZ_2^n$.
Then we see that $q^{(1)}(v; R,a,b) \equiv 0$ (mod $2^{\ell} = 4$), and hence the resulting expression $\tau_R^{(\ell)} E(a,b) (\tau_R^{(\ell)})^{\dagger} = E([a,b] \Gamma_R)$ matches exactly with the formula derived for $t_R \in \text{Cliff}_N$ in Section~\ref{sec:elem_symp}.
\end{remark}

Note that the matrices $\Gamma$ also satisfy 
\begin{align}
\label{eq:integer_symp_mat}
\Gamma \Omega \Gamma^T = \Omega\ (\bmod\ 2),
\end{align}
so they are integer symplectic matrices. 
Hence this generalizes from $\MZ_2$ the third elementary symplectic matrix in Table~\ref{tab:std_symp}.

Lemma~\ref{lem:tauR_conjugate} describes the result of conjugating a Pauli matrix with a diagonal unitary by its action on the (computational) basis states $\dket{v}$.
It is clear that this action can be expressed, without explicitly writing these basis states, as
\begin{align}
\label{eq:tauR_conjugate}
\tau_R^{(\ell)} & E(a,b) (\tau_R^{(\ell)})^{\dagger} 
      = E([a_0,b_0] \Gamma_R) \, \text{diag}\left( \xi^{q^{(\ell-1)}(v; R,a,b) \bmod 2^{\ell}} \right).
\end{align}
Next we prove a simple corollary that provides a more succinct and recursive description of the above result, using the binary diagonal matrices $D_x$ introduced just before Lemma~\ref{lem:tauR_conjugate}.

\begin{corollary}
\label{cor:tauR_conjugate}
The result of conjugating a Pauli matrix $E(a,b)$ with a diagonal unitary $\tau_R^{(\ell)}$ can be expressed as
\begin{align}
\tau_R^{(\ell)} E(a,b) (\tau_R^{(\ell)})^{\dagger} & = \xi^{\phi(R,a,b,\ell)} E([a_0,b_0] \Gamma_R) \, \tau_{\tilde{R}(R,a,\ell)}^{(\ell-1)},
\end{align}
where the global phase $\phi(R,a,b,\ell)$ and the new symmetric matrix $\tilde{R}(R,a,\ell)$ over $\MZ_{2^{\ell-1}}$ are
\begin{align}
\phi(R,a,b,\ell) & \coloneqq (1 - 2^{\ell-2}) a_0 R a_0^T + 2^{\ell-1} (a_0 b_1^T + b_0 a_1^T), \\
\tilde{R}(R,a,\ell) & \coloneqq (1 + 2^{\ell-2}) D_{a_0 R} - (D_{\bar{a}_0} R D_{a_0} + D_{a_0} R D_{\bar{a}_0} + 2 D_{a_0 R D_{a_0}}).
\end{align}
Therefore, up to a deterministic global phase, we have
\begin{align}
\label{eq:tau_recursion}
\tau_R^{(\ell)} E(a,b) (\tau_R^{(\ell)})^{\dagger} \equiv E([a_0,b_0] \Gamma_R) \, \tau_{\tilde{R}(R,a,\ell)}^{(\ell-1)} = E(a_0,b_0 + a_0 R) \, \tau_{\tilde{R}(R,a,\ell)}^{(\ell-1)},
\end{align}
thereby yielding a natural recursion in $\ell$.
\begin{proof}
Since $vRa_0^T = v \, D_{Ra_0^T} \, v^T = v \, D_{a_0 R} \, v^T, 2v \, D_{a_0} R D_{\bar{a}_0} \, v^T = v \, (D_{\bar{a}_0} R D_{a_0} + D_{a_0} R D_{\bar{a}_0}) \, v^T$,
\begin{align}
& q^{(\ell-1)}(v; R,a,b) \nonumber \\
  & = (1 - 2^{\ell-2}) a_0 R a_0^T + 2^{\ell-1} (a_0 b_1^T + b_0 a_1^T) + (2 + 2^{\ell-1}) vRa_0^T - 4 \eta(v; R,a_0) \\
  & = (1 - 2^{\ell-2}) a_0 R a_0^T + 2^{\ell-1} (a_0 b_1^T + b_0 a_1^T) + (2 + 2^{\ell-1}) v D_{Ra_0^T} v^T \nonumber \\
  & \hspace*{8cm} - 4 v [D_{a_0} R D_{\bar{a}_0} + D_{a_0 R D_{a_0}}] v^T \\
  & = (1 - 2^{\ell-2}) a_0 R a_0^T + 2^{\ell-1} (a_0 b_1^T + b_0 a_1^T) \nonumber \\
  & \hspace*{2.5cm} + v \, \left[ (2 + 2^{\ell-1}) D_{a_0 R} - 4 (D_{a_0} R D_{\bar{a}_0} + D_{a_0 R D_{a_0}}) \right] \, v^T \\
  & = (1 - 2^{\ell-2}) a_0 R a_0^T + 2^{\ell-1} (a_0 b_1^T + b_0 a_1^T) \nonumber \\
  & \hspace*{2.5cm} + 2 v \, \left[ (1 + 2^{\ell-2}) D_{a_0 R} - (D_{\bar{a}_0} R D_{a_0} + D_{a_0} R D_{\bar{a}_0} + 2 D_{a_0 R D_{a_0}}) \right] \, v^T \\
  & = \phi(R,a,b,\ell) + 2 v \, \tilde{R}(R,a,\ell) \, v^T.
\end{align}
Therefore, we can write
\begin{align}
\tau_R^{(\ell)} E(a,b) (\tau_R^{(\ell)})^{\dagger} & = E([a_0,b_0] \Gamma_R) \, \text{diag}\left( \xi^{q^{(\ell-1)}(v; R,a,b) \bmod 2^{\ell}} \right) \\
  & = \xi^{\phi(R,a,b,\ell)} E([a_0,b_0] \Gamma_R) \, \text{diag}\left( (\xi^2)^{v \tilde{R}(R,a,\ell) v^T \bmod 2^{\ell-1}} \right) \\
  & = \xi^{\phi(R,a,b,\ell)} E([a_0,b_0] \Gamma_R) \, \tau_{\tilde{R}(R,a,\ell)}^{(\ell-1)}.   \tag*{\qedhere}
\end{align}
\end{proof}
\end{corollary}

\subsection{The $T$ Gate and Other Standard QFD Gates}

\begin{example}
\label{eg:T_gate}
Let $n = 1, \ell = 3$, and consider the ``$\pi/8$''-gate defined by 
$T \coloneqq
\begin{bmatrix}
1 & 0 \\
0 & e^{\imath \pi/4}
\end{bmatrix}$.
Since $\xi = e^{\imath \pi/4}$ in this case, it is clear that $R = [\, 1\, ]$.
It is well-known, and direct calculation shows, that $TXT^{\dagger} = \frac{1}{\sqrt{2}} (X + Y)$.
This result can be cast in the form obtained in the above lemma as follows.
For $X = E(1,0)$ we have $a = 1, b = 0$.
So for $v = 0$ we get $q^{(\ell-1)}(v; R,a,b) = -1$,
\begin{align}
TXT^{\dagger} \dket{0} & = \tau_R^{(3)} E(1,0) (\tau_R^{(3)})^{\dagger} \dket{0} 
                         = \xi^{-1} E(1,0+1) \dket{0} 
                         = e^{-\imath\pi/4} Y \dket{0}.
\end{align}
For $v = 1$ we get $q^{(\ell-1)}(v; R,a,b) = -1 + 6 - 4 = 1$,
\begin{align}
TXT^{\dagger} \dket{1} 
                        & = \xi^{+1} E(1,0+1) \dket{1} 
                          = e^{\imath\pi/4} Y \dket{1}.
\end{align}
These two actions can be simplified as shown below, where we use $Z \dket{0} = \dket{0}, Z \dket{1} = -\dket{1}$. 
\begin{align}
e^{-\imath\pi/4} Y \dket{0} & = \frac{(1 - \imath)}{\sqrt{2}} Y \dket{0} = \frac{Y - \imath \times \imath XZ}{\sqrt{2}} \dket{0} = \frac{Y+X}{\sqrt{2}} \dket{0}, \\
e^{\imath\pi/4} Y \dket{1} & = \frac{(1 + \imath)}{\sqrt{2}} Y \dket{1} = \frac{Y + \imath \times \imath XZ}{\sqrt{2}} \dket{1} = \frac{Y+X}{\sqrt{2}} \dket{1}. 
\end{align}
In this case, the action of $T$ can be unified for both basis vectors $\dket{0}$ and $\dket{1}$ as $\frac{1}{\sqrt{2}} (X+Y)$. \hfill \qedhere
Finally, we have $\phi(R,a,b,\ell) = -1, \tilde{R}(R,a,\ell) = [\, 1\, ]$ which implies 
\begin{align*}
TXT^{\dagger} = \xi^{-1} E(1,1) \, \text{diag}(1,\imath) = e^{-\imath\pi/4} Y \, P.  
\end{align*}
\end{example}

\begin{example}
\label{ex:diag_m_1_2}
Consider $n = 1, \ell = 3$.
The matrices $R$ corresponding to standard single-qubit gates in $\MCC_d^{(3)}$ are:
\begin{IEEEeqnarray*}{rClCrCl}
I_2 & = &
\begin{bmatrix}
1 & 0 \\
0 & 1
\end{bmatrix} \colon R = [ 0 ] & , &\ 
P & = &
\begin{bmatrix}
1 & 0 \\
0 & \imath
\end{bmatrix} \colon R = [ 2 ], \\ 
Z & = &
\begin{bmatrix}
1 & 0 \\
0 & -1
\end{bmatrix} \colon R = [ 4 ] & , &\ 
P^{\dagger} & = &
\begin{bmatrix}
1 & 0 \\
0 & -\imath
\end{bmatrix} \colon R = [ 6 ], \\
T & = &
\begin{bmatrix}
1 & 0 \\
0 & e^{\imath\pi/4} 
\end{bmatrix} \colon R = [ 1 ] & , &\ 
TZ & = &
\begin{bmatrix}
1 & 0 \\
0 & -e^{\imath\pi/4}
\end{bmatrix} \colon R = [ 5 ], \\ 
T^{\dagger} & = &
\begin{bmatrix}
1 & 0 \\
0 & e^{-\imath\pi/4}
\end{bmatrix} \colon R = [ 7 ] & , &\ 
T^{\dagger} Z & = &
\begin{bmatrix}
1 & 0 \\
0 & -e^{-\imath\pi/4} 
\end{bmatrix} \colon R = [ 3 ].
\end{IEEEeqnarray*}
Similarly, for two-qubit gates ($n = 2$) in $\MCC_d^{(3)}$ we have: (C$P$: Controlled-Phase)
\begin{align*}
\text{CZ} &=
\begin{bmatrix}
1 & 0 & 0 & 0 \\
0 & 1 & 0 & 0 \\
0 & 0 & 1 & 0 \\
0 & 0 & 0 & -1
\end{bmatrix} \colon R = 
\begin{bmatrix}
0 & 2 \\
2 & 0
\end{bmatrix}, \ \ \, 
\text{C}P =  
\begin{bmatrix}
1 & 0 & 0 & 0 \\
0 & 1 & 0 & 0 \\
0 & 0 & 1 & 0 \\
0 & 0 & 0 & \imath
\end{bmatrix} \colon R = 
\begin{bmatrix}
0 & 1 \\
1 & 0
\end{bmatrix}, \\ 
I_2 \otimes P & = 
\begin{bmatrix}
1 & 0 & 0 & 0 \\
0 & \imath & 0 & 0 \\
0 & 0 & 1 & 0 \\
0 & 0 & 0 & \imath
\end{bmatrix} \colon R = 
\begin{bmatrix}
0 & 0 \\
0 & 2
\end{bmatrix}, \ 
I_2 \otimes Z =
\begin{bmatrix}
1 & 0 & 0 & 0 \\
0 & -1 & 0 & 0 \\
0 & 0 & 1 & 0 \\
0 & 0 & 0 & -1
\end{bmatrix} \colon R = 
\begin{bmatrix}
0 & 0 \\
0 & 4
\end{bmatrix}, \\ 
P \otimes I_2 & =
\begin{bmatrix}
1 & 0 & 0 & 0 \\
0 & 1 & 0 & 0 \\
0 & 0 & \imath & 0 \\
0 & 0 & 0 & \imath
\end{bmatrix} \colon R = 
\begin{bmatrix}
2 & 0 \\
0 & 0
\end{bmatrix}, \  
Z \otimes I_2 =
\begin{bmatrix}
1 & 0 & 0 & 0 \\
0 & 1 & 0 & 0 \\
0 & 0 & -1 & 0 \\
0 & 0 & 0 & -1
\end{bmatrix} \colon R = 
\begin{bmatrix}
4 & 0 \\
0 & 0
\end{bmatrix}.
\end{align*}
\end{example}


Next we prove a simple result that determines the symmetric matrix $R$ for a given diagonal unitary that is a tensor product of diagonal unitaries.

\begin{lemma}
\label{lem:diag_tensor}
Let $\ell, k \in \MZ_{>0}$ such that $\ell < k$, and define $\xi_{\ell} \coloneqq \exp(\frac{2\pi \imath}{2^{\ell}}), \xi_k \coloneqq \exp(\frac{2\pi \imath}{2^k})$.
Suppose that $\tau_{R_1,m}^{(k)}$ and $\tau_{R_2,n}^{(\ell)}$ are two diagonal unitaries, where $R_1 \in \MZ_{2^k}^{m \times m}$ and $R_2 \in \MZ_{2^{\ell}}^{n \times n}$ are symmetric, and $m,n$ represent the number of qubits on which the unitaries are defined.
Then the symmetric matrix $R \in \MZ_{2^k}^{(m+n) \times (m+n)}$ corresponding to $\tau_{R,m+n}^{(k)} \coloneqq \tau_{R_1,m}^{(k)} \otimes \tau_{R_2,n}^{(\ell)}$ is given by $R = 
\begin{bmatrix}
R_1 & 0 \\
0 & 2^{k-\ell} R_2
\end{bmatrix}$.
\begin{proof}
We can simplify the tensor product as follows:
\begin{align}
\tau_{R_1,m}^{(k)} \otimes \tau_{R_2,n}^{(\ell)} & = \sum_{v \in \MZ_2^m} \xi_k^{v R_1 v^T \bmod 2^k} \dketbra{v} \otimes \sum_{w \in \MZ_2^n} \xi_{\ell}^{w R_2 w^T \bmod 2^{\ell}} \dketbra{w} \\
& = \sum_{\substack{v \in \MZ_2^m\\w \in \MZ_2^n}} \xi_k^{(v R_1 v^T + 2^{k-\ell} w R_2 w^T) \bmod 2^k} (\dket{v} \otimes \dket{w}) (\dbra{v} \otimes \dbra{w}) \\
& = \sum_{[v,w] \in \MZ_2^{m+n}} \xi_k^{\small
\begin{bmatrix}
v & w
\end{bmatrix} 
\begin{bmatrix}
R_1 & 0 \\
0 & 2^{k-\ell} R_2
\end{bmatrix}
\begin{bmatrix}
v^T \\ w^T
\end{bmatrix}} \dketbra{v,w} \\
& = \sum_{u \in \MZ_2^{m+n}} \xi_k^{u R u^T} \dketbra{u} 
  = \tau_{R,m+n}^{(k)}.   \tag*{\qedhere}
\end{align}
\end{proof}
\end{lemma}

Following the symmetric matrices given previously for the single-qubit case, the above result can be used to produce the symmetric matrices for the two-qubit tensor product unitaries in Example~\ref{ex:diag_m_1_2}.
Now, we show that not all $3$-local diagonal unitaries are QFD by producing a $3$-local diagonal unitary that cannot be characterized by any symmetric matrix $R$.

\subsection{CCZ is not QFD}

\begin{example}
Consider the Controlled-Controlled-$Z$ (CCZ) gate on $n = 3$ qubits represented by the unitary $\text{CCZ} = \text{diag}\, (1,1,1,1,1,1,1,-1)$.
It can be checked that this unitary belongs to level $\ell = 3$ of the Clifford hierarchy.
Let $R = 
\begin{bmatrix}
a & b & c \\
b & d & e \\
c & e & f
\end{bmatrix}$ be a symmetric matrix with entries in $\MZ_8$.
Equating $\text{CCZ} = \tau_R^{(3)}$, we see that the exponent of $\xi = \exp(\frac{2\pi\imath}{8})$ is $0$ for the first $7$ entries in the diagonal and $-4 \equiv 4$ (mod $8$) for the last entry.
Solving $v R v^T = 0$ for the first $7$ entries, we find that all entries in $R$ have to be $0$.
Thus, there are not enough degrees of freedom in $R$ and we can only produce the identity $I_8$.
\end{example}

Therefore, we have the following result about the diagonal unitaries we characterize in each level of the Clifford hierarchy~\footnote{We would like to thank Theodore Yoder for pointing out that the QFD framework does not include all $d$-local diagonal gates in the hierarchy, for $d > 2$, as we had originally thought.}.

\subsection{All $1$- and $2$-Local Diagonal Gates are QFD}

\begin{theorem}
\label{thm:diagonals}
For any symmetric $R \in \MZ_{2^{\ell}}^{n \times n}$, the matrix $\tau_R^{(\ell)} \in \MCC_d^{(\ell)}$. 
All two-local diagonal unitaries in the Clifford hierarchy can be expressed in the form $\tau_R^{(\ell)}$ for some $\ell \in \mathbb{N}$ and symmetric $R \in \MZ_{2^{\ell}}^{n \times n}$, up to a global phase.
\begin{proof}
We will prove the first part by induction.
For $\ell = 1$, $R$ has binary entries and since $\xi = \exp(\frac{2\pi\imath}{2}) = -1$, only the diagonal $d_R$ contributes non-trivially to $vRv^T = \sum_i R_{ii} v_i + 2 \sum_{i < j} R_{ij} v_i v_j$.
So the diagonal entries of $\tau_R^{(1)}$ are $(-1)^{v d_R^T}$ (since $v_i^2 = v_i$), i.e., $\tau_R^{(1)} \dket{v} = (-1)^{v d_R^T} \dket{v}$, and hence $\tau_R^{(1)} = E(0,d_R) \in \MCC_d^{(1)}$. 
Suppose that we have shown $\tau_R^{(\ell)} \in \MCC_d^{(\ell)}$ for $\ell \geq 1$ and any symmetric matrix $R \in \MZ_{2^{\ell}}^{n \times n}$.
For level $(\ell+1)$, we have
\begin{align}
\tau_R^{(\ell+1)} E(a,b) (\tau_R^{(\ell+1)})^{\dagger} & = \xi^{\phi(R,a,b,\ell+1)} E([a_0,b_0] \Gamma_R) \, \tau_{\tilde{R}(R,a,\ell+1)}^{(\ell)}.
\end{align}
Since the global phase can be safely ignored and $\tilde{R}(R,a,\ell+1) \in \MZ_{2^{\ell}}^{n \times n}$ is symmetric, by the induction hypothesis, $\tau_{\tilde{R}(R,a,\ell+1)}^{(\ell)} \in \MCC_d^{(\ell)}$. 
(Note that $\tau_R^{(0)} = I_N$ for all $R$).
Using the fact that the first two levels of the hierarchy are unaffected by multiplication by Paulis, a simple induction shows that if $V \in \MCC^{(\ell)}$ (not necessarily diagonal) then $E(c,d) V \in \MCC^{(\ell)}$ as well, for any $c,d \in \MZ_2^n$.
(Note that it is easier to show that $V E(c,d) \in \MCC^{(\ell)}$ by just using the definition of the hierarchy and the fact that Paulis commute or anti-commute).
Therefore, by the definition of the Clifford hierarchy we have $\tau_R^{(\ell+1)} \in \MCC_d^{(\ell+1)}$.
This completes the proof for the first part.

A two-local diagonal unitary $U$ is a tensor product of single- and two-qubit diagonal unitaries.
For $n = 1$, consider a diagonal unitary $W \in \MCC_d^{(\ell)}$ for any $\ell \geq 1$.
Then, up to a global phase, there is only one degree of freedom given by the second diagonal entry of $W$ and this must be of the form $\xi^a$ for some $a \in \MZ_{2^{\ell}}$~\cite{Cui-physreva17}.
In this case, we can take $R = [\, a\, ]$ so that $W \equiv \tau_R^{(\ell)}$.
Similarly, for $n = 2$, any diagonal unitary $W$ in the hierarchy has $3$ degrees of freedom with diagonal entries of the form $\xi_{\ell}^{\alpha}, \xi_{\ell}^{\beta}, \xi_{\ell}^{\gamma}$ for some $\ell \geq 1$, $\xi_{\ell} = \exp(\frac{2\pi\imath}{2^{\ell}})$, and $\alpha, \beta, \gamma \in \MZ_{2^{\ell}}$.
Let $R = 
\begin{bmatrix}
a & b \\
b & c
\end{bmatrix}$ so that the diagonal entries of $\tau_R^{(\ell)}$ are $\xi_{\ell}^c, \xi_{\ell}^a, \xi_{\ell}^{a+2b+c}$.
Then we can directly set $c = \alpha, a = \beta$ and attempt to solve for $2b = \gamma - a - c$.
If $(\gamma - a - c)$ is even then there exists a $b \in \MZ_{2^{\ell}}$, but if $(\gamma - a - c)$ is odd then we can move to level $\ell + 1$ so that we map $\gamma \mapsto 2 \gamma, a \mapsto 2a, c \mapsto 2c$ (with respect to $\xi_{\ell+1}$) and then there exists a solution for $b \in \MZ_{2^{\ell+1}}$.
Hence we satisfy $W \equiv \tau_R^{(k)}$ for $k = \ell$ or $\ell + 1$.
Since $U$ is a tensor product of such unitaries, Lemma~\ref{lem:diag_tensor} implies that we can determine the exact symmetric matrix corresponding to $U$.
This completes the proof.
\end{proof}
\end{theorem}


\begin{example}
Consider the diagonal unitary $U = \text{diag}(1,\imath,\imath,\imath)$.
By the argument in the above proof, we choose $\ell = 2$ since $\imath = \exp(\frac{2\pi\imath}{2^2})$.
Then using the form of $R$ as in the above proof, we see that $c = a = 1$ given the second and third diagonal entries of $U$.
This implies that we need to find $b$ such that $a + 2b + c = 1 \Rightarrow 2b = -1 \equiv 3$.
Since this does not have a solution in $\MZ_{2^2}$, we move to $\ell = 3$.
Then we get $c = a = 2$, $2b = 2 - 4 \equiv 6$ and this implies $b = 3$.
Hence, we find that $U = \tau_R^{(3)}$.
\end{example}


\subsection{M{\o}lmer-S{\o}rensen Gates are QFD}

\begin{example}
Since we can produce all $2$-local diagonal unitaries in the hierarchy, the gate 
\begin{align}
\text{ZZ}(\theta) \coloneqq \exp(-\imath \theta (Z \otimes Z)) = \cos\theta\, I_4 - \imath \sin\theta\, (Z \otimes Z) = e^{-\imath \theta}\, \text{diag}\, (1, e^{\imath 2\theta}, e^{\imath 2\theta}, 1) 
\end{align}
can be represented as $\tau_R^{(\ell)}$ with $R = 
\begin{bmatrix}
1 & -1 \\
-1 & 1
\end{bmatrix}$, where $\theta = \frac{\pi}{2^{\ell}}$ for some $\ell \geq 1$.
Hence, when combined with Hadamard gates, we can incorporate the M{\o}lmer-S{\o}rensen family of gates $\text{XX}_{ij}(\theta) \coloneqq \exp(-\imath \theta\, X_i X_j) = \cos\theta\, I_4 - \imath \sin\theta\, X_i X_j$ in our framework, where the subscripts $i$ and $j$ denote the qubits involved in the gate.
Since these gates are the native operations in trapped-ion quantum computers, this observation can potentially lead to applications such as efficient circuit optimization for such systems.
\end{example}

\begin{remark}
\label{rem:u_ccz}
\normalfont
The result in Theorem~\ref{thm:diagonals} only implies that we cannot represent ``all'' $d$-local unitaries for $d > 2$ via a symmetric matrix in our framework.
However, since $\tau_R^{(\ell)} \in \mathcal{C}_d^{(\ell)}$ for symmetric $R \in \MZ_{2^{\ell}}^{n \times n}$, our framework can generate $2^{nk} 2^{(\ell-1)n(n-1)/2}$ diagonal gates at the $k$-th level (see Theorem~\ref{thm:symmetric_representation} for the reason behind this count), and this includes a large set of $d$-local unitaries with $d > 2$.
For example, consider the gate $U = \exp(\imath \frac{\pi}{8} (Z \otimes Z \otimes Z)) = \cos\frac{\pi}{8} \, I_8 + \imath \sin\frac{\pi}{8}\, (Z \otimes Z \otimes Z) \in \MCC_d^{(3)}$.
Clearly this gate is $3$-local.
Since $\xi = \exp(\frac{2\pi\imath}{8}) = \exp(\frac{\imath\pi}{4})$, we have $U = \exp(\frac{\imath \pi}{8}) \, \text{diag}\, (\xi^0, \xi^7, \xi^7, \xi^0, \xi^7, \xi^0, \xi^0, \xi^7)$.
Considering $R = 
\begin{bmatrix}
a & b & c \\
b & d & e \\
c & e & f
\end{bmatrix}$ and solving for the entries by setting $v R v^T$ to the above given entries of $U$ (ignoring the global phase), we find that the first seven entries imply $a = d = f = 7, b = c = e = -3 \equiv 5$ (mod $8$).
Therefore, the exponent of the last diagonal entry of $\tau_R^{(3)}$ must be $a + 2b + 2c + d + 2e + f \equiv 3$ whereas the last entry of $U$ is $\xi^7$.
Interestingly, the difference is exactly the factor $\xi^4 = -1$, which means that $\tau_R^{(3)} = U \times \text{CCZ}$ has the above representation $R$ in our framework although it is not $2$-local.
Note that taking $b = c = e = 1$ does not change the diagonal gate.
\end{remark}

The action of $\tau_R^{(\ell)}$ on the Pauli matrices directly implies the following result.

\begin{lemma}
\label{lem:isomorphism}
For a fixed $\ell \in \MZ$ and symmetric $R \in \MZ_{2^{\ell}}^{n \times n}$, the map 
\begin{align}
\varphi \colon E(a,b) \mapsto \tau_R^{(\ell)} E(a,b) (\tau_R^{(\ell)})^{\dagger}
\end{align}
is a group isomorphism.
\end{lemma}
%
%
%

Next we discuss some properties of the objects defined above.

\subsection{Some Properties of QFD Gates and an Isomorphism}


\begin{lemma}
\label{lem:properties}
For $v \in \MZ_2^n$, any $a,b,c,d \in \MZ^n$, and any symmetric $R \in \MZ_{2^{\ell}}^{n \times n}$ the following properties hold.
\begin{enumerate}

\item[(a)] The diagonal unitaries defined by $\xi$ and $q^{(\ell-1)}(v; R,a,b)$ satisfy, for any $e,f \in \MZ^n$,
\begin{align}
\text{diag}\left( \xi^{q^{(\ell-1)}(v \oplus e_0 ; R,a,b) \bmod 2^{\ell}} \right) & = E(e_0,f)\ \text{diag}\left( \xi^{q^{(\ell-1)}(v; R,a,b) \bmod 2^{\ell}} \right) E(e_0,f).
\end{align}

\item[(b)] The function $q^{(\ell-1)}(v; R, \cdot, \cdot)$ satisfies (modulo $2^{\ell}$)
\begin{align}
& q^{(\ell-1)}(v \oplus c_0; R,a,b) + q^{(\ell-1)}(v; R,c,d) \nonumber \\
\label{eq:propertiesb1}
  & \equiv q^{(\ell-1)}(v; R,a,b) + q^{(\ell-1)}(v \oplus a_0; R,c,d) \\
\label{eq:propertiesb2}  
  & \equiv q^{(\ell-1)}(v; R,a+c,b+d) + 2^{\ell-1} (b_0 c_1^T + b_1 c_0^T - a_0 d_1^T - a_1 d_0^T).
\end{align}

\item[(c)] The action of $\tau_R^{(\ell)}$ satisfies 
\begin{align}
\tau_R^{(\ell)} E(c,d) (\tau_R^{(\ell)})^{\dagger} \times \tau_R^{(\ell)} E(a,b) (\tau_R^{(\ell)})^{\dagger}  & = E(a_0,e) \left[ \tau_R^{(\ell)} E(c,d) (\tau_R^{(\ell)})^{\dagger} \right] E(a_0,e) \nonumber \\
 & \quad \times E(c_0,f) \left[ \tau_R^{(\ell)} E(a,b) (\tau_R^{(\ell)})^{\dagger} \right] E(c_0,f),
\end{align}
for any $e,f \in \MZ^n$ such that $\syminn{[a_0,b_0]}{[c_0,d_0]} = \syminn{[a_0,e_0]}{[c_0,f_0]}$, and in particular for $e = b_0 + a_0 R, f = d_0 + c_0 R$.

\end{enumerate}
\begin{proof}
We use identities related to these quantities to complete the proof.
\begin{enumerate}

\item[(a)] Recall that $E(e_0,f) = \imath^{e_0 f^T} E(e_0,0) E(0,f),\ E(0,f) = D(0,f)$ is diagonal, and $E(e_0,0) = D(e_0,0)$ is a permutation matrix corresponding to $\dket{v} \mapsto \dket{v \oplus e_0}$ (\eqref{eq:Eab_expansion}).

\item[(b)] This can be verified by explicitly enumerating and matching terms on each side of the equality (see Section~\ref{sec:proof_prop1}).
Here we illustrate a more elegant approach.
Using the result of part (a) we calculate
\begin{align}
& \tau_R^{(\ell)} E(a,b) (\tau_R^{(\ell)})^{\dagger} \times \tau_R^{(\ell)} E(c,d) (\tau_R^{(\ell)})^{\dagger} \nonumber \\
  & = \left[ E([a_0,b_0] \Gamma_R)\ \text{diag}\left( \xi^{q^{(\ell-1)}(v; R,a,b)} \right) \right] \times \left[ E([c_0,d_0] \Gamma_R)\  \text{diag}\left( \xi^{q^{(\ell-1)}(v; R,c,d)} \right) \right] \\
  & = E([a_0,b_0] \Gamma_R) E([c_0,d_0] \Gamma_R)\ \text{diag}\left( \xi^{q^{(\ell-1)}(v \oplus c_0 ; R,a,b)} \right)\ \text{diag}\left( \xi^{q^{(\ell-1)}(v; R,c,d)} \right) \\
\label{eq:prop2}
  & = (-1)^{\syminn{[a_0,b_0] \Gamma_R}{[c_0,d_0] \Gamma_R}} E([c_0,d_0] \Gamma_R) E([a_0,b_0] \Gamma_R) \nonumber \\
  & \hspace*{4cm} \text{diag}\left( \xi^{q^{(\ell-1)}(v; R,c,d)} \right)\  \text{diag}\left( \xi^{q^{(\ell-1)}(v \oplus c_0 ; R,a,b)} \right) \\
\label{eq:prop2b}
  & \overset{\text{(or)}}{=} \imath^{(b_0 + a_0 R) c_0^T - a_0 (d_0 + c_0 R)^T} E([a_0 + c_0, b_0 + d_0] \Gamma_R) \nonumber \\
  & \hspace*{4cm} \text{diag}\left( \xi^{q^{(\ell-1)}(v \oplus c_0 ; R,a,b)} \right)\ \text{diag}\left( \xi^{q^{(\ell-1)}(v; R,c,d)} \right). 
\end{align}
The first equality uses~\eqref{eq:tauR_conjugate}, the second equality follows from (a), and the last two equalities use the properties given in~\eqref{eq:Eab_multiply}.
Note that we have slightly abused notation since the symplectic inner product is defined only for binary vectors.
However, this can be generalized to integer vectors since only their modulo $2$ components play a role in the exponent of $(-1)$.
Once again using the results referenced above, we have
\begin{align}
& \tau_R^{(\ell)} E(a,b) (\tau_R^{(\ell)})^{\dagger} \times \tau_R^{(\ell)} E(c,d) (\tau_R^{(\ell)})^{\dagger} \nonumber \\
  & = (-1)^{\syminn{[a_0,b_0]}{[c_0,d_0]}} \tau_R^{(\ell)} E(c,d) (\tau_R^{(\ell)})^{\dagger} \times \tau_R^{(\ell)} E(a,b) (\tau_R^{(\ell)})^{\dagger} \\
  & = (-1)^{\syminn{[a_0,b_0]}{[c_0,d_0]}} \left[ E([c_0,d_0] \Gamma_R)\ \text{diag}\left( \xi^{q^{(\ell-1)}(v; R,c,d)} \right) \right] \nonumber \\
  & \hspace*{4cm} \times \left[ E([a_0,b_0] \Gamma_R)\ \text{diag}\left( \xi^{q^{(\ell-1)}(v; R,a,b)} \right) \right] \\
  & = (-1)^{\syminn{[a_0,b_0]}{[c_0,d_0]}} E([c_0,d_0] \Gamma_R) E([a_0,b_0] \Gamma_R) \nonumber \\
  & \hspace*{4cm} \text{diag}\left( \xi^{q^{(\ell-1)}(v \oplus a_0; R,c,d)} \right)\ \text{diag}\left( \xi^{q^{(\ell-1)}(v ; R,a,b)} \right). 
\end{align}
This must be equal to~\eqref{eq:prop2} and, using~\eqref{eq:integer_symp_mat}, we verify
\begin{align}
\syminn{[a_0,b_0] \Gamma_R}{[c_0,d_0] \Gamma_R} & = [a_0,b_0] \Gamma_R\ \Omega\ \Gamma_R^T [c_0, d_0]^T \nonumber \\
%
%
   & = [a_0, b_0] \, \Omega \, [c_0, d_0]^T \nonumber \\
   & = \syminn{[a_0,b_0]}{[c_0,d_0]}
\end{align}
as required (all modulo $2$).
Hence, the equality~\eqref{eq:propertiesb1} in the lemma must be true.
Similarly, we have
\begin{align}
& \tau_R^{(\ell)} E(a,b) (\tau_R^{(\ell)})^{\dagger} \times \tau_R^{(\ell)} E(c,d) (\tau_R^{(\ell)})^{\dagger} \nonumber \\ 
  & = \tau_R^{(\ell)} \left[ \imath^{bc^T - ad^T} E(a+c,b+d) \right] (\tau_R^{(\ell)})^{\dagger} \\
  & = \xi^{2^{\ell-2}(bc^T - ad^T)} E([a_0 + c_0, b_0 + d_0] \Gamma_R) \times \text{diag}\left( \xi^{q^{(\ell-1)}(v; R,a+c,b+d)} \right).
\end{align}
Comparing this with~\eqref{eq:prop2b}, and observing that $bc^T - ad^T = b_0 c_0^T - a_0 d_0^T + 2(b_0 c_1^T + b_1 c_0^T - a_0 d_1^T - a_1 d_0^T)\ (\bmod\ 4)$, proves the second equality~\eqref{eq:propertiesb2}.

\item[(c)] This follows from the previous properties as shown below.
\begin{align}
& E(a_0,e) \left[ \tau_R^{(\ell)} E(c,d) (\tau_R^{(\ell)})^{\dagger} \right] E(a_0,e) \times E(c_0,f) \left[ \tau_R^{(\ell)} E(a,b) (\tau_R^{(\ell)})^{\dagger} \right] E(c_0,f) \\
 & = E(a_0,e) E(c_0,d_0 + c_0 R)\ \text{diag}\left( \xi^{q^{(\ell-1)}(v; R,c,d)} \right) E(a_0,e) \nonumber \\
 & \hspace*{1.5cm}  \times E(c_0,f) E(a_0,b_0 + a_0 R)\ \text{diag}\left( \xi^{q^{(\ell-1)}(v; R,a,b)} \right) E(c_0,f) \\
 & = (-1)^{a_0 (d_0 + c_0 R)^T + e_0 c_0^T} E(c_0, d_0 + c_0 R)\ \text{diag}\left( \xi^{q^{(\ell-1)}(v \oplus a_0; R,c,d)} \right) \nonumber \\
 & \hspace*{1.5cm} \times (-1)^{c_0 (b_0 + a_0 R)^T + f_0 a_0^T} E(a_0, b_0 + a_0 R)\ \text{diag}\left( \xi^{q^{(\ell-1)}(v \oplus c_0; R,a,b)} \right)  \\
 & = (-1)^{\syminn{[a_0,b_0]}{[c_0,d_0]} + \syminn{[a_0,e_0]}{[c_0,f_0]}} E(c_0, d_0 + c_0 R) E(a_0, b_0 + a_0 R) \nonumber \\
 & \hspace*{2.5cm} \times \text{diag}\left( \xi^{q^{(\ell-1)}(v; R,c,d)} \right) \text{diag}\left( \xi^{q^{(\ell-1)}(v \oplus c_0; R,a,b)} \right) \\
 & = E(c_0, d_0 + c_0 R) E(a_0, b_0 + a_0 R) \ \text{diag}\left( \xi^{q^{(\ell-1)}(v \oplus a_0; R,c,d)} \right) \text{diag}\left( \xi^{q^{(\ell-1)}(v; R,a,b)} \right) \\
 & = E(c_0, d_0 + c_0 R)\ \text{diag}\left( \xi^{q^{(\ell-1)}(v; R,c,d)} \right) \times E(a_0, b_0 + a_0 R)\ \text{diag}\left( \xi^{q^{(\ell-1)}(v; R,a,b)} \right) \\
 & = \tau_R^{(\ell)} E(c,d) (\tau_R^{(\ell)})^{\dagger} \times \tau_R^{(\ell)} E(a,b) (\tau_R^{(\ell)})^{\dagger}. 
\end{align}
Again, the first equality uses~\eqref{eq:tauR_conjugate}. 
The second equality uses the properties in~\eqref{eq:Eab_multiply} to swap the order of Paulis, then uses the result of (a) to pass $E(a_0,e)$ and $E(c_0,f)$ through the diagonals, and then observes the property that $E(a_0,e)^2 = E(c_0,f)^2 = I_N$.
The third equality collects exponents by noting that $a_0 R c_0^T = c_0 R a_0^T$ (since $R$ is symmetric), and then uses the result of (a) to pass $E(a_0, b_0 + a_0 R)$ through the diagonal on its left.
The fourth equality utilizes the condition assumed in the hypothesis as well as the result of (b).
The fifth equality once again uses (a) to pass back $E(a_0, b_0 + a_0 R)$, and finally the last step follows from~\eqref{eq:tauR_conjugate}.

\end{enumerate}
This completes the proof. \hfill \qedhere
\end{proof}
\end{lemma}




\begin{theorem}
\label{thm:symmetric_representation}
Fix $\ell \geq 1$.
Define $\MCC_{d,\text{sym}}^{(\ell)}$ to be the set of diagonal unitaries $\tau_R^{(\ell)}$ for all matrices $R \in \MZ_{2^{\ell}, \text{sym}}^{n \times n}$, where the subscript ``sym'' represents symmetric matrices whose diagonal entries are in $\MZ_{2^{\ell}}$ and off-diagonal entries are in $\MZ_{2^{\ell-1}}$.
Also, addition in $\MZ_{2^{\ell}, \text{sym}}^{n \times n}$ is defined as addition over $\MZ_{2^{\ell}}$ for the diagonal entries and addition over $\MZ_{2^{\ell-1}}$ for the off-diagonal entries.
Then $\MCC_{d,\text{sym}}^{(\ell)}$ is a subgroup of $\MCC_d^{(\ell)}$.
Furthermore, the map $\gamma \colon \MCC_{d,\text{sym}}^{(\ell)} \rightarrow \MZ_{2^{\ell},\text{sym}}^{n \times n}$ defined by $\gamma(\tau_{R}^{(\ell)}) \coloneqq R$ is an isomorphism. 
\begin{proof}
From Theorem~\ref{thm:diagonals} we know that $\tau_R^{(\ell)} \in \MCC_d^{(\ell)}$. 
Then 
\begin{align}
\gamma\left( \tau_{R_1}^{(\ell)} \times \tau_{R_2}^{(\ell)} \right) & = \gamma\left(\text{diag}\left( \xi^{vR_1v^T} \right) \times \text{diag}\left( \xi^{vR_2v^T} \right) \right) \\
  & = \gamma\left( \tau_{R_1 + R_2}^{(\ell)} \right) \\
  & = R_1 + R_2 \\
  & = \gamma\left( \tau_{R_1}^{(\ell)} \right) + \gamma\left( \tau_{R_2}^{(\ell)} \right).
\end{align}
As discussed in the proof of Theorem~\ref{thm:diagonals}, since $vRv^T = \sum_i R_{ii} v_i + 2 \sum_{i < j} R_{ij} v_i v_j$, when $2^{\ell-1}$ is added to any off-diagonal entry $R_{ij}$, the factor of $2$ produces $2^{\ell} R_{ij} v_i v_j$ which vanishes modulo $2^{\ell}$ (see Remark~\ref{rem:u_ccz} for an example).
Therefore, only when the off-diagonal entries are restricted to values in the ring $\MZ_{2^{\ell-1}}$, the vectors $[v R_1 v^T]_{v \in \MZ_2^n}$ and $[v R_2 v^T]_{v \in \MZ_2^n}$ are distinct for distinct $R_1, R_2$ and $\ell \geq 1$. 
Here, the sum $R_1 + R_2$ is taken over $\MZ_{2^{\ell}}$ for the diagonal entries and over $\MZ_{2^{\ell-1}}$ for the off-diagonal entries.
Hence, the closure implies that $\MCC_{d,\text{sym}}^{(\ell)}$ is clearly a subgroup of $\MCC_d^{(\ell)}$. 
Moreover, by definition $\MCC_{d,\text{sym}}^{(\ell)}$ does not include global phases, so the map $\gamma$ is an isomorphism.
\end{proof} 
\end{theorem}

\subsection{Alternate Proof of Lemma~\ref{lem:properties}(b)}
\label{sec:proof_prop1}

We ignore the common terms $q^{(\ell-1)}(v; R,a,b) + q^{(\ell-1)}(v; R,c,d)$ on both sides of the equality and consider only the remainder.
Note that the calculation is modulo $2^{\ell}$.
Let $\tilde{c}_0 = c_0 - 2 (v \ast c_0)$.
For the left hand side, by first ignoring $q^{(\ell-1)}(v; R,c,d)$ and then $q^{(\ell-1)}(v; R,a,b)$,
\begin{align}
& q^{(\ell-1)}(v \oplus c_0; R,a,b) \nonumber \\
& = (1 - 2^{\ell-2}) a_0 R a_0^T + 2^{\ell-1} (a_0 b_1^T + b_0 a_1^T) + (2 + 2^{\ell-1}) (v \oplus c_0) Ra_0^T \nonumber \\
& \hspace*{1cm} - 4 [((v \oplus c_0) + a_0) - ((v \oplus c_0) \ast a_0)] R ((v \oplus c_0) \ast a_0)^T \\
& = (1 - 2^{\ell-2}) a_0 R a_0^T + 2^{\ell-1} (a_0 b_1^T + b_0 a_1^T) + (2 + 2^{\ell-1}) (v + \tilde{c}_0) Ra_0^T \nonumber \\
& \hspace*{1cm} - 4 [((v + \tilde{c}_0) + a_0) - ((v + \tilde{c}_0) \ast a_0)] R ((v + \tilde{c}_0) \ast a_0)^T \\
& = q^{(\ell-1)}(v; R,a,b) + (2 + 2^{\ell-1}) \tilde{c}_0 R a_0^T - 4 \biggr[ (v + a_0 - v \ast a_0) R (\tilde{c}_0 \ast a_0)^T \nonumber \\
 & \hspace*{1cm} + (\tilde{c}_0 - \tilde{c}_0 \ast a_0) R (v \ast a_0)^T \biggr] + (\tilde{c}_0 - \tilde{c}_0 \ast a_0) R (\tilde{c}_0 \ast a_0)^T \biggr] \\
& \equiv (2 + 2^{\ell-1}) c_0 R a_0^T - 4 (v \ast c_0) R a_0^T - 4 (v + a_0 - v \ast a_0) R (c_0 \ast a_0)^T \nonumber \\
& \quad + 8 (v + a_0 - v \ast a_0) R (v \ast c_0 \ast a_0)^T - 4 (c_0 - 2(v \ast c_0)) R (v \ast a_0)^T \nonumber \\
& \quad + 4 ((c_0 \ast a_0) - 2 v \ast c_0 \ast a_0) R (v \ast a_0)^T - 4 (c_0 - 2 v \ast c_0) R ((c_0 - 2 v \ast c_0) \ast a_0)^T \nonumber \\
& \quad + 4 (c_0 \ast a_0 - 2 v \ast c_0 \ast a_0) R (c_0 \ast a_0 - 2 v \ast c_0 \ast a_0)^T \\
& = [(2 + 2^{\ell-1}) c_0 R a_0^T]_1 - [4 (v \ast c_0) R a_0^T]_2 - [4 v R (c_0 \ast a_0)^T]_3 - [4 a_0 R (c_0 \ast a_0)^T]_4 \nonumber \\
& \quad + [4 (v \ast a_0) R (c_0 \ast a_0)^T]_5 + [8 v R (v \ast c_0 \ast a_0)^T]_6 + [8 a_0 R (v \ast c_0 \ast a_0)^T]_7 \nonumber \\
& \quad - [8 (v \ast a_0) R (v \ast c_0 \ast a_0)^T]_8 - [4 c_0  R (v \ast a_0)^T]_2 + [8 (v \ast c_0)  R (v \ast a_0)^T]_9 \nonumber \\
& \quad + [4 (c_0 \ast a_0) R (v \ast a_0)^T]_5 - [8 (v \ast c_0 \ast a_0) R (v \ast a_0)^T]_8 - [4 c_0 R (c_0 \ast a_0)^T]_4 \nonumber \\
& \quad + [8 c_0 R (v \ast c_0 \ast a_0)^T]_7 + [8 (v \ast c_0) R (c_0 \ast a_0)^T]_5 - [16 (v \ast c_0) R (v \ast c_0 \ast a_0)^T]_8 \nonumber \\
& \quad + [4 (c_0 \ast a_0) R (c_0 \ast a_0)^T]_{10} - [16 (c_0 \ast a_0) R (v \ast c_0 \ast a_0)^T]_{11} \nonumber \\
& \quad + [16 (v \ast c_0 \ast a_0) R (v \ast c_0 \ast a_0)^T]_{12}.
\end{align}
Observe that using the same strategy as above, the terms for the right hand side (of the first equality in Lemma~\ref{lem:properties}(b)) will simply be the above expression with $a_0$ and $c_0$ swapped.
The numbers in the subscript are given to facilitate matching the terms obtained by swapping $a_0$ and $c_0$.
A quick inspection shows that every term is either symmetric about $a_0$ and $c_0$ or has a pair under the swap, and hence the overall expression remains the same.
Therefore the two sides are equal and this completes the proof (of the first equality). \hfill \qed

\section{Discussion on Applications}
\label{sec:discuss}

In this section, we describe how we might apply our new characterization to classical simulation of quantum circuits, synthesis of logical diagonal unitaries, and decomposition of unitaries into Cliffords and diagonal gates.

The classical simulation problem can be succinctly described as follows.
Given a unitary operator $U$ acting on $\dket{0}^{\otimes n}$ to produce the state $\dket{\psi} = U \dket{0}^{\otimes n}$, efficiently sample from the distribution $P_{\psi}(x) = |\dbraket{x}{\psi}|^2$, where $x \in \mathbb{Z}_2^n$.
We know that the stabilizer for the initial state $\dket{0}^{\otimes n}$ is $Z_N \coloneqq \{ E(0,b) \colon b \in \mathbb{Z}_2^n \}$.
Note that this is a maximal commutative subgroup of the Pauli group as it has $n$ generators.
If $U \in \text{Cliff}_N$, we can track the stabilizer of the state $\dket{\psi}$ as $U Z_N U^{\dagger}$, which can be done efficiently using the symplectic representation of $U$ and the identity~\eqref{eq:symp_action}.
More generally, any unitary $U$ can be decomposed as
\begin{align}
U = C_n D_n C_{n-1} D_{n-1} \cdots C_1 D_1 C_0,
\end{align}
where $C_i \in \text{Cliff}_N$ and $D_i \in \mathcal{C}_d^{(\ell_i)}$ for $\ell_i \in \{3,4,\ldots\}$~\cite{Bravyi-arxiv18}.
For simplicity, assume $\ell_i = \ell$ for all $i$. 
First, let $n = 1$ and let the stabilizer before $C_0$ be $S = \langle E(a_j,b_j); j=1,\ldots,n \rangle$ to keep the initial state generic.
(Each $E(a_j,b_j)$ can also have an overall $(-1)$ factor, but we ignore this since it does not provide any new insight.)
Let $F_0$ be the symplectic matrix corresponding to $C_0$.
Then the new stabilizer can be expressed as
\begin{align}
S_0 & = \langle C_0 E(a_j,b_j) C_0^{\dagger}; j=1,\ldots,n \rangle \\
    & = \langle \pm E([a_j,b_j]F_0); j=1,\ldots,n \rangle.
\end{align}
The \texttt{CHP} simulator of Aaronson and Gottesman~\cite{Aaronson-pra04} indeed keeps track of the stabilizer in this manner and the stabilizer rank approach of Bravyi et al. builds on this~\cite{Bravyi-arxiv18}.
Define $[a_{0,j}, b_{0,j}] \coloneqq [a_j,b_j]F_0$.
Suppose $D_1 = \tau_{R_1}^{(\ell)}$ for some symmetric $R_1$ and let $\Gamma_1 = 
\begin{bmatrix}
I_n & R_1 \\
0 & I_n
\end{bmatrix}$. 
Then, using Corollary~\ref{cor:tauR_conjugate}, we can track the new stabilizer after $D_1$ as
\begin{align}
S_1' &= \langle \pm \tau_{R_1}^{(\ell)} E(a_{0,j},b_{0,j}) (\tau_{R_1}^{(\ell)})^{\dagger}; j=1,\ldots,n \rangle \\
  & = \langle \pm \xi^{\phi(R_1,a_{0,j},b_{0,j},\ell)} E([a_j,b_j]F_0 \Gamma_1) \times \tau_{\tilde{R}_1(R_1,a_{0,j},\ell)} ; j=1,\ldots,n \rangle.
\end{align}
At this point, note that each stabilizer generator is completely determined by $a_j,b_j,F_0$ and $\Gamma_1$ (or equivalently $R_1$), whose sizes grow only as $O(n^2)$.
Next, let $F_1$ be the binary symplectic matrix corresponding to $C_1$.
Then the new stabilizer is
\begin{align}
S_1 & = \langle \pm \xi^{\phi(R_1,a_{0,j},b_{0,j},\ell)} C_1 E([a_j,b_j]F_0 \Gamma_1) C_1^{\dagger} \times C_1 \tau_{\tilde{R}_1(R_1,a_{0,j},\ell)} C_1^{\dagger} ; j=1,\ldots,m \rangle \\
  & = \langle \pm \xi^{\phi(R_1,a_{0,j},b_{0,j},\ell)} E([a_j,b_j]F_0 \Gamma_1 F_1) \times \left( C_1 \tau_{\tilde{R}_1(R_1,a_{0,j},\ell)} C_1^{\dagger} \right) ; j=1,\ldots,m \rangle.
\end{align}
We could expand the second term in each generator as follows.
For simplicity, just consider some $g \in \text{Cliff}_N$ and a $\tau_R^{(\ell)} \in \mathcal{C}_d^{(\ell)}$.
\begin{align}
g \tau_R^{(\ell)} g^{\dagger} & = g \left( \sum_{v \in \mathbb{Z}_2^n} \xi^{vRv^T \bmod 2^{\ell}} \dketbra{v} \right) g^{\dagger} \\
                           & = \sum_{v \in \mathbb{Z}_2^n} \xi^{vRv^T \bmod 2^{\ell}} g \dketbra{v} g^{\dagger}.
\end{align}
So now the stabilizer involves operators that are diagonal in an eigenbasis of stabilizer states $\{ g\dket{v} \}$.
If we proceed as before to apply another diagonal gate $D_2$ then the interactions become more complicated as we might expect, since arbitrary stabilizers are indeed hard to track and this is one way to see the gap between quantum and classical computation.
However, we see that our perspective enables to continue this recursion and shows that every stabilizer generator is \emph{structured}: it always involves a Hermitian Pauli matrix, that can be \emph{efficiently} tracked using the symplectic matrices $F_i$ and $\Gamma_i$, and additional terms that become more complex with the depth of the decomposition of $U$.

Although we did this calculation in the context of classical simulation, it captures the calculations in the other two applications as well.
For logical Clifford operations, once we generate logical Paulis using Gottesman's~\cite{Gottesman-phd97} or Wilde's~\cite{Wilde-physreva09} algorithm, we need to perform the above type of calculations to impose linear constraints on the target symplectic matrix that represents the physical realization of the logical operator (as discussed in Chapter~\ref{ch:ch5_lcs_algorithm}).
Although the same approach can be attempted for logical diagonal unitaries, the fact that we need to fix the code by normalizing the stabilizer introduces complications.
In other words, when the (Pauli) stabilizer of the code is conjugated by a non-Clifford operator, the stabilizer generators are no more purely Paulis and hence the code space might be disturbed.
This is the challenge overcome by magic state distillation~\cite{Bravyi-pra05}, but since that procedure is usually expensive, we think it will be interesting to explore if our unification via symplectic matrices produces alternative strategies for non-Clifford (diagonal) logical operations.
We make some progress in this direction in Chapter~\ref{ch:ch7_stabilizer_codes_qfd}.

Similarly, Clifford unitaries are decomposed by suitably multiplying elementary symplectic matrices from Table~\ref{tab:std_symp} (see~\cite{Dehaene-physreva03},\cite[Appendix I]{Rengaswamy-isit18}).
In order to produce decompositions of the form shown above for a general unitary $U$, we need to understand the interaction between binary symplectic matrices $F_i$ and integer symplectic matrices $\Gamma_i$.
Such an understanding might enable us to develop decomposition algorithms that take advantage of \emph{native} operations in quantum technologies such as arbitrary angle $X$- and $Z$-rotations, and M{\o}lmer-S{\o}rensen gates, in trapped-ion architectures~\cite{Linke-nas17}.
For these purposes, it will be interesting to see if Lemma~\ref{lem:properties} can be effectively put to use.

%% file: ch7_stabilizer_codes_qfd.tex

\label{ch:ch7_stabilizer_codes_qfd}


\section{Motivation and Contributions}

Given a stabilizer code, we need to develop a scheme to perform fault-tolerant universal quantum computation on the logical qubits protected by the code, because otherwise the code can only be used as a quantum memory\footnote{Part of this work has been accepted to the 2020 IEEE International Symposium on Information Theory~\cite{Rengaswamy-arxiv20*2}.}.
In Chapter~\ref{ch:ch5_lcs_algorithm}, we produced a systematic algorithm that can synthesize all (equivalence classes of) Clifford circuits on the physical qubits that realize a given logical Clifford operator (on the logical qubits).
As mentioned before, for universal quantum computation, we also need to determine a way to synthesize circuits that realize at least one non-Clifford logical operator.
Moreover, the simplest example of a fault-tolerant circuit is a \emph{transversal} operator, that splits as a tensor product of individual single-qubit operators on the physical qubits of the code, since errors on individual qubit lines do not spread to other qubits.
However, the Eastin-Knill theorem shows that there is no QECC that detects at least $1$ error and possesses a universal set of logical gates that can be realized via transversal operations~\cite{Eastin-prl09, Zeng-it07}.
Therefore, there is a tradeoff between the operations that can be implemented transversally and the operations for which other fault-tolerant mechanisms must be devised.
Since fault-tolerant realizations of logical \emph{non-Clifford} operators are harder to produce, we begin by asking a question that is motivated by the easiest non-Clifford operator to engineer.

\begin{center}
What kind of stabilizer codes support a transversal operator composed of $T$ and $T^{\dagger}$ gates on the physical qubits?
\end{center}

In other words, what structure is needed in the stabilizer $S$ so that the code subspace $V(S)$ is preserved under the application of a given pattern of $T$ and $T^{\dagger}$ (and identity) gates on the physical qubits?
This reverses the strategy employed in the LCS algorithm, where we translated a logical operator to a physical operator. 
The above question is more practically motivated because single-qubit $Z$-rotations are some of the easiest examples of non-Clifford gates that can be performed in the lab, e.g., for trapped ion systems these gates are actually \emph{native} operations~\cite{Linke-nas17}, and we want to make the maximum use of them.
Assuming the stabilizer has the necessary structure, we also need to determine what logical operator is realized by the given pattern of $T$ and $T^{\dagger}$ gates.
We also address this question for some codes and logical operators, e.g., we consider the case when a transversal application of the $T$ gate realizes the logical transversal $T$ on all the $k$ logical qubits encoded by a CSS($X,C_2 ; Z, C_1^{\perp}$) code.
This establishes a tight connection with \emph{triorthogonal codes} defined by Bravyi and Haah~\cite{Bravyi-pra12}. 
Subsequently, we provide a partial answer to the extension of the above question to $Z$-rotations above level $3$ of the Clifford hierarchy~\cite{Landahl-arxiv13,Haah-quantum17b,Campbell-pra17,Campbell-prl17,Vuillot-arxiv19}.
Finally, we produce a family of $\llbr 2^m, \binom{m}{r}, 2^r \rrbr$ quantum Reed-Muller codes, where $1 \leq r \leq m/2$ and $r$ divides $m$, and show that the transversal $\pi/2^{m/r}$ $Z$-rotation is a logical operator on these codes.
Furthermore, we also derive the exact logical operation realized by this transversal gate on these codes.

\subsection{Distinction from Prior Works}

In general, logical Clifford gates are easier to implement than logical $T$ gates.
This is because self-dual CSS codes, i.e., CSS codes where the pure $X$-type and pure $Z$-type stabilizers are constructed from the same classical code, admit a transversal implementation of the logical Clifford group, but not of the logical $T$ gate.
On the other hand, there are some code families such as \emph{triorthogonal codes}~\cite{Bravyi-pra12} and \emph{color codes}~\cite{Kubica-pra15} that realize the logical $T$ gate transversally.
Therefore, a common strategy is to utilize these codes to perform magic state distillation and state injection to apply the logical $T$ gate on the data ~\cite{Bravyi-pra12,Gottesman-nature99,Bravyi-pra05}.
By this approach, circuits on the error-corrected quantum computer will only consist of Clifford operations, augmented by ancillary magic states, and these operations can be realized transversally.

There has also been interest in employing smaller angle rotations, compared to the $\pi/8$ rotation of the $T$ gate, logically~\cite{Landahl-arxiv13}.  
This poses heavier requirements on the distillation code, but can also result in shorter gate sequences during compilation.  
In contrast to the difficulty of small-angle logical rotations, the fidelity of physical rotations can increase at finer angles~\cite{Nam-arxiv19}, helping to mitigate the burden of magic state distillation with cumbersome codes.  
Thus, it may be profitable to further understand codes supporting smaller-angle transversal $Z$-rotations as well.

Several works have studied the problem of realizing non-trivial logical operators via physical $Z$-rotations~\cite{Bravyi-pra12,Haah-quantum17b,Campbell-pra17,Campbell-prl17,Vuillot-arxiv19}.
These works approach this problem by restricting themselves to \emph{Calderbank-Shor-Steane (CSS)} codes and then examining the action of these gates on the basis states of these codes.
When a $\pi/2^{\ell}$ $Z$-rotation
\begin{align}
\label{eq:Zrotations}
\exp\left( \frac{-\imath\pi}{2^{\ell}} Z \right) = \cos\frac{\imath\pi}{2^{\ell}} \cdot I_2 - \imath \sin\frac{\imath\pi}{2^{\ell}} \cdot Z \equiv \text{diag}\left( 1, \exp\left( \frac{2\pi\imath}{2^{\ell}} \right) \right)
\end{align}
acts on a qubit in the computational basis $\{ \ket{0}, \ket{1} \}$, it picks up a phase of $\exp\left( \frac{2\pi\imath}{2^{\ell}} \right)$ when acting on $\ket{1}$ and it leaves $\ket{0}$ undisturbed.
Hence, a transversal application of this gate on an $n$-qubit state $\ket{v}, v \in \mathbb{Z}_2^n$, picks up the phase $\exp\left( \frac{2\pi\imath}{2^{\ell}} w_H(v) \right)$, where $w_H(v) \coloneqq \sum_{i=1}^{n} v_i$ denotes the Hamming weight of $v$.
Therefore, by engineering the Hamming weights of the binary vectors describing the superposition in the CSS basis states, these works determined \emph{sufficient} conditions for such transversal $Z$-rotations to realize logical operators on these codes.

In contrast to these previous works, we take a Heisenberg approach to this problem by examining the action of the physical operation on the stabilizer group defining the code, naturally generalizing the aforementioned strategy.
Consequently, we are able to derive necessary and sufficient conditions for any stabilizer code to support a physical transversal $T, T^{\dagger}$ gate, without restricting ourselves to CSS codes (see Theorems~\ref{thm:transversal_T},~\ref{thm:transversal_T_Tinv}).
When applied to CSS codes, these conditions translate to constructing a pair of classical codes $C_X$ and $C_Z$ such that $C_Z$ contains a self-dual code supported on each codeword in $C_X$.
Concretely, this result allows us to prove the following corollaries which broadly form ``converses'' to the sufficient conditions derived in the aforementioned works.
\begin{enumerate}
    \item Given an $\llbr n,k,d \rrbr$ \emph{non-degenerate} stabilizer code supporting a physical transversal $T, T^{\dagger}$ gate, there exists an $\llbr n,k,d \rrbr$ CSS code supporting the same operation (see Corollary~\ref{cor:csst_sufficient}).
    An $\llbr n,k,d \rrbr$ stabilizer code is non-degenerate if every stabilizer element has weight at least $d$.
    For degenerate stabilizer codes, this statement holds under an additional assumption on the stabilizer generators.
    
    \item Triorthogonal codes form the most general family of CSS codes that realize logical transversal $T$ from physical transversal $T$ (see Theorem~\ref{thm:logical_trans_T} and Corollary~\ref{cor:triortho_general}).
    
    \item Triorthogonality is necessary for physical transversal $T$ on a CSS code to realize the logical identity (see Theorem~\ref{thm:logical_identity}).
    An additional condition on the logical $X$ operators distinguishes this case from triorthogonal codes where the logical operation is also a transversal $T$.
\end{enumerate}
These results suggest that, for the problem of distilling magic states using physical transversal $T$, CSS codes might indeed be optimal.
We emphasize that we are able to make such conclusions because we focus on the effects of physical operations directly on the stabilizer (and logical Pauli) group(s), rather than just on the basis states of CSS codes, i.e., by taking a ``Heisenberg'' perspective rather than a ``Schr{\"o}dinger'' perspective.

\subsection{Connections to Classical Coding Theory}

When our main result (Theorem~\ref{thm:transversal_T}) is specialized to CSS codes we obtain new classical coding problems, and the general case is quite similar.
Since this is a self-contained problem that classical coding theorists can analyze, we describe it here.

\textbf{CSS-T Codes:} A pair $(C_1,C_2)$ of binary linear codes with parameters $[n,k_1,d_1]$ and $[n,k_2,d_2]$, respectively, such that $C_2 \subset C_1$ and the following properties hold:
\begin{enumerate}
    \item $C_2$ is an even code, i.e., $w_H(x) \equiv 0$ (mod $2$) for all $x \in C_2$, where $w_H(x)$ is the Hamming weight of $x$.
    
    \item For each $x \in C_2$, there exists a dimension $w_H(x)/2$ self-dual code in $C_1^{\perp}$ that is supported on $x$, i.e., there exists $ C_x \subseteq C_1^{\perp}$ s.t. $|C_x| = 2^{w_H(x)/2}, C_x = C_x^{\perp}$, and $z \in C_x \Rightarrow z \preceq x$, i.e., $\text{supp}(z) \subseteq \text{supp}(x)$, where $C_1^{\perp}$ is the code dual to $C_1$ and $\text{supp}(x)$ is the support of $x$.
\end{enumerate}
\textbf{Open Problem:} Find an infinite family of $\llbr n, k_1-k_2, \text{min}(d_1,d_2^{\perp}) \rrbr$ CSS-T codes such that $\frac{(k_1-k_2)}{n} = \Omega(1)$ and (ideally) $\frac{\text{min}(d_1,d_2^{\perp})}{n} = \Omega(1)$, where $d_2^{\perp}$ is the minimum distance of $C_2^{\perp}$.

The specific requirements arise when the $T$ gate is applied transversally, but different patterns of $T$ and $T^{\dagger}$ gates also produce variants of it~\cite{Rengaswamy-arxiv19c} (see Theorem~\ref{thm:transversal_T_Tinv}).

We believe this result opens the way to leverage the rich classical literature on self-dual codes~\cite{Rains-arxiv02,Nebe-2006}, the MacWilliams identities~\cite{Macwilliams-1977}, and the McEliece theorems on divisibility of weights~\cite{McEliece-jpl72,McEliece-dm72}, to potentially construct new stabilizer codes with transversal gates. 
Furthermore, this perspective is a new tool for arguing about the best possible scaling achievable for rates and distances of stabilizer codes supporting transversal $T$ gates, or even general $\pi/2^{\ell}$ $Z$-rotations. 

Among several examples, we construct a $\llbr 16,3,2 \rrbr$ code where transversal $T$ realizes the logical CCZ (up to Pauli corrections; see Section~\ref{sec:logical_ccz}). 
This code belongs to the compass code family studied in \cite{Li-prx19}.
This is also closely related to Campbell's $\llbr 8,3,2 \rrbr$ color code~\cite{Campbell-blog16} that is defined on a $3$-dimensional cube, and it can be interpreted as three such cubes in a chain.
(The construction can be extended to a chain of arbitrary number of cubes.)
As we show in Example~\ref{eg:Campbell}, the $\llbr 8,3,2 \rrbr$ code belongs to a family of $\llbr 2^m,m,2 \rrbr$ quantum Reed-Muller codes defined on $m$-dimensional cubes.
However, as we discuss in Example~\ref{eg:decreasing_monomial}, the $\llbr 16,3,2 \rrbr$ code can be constructed using the (classical) formalism of \emph{decreasing monomial codes} that was introduced by Bardet et al.~\cite{Bardet-isit16,Bardet-arxiv16}.
This formalism generalizes Reed-Muller and polar codes~\cite{Arikan-it09}, and provides a general framework for synthesizing a large family of codes via evaluations of polynomials.
Recently, Krishna and Tillich~\cite{Krishna-arxiv18} have exploited this framework to construct triorthogonal codes from punctured polar codes for magic state distillation.
Thus, the $\llbr 16,3,2 \rrbr$ code forms an interesting example because it points towards a general application of the formalism of decreasing monomial codes for transversal $Z$-rotations, where the logical $X$ and $Z$ strings are not necessarily identical as in the standard presentation of triorthogonal codes.
Such asymmetry in logical operators and hence the $X$- and $Z$-distances of the codes, which can also exist in triorthogonal codes, might be useful in scenarios of biased noise as well~\cite{Tuckett-prl18}. 
Hence, this formalism provides more flexibility in designing codes as well as analyzing them.

Finally, we extend this approach beyond $T$ gates and establish conditions for a stabilizer code to support a transversal $\pi/2^{\ell}$ $Z$-rotation (see Theorem~\ref{thm:transversal_Z_rot}).
However, the conditions we derive involve trigonometric quantities on the weights of vectors describing the stabilizer, and we are unable to distill finite geometric conditions without making a simplifying assumption.
Therefore, we have yet to establish a full generalization of Theorems~\ref{thm:transversal_T} and~\ref{thm:transversal_T_Tinv} to general $\pi/2^{\ell}$ $Z$-rotations.
Note that we only discuss $Z$-rotations of the form in~\eqref{eq:Zrotations} because non-trivial error-detecting stabilizer codes only support rotations belonging to the Clifford hierarchy~\cite{JochymOconnor-prx17,Cui-physreva17}.
However, we are able to study a family of $\llbr 2^m, \binom{m}{r}, 2^r \rrbr$ quantum Reed-Muller codes, where $1 \leq r \leq m/2$ and $r$ divides $m$, and provide an alternative perspective that highlights the logical operation realized by a transversal $\exp\left( \frac{-\imath\pi}{2^{m/r}} Z \right)$ gate.
This recovers the well-known $\llbr 8,3,2 \rrbr$ code of Campbell~\cite{Campbell-blog16}, and also provides new information about the family of codes discussed in~\cite{Haah-quantum17b,Campbell-pra17}.
By the ``CSS sufficiency'' intuition above, this ties back to the ``Schr{\"o}dinger'' perspective of past works.

We begin by outlining the general strategy for understanding when a physical QFD gate preserves a stabilizer code subspace.

\section{General Approach for QFD Gates}
\label{sec:general_QFD}

The key idea in addressing the above question is the following: a physical operator $U \in \mathbb{U}_N$ preserves the code subspace of a stabilizer code defined by a stabilizer group $S$ if and only if $U \Pi_S U^{\dagger} = \Pi_S$.
This is because two operators preserve each others' eigenspaces if and only if they commute.
Here, we say an operator $A$ preserves the eigenspace of another operator $B$ if it holds that for any eigenvector $v$ of $B$ with eigenvalue $b$, $Av$ is also an eigenvector of $B$ with eigenvalue $b$.
Thus, $U$ is a valid logical operator for $S$ if and only if
\begin{align}
\label{eq:U_logical_op}
U \Pi_S U^{\dagger} = \frac{1}{2^r} \sum_{j = 1}^{2^r} \epsilon_j U E(a_j,b_j) U^{\dagger} = \frac{1}{2^r} \sum_{j = 1}^{2^r} \epsilon_j E(a_j,b_j) = \Pi_S.
\end{align}

Recollect that the Pauli operators $\{ \frac{1}{\sqrt{N}} E(a,b),\, a,b \in \mathbb{Z}_2^n \}$ form an orthonormal basis for all unitary matrices under the trace inner product $\langle A,B \rangle_{\text{Tr}} \coloneqq \text{Tr}(A^{\dagger} B)$, where $A^{\dagger}$ represents the Hermitian transpose of $A$ (see Theorem~\ref{thm:Pauli_basis}).
Therefore, given any matrix $U \in \mathbb{U}_N$, where $\mathbb{U}_N$ denotes the group of $N \times N$ unitary matrices, we can express it as
\begin{align}
U = \sum_{a,b \in \mathbb{Z}_2^n} \text{Tr}\left( \frac{1}{\sqrt{N}} E(a,b) \cdot U \right) \cdot \frac{1}{\sqrt{N}} E(a,b) = \frac{1}{N} \sum_{a,b \in \mathbb{Z}_2^n} \text{Tr}(E(a,b) U) E(a,b).
\end{align}
Note that $\text{Tr}(E(a,b)) = 0$ unless $E(a,b) = E(0,0) = I_N$, in which case $\text{Tr}(E(0,0)) = N$.
If $U = \sum_{v \in \mathbb{Z}_2^n} \phi_v \dketbra{v}$ is diagonal (in the standard coordinate basis), then for $a \neq 0$,
\begin{align}
\text{Tr}(E(a,b) U) & = \text{Tr}\left[ \imath^{ab^T} \sum_{v \in \mathbb{Z}_2^n} \dketbra{v \oplus a}{v} \cdot \sum_{v' \in \mathbb{Z}_2^n} (-1)^{v'b^T} \dketbra{v'} \cdot U \right] \\
  & = \imath^{ab^T} \sum_{v \in \mathbb{Z}_2^n} (-1)^{vb^T} \dbra{v} U \dket{v \oplus a} \\
  & = 0.
%
\end{align}
Hence, $\text{Tr}(E(a,b) U) \neq 0$ if and only if $a = 0$, and $\text{Tr}(E(0,b) U) = \sum_{v \in \mathbb{Z}_2^n} (-1)^{vb^T} \phi_{v}$.

Using this observation, let us expand $\tau_R^{(\ell)}$ in the Pauli basis. 
For $x \in \mathbb{Z}_2^n$ we define 
\begin{align}
\label{eq:tau_coeff}
c_{R,x}^{(\ell)} & \coloneqq \frac{1}{\sqrt{2^n}} \text{Tr}\left[ E(0,x) \tau_R^{(\ell)} \right] = \frac{1}{\sqrt{2^n}} \sum_{v \in \mathbb{Z}_2^n} (-1)^{vx^T} \xi^{vRv^T} \\
\Rightarrow \tau_R^{(\ell)} & = \frac{1}{\sqrt{2^n}} \sum_{x \in \mathbb{Z}_2^n} c_{R,x}^{(\ell)} E(0,x).
\end{align}
Using this Pauli expansion for $\tau_{\tilde{R}(R,a,\ell)}^{(\ell-1)}$ in Corollary~\ref{cor:tauR_conjugate} and assuming $a,b \in \mathbb{Z}_2^n$, we get
\begin{align}
\tau_R^{(\ell)} E(a,b) ( \tau_R^{(\ell)} )^{\dagger} & = \xi^{\phi(R,a,b,\ell)} E(a, b + a R) \, \tau_{\tilde{R}(R,a,\ell)}^{(\ell-1)} \\
  & = \xi^{\phi(R,a,b,\ell)} E(a, b + a R) \cdot \frac{1}{\sqrt{2^n}} \sum_{x \in \mathbb{Z}_2^n} c_{\tilde{R}(R,a,\ell),x}^{(\ell-1)} E(0,x) \\
\label{eq:tau_Eab_expand}
  & = \frac{1}{\sqrt{2^n}} \xi^{\phi(R,a,b,\ell)} \sum_{x \in \mathbb{Z}_2^n} c_{\tilde{R}(R,a,\ell),x}^{(\ell-1)} \imath^{-ax^T} E(a, b + aR + x).
\end{align}
%
%
Therefore, if $U = \tau_R^{(\ell)}$ in~\eqref{eq:U_logical_op}, for some $\ell \geq 2$ and $R$ symmetric over $\MZ_{2^\ell}$, then for $\tau_R^{(\ell)}$ to be a valid logical operator we need
\begin{align}
\tau_R^{(\ell)} \Pi_S (\tau_R^{(\ell)})^{\dagger} & = \frac{1}{2^r} \sum_{j = 1}^{2^r} \epsilon_j \tau_R^{(\ell)} E(a_j,b_j) (\tau_R^{(\ell)})^{\dagger} \\
  & = \frac{1}{2^r} \sum_{j = 1}^{2^r} \epsilon_j \frac{1}{\sqrt{2^n}} \xi^{\phi(R,a_j,b_j,\ell)} \sum_{x \in \Fn} c_{\tilde{R}(R,a_j,\ell),x}^{(\ell-1)} \imath^{-a_j x^T} E(a_j, b_j + a_j R + x) \\
  & = \frac{1}{2^r} \sum_{j = 1}^{2^r} \epsilon_j E(a_j,b_j).
\end{align}
This shows one important use of the formula in Corollary~\ref{cor:tauR_conjugate} that we derived in~\cite{Rengaswamy-pra19}.
The primary challenge here is to determine which coefficients are non-zero for given $R,a,\ell$, and also their values.
In principle, we can solve for all the conditions on $S$ that are necessary (and sufficient) for this equality. 
So, if we want to take advantage of an operation that we can do ``easily'' in the lab, then we can use the above approach to derive codes accordingly; if the operation is also a QFD gate, then we can exactly use the above equations. 
However, solving the above equality for arbitrary $R,\ell$ might be hard.
In this chapter, we solve the equality completely when $\tau_R^{(3)}$ is a transversal combination of $T$ and $T^{\dagger}$ gates. 
We also provide a nearly complete solution for the transversal application of higher level $Z$-rotations.

Fault-tolerance makes it natural to partition the physical qubits into small groups and employ ``generalized'' transversal gates that split into operations on these individual groups.
Indeed, such a scheme has been recently explored by Jochym-O'Connor et al.~\cite{JochymOconnor-prx17}, and can be used to construct a universal set of fault-tolerant gates~\cite{JochymOconnor-prl14}.
In fact, they showed that if we allow the partition to change during computation, then we can obtain a universal set of logical gates through transversal operations alone.
Therefore, our general approach to analyze QFD gates allows one to investigate codes that support transversal $1$- and $2$-local diagonal gates, on a partition of qubits into groups of at most two.
This work is a proof-of-concept for the important case of $Z$-rotations.

\section{Stabilizer Codes that Support $T$ and $T^{\dagger}$ Gates}
\label{sec:stab_codes_T_Tinv}

We begin with a formula for the physical transversal $T$ gate which, given several applications, is of independent interest. 

\begin{lemma}
\label{lem:conj_by_trans_T}
Let $E(a,b) \in HW_N, N = 2^n$, for some $a,b \in \mathbb{Z}_2^n$.
Then the transversal $T$ gate acts on $E(a,b)$ as
\begin{align}
T^{\otimes n} E(a,b) \left( T^{\otimes n} \right)^{\dagger} = \frac{1}{2^{w_H(a)/2}} \sum_{y \preceq a} (-1)^{b y^T} E(a, b \oplus y),
\end{align}
where $w_H(a) = aa^T$ is the Hamming weight of $a$, and $y \preceq a$ denotes that $y$ is contained in the support of $a$.
\begin{proof}
This result is a special case of Lemma~\ref{lem:conj_by_trans_T_Tinv}, which we prove in Section~\ref{sec:proof_conj_by_trans_T_Tinv}.
\end{proof}
\end{lemma}

Using this lemma, we state our first result which partially answers the above question.

\begin{theorem}[Transversal $T$]
\label{thm:transversal_T}
%
%
%
%
%
Let $S = \langle \nu_i E(c_i,d_i) ; i = 1,\ldots,r \rangle$ define an $\llbr n,n-r,d \rrbr$ stabilizer code, with arbitrary $\nu_i \in \{ \pm 1 \}$, and denote the elements of $S$ by $\epsilon_j E(a_j, b_j), j = 1,2,\ldots,2^r$. 
If the transversal application of the $T$ gate  preserves the code space $V(S)$ and hence realizes a logical operation on $V(S)$, then the following are true.
\begin{enumerate}

\item For any $\epsilon_j E(a_j,b_j) \in S$ with non-zero $a_j$, $w_H(a_j)$ is even, where $w_H(a_j)$ represents the Hamming weight of $a_j \in \mathbb{Z}_2^n$.

\item  For any $\epsilon_j E(a_j,b_j) \in S$ with non-zero $a_j$, define $Z_j \coloneqq \{ z \preceq a_j \colon \epsilon_z E(0,z) \in S\ \text{for\ some}\ \epsilon_z \in \{ \pm 1 \} \}$%
	  \footnote{Although using the above notation it is true that $\epsilon_z E(0,z) = \epsilon_{j'} E(a_{j'},b_{j'})$ for some $j' \in \{ 1,2,\ldots,2^r \}$ with $a_{j'} = 0$, we use the notation with $z$ for convenience and also because it actually refers to pure $Z$-type stabilizers.}.
      Then $Z_j$ contains its dual computed only on the support of $a_j$, i.e., on the ambient dimension $w_H(a_j)$.
      Equivalently, $Z_j$ contains a dimension $w_H(a_j)/2$ self-dual code $A_j$ that is supported on $a_j$, i.e., there exists a subspace $A_j \subseteq Z_j$ such that $y z^T = 0$ (mod 2) for any $y,z \in A_j$ (including $y = z$) and $\text{dim}(A_j) = w_H(a_j)/2$.
      
\item Let $\tilde{Z}_j \subseteq \mathbb{Z}_2^{w_H(a_j)}$ denote the subspace $Z_j$ where all positions outside the support of $a_j$ are punctured (dropped).
      Then, for each $z \in \mathbb{Z}_2^n$ such that $\tilde{z} \in (\tilde{Z}_j)^{\perp}$ for some $j \in \{ 1,\ldots,2^r \}$, we have $\epsilon_z = \imath^{zz^T}$, i.e., $\imath^{zz^T} E(0,z) \in S$.
      Here, $(\tilde{Z}_j)^{\perp}$ denotes the dual of $Z_j$ taken over this punctured space with ambient dimension $w_H(a_j)$.
      (Also, $Z_j \supseteq (\tilde{Z}_j)^{\perp}$ with zeros added outside the support of $a_j$.)

\end{enumerate} 
Conversely, if the first two conditions above are satisfied, and if the third condition holds for all $z \in A_j$ instead of just the dual of (the punctured) $Z_j$, then transversal $T$ preserves the code space $V(S)$ and hence induces a logical operation.
\begin{proof}
See Section~\ref{sec:proof_transversal_T}.
\end{proof}
\end{theorem}

\begin{remark}
\label{rem:transversal_T_Pauli_correction}
\normalfont
Note that, since $A_j$ is supported on $a_j$, the ambient dimension of vectors in $A_j$ is essentially $w_H(a_j)$.
So $A_j$ is a $[w_H(a_j), w_H(a_j)/2]$ self-dual code embedded in $\mathbb{Z}_2^n$.
The last point is requiring that the $Z$-stabilizers arising from vectors in the subspaces $A_j$ have the correct sign, given by $\imath^{w_H(z)} = \imath^{zz^T} \in \{ \pm 1 \}$.
If this is not taken care of, then an appropriate Pauli operator has to be applied before and after transversal $T$ in order to make a valid logical operator.
Indeed, this Pauli operator is essentially fixing the signs of the $Z$-stabilizers as required.
Hence, although the necessary and sufficient directions of the theorem differ in the last sign condition, for all practical purposes one can take the signs to be imposed on all of $A_j$ instead of just its subspace that is identified with $(\tilde{Z}_j)^{\perp}$.
These Pauli corrections preserve the code parameters $\llbr n,n-r,d \rrbr$.
\end{remark}


Let us now look at a simple example constructed using this theorem that will clarify the requirements above.

\begin{example}
\label{eg:cssT_622}
\normalfont
Define a $\llbr 6,2,2 \rrbr$ CSS code by the following stabilizer generator matrix:
\begin{align}
\setlength\aboverulesep{0pt}\setlength\belowrulesep{0pt}
    \setlength\cmidrulewidth{0.5pt}
G_S = 
\left[
\begin{array}{cccccc|cccccc}
1 & 1 & 1 & 1 & 1 & 1 & 0 & 0 & 0 & 0 & 0 & 0 \\
\hline 
0 & 0 & 0 & 0 & 0 & 0 & 1 & 1 & 0 & 0 & 0 & 0 \\
0 & 0 & 0 & 0 & 0 & 0 & 0 & 0 & 1 & 1 & 0 & 0 \\
0 & 0 & 0 & 0 & 0 & 0 & 0 & 0 & 0 & 0 & 1 & 1 \\
\end{array}
\right].
\end{align}
The right half of the last $3$ rows form the generators of $Z_S$ for this code.
Since there is only one non-trivial $a_j$ for this code, we see that $Z_S = A_1$ with $a_1 = [1,1,1,1,1,1]$.
Hence, the stabilizer generators are $X^{\otimes 6} = X_1 X_2 \cdots X_6, - Z_1 Z_2, - Z_3 Z_4, - Z_5 Z_6$, since the generators of $Z_S$ have weight $2$.
Multiplying $X^{\otimes 6}$ and the product of these three $Z$-stabilizers, we see that $Y^{\otimes 6} \in S$.
We can define the logical $X$ operators for this code to be $\bar{X}_1 = X_1 X_2, \bar{X}_2 = X_3 X_4$, since these are linearly independent and commute with all stabilizers.
Using the identity we observed in Example~\ref{eg:T_gate}, we see that 
\begin{align}
T^{\otimes 6} X_1 X_2 (T^{\otimes 6})^{\dagger} = e^{-\imath \cdot 2\pi/4} (Y_1 P_1) (Y_2 P_2) = -\imath \cdot (\imath X_1 Z_1 P_1) (\imath X_2 Z_2 P_2) \equiv -\imath (X_1 X_2) (P_1 P_2),
\end{align}
since $-Z_1 Z_2 \in S$.
We observe that $(P_1 P_2) X^{\otimes 6} (P_1 P_2)^{\dagger} = Y_1 Y_2 X_3 X_4 X_5 X_6 \equiv X^{\otimes 6}$ up to the stabilizer $-Z_1 Z_2$, so $P_1 P_2$ indeed preserves $V(S)$.
But $(P_1 P_2) (X_1 X_2) (P_1 P_2)^{\dagger} = Y_1 Y_2 = (X_1 X_2) (-Z_1 Z_2) \equiv X_1 X_2$, and $P_1 P_2$ obviously commutes with $\bar{X}_2$, so $P_1 P_2$ is essentially the logical identity gate.
A similar reasoning holds for $P_3 P_4$.
Therefore, up to a global phase, the transversal $T$ preserves the logical operators $\bar{X}_1$ and $\bar{X}_2$, so in this case the transversal $T$ gate realizes just the logical identity (up to a global phase).
This can also be checked explicitly by writing the logical basis states $\dket{x_1 x_2}_L$ for $x_i \in \mathbb{Z}_2$:
\begin{align}
\dket{x_1 x_2}_L = \bar{X}_1^{x_1} \bar{X}_2^{x_2} \cdot \frac{1}{\sqrt{2}} \left( \dket{010101} + \dket{101010} \right).
\end{align}
If the $Z$-stabilizer generators were instead taken to be $Z_1 Z_2, Z_3 Z_4, Z_5 Z_6$, then the superposition above in $\dket{00}_L$ will be $(\dket{000000} + \dket{111111})$.
Therefore, $T^{\otimes 6} X_1 X_3 X_5$ will be a valid logical operator (that still implements the logical identity).

Given that $S$ has the necessary structure given by Theorem~\ref{thm:transversal_T}, note that we can freely add another $Z$-stabilizer generator that commutes with $X^{\otimes 6}$, e.g., $Z_1 Z_3 Z_4 Z_6 \leftrightarrow [1,0,1,1,0,1] \notin Z_S$.
This does not affect the transversal $T$ property: once $T^{\otimes n} \Pi_S (T^{\otimes n})^{\dagger} = \Pi_S$, mapping $\Pi_S \mapsto \Pi_S \cdot \frac{(I_N + E(0,z))}{2}$ preserves the equality since $(I_N + E(0,z))$ is diagonal.
\end{example}

Now we generalize Lemma~\ref{lem:conj_by_trans_T} and Theorem~\ref{thm:transversal_T} to $T$ and $T^{\dagger}$ gates, which addresses the initial question completely.

\begin{lemma}
\label{lem:conj_by_trans_T_Tinv}
For $N = 2^n$ and $a,b \in \mathbb{Z}_2^n$, let $E(a,b) \in HW_N$ and choose $t_1, t_7 \in \mathbb{Z}_2^n$ such that $t_1 \ast t_7 = 0$ (i.e., $\text{supp}(t_1)\, \cap\, \text{supp}(t_7) = \emptyset$).
Define $t = t_1 + 7 t_7 \in \{0,1,7\}^n , t' = t_1 + t_7 \in \mathbb{Z}_2^n$.
Then the physical operation $T^{\otimes t}$ acts on $E(a,b)$ as
\begin{align}
T^{\otimes t} E(a,b) \left( T^{\otimes t} \right)^{\dagger} = \frac{1}{2^{w_H(a \ast t')/2}} \sum_{y \preceq (a \ast t')} (-1)^{(b + t_7) y^T} E(a, b \oplus y),
\end{align}
where $T^{\otimes t}$ denotes that $T$ (resp. $T^{\dagger} = T^7$) is applied to qubits in the support of $t_1$ (resp. $t_7$).
\begin{proof}
See Section~\ref{sec:proof_conj_by_trans_T_Tinv}.
\end{proof}
\end{lemma}

\begin{corollary}
\label{cor:conj_by_trans_T_Tinv}
Let $E(a,b) \in HW_N$ for $N = 2^n$ and some $a,b \in \mathbb{Z}_2^n$.
For $j,j' \in \{ 1,2,\ldots,7 \}$, let $t_j \in \mathbb{Z}_2^n$ and assume that $t_j \ast t_{j'} = 0$ (i.e., $\text{supp}(t_j) \cap \text{supp}(t_{j'}) = \emptyset$) for $j \neq j'$.
Define $t \coloneqq \sum_{j=1}^{7} j t_j \in \mathbb{Z}_8^n , \tilde{t}_1 \coloneqq t_1 + t_5, \tilde{t}_2 \coloneqq t_2 + t_6, \tilde{t}_3 \coloneqq t_3 + t_7 \in \mathbb{Z}_2^n$.
Then the physical operation $T^{\otimes t}$ acts on $E(a,b)$ as
\begin{align}
T^{\otimes t} E(a,b) \left( T^{\otimes t} \right)^{\dagger} = \frac{ (-1)^{a (t_3 + t_4 + t_5 + t_6)^T} }{2^{w_H(a \ast (\tilde{t}_1 + \tilde{t}_3))/2}} \sum_{(a \ast \tilde{t}_2) \preceq z \preceq (a \ast (\tilde{t}_1 + \tilde{t}_2 + \tilde{t}_3))}  (-1)^{(b + \tilde{t}_3) z^T} E\left( a, b \oplus z \right),
\end{align}
where $T^{\otimes t}$ denotes that $T^j$ is applied to the qubits in the support of $t_j$.
\begin{proof}
See Section~\ref{sec:proof_cor_conj_by_trans_T_Tinv}.
\end{proof}
\end{corollary}


\begin{theorem}[Transversal $T(t)$]
\label{thm:transversal_T_Tinv}
Let $S = \langle \nu_i E(c_i,d_i) ; i = 1,\ldots,r \rangle$ define an $\llbr n,n-r,d \rrbr$ stabilizer code as in Theorem~\ref{thm:transversal_T}.
Let $t = t_1 + 7 t_7, t_1 \ast t_7 = 0,$ with supports of $t_1, t_7 \in \mathbb{Z}_2^n$ indicating the qubits on which $T$ and $T^{\dagger} = T^7$ are applied, respectively.
Define $t' = t_1 + t_7 \in \mathbb{Z}_2^n$.
If the application of the $T^{\otimes t}$ gate realizes a logical operation on $V(S)$, then the following are true.


      

\begin{enumerate}

\item For any $\epsilon_j E(a_j,b_j) \in S$ with non-zero $a_j$, $w_H(a_j \ast t')$ is even, where $w_H(a_j \ast t')$ represents the Hamming weight of $(a_j \ast t') \in \mathbb{Z}_2^n$.

\item  For any $\epsilon_j E(a_j,b_j) \in S$ with non-zero $a_j$, define $Z_{j,t'} \coloneqq \{ z \preceq (a_j \ast t') \colon \epsilon_z E(0,z) \in S\ \text{for\ some}\ \epsilon_z \in \{ \pm 1 \} \}$. 
      Then $Z_{j,t'}$ contains its dual computed only on the support of $(a_j \ast t')$, i.e., on the ambient dimension $w_H(a_j \ast t')$.
      Equivalently, $Z_j$ contains a dimension $w_H(a_j \ast t')/2$ self-dual code $A_{j,t'}$ that is supported on $(a_j \ast t')$, i.e., there exists a subspace $A_{j,t'} \subseteq Z_{j,t'}$ such that $y z^T = 0$ (mod 2) for any $y,z \in A_{j,t'}$ (including $y = z$) and $\text{dim}(A_{j,t'}) = w_H(a_j \ast t')/2$.
      
\item Let $\tilde{Z}_{j,t'} \subseteq \mathbb{Z}_2^{w_H(a_j \ast t')}$ denote the subspace $Z_{j,t'}$ where all positions outside the support of $(a_j \ast t')$ are punctured (dropped).
      Then, for each $z \in \mathbb{Z}_2^n$ such that $\tilde{z} \in (\tilde{Z}_{j,t'})^{\perp}$ for some $j \in \{ 1,\ldots,2^r \}$, we have $\epsilon_z = \imath^{zz^T + 2t_7 z^T}$, i.e., $\imath^{zz^T + 2t_7 z^T} E(0,z) \in S$.
      Here, $(\tilde{Z}_{j,t'})^{\perp}$ denotes the dual of $Z_{j,t'}$ taken over this punctured space with ambient dimension $w_H(a_j \ast t')$.
      ($Z_{j,t'} \supseteq (\tilde{Z}_{j,t'})^{\perp}$ with zeros added outside the support of $(a_j \ast t')$.)

\end{enumerate} 
Conversely, if the first two conditions above are satisfied, and if the third condition holds for all $z \in A_{j,t'}$ instead of just the dual of (the punctured) $Z_{j,t'}$, then transversal $T$ preserves the code space $V(S)$ and hence induces a logical operation.
\begin{proof}
The proof is along the same lines as for Theorem~\ref{thm:transversal_T}, but adapted suitably to the general case in Lemma~\ref{lem:conj_by_trans_T_Tinv}.
\end{proof}
\end{theorem}

Notice that the above two results reduce to Lemma~\ref{lem:conj_by_trans_T} and Theorem~\ref{thm:transversal_T}, respectively, when $t_1 = [1,1,\ldots,1]$ and $t_j = [0,0,\ldots,0]$ for $j = 2,3,\ldots,7$.
The main difference is that in this general scenario, the conditions in Theorem~\ref{thm:transversal_T} are applied to the intersection of the support of $a_j$ and $(t_1 + t_7)$.

\begin{example*}[contd.]
\normalfont
Assume that now we want to apply $T$ and $T^{\dagger}$ according to $t_1 = [1,0,1,0,1,0]$ and $t_7 = [0,1,0,1,0,1]$, respectively.
Since $t' = t_1 + t_7 = [1,1,\ldots,1]$, $a_j \ast (t_1 + t_7) = a_j$ always and so the first two conditions of Theorem~\ref{thm:transversal_T_Tinv} reduce to the transversal $T$ case.
However, the last condition needs the sign for the $Z$-stabilizer generators to be $\imath^{2 + 2} = 1$, so we need to change the stabilizer to be $S = \langle X^{\otimes 6}, Z_1 Z_2, Z_3 Z_4, Z_5 Z_6 \rangle$.
Then the superposition for $\dket{00}_L$ will indeed be $(\dket{000000} + \dket{111111})$, and it is easy to verify that $T^{\otimes t}$ fixes the basis states $\dket{00}_L, \dket{01}_L, \dket{10}_L, \dket{11}_L$, so it also realizes the logical identity.
\end{example*}

In principle, we can generalize Theorem~\ref{thm:transversal_T_Tinv} to the case of arbitrary powers of $T$ by using Corollary~\ref{cor:conj_by_trans_T_Tinv}.
However, the derivation is more complicated and the final conditions are not fully clear because the summation in Corollary~\ref{cor:conj_by_trans_T_Tinv} is over a coset and not a subspace as in Lemma~\ref{lem:conj_by_trans_T_Tinv}.
Hence, this generalization still remains open.

Using these results, we can refine the CSS construction to produce codes that support a desired pattern of $T$ and $T^{\dagger}$ gates.
Note that the first two conditions in Theorem~\ref{thm:transversal_T_Tinv} only depend on $(t_1 + t_7)$ and not individually on $t_1$ and $t_7$, i.e., on the union of their supports. 
Hence, any pattern of $T$ and $T^{\dagger}$ on the support of $(t_1 + t_7)$ will preserve the code subspace, up to an initial Pauli application that produces the right signs for the $Z$-stabilizers as prescribed by the last condition in Theorem~\ref{thm:transversal_T_Tinv}.

\begin{corollary}[CSS-T Codes]
\label{cor:css-t_codes}
Let $t_1, t_7 \in \mathbb{Z}_2^n$ be such that $t_1 \ast t_7 = 0$, and define $t = t_1 + 7 t_7, t' = t_1 + t_7$.
Consider a code CSS($X, C_2 ; Z, C_1^{\perp}$) with stabilizer $S$, such that $w_H(x \ast t')$ is even for all $x \in C_2$.
For each $x \in C_2$, let $C_1^{\perp}$ contain a dimension $w_H(x \ast t')/2$ self-dual code $A_{x,t'}$ supported on $(x \ast t')$. 
Moreover, for all $x \in C_2$ and for each $z \in A_{x,t'}$, let $\imath^{w_H(z) + 2 t_7 z^T} E(0,z) \in S$.
Notice that this means $C_2 \subset C_1^{\perp} \subset C_2^{\perp}$, since $x \in A_{x,t'}$.
Then $T^{\otimes t}$ is a valid logical operator for CSS($X, C_2 ; Z, C_1^{\perp}$).
If for all $x \in C_2$ and for all $z \in A_{x,t'}$ we have $t_7 z^T \equiv 0$ (mod $2$) , then $T^{\otimes t'}$ (which is composed of only $T$ gates) is also a valid logical operator, as the sign constraints for $E(0,z)$ are now independent of $t_7$.
\end{corollary}

\begin{remark}
\normalfont
Intuitively, a CSS-T code (for transversal $T$) is determined by two classical codes $C_2 \subset C_1$ such that for every codeword $x \in C_2$, there exists a dimension $w_H(x)/2$ self-dual code in $C_1^{\perp}$ supported on $x$.
This also means that $C_1 \ast C_2 \subseteq C_1^{\perp}$ for the following reason.
For $a \in C_1, x \in C_2$, it is clear that $a$ is orthogonal to every vector in $C_1^{\perp}$.
In particular, $a$ is orthogonal to the self-dual code $C_x \subset C_1^{\perp}$ supported on $x$.
But, for any $z \in C_x$, we have $az^T = (a \ast x)z^T = 0$.
This means $a \ast x \in C_x \subset C_1^{\perp}$ since $C_x$ is self-dual.
We think that this observation might make it more convenient to derive some properties of CSS-T codes, since there is a good literature on the star product~\cite{Randriambololona-aagct15}.
\end{remark}

Corollary~\ref{cor:css-t_codes} also suggests that there might not be a significant advantage in working with general stabilizer codes, rather than just CSS codes, as far as $T$ and $T^{\dagger}$ gates are concerned.
This is because, by Theorem~\ref{thm:transversal_T_Tinv}, there is always a large asymmetry required between the number of stabilizer elements that have at least one $X$ (or $Y$) in them, and the number of purely $Z$-type stabilizer elements. 
Hence, altering the pure $X$-type stabilizers into $X,Y$-type stabilizers might not provide much gain, say, in terms of the distance of the code.
The next corollary confirms this intuition for non-degenerate stabilizer codes.

\begin{definition}
An $\llbr n,k,d \rrbr$ stabilizer code is non-degenerate if every stabilizer element has weight at least $d$, i.e., it acts non-trivially (as $X, Y$ or $Z$) on at least $d$ qubits.
\end{definition}

\begin{corollary}[Sufficiency of CSS-T Codes]
\label{cor:csst_sufficient}
Consider an $\llbr n, k, d \rrbr$ non-degenerate stabilizer code generated by the matrix 
$G_S = 
\begin{bmatrix}
 A & B \\ 
 C & 0 \\
 0 & D 
\end{bmatrix}$
that satisfies the transversal $T(t)$ property (Theorem~\ref{thm:transversal_T_Tinv}). 
Then the CSS code generated by 
$G_S = 
\begin{bmatrix}
 A & 0 \\ 
 C & 0 \\
 0 & D 
\end{bmatrix}$
has parameters $\llbr n, \geq k, \geq d \rrbr$ and also satisfies the transversal $T(t)$ property for the same $t \in \{0,1,7\}^n$.
\begin{proof}
For convenience, we will use the notation $A,B,C,D$ to also refer to the subspaces $\langle A \rangle, \langle B \rangle, \langle C \rangle, \langle D \rangle$ generated by the matrices $A,B,C,D$, respectively.
We assume that $A$ and $C$ are disjoint, that $B$ and $D$ are disjoint, and that $C$ and $D$ have full rank, all without loss of generality.
(Note that if some of these are not satisfied, then we can perform suitable row operations to subsume rows $\begin{bmatrix} a & 0 \end{bmatrix}$ into $\begin{bmatrix} C & 0 \end{bmatrix}$ and rows $\begin{bmatrix} 0 & b \end{bmatrix}$ into $\begin{bmatrix} 0 & D \end{bmatrix}$.)
We will prove the result for $t = [1,1,\ldots,1]$, i.e., transversal $T$, but the extension to any $t \in \{0,1,7\}^n$ is straightforward as we comment at the end of the proof.
Firstly, it is clear that the CSS code has the transversal $T$ property because this depends on the (binary) subspace $\langle A,C \rangle$ being even and the existence of a self-dual code in $D$ within the support of each vector in the subspace $\langle A,C \rangle$. 
Dropping $B$ does not affect these properties. 
It is also clear that the CSS code still encodes $k$ qubits if $A$ has full rank, but if this is violated then some rows of the stabilizer matrix are removed to provide room for more than $k$ logical qubits (without affecting the transversal $T$ property).

Secondly, the distance of the CSS code is lower bounded by the minimum of the minimum weights of $\langle A,C \rangle^{\perp}$ and $D^{\perp}$. 
Since $\begin{bmatrix} 0 & \langle A,C \rangle^{\perp} \end{bmatrix}$ belongs to the normalizer of the given stabilizer code, and $D$ has minimum weight at least $d$ by non-degeneracy, we know that the minimum weight of $\langle A,C \rangle^{\perp}$ is at least $d$. 
Originally, $\begin{bmatrix} \langle B,D \rangle^{\perp} & 0 \end{bmatrix}$ is in the normalizer of the given stabilizer, so the minimum weight of $\langle B,D \rangle^{\perp}$ is at least $d$ as well. 
Since $\begin{bmatrix} C & 0 \end{bmatrix}$ was initially in the stabilizer, we know that $C \subset \langle B,D \rangle^{\perp}$ and minimum weight of $C$ must be at least $d$ by non-degeneracy. 
However, $A \subset D^{\perp}$ and $\begin{bmatrix} A & 0 \end{bmatrix}$ was not originally part of the stabilizer, so it appears that $A$ might have vectors of weight less than $d$. 
But by non-degeneracy, minimum weight of $D$ is $d$. 
So the minimum weight of any self-dual code $C_x \subset D$ is $d$, for any $x \in \langle A,C \rangle$. 
Hence, this means that $w_H(x) \geq d$ since $x \in C_x$
\footnote{Strictly speaking, we are only concerned with the minimum weight of $D^{\perp} \setminus \langle A, C \rangle$ and not of $D^{\perp}$, i.e., we do not require that the new CSS code is also non-degenerate. 
However, the above shows that the minimum weight of $\langle A,C \rangle$ is already at least $d$ due to Theorem~\ref{thm:transversal_T}.}.

Now consider a $z \in D^{\perp}$. 
We want to show that a minimal weight $z$ has weight at least $d$. 
Suppose, for the sake of contradiction, that $w_H(z)$ is minimal but is strictly less than $d$.
Now consider any $x \in \langle A,C \rangle$ and look at the projection $(x \ast z)$. 
By assumption, $z$ is orthogonal to $D$ and hence is orthogonal to $C_x$. 
But the inner product of $z$ with any vector in $C_x$ only depends on the projection $(x \ast z)$. 
Therefore, $(x \ast z)$ is orthogonal to $C_x$, and hence belongs to $C_x$ as $C_x$ is self-dual.
Observe that $w_H(x \ast z) \leq w_H(z) < d$, which implies that we have found an element $(x \ast z) \in C_x$ that has weight less than $d$. 
This is a contradiction since minimum weight of $C_x$ is $d$, and this completes the proof for transversal $T$.
Note that for any other $t = t_1 + 7t_7$, as Theorem~\ref{thm:transversal_T_Tinv} suggests, we simply replace $x \in \langle A,C \rangle$ in the above argument with $(x \ast t')$, where $t' = t_1 + t_7$.
\end{proof}
\end{corollary}

\begin{remark}[Degenerate Codes]
\normalfont
Observe that the arguments above can be extended to the case when the given stabilizer code is degenerate, but now the distance of the new CSS code constructed above is lower bounded only by the minimum weight of $D$, which can be strictly less than $d$.
More explicitly, the minimum weight of $\langle A,C \rangle^{\perp} \setminus D$ is still $d$ since this space is strictly outside the stabilizer but in the normalizer of the given stabilizer code.
So the distance of the CSS code mainly depends on the minimum weight of $D^{\perp} \setminus \langle A,C \rangle$ and the vector $z$ at the end can be assumed to be taken from this subspace.
As a result, such a $z$ with weight less than $d$ cannot also belong to $B^{\perp}$ since otherwise this would contradict the assumption that the given stabilizer code has distance $d$.
Therefore, under the assumption that for the given stabilizer code any vector $z \in D^{\perp} \setminus (\langle A,C \rangle \cup B^{\perp})$ has weight at least $d$, the above corollary can be extended to the degenerate case.
We leave the more general problem of addressing the full extension of the above corollary to the degenerate case for future work.
\end{remark}

Motivated by the above corollary, all our examples in this chapter are CSS-T codes (including the $\llbr 6,2,2 \rrbr$ code in Example~\ref{eg:cssT_622}).
The $\llbr 6,2,2 \rrbr$ code is not just a corner case where the transversal $T$ gate realizes the logical identity.
The following result provides necessary and sufficient conditions for this to happen.

\begin{theorem}[Logical Identity]
\label{thm:logical_identity}
Let $S$ be the stabilizer for an $\llbr n,k,d \rrbr$ CSS-T code that we denote as CSS($X, C_2 ; Z, C_1^{\perp}$).
Let the logical Pauli $X$ group be $\bar{X} = \langle E(x_i,0) ; i = 1,\ldots,k \rangle$. 
Then the transversal $T$ gate on the $n$ physical qubits realizes the logical identity operation if and only if the following are true.
\begin{enumerate}

\item For each $E(x,0) \in \bar{X}$, $\imath^{w_H(x)} E(0,x)$ must be a stabilizer.

\item For each $E(x,0) \in \bar{X}$ and $\epsilon_a E(a,0) \in S$, $\imath^{w_H(x \ast a)} E(0,x \ast a)$ must be a stabilizer.

\item For any two logical Paulis $E(x,0), E(y,0) \in \bar{X}$, $\imath^{w_H(x \ast y)} E(0, x \ast y)$ must be a stabilizer.

\item For any two $X$-type stabilizers $E(a,0), E(b,0) \in S$, $\imath^{w_H(a \ast b)} E(0, a \ast b)$ must be a stabilizer.

\end{enumerate}
\begin{proof}
See Section~\ref{sec:proof_logical_identity}.
We will see shortly that the last three conditions essentially constitute the property of \emph{triorthogonality} for the generator matrix $G_1$ for the classical binary code $C_1$.
See the proof for a more detailed argument.
\end{proof}
\end{theorem}

\subsection{Classical Reed-Muller Codes}
\label{sec:rm_codes}

Given an integer $m \geq 1$, let $x_1, x_2, \ldots, x_m$ be binary variables and we adopt the convention that $x_1$ represents the least significant bit (LSB) and $x_m$ represents the most significant bit (MSB).
These variables can also be interpreted as \emph{monomials} of degree $1$, and we can construct degree $t$ monomials $x_{i_1} x_{i_2} \cdots x_{i_t}$ where $i_j \in \{ 1,\ldots,m \}$.
The set of all monomials in $m$ variables is denoted by $\mathcal{M}_m$.
A degree $t$ polynomial $f$ on $m$ variables is a binary linear combination of monomials such that the maximum degree term(s) has (have) degree $t$.
Any polynomial $f \in \mathbb{F}_2[x_1,\ldots,x_m]$ can be associated one-to-one to its evaluation vector $\text{ev}(f) \coloneqq [\, f(x_m,\ldots,x_1)\, ]_{(x_m,\ldots,x_1) \in \mathbb{F}_2^m} \in \mathbb{F}_2^{2^m}$.
Note that the unique degree $0$ monomial is taken to be $1$ whose evaluation vector is the all-$1$s vector. 

For $0 \leq r \leq m$, the binary Reed-Muller code RM($r,m$) is generated by evaluation vectors of all monomials on $m$ binary variables with degree at most $r$, i.e.,
\begin{align}
\text{RM}(r,m) & \coloneqq \{ \text{ev}(f) \in \mathbb{F}_2^{2^m} \colon f \in \mathbb{F}_2[x_1,\ldots,x_m], \text{deg}(f) \leq r \} \\
  & = \langle \text{ev}(f) \in \mathbb{F}_2^{2^m} \colon f \in \mathcal{M}_m, \text{deg}(f) \leq r \rangle.
\end{align}
Hence, the dimension of RM($r,m$) is given by $k = \sum_{t=0}^{r} \binom{m}{t}$.
It is well-known that the minimum distance of RM($r,m$) is $2^{m-r}$ and that the dual of RM($r,m$) is RM($m-r-1,m$)~\cite{Macwilliams-1977}.
If $\text{ev}(f) \in \text{RM}(r,m)$, then we also write $f \in \text{RM}(r,m)$.

\subsection{Realizing Logical $T$ Gates with Transversal $T$}
\label{sec:logical_T}

Let us begin by constructing the well-known $\llbr 15,1,3 \rrbr$ (punctured) quantum Reed-Muller code~\cite{Anderson-prl14,Quan-jpmt18} that supports a transversal $T$, using the conditions in Theorem~\ref{thm:transversal_T_Tinv}.
The construction is shown in Fig.~\ref{fig:RM15}.

\begin{figure}
\begin{center}

\begin{tikzpicture}

\node (Z) at (-2,0) {$\{ 0 \}$};
\node (C2) at (-2,1.5) {$C_2$};
\node (C1) at (-2,3) {$C_1$};
\node (F2m) at (-2,4.5) {$\mathbb{F}_2^{15}$};

\path[draw] (Z) -- (C2) node[midway,left] {{$4$}} -- (C1) node[midway,left] {{$1$}} -- (F2m) node[midway,left] {{$10$}};

\node[text width=5cm] at (0.9,3) {$=$ Punctured RM($1,4$)};

\node[text width=5cm] at (0.9,1.5) {$=$ Simplex code};

\node (Zp) at (4,0) {$\{ 0 \}$};
\node (C1p) at (4,1.5) {$C_1^{\perp}$};
\node (C2p) at (4,3) {$C_2^{\perp}$};
\node (F2m) at (4,4.5) {$\mathbb{F}_2^{15}$};

\path[draw] (Zp) -- (C1p) node[midway,left] {{$10$}} -- (C2p) node[midway,left] {{$1$}} -- (F2m) node[midway,left] {{$4$}};

\node[text width=5cm] at (6.9,3) {$=$ Hamming code};

\node[text width=5cm] at (6.9,1.5) {$=$ Even weight subcode};

\node at (2,-5.5) {$\setlength\aboverulesep{0pt}\setlength\belowrulesep{0pt}
    \setlength\cmidrulewidth{0.5pt}
G_1^{\perp} \coloneqq 
\begin{blockarray}{cccccccccccccccc}
 \\
\begin{block}{[ccccccccccccccc]c}
 \\
1 & 0 & 1 & 0 & 1 & 0 & 1 & 0 & 1 & 0 & 1 & 0 & 1 & 0 & 1 & x_1 \\
0 & 1 & 1 & 0 & 0 & 1 & 1 & 0 & 0 & 1 & 1 & 0 & 0 & 1 & 1 & x_2 \\
0 & 0 & 0 & 1 & 1 & 1 & 1 & 0 & 0 & 0 & 0 & 1 & 1 & 1 & 1 & x_3 \\
0 & 0 & 0 & 0 & 0 & 0 & 0 & 1 & 1 & 1 & 1 & 1 & 1 & 1 & 1 & x_4 \\
\cmidrule(lr){1-15}
0 & 0 & 1 & 0 & 0 & 0 & 1 & 0 & 0 & 0 & 1 & 0 & 0 & 0 & 1 & x_1 x_2 \\
0 & 0 & 0 & 0 & 1 & 0 & 1 & 0 & 0 & 0 & 0 & 0 & 1 & 0 & 1 & x_1 x_3 \\
0 & 0 & 0 & 0 & 0 & 0 & 0 & 0 & 1 & 0 & 1 & 0 & 1 & 0 & 1 & x_1 x_4 \\
0 & 0 & 0 & 0 & 0 & 1 & 1 & 0 & 0 & 0 & 0 & 0 & 0 & 1 & 1 & x_2 x_3 \\
0 & 0 & 0 & 0 & 0 & 0 & 0 & 0 & 0 & 1 & 1 & 0 & 0 & 1 & 1 & x_2 x_4 \\
0 & 0 & 0 & 0 & 0 & 0 & 0 & 0 & 0 & 0 & 0 & 1 & 1 & 1 & 1 & x_3 x_4 \\
 \\
\end{block}
\end{blockarray}$};

\node[align=center] at (2,-11) {The first $4$ rows of $G_1^{\perp}$ form $G_2$, so $C_2 \subset C_1^{\perp}$};

\end{tikzpicture}
\caption{\label{fig:RM15}The CSS($X, C_2 ; Z, C_1^{\perp}$) construction for the $\llbr 15,1,3 \rrbr$ quantum (punctured) Reed-Muller code.}

\end{center}
\end{figure}

The generator matrix $G_2$ for the simplex code $C_2$ that produces all $X$-type stabilizers (which are all weight $8$) is formed by the first $4$ rows of $G_1^{\perp}$, as shown in Fig.~\ref{fig:RM15}.
\setcounter{MaxMatrixCols}{20}
Notice that this is obtained by \emph{shortening} RM($1,4$): take the generator matrix for the Reed-Muller code RM($1,4$), remove the first row of all $1$s, and then remove the first column which is all $0$s in the remaining matrix.
In other words, let $x_1,x_2,x_3,x_4$ be binary variables that also represent degree-$1$ monomials, with $x_1$ being the least significant bit and $x_4$ being the most significant bit.
Then, the rows of $G_2$ from the top are $x_1,x_2,x_3,x_4$ respectively, with the first coordinate removed.
Similarly, since the dual of RM($1,4$) is the $[16,11,4]$ extended Hamming code RM($2,4$), the dual $C_1^{\perp}$ of the \emph{punctured} RM($1,4$) code $C_1$ is obtained by shortening RM($2,4$).
Therefore, the rows of $G_1^{\perp}$ must be the degree-$1$ monomials $x_1,x_2,x_3,x_4$ and the degree-$2$ monomials $x_i x_j$ for $i < j$, with the first coordinate removed.

Since all vectors in $C_2$ have weight $8$, the first condition of Theorem~\ref{thm:transversal_T} is satisfied.
Now consider, say, the $X$-stabilizer arising from the monomial $x_1$ belonging to $C_2$.
By direct observation of the rows of $G_1^{\perp}$, we see that the monomials $x_1, x_1 x_2, x_1 x_3$, and $x_1 x_4$ are linearly independent vectors contained in the support of $x_1$.
If we project these vectors onto just the support of $x_1$, i.e., drop the $x_1$ in the description of the monomials, then these $4$ vectors form the monomials $1, \tilde{x}_1, \tilde{x}_2, \tilde{x}_3$ in the space of $3$ binary variables.
By definition, these generate the Reed-Muller code RM($1,3$), which is also the $[8,4,4]$ extended Hamming code that is self-dual.
Since all other codewords in $C_2$ are also degree-$1$ polynomials, the same argument as above can be applied to them.
Therefore, the second condition of Theorem~\ref{thm:transversal_T} is satisfied as well.
Finally, since all generating codewords of $C_1^{\perp}$ have Hamming weight $4$ or $8$, the last condition of Theorem~\ref{thm:transversal_T} produces no negative signs for the $Z$-stabilizers.
Hence, the $\llbr 15,1,3 \rrbr$ quantum Reed-Muller code supports the transversal $T$ gate, and it can be checked that this realizes the logical $T^{\dagger}$ gate on the encoded qubit.

We can also construct CSS codes where the physical transversal $T$ realizes logical transversal $T$.
In fact, \emph{triorthogonal codes} introduced by Bravyi and Haah~\cite{Bravyi-pra12} serve exactly this purpose, although they allow for an additional Clifford correction beyond the strict (physical) transversal $T$ operation.
As our next result, using our methods we show a ``converse'' that triorthogonality is not only sufficient but also necessary if we desire to realize logical transversal $T$ via (strict) physical transversal $T$ (using a CSS-T code).
We first repeat the definition of a triorthogonal matrix for clarity.

\begin{definition}[Triorthogonality~\cite{Bravyi-pra12}]
\label{def:triorthogonality}
A $p \times q$ binary matrix $G$ is said to be triorthogonal if and only if the support of any pair and triple of its rows has even overlap, i.e., $w_H(G_a \ast G_b) \equiv 0$ (mod $2$) for any two rows $G_a$ and $G_b$ for $1 \leq a < b \leq p$, and $w_H(G_a \ast G_b \ast G_c) \equiv 0$ (mod $2$) for all triples of rows $G_a, G_b, G_c$ for $1 \leq a < b < c \leq p$.
\end{definition}

Before we state our next result, let us clarify some unfortunate ambiguity in the terminology of triorthogonal codes.
First, a \emph{binary triorthogonal code} is one that has a generator matrix that satisfies the above triorthogonality property.
Second, a \emph{quantum CSS triorthogonal code} as defined by Bravyi and Haah starts with a binary triorthogonal code and adds constraints to some of its rows that are used to describe generators for the logical $X$ operators of the resulting CSS code.
This family of codes satisfy the property that when transversal $T$ is applied to the physical qubits of the triorthogonal code, along with possibly an additional Clifford operation (correction), then it induces a transversal $T$ on the logical qubits.
But, in the literature, sometimes the terminology is more casual where triorthogonal codes are described as codes that realize logical transversal $T$ via physical transversal $T$.
Finally, we note that the notion of triorthogonality is closely related to \emph{triply-even} codes, but they are distinct objects~\cite{Haah-pra18}.

\begin{theorem}[Logical Transversal $T$]
\label{thm:logical_trans_T}
Let $S$ be the stabilizer for an $\llbr n,k,d \rrbr$ CSS-T code CSS($X, C_2 ; Z, C_1^{\perp}$).
Let $G_1 = 
\begin{bmatrix}
G_{C_1/C_2} \\
G_2
\end{bmatrix}$ be a generator matrix for the classical code $C_1 \supset C_2$ such that the rows $x_i, i = 1,\ldots,k$, of $G_{C_1/C_2}$ form a generating set for the coset space $C_1/C_2$ that produces the logical $X$ group of the CSS-T code, i.e., $\bar{X} = \langle E(x_i,0) ; i = 1,\ldots,k \rangle$.
Then the physical transversal $T$ gate realizes the logical transversal $T$ gate, without any Clifford correction as in~\cite{Bravyi-pra12}, if and only if the matrix $G_1$ is triorthogonal and the following condition holds true:
\begin{align}
x = \bigoplus_{i=1}^{k} c_i x_i,\ c_i \in \{0,1\} \Rightarrow w_H(x \oplus a) \equiv w_H(c) \ (\bmod\ 8)\ \text{for\ all}\ a \in C_2.
\end{align}
\begin{proof}
See Section~\ref{sec:proof_logical_trans_T}.
\end{proof}
\end{theorem}

\begin{corollary}
\label{cor:triortho_general}
The family of triorthogonal codes introduced by Bravyi and Haah~\cite{Bravyi-pra12} is the most general CSS-T family that realizes logical transversal $T$ from physical transversal $T$.
\begin{proof}
The Bravyi-Haah construction allows for a Clifford correction after the transversal $T$ gate in order to exactly realize logical transversal $T$.
In order to prove the equivalence of their construction to Theorem~\ref{thm:logical_trans_T}, we need to show that, by requiring their Clifford correction to be trivial, we arrive at the same conditions as listed above.
Since triorthogonality of $G_1$ is a common constraint in both Theorem~\ref{thm:logical_trans_T} and the Bravyi-Haah construction, we are left to verify that the Hamming weight condition above coincides with the condition for their Clifford correction to be trivial.

Let $C_2$ be an $[n,k_2]$ code so that the number of rows in $G_2$ is $k_2$.
Let $y = \bigoplus_{i=1}^{k + k_2} d_i y_i$ with $d_i = c_i, y_i = x_i$ for $i = 1,\ldots,k$ and $y_i = a_{i-k}$ for $i > k$, where $a_i$ are the rows of $G_2$.
The Clifford correction depends on the phase $\imath^{Q(d)}$ and is trivial when
\begin{align}
Q(d) \coloneqq \sum_{i=1}^{k + k_2} \Gamma_i d_i - 2 \sum_{i < j} \Gamma_{ij} d_i d_j \equiv 0\ (\text{mod}\ 4),
\end{align}
where $w_H(y_i) = 
\begin{cases}
2 \Gamma_i + 1 & \text{for}\ 1 \leq i \leq k, \\
2 \Gamma_i & \text{for}\ i > k,
\end{cases}$
and $y_i y_j^T = 2 \Gamma_{ij}$.
This is because, by their construction $w_H(x_i)$ is odd.
Substituting for $\Gamma_i$ and $2 \Gamma_{ij}$, we get
\begin{align}
\sum_{i=1}^{k} c_i \frac{(w_H(x_i) - 1)}{2} + \sum_{j=1}^{k_2} d_{k+j} \frac{w_H(a_j)}{2} - \sum_{\substack{i < j\\i,j = 1}}^{k+k_2} d_i d_j (y_i y_j^T) & \equiv 0\ (\text{mod}\ 4) \\
\Leftrightarrow \sum_{i=1}^{k} c_i w_H(x_i) - w_H(c) + \sum_{j = 1}^{k_2} d_{k+j} w_H(a_j) - 2 \sum_{\substack{i < j\\i,j = 1}}^{k+k_2} d_i d_j (y_i y_j^T) & \equiv 0\ (\text{mod}\ 8) \\
%
%
\Leftrightarrow w_H(x \oplus a) & \equiv w_H(c)\ (\text{mod}\ 8).
\end{align}
For the last step, using the fact that $w_H(x \oplus a) = w_H(x) + w_H(a) - 2 xa^T$ recursively for $x = \bigoplus_{i=1}^{k} c_i x_i, a = \bigoplus_{j=1}^{k_2} d_{k+j} a_j$, it can be verified that 
\begin{align}
w_H(x \oplus a) & = w_H\left( \bigoplus_{i=1}^{k} c_i x_i \oplus \bigoplus_{j=1}^{k_2} d_{k+j} a_j \right) \\
  & = w_H\left( \bigoplus_{i=1}^{k} c_i x_i \right) + w_H\left( \bigoplus_{j=1}^{k_2} d_{k+j} a_j \right) - 2 \left( \bigoplus_{i=1}^{k} c_i x_i \right) \left( \bigoplus_{j=1}^{k_2} d_{k+j} a_j \right)^T \\
  & = \left[ c_1 w_H(x_1) + w_H\left( \bigoplus_{i=2}^{k} c_i x_i \right) - 2 c_1 x_1 \left( \bigoplus_{i=2}^{k} c_i x_i \right)^T \right] \nonumber \\
  & \qquad + \left[ d_{k+1} w_H(a_1) + w_H\left( \bigoplus_{j=2}^{k_2} d_{k+j} a_j \right) - 2 d_{k+1} a_1 \left( \bigoplus_{j=2}^{k_2} d_{k+j} a_j \right)^T \right] \nonumber \\
  & \qquad - 2 \left( \bigoplus_{i=1}^{k} d_i y_i \right) \left( \bigoplus_{j=k+1}^{k+k_2} d_j y_j \right)^T \\
  & = \left[ c_1 w_H(x_1) + w_H\left( \bigoplus_{i=2}^{k} c_i x_i \right) + d_{k+1} w_H(a_1) + w_H\left( \bigoplus_{j=2}^{k_2} d_{k+j} a_j \right) \right] \\
  & \qquad - 2 \left[ d_1 y_1 \left( \bigoplus_{i=2}^{k} d_i y_i \right)^T + d_{k+1} y_{k+1} \left( \bigoplus_{j=k+2}^{k+k_2} d_j y_j \right)^T \right. \nonumber \\
  & \hspace{5.5cm} \left. + \left( \bigoplus_{i=1}^{k} d_i y_i \right) \left( \bigoplus_{j=k+1}^{k+k_2} d_j y_j \right)^T \right] \\
  & \ \ {\vdots} \ (\text{continue\ recursion\ for}\ i=2,\ldots,k-1\ \text{and}\ j=k+2,\ldots,k+k_2-1) \nonumber \\
  & = \sum_{i=1}^{k} c_i w_H(x_i) + \sum_{j = 1}^{k_2} d_{k+j} w_H(a_j) - 2 \sum_{\substack{i < j\\i,j = 1}}^{k+k_2} d_i d_j (y_i y_j^T).
\end{align}
This completes the proof. 
\end{proof}
\end{corollary}

While the Hamming weight condition above can be hard to check in practice, using the Bravyi-Haah recipe still implies that one has to calculate a final Clifford correction.
We suspect that CSS-T codes constructed using classical monomial codes, such as Reed-Muller codes or more general decreasing monomial codes~\cite{Bardet-isit16,Krishna-arxiv18}, might possess simple ways to check the Hamming weight condition above, since the weight distribution of some of these codes are known.

Observe that triorthogonality is a common condition for realizing either logical transversal $T$ or logical identity from physical transversal $T$, since the last three conditions in Theorem~\ref{thm:logical_identity} constitute the property of triorthogonality.
Indeed, if $\imath^{xy^T} E(0,x \ast y) \in S$ for any $x,y \in C_1/C_2$, then we need $xy^T \equiv 0$ (mod $2$) for $x \neq y$ and $w_H(a \ast (x \ast y)) \equiv 0, w_H(z \ast (x \ast y)) \equiv 0$ (mod $2$) for any $a \in C_2, z \in C_1/C_2, z \notin \{ x,y \}$, all because $E(0,x \ast y)$ needs to commute with $X$-type stabilizers and logical $X$ operators.
Similarly, the other conditions of triorthogonality can be derived from Theorem~\ref{thm:logical_identity}.

Therefore, the essential difference between transversal $T$ realizing logical transversal $T$ or logical identity is the following: for the former we need the Hamming weight condition above which in part implies $w_H(x_i) \equiv 1$ (mod $8$), while for the latter we need $\imath^{w_H(x)} E(0,x) \in S$ which implies $w_H(x) \equiv 0$ (mod $2$), and these are mutually contradictory.
Note that even if we permit a Clifford correction and omit the Hamming weight condition above, the proof of Theorem~\ref{thm:logical_trans_T} implies that the constraint $w_H(x_i) \equiv 1$ (mod $8$) is still necessary, so the contradiction remains.
Even in the Bravyi-Haah recipe, they impose that $w_H(x_i) \equiv 1$ (mod $2$).
We will construct a Reed-Muller family of CSS-T codes shortly, where we explicitly state a condition that differentiates between when the physical transversal $T$ realizes the logical identity and when it realizes some non-trivial logical operator.



\subsection{Realizing Logical CCZ via Transversal $T$}
\label{sec:logical_ccz}

The controlled-controlled-$Z$ (CCZ) gate defined in Chapter~\ref{ch:ch2_background} is a $3$-qubit gate that applies the Pauli $Z$ operator on the third qubit if and only if the first two qubits are in state $\dket{1}$.
Similar to the CZ gate, this unitary is symmetric with respect to all the three qubits.

\begin{example}[$\llbr 8,3,2 \rrbr$ Color Code]
\label{eg:Campbell}
\normalfont
First we revisit the construction of the $\llbr 8,3,2 \rrbr$ color code of Campbell~\cite{Campbell-blog16}, since it is now well-known, and show how it satisfies Theorem~\ref{thm:transversal_T}.
The code can be defined by considering the $8$ physical qubits to be the vertices of a cube.
There is a single $X$-type stabilizer generator that is defined by $X$ on all the vertices.
There are $4$ independent $Z$-type generators that are defined by $Z$ on the vertices of ($4$ independent) faces of the cube.
So the $X$-type stabilizers come from the $[8,1,8]$ classical repetition code, which can be written as the Reed-Muller code RM($0,3$).
It is easy to verify that the $Z$-type stabilizers come from the $[8,4,4]$ extended Hamming code, which is also the self-dual Reed-Muller code RM($1,3$).
By appropriately defining the logical $X$ strings from faces of the cube, it can be shown that transversal $T$ realizes logical CCZ on this code.
This code is also a special case of Theorem~\ref{thm:QRM_family} for $m = 3, r = 1$, which generalizes to any $m$ (and $r = 1$) by the conditions of the theorem.
Thus, this is a family of $\llbr 2^m, m, 2 \rrbr$ CSS($X, C_2 ; Z, C_1^{\perp}$) codes defined on $m$-dimensional cubes as $C_2 = \text{RM}(0,m), C_1 = \text{RM}(1,m)$, similar to the 3D code above.
\end{example}

\begin{example}[$\llbr 16,3,2 \rrbr$ Bacon-Shor-like Code]
\normalfont
Now we construct a $\llbr 16,3,2 \rrbr$ Bacon-Shor-like code using the conditions of Theorem~\ref{thm:transversal_T} and show that the transversal $T$ realizes the logical CCZ gate (up to Paulis).  
In particular, this code belongs to the compass code family studied in \cite{Li-prx19}.
Although the $\llbr 8,3,2 \rrbr$ code is smaller while having essentially the same properties, we will demonstrate shortly that the $\llbr 16,3,2 \rrbr$ code can be constructed from \emph{decreasing monomial codes}~\cite{Bardet-isit16,Bardet-arxiv16}.
While this framework has been recently used by Krishna and Tillich~\cite{Krishna-arxiv18} to construct triorthogonal codes from punctured polar codes, this example has non-identical logical $X$ and $Z$ generators unlike the standard presentation of triorthogonal codes~\cite{Bravyi-pra12}.

\begin{figure}

\centering

\includegraphics[scale=0.135,keepaspectratio]{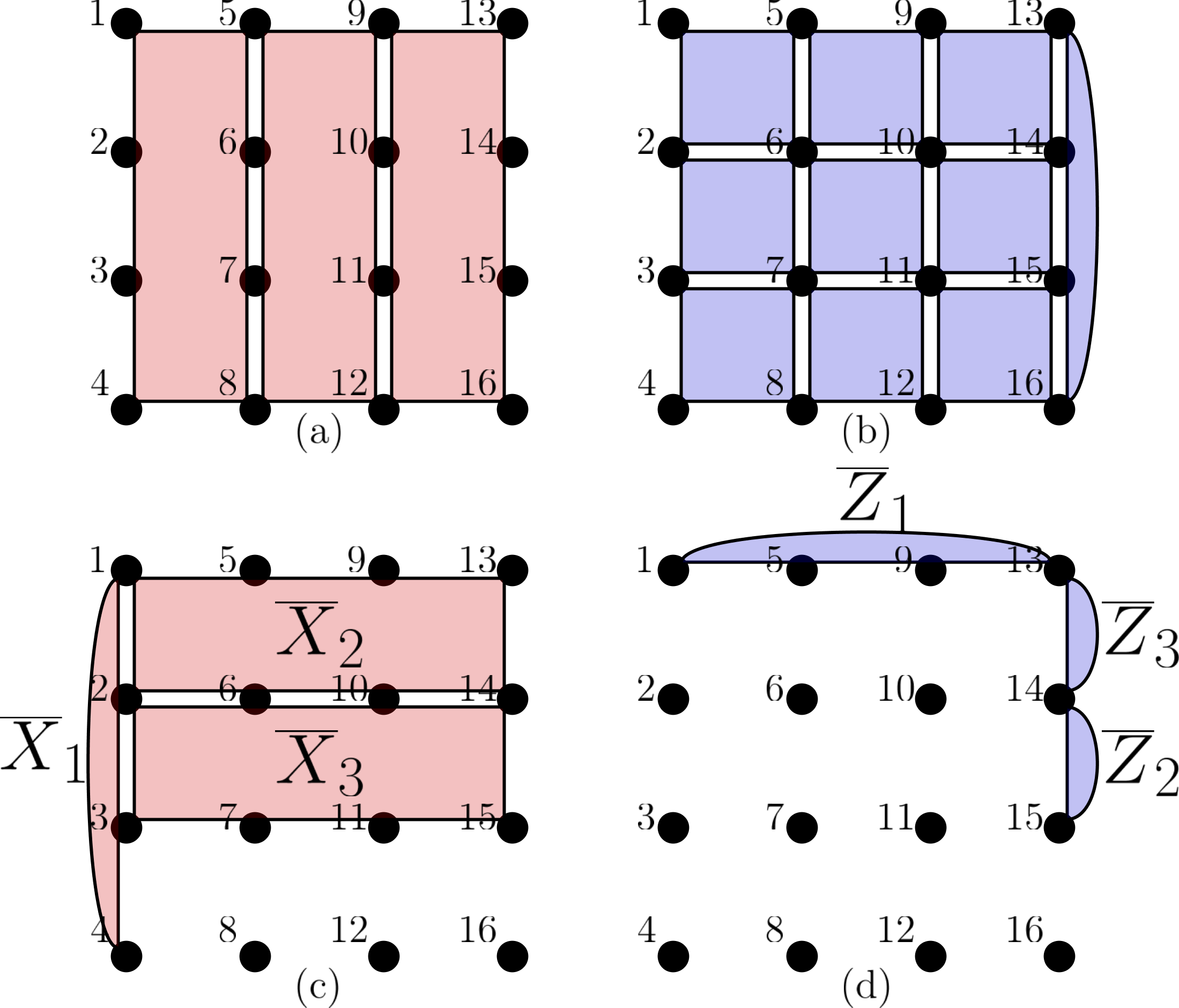}

\caption[A $\llbr 16,3,2 \rrbr$ CSS-T code where transversal $T$ realizes logical CCZ (up to logical Paulis).]{\label{fig:CSST_16_3_2}A $\llbr 16,3,2 \rrbr$ CSS-T code where transversal $T$ realizes logical CCZ (up to logical Paulis). (a) The $3$ weight-$8$ $X$-type stabilizer generators. (b) The $10$ weight-$4$ $Z$-type stabilizer generators. (c) The $3$ $X$-type logical Pauli generators. (d) The corresponding $3$ $Z$-type logical Pauli generators.}

\end{figure}

The construction of the $\llbr 16,3,2 \rrbr$ CSS-T code CSS($X, C_2 ; Z, C_1^{\perp}$) is shown in Fig.~\ref{fig:CSST_16_3_2}.
The code $C_2$ is generated by three weight-$8$ vectors that are represented as the vertical rectangles in Fig.~\ref{fig:CSST_16_3_2}(a).
It is easy to check that all but one non-zero vector in $C_2$ are weight-$8$ and there is one vector which is all $1$s.
The code $C_1^{\perp}$ is generated by $10$ weight-$4$ vectors, $9$ of which are represented as plaquette operators and the last one corresponds to the vertical string $Z_5 Z_6 Z_7 Z_8$ (or $Z_{13} Z_{14} Z_{15} Z_{16}$ in Fig.~\ref{fig:CSST_16_3_2}(b)).
Consider the first $X$-type generator $X_1 X_2 \cdots X_8$.
In its support, there are $3$ plaquette weight-$4$ strings and $1$ vertical weight-$4$ string, all of which are linearly independent and have mutually even overlap,
Hence, these clearly form a self-dual code which is in fact the $[8,4,4]$ extended Hamming code.
This can be checked for all the other vectors in $C_2$, so the first two conditions of Theorem~\ref{thm:transversal_T} are satisfied.
The last condition imposes no negative signs to the $Z$-type stabilizers since all of them have weight that is a multiple of $4$.
Therefore, transversal $T$ preserves the code subspace.

To see that the realized logical operator is CCZ, consider the action of $T^{\otimes 16}$ on $\bar{X}_1 = X_1 X_2 X_3 X_4$.
Recollect that CCZ on qubits $a,b,c$ maps $X_a \mapsto X_a\, \text{CZ}_{bc}, X_b \mapsto X_b\, \text{CZ}_{ac}, X_c \mapsto X_c\, \text{CZ}_{ab}$, and CZ on qubits $e,f$ maps $X_e \mapsto X_e Z_f, X_f \mapsto X_f Z_e$.
Then,
\begin{align}
T^{\otimes 16} \bar{X}_1 \left( T^{\otimes 16} \right)^{\dagger} = e^{-\frac{\imath\pi}{4} \cdot 4} (Y_1 Y_2 Y_3 Y_4) (P_1 P_2 P_3 P_4) = - \bar{X}_1 (P_1^{\dagger} P_2^{\dagger} P_3^{\dagger} P_4^{\dagger}).
\end{align}
We need to show that $U \coloneqq P_1^{\dagger} P_2^{\dagger} P_3^{\dagger} P_4^{\dagger} \equiv \bar{\text{CZ}}_{23}$.
Recollecting that $P^{\dagger} X P = -Y$, we notice 
\begin{align}
U \bar{X}_2 U^{\dagger} & = Y_1 Y_2 X_5 X_6 X_9 X_{10} X_{13} X_{14} = - \bar{X}_2 (Z_1 Z_2) \equiv - \bar{X}_2 \bar{Z}_3, \\
U \bar{X}_3 U^{\dagger} & = Y_2 Y_3 X_6 X_7 X_{10} X_{11} X_{14} X_{15} = - \bar{X}_3 (Z_2 Z_3) \equiv - \bar{X}_3 \bar{Z}_2,
\end{align}
since $Z_1 Z_2 Z_{13} Z_{14}, Z_2 Z_3 Z_{14} Z_{15} \in S$.
Thus, up to signs, we have verified that $P_1^{\dagger} P_2^{\dagger} P_3^{\dagger} P_4^{\dagger} \equiv \bar{\text{CZ}}_{23}$.
Hence, $T^{\otimes 16}$ acts like logical CCZ on $\bar{X}_1$, and similar calculations can be done to verify the other relations for CCZ.
In this case, the signs can be fixed by checking the relations for $(\bar{X}_2 \bar{X}_3)\, \bar{\text{CCZ}}\, (\bar{X}_2 \bar{X}_3)$, since this is the logical operator realized by $T^{\otimes 16}$.

\end{example}

\begin{example}[$\llbr 16,3,2 \rrbr$ Decreasing Monomial Code]
\label{eg:decreasing_monomial}
\normalfont
An equivalent $\llbr 16,3,2 \rrbr$ code can be constructed as a decreasing monomial code as follows, using the monomial description of Reed-Muller codes we discussed in Section~\ref{sec:rm_codes}.
Define the code $C_2$ as the space generated by the monomials $G_2 = \{1, x_1, x_2\}$, and the code $C_1$ as the space generated by $G_1 = G_2 \cup \{ x_3, x_4, x_1 x_2 \}$.
Hence, the logical $X$ group is generated by $G_X = \{ x_3, x_4, x_1 x_2 \}$.
While Reed-Muller codes always include all monomials up to some degree as the generators, decreasing monomial codes with maximum degree $r$ might include only some of the degree $r$ monomials among the generators.
However, the code must include all monomials of degree up to $r-1$, and the degree $r$ terms must be chosen according to a partial order as described in~\cite{Bardet-arxiv16}.
In the construction above, both $C_2$ and $C_1$ are decreasing monomial codes.
Using the formalism in~\cite{Bardet-arxiv16}, it is easy to see that the dual codes $C_1^{\perp}$ and $C_2^{\perp}$ are generated respectively by $G_1^{\perp} = \{ 1,x_1,x_2,x_3,x_4,x_1 x_2,x_1 x_3,x_1 x_4,x_2 x_3,x_2 x_4 \}$ and $G_2^{\perp} = G_1^{\perp} \cup \{ x_3 x_4,x_1 x_2 x_3,x_1 x_2 x_4 \}$.
So the logical $Z$ group is generated by $G_Z = \{ x_1 x_2 x_4, x_1 x_2 x_3, x_3 x_4 \}$, where we have rewritten the generators in an order such that they form corresponding pairs with logical $X$ generators in $G_X$.
In other words, we see that the corresponding entries in $G_X$ and $G_Z$ multiply to the full monomial $x_1 x_2 x_3 x_4$, which is the only monomial whose evaluation has odd weight, and hence the pairs anti-commute as required.
Similarly, multiplying terms from $G_X$ and $G_Z$ that are not pairs does not yield the full monomial, thereby ensuring they have even overlap.
Finally, we see that the product of the three logical $X$ generators produces the full monomial, which means their triple product has odd weight.
This is precisely one of the requirements in the \emph{generalized triorthogonality} conditions established by Haah and Hastings~\cite{Haah-quantum17b}, which is a special case of \emph{quasitransversality} established earlier by Campbell and Howard~\cite{Campbell-pra17}, in order to ensure that transversal $T$ performs a logical CCZ on a CSS code.
The other requirements in their conditions can also be quickly verified simply using the fact that the only monomial of odd weight is the full monomial.

To see that this also satisfies Theorem~\ref{thm:transversal_T}, consider for example the $X$-stabilizer corresponding to the monomial $x_1 \in G_2$. 
We observe that the elements $x_1, x_1 x_2, x_1 x_3, x_1 x_4 \in G_1^{\perp}$ are supported on $x_1$.
When we project down to $x_1$, i.e., consider only the support of $x_1$, we get the monomials $1, \tilde{x}_1 = x_2, \tilde{x}_2 = x_3, \tilde{x}_3 = x_4$ that precisely generate the code RM($1,3$) that is self-dual.
A similar analysis can be made for other elements in $C_2$.
Moreover, since the elements in $G_1^{\perp}$ have weights $4,8$, or $16$, the last condition of Theorem~\ref{thm:transversal_T} does not introduce any negative signs for the $Z$-stabilizers.
Therefore, we have used the decreasing monomial codes formalism to produce an equivalent $\llbr 16,3,2 \rrbr$ code, where only the logical $X$ and $Z$ generators have changed in comparison to the construction above.
We believe this is not just one special case but points to a general construction of CSS codes using this formalism that support transversal $Z$-rotations.
\end{example}



\subsection{Realizing Products of $\mathcal{C}^{(3)}$ gates with Transversal $T$}
\label{sec:product_of_CCZs}

We demonstrate two examples where transversal $T$ realizes a logical diagonal gate at the $3$rd level that is a product of elementary gates.
These codes have been partially discussed in recent works~\cite{Haah-quantum17b,Campbell-pra17} but we describe the general family and later derive the exact logical operation realized by transversal $T$ on these codes.

\begin{example}[$\llbr 64,15,4 \rrbr$ Reed-Muller Code]
\normalfont
Consider a CSS($X, C_2 ; Z, C_1^{\perp}$) code where $C_2 = \text{RM}(1,6) \subset C_1 = \text{RM}(2,6)$ and therefore $C_1^{\perp} = \text{RM}(3,6) \subset C_2^{\perp} = \text{RM}(4,6)$.
The distance of this code is the minimum of the minimum distances of $C_1$ and $C_2^{\perp}$.
It is well-known that the minimum distance of RM($r,m$) is $2^{m-r}$, so the distance of this CSS-T code is $4$.
Therefore, this gives a $\llbr 64,15,4 \rrbr$ code.
Let us quickly check the conditions in Theorem~\ref{thm:transversal_T}.
The code $C_2$ is generated by degree-$1$ monomials in $6$ binary variables $x_1,x_2,\ldots,x_6$, and all of its codewords have even weight.
Using the same strategy that we used for the $\llbr 15,1,3 \rrbr$ RM code, consider the monomial $x_1$ in $C_2$.
Since $C_1^{\perp}$ is generated by all monomials of degree less than or equal to $3$, it contains the monomials $x_1 (1), x_1 (x_i), x_1 (x_i x_j)$ for $i,j \in \{ 2,3,4,5,6 \}$ and $i < j$.
If we project down to $x_1$, then the monomials $1, \tilde{x}_i \tilde{x}_i \tilde{x}_j$ for $i,j \in \{1,2,3,4,5\}$ and $i < j$ exactly generate the code RM($2,5$) which is self-dual.
A similar analysis holds for all the other codewords in $C_2$ which are degree-$1$ polynomials as well.
Finally, the generators of $C_1^{\perp}$ have Hamming weights $8,16,32$ and $64$, so there are no negative signs introduced by the last condition of Theorem~\ref{thm:transversal_T}. 

In order to determine the logical operation realized by transversal $T$, we initially wrote a computer program to generate the vectors in the superposition of all the $2^{15}$ logical computational basis states. 
(Later, in Theorem~\ref{thm:QRM_family}, we derive the exact logical operation analytically.)
Then we calculated the action of $T^{\otimes 64}$ on them by just computing the Hamming weights of the vectors in the superposition.
The effective logical operation was a diagonal unitary with entries $\pm 1$, and there were $13,888$ entries that were $(-1)$ in the diagonal (out of the $2^{15} = 32,768$).
We determined the Boolean function encoding the locations of $1$ and $-1$ in the diagonal, and simplified the naive $13,888$ term sum-of-products (SOP) expression into a $1991$ term SOP expression using the software ``Logic Friday''.
Subsequently, we used ``Mathematica'' to convert this Boolean function into its \emph{algebraic normal form (ANF)} and obtained the following polynomial.
\begin{align}
\label{eq:QRM26_poly}
q(v_1,\ldots,v_{15}) & = v_1 v_{10} v_{15} + v_1 v_{11} v_{14} + v_1 v_{12} v_{13}  \nonumber \\
                     & \quad + v_2 v_7 v_{15} + v_2 v_8 v_{14} + v_2 v_9 v_{13} \nonumber \\
                     & \quad + v_3 v_6 v_{15} + v_3 v_8 v_{12} + v_3 v_9 v_{11}  \nonumber \\ 
                     & \quad + v_4 v_6 v_{14} + v_4 v_7 v_{12} + v_4 v_9 v_{10}  \nonumber \\ 
                     & \quad + v_5 v_6 v_{13} + v_5 v_7 v_{11} + v_5 v_8 v_{10}.
\end{align}
Therefore, the logical diagonal gate can be represented as 
\begin{align}
U^L \dket{v_1 \cdots v_{15}}_L = (-1)^{q(v_1,\ldots,v_{15})} \dket{v_1 \cdots v_{15}}_L.
\end{align}
This implies that the gate decomposes into exactly $15$ CCZ gates on the logical qubits, and hence belongs to the $3$rd level of the Clifford hierarchy.
More interestingly, note that the CSS superposition of a given logical computational basis state $\dket{v_1 \cdots v_{15}}_L$ consists of all vectors in the corresponding coset of $C_2$ in $C_1$ generated by $\prod_{i=1}^{15} \bar{X}_i^{v_i}$, i.e., the binary vector representations $x_i$ of the logical operators $\bar{X}_i = E(x_i,0)$.
Therefore, the diagonal of $U^L$ encodes exactly which cosets of RM($1,6$) in RM($2,6$) have all vectors of weight exactly $4$ mod $8$ (diagonal entry $-1$) and which cosets have all vectors of weight $0$ mod $8$ (diagonal entry $1$).
Since the above phase polynomial is a codeword in RM($3,15$) of degree $3$, this suggests a deeper connection to RM codes where the coset weight distribution modulo $8$ is encoded exactly by a codeword in RM($3,15$).
Theorem~\ref{thm:QRM_family} explores this connection more rigorously than the empirical approach described above.
\end{example}

\begin{example}[$\llbr 128,21,4 \rrbr$ Reed-Muller Code]
\normalfont
Similarly, we constructed a $\llbr 128,21,4 \rrbr$ Reed-Muller CSS-T code by setting $C_2 = \text{RM}(1,7) \subset C_1 = \text{RM}(2,7)$, and hence $C_1^{\perp} = \text{RM}(4,7) \subset C_2^{\perp} = \text{RM}(5,7)$.
The $X$-stabilizers are generated by degree-$1$ monomials, and logical $X$ operators are given by the coset representatives for $C_1/C_2$, which are degree-$2$ polynomials.
This implies, for any $a,b \in C_2$ and $x,y \in C_1/C_2$, $a \ast b, a \ast x, x \ast y$ are either degree $2,3$, or $4$ polynomials, all of which belong to $C_1^{\perp}$ by the definition of RM codes.
So $C_1$ is a triorthogonal code that also satisfies the first condition of Theorem~\ref{thm:logical_identity}, and hence transversal $T$ realizes the logical identity.
However, the code supports the application of the $T$ gate on the physical qubits corresponding to the pattern prescribed by any degree-$1$ polynomial, as can be verified from the conditions in Theorem~\ref{thm:transversal_T_Tinv} by setting $t_7 = 0$ and $t_1$ a degree-$1$ polynomial.
Although this example does not fit in Theorem~\ref{thm:QRM_family} that concerns strictly with transversal $Z$-rotations, we verified computationally that the logical gate is non-trivial in this case.
\end{example}

\begin{example}[Reed-Muller Family]
\label{ex:QRM_family}
\normalfont
We can generalize this construction as a Reed-Muller family of $\llbr n = 2^m, k = \binom{m}{r}, d = 2^{r} \rrbr$ CSS-T codes defined by $C_2 = \text{RM}(r-1,m), C_1 = \text{RM}(r,m)$, and hence $C_1^{\perp} = \text{RM}(m-r-1,m) \subset C_2^{\perp} = \text{RM}(m-r,m)$.
Using the conditions in Theorem~\ref{thm:transversal_T} and Theorem~\ref{thm:logical_identity}, we see that we need $r \leq \frac{m}{3}$ for transversal $T$ to be supported, and also $r > \frac{m-1}{3}$ for transversal $T$ to not realize the logical identity.
This appears to imply that there is exactly one integer value of $r$ that provides a valid code, but this need not be true since decreasing monomial codes correspond precisely to non-integer values of $r$.
For example, one can take $m = 9, r = 3$ to obtain a valid $\llbr 512,84,8 \rrbr$ code.
Indeed, notice that $C_1^{\perp} = \text{RM}(5,9)$ contains the code $\text{RM}(4,8)$ in the support of any degree-$1$ polynomial, and $\text{RM}(4,8)$ contains the self-dual code ``$\text{RM}(3.5,8)$'', which is generated by all degree at most $3$ monomials as well as the first half of all degree $4$ monomials when they are arranged in lexicographic order.
Once again, Theorem~\ref{thm:QRM_family} provides the logical gate realized by transversal $T$ on this $\llbr 512,84,8 \rrbr$ code.
\end{example}

The family of quantum Reed-Muller codes in Example~\ref{ex:QRM_family} appears in recent work by Haah and Hastings~\cite{Haah-quantum17b}, and Campbell and Howard~\cite{Campbell-pra17,Campbell-prl17}, where in~\cite{Haah-quantum17b} they focused on distilling CCZ magic states from these codes via physical transversal $T$.
For this reason, they needed the logical CCZs to be on \emph{distinct} triples of (logical) qubits and they provided an analytic construction that guarantees a $\llbr 2^m, 3(2^{m/3}-2), 2^{m/3} \rrbr$ quantum Reed-Muller code satisfying this constraint.
However, using a computational search strategy, they show that in certain cases the number of logical qubits can be increased to produce more disjoint CCZs.
For example, for $m = 9$, they expand the $\llbr 512,18,8 \rrbr$ code from the analytic construction that yields $6$ logical CCZs into a $\llbr 512,30,8 \rrbr$ code that yields $10$ logical CCZs on disjoint triples of logical qubits.

In Theorem~\ref{thm:QRM_family}, we generalize the quantum Reed-Muller family from Example~\ref{ex:QRM_family} for general $\pi/2^{\ell}$ $Z$-rotations, and also prove the exact logical operator realized on these codes.
We believe that, when applied to the transversal $T$ scenario, this result allows one to \emph{analytically} derive the above codes in~\cite{Haah-quantum17b} by using combinatorial arguments to carefully ``peel off'' the additional CCZs that either overlap on qubits involved in existing ones or violate the \emph{generalized triorthogonality} constraints~\cite{Haah-quantum17b}.
This peeling procedure effectively drops all logical qubits that are not involved in the maximum number of disjoint CCZs, $k_{\text{CCZ}}^{\text{max}}$, obtainable on these codes.
Moreover, this approach might lead to an exact characterization of $k_{\text{CCZ}}^{\text{max}}$ for any $m$, without involving the \emph{Lovasz Local Lemma} that appears to provide guarantees only for large $m$.
We leave this investigation for future work.

\section{Transversal Finer Angle $Z$-Rotations}
\label{sec:stab_codes_Z_rotations}

We first generalize Lemma~\ref{lem:conj_by_trans_T} to transversal $\pi/2^{\ell}$ $Z$-rotations, that again could be of independent interest.

\begin{lemma}
\label{lem:conj_by_trans_Z_rot}
Let $E(a,b) \in HW_N, N = 2^n$, for some $a,b \in \mathbb{Z}_2^n$.
Then transversal $\tau_{[ 1 ]}^{(\ell)} = \exp\left( \frac{\imath\pi}{2^{\ell}} Z \right), \ell \geq 2$, acts on $E(a,b)$ as
\begin{align}
\tau_{I_n}^{(\ell)} E(a,b) \left( \tau_{I_n}^{(\ell)} \right)^{\dagger} = \frac{1}{\left( \sec\frac{2\pi}{2^{\ell}} \right)^{w_H(a)}} \sum_{y \preceq a} \left( \tan\frac{2\pi}{2^{\ell}} \right)^{w_H(y)} (-1)^{b y^T} E(a, b \oplus y),
\end{align}
where $w_H(a) = aa^T$ is the Hamming weight of $a$, and $y \preceq a$ denotes that $y$ is contained in the support of $a$.
\begin{proof}
See Section~\ref{sec:proof_conj_by_trans_Z_rot}.
\end{proof}
\end{lemma}

For $\ell = 3$ (transversal $T$), the cosine term produced a $2^{-w_H(a_j)/2}$ factor which we were able to ensure was an integer by enforcing $a_j$ to have even Hamming weight.
Then we produced $2^{w_H(a_j)/2}$ copies of each stabilizer element in order to cancel this factor and thereby reproduced the code projector.
However, for $\ell > 3$, extending this idea requires that $\left( \sec\frac{2\pi}{2^{\ell}} \right)^{w_H(a)}$ cancel the sum of (signed) tangents acquired for each copy of the stabilizer element.
This leads us to an extension of Theorem~\ref{thm:transversal_T}.

\begin{theorem}[Transversal $Z$-rotations]
\label{thm:transversal_Z_rot}
Let $S = \langle \nu_i E(c_i,d_i) ; i = 1,\ldots,r \rangle$ define an $\llbr n,n-r \rrbr$ stabilizer code as in Theorem~\ref{thm:transversal_T}.
Let $Z_S \coloneqq \{ z \in \mathbb{Z}_2^n \colon \epsilon_z E(0,z) \in S \}$ for some $\{ \epsilon_z \}$. 
For any $\epsilon_j E(a_j,b_j) \in S$ with non-zero $a_j$, define the subspace $Z_j \coloneqq \{ v \in Z_S \colon v \preceq a_j \}$ and the set $W_j \coloneqq \{ y \in \mathbb{Z}_2^n \colon y \preceq a_j, y \notin Z_j \}$.
Then the transversal application of the $\exp\left( \frac{\imath\pi}{2^{\ell}} Z \right)$ gate realizes a logical operation on $V(S)$ if and only if the following are true for all such $a_j \neq 0$:
\begin{align}
\sum_{v \in Z_j} \epsilon_v \left( \imath \tan\frac{2\pi}{2^{\ell}} \right)^{w_H(v)} & = \left( \sec\frac{2\pi}{2^{\ell}} \right)^{w_H(a_j)}, \\
\sum_{v \in Z_j} \epsilon_v \left( \imath \tan\frac{2\pi}{2^{\ell}} \right)^{w_H(v \oplus y)} & = 0 \quad \text{for\ all}\ y \in W_j,
\end{align}      
where $\epsilon_v \in \{ \pm 1 \}$ is the sign of $E(0,v)$ in the stabilizer group $S$.
\begin{proof}
See Section~\ref{sec:proof_transversal_Z_rot}.
\end{proof}
\end{theorem}

The extension here is only partial in the sense that the conditions on the stabilizer involve trigonometric quantities and we still have to distill finite geometric constraints on the vectors describing the stabilizer elements, similar to Theorems~\ref{thm:transversal_T} and~\ref{thm:transversal_T_Tinv}.
However, under the assumption that $Z_j$ is a self-dual code and $\epsilon_v = 1$ for all $v \in Z_j$, we are able to deduce the following condition on the Hamming weights of $v$ and $a_j$.

\begin{lemma}
Let $C$ be an $[m,m/2]$ self-dual code and $\ell \geq 2$.
Then $\sum_{v \in C} \left( \imath \tan\frac{2\pi}{2^{\ell}} \right)^{w_H(v)} = \left( \sec\frac{2\pi}{2^{\ell}} \right)^{m}$ if and only if $(m - 2 w_H(v))$ is divisible by $2^{\ell}$ for all $v \in C$.
\begin{proof}
The weight enumerator of the code $C$ is 
\begin{align}
W_C(x,y) = \sum_{i=0}^{m} A_i x^{m-i} y^i = \sum_{v \in C} x^{m - w_H(v)} y^{w_H(v)},
\end{align}
where $A_i$ is the number of codewords in $C$ of Hamming weight $i$.
The MacWilliams identities for a self-dual code are $W_C(x,y) = \frac{1}{|C|} W_C(x+y, x-y)$, where $|C|$ is the number of codewords in $C$.
Then we observe that
\begin{align}
\sum_{v \in C} \left( \imath \tan\frac{2\pi}{2^{\ell}} \right)^{w_H(v)} & = \left( \sec\frac{2\pi}{2^{\ell}} \right)^{m} \\
\Rightarrow \sum_{v \in C} \left( \imath \tan\frac{2\pi}{2^{\ell}} \right)^{w_H(v)} \left( \cos\frac{2\pi}{2^{\ell}} \right)^{m} & = 1 \\
\Rightarrow \sum_{v \in C} \left( \imath \sin\frac{2\pi}{2^{\ell}} \right)^{w_H(v)} \left( \cos\frac{2\pi}{2^{\ell}} \right)^{m - w_H(v)} & = 1 \\
\Rightarrow W_C\left( \cos\frac{2\pi}{2^{\ell}}, \imath\sin\frac{2\pi}{2^{\ell}} \right) & = 1 \\
\Rightarrow \frac{1}{|C|} \sum_{v \in C} \left( \cos\frac{2\pi}{2^{\ell}} + \imath \sin\frac{2\pi}{2^{\ell}} \right)^{m - w_H(v)} \left( \cos\frac{2\pi}{2^{\ell}} - \imath \sin\frac{2\pi}{2^{\ell}} \right)^{w_H(v)} & = 1 \\
\Rightarrow \frac{1}{|C|} \sum_{v \in C} \left( \cos\frac{2\pi}{2^{\ell}} + \imath \sin\frac{2\pi}{2^{\ell}} \right)^{m - 2w_H(v)} & = 1,
\end{align}
where the last step follows from the fact that $\exp\left( \frac{-2\pi\imath}{2^{\ell}} \right) = \cos\frac{2\pi}{2^{\ell}} - \imath \sin\frac{2\pi}{2^{\ell}}$.
We note that $w_H(v)$ is even for all $v \in C$ and observe three cases.
\begin{enumerate}

\item If $w_H(v) = m/2$, then the term $v$ contributes $1$ to the sum.

\item If $w_H(v) < m/2$, then the term $v$ contributes $\left( \cos\frac{2 (m - 2 w_H(v)) \pi}{2^{\ell}} + \imath \sin\frac{2 (m - 2 w_H(v)) \pi}{2^{\ell}} \right)$.

\item If $w_H(v) > m/2$, then the term $v$ contributes $\left( \cos\frac{2 (2 w_H(v) - m) \pi}{2^{\ell}} - \imath \sin\frac{2 (2 w_H(v) - m) \pi}{2^{\ell}} \right)$.

\end{enumerate}
Since the all-$1$s vector is always present in a self-dual code, we pair the terms $v$ and $w = \vecnot{1} \oplus v$ such that $w_H(v) < m/2$ and $w_H(w) = m - w_H(v)$.
Hence, the term $w$ contributes $\left( \cos\frac{2 (m - 2 w_H(v)) \pi}{2^{\ell}} - \imath \sin\frac{2 (m - 2 w_H(v)) \pi}{2^{\ell}} \right)$.
Therefore, we have the condition 
\begin{align}
\frac{1}{|C|} \sum_{v \in C} \cos\frac{2 (m - 2 w_H(v)) \pi}{2^{\ell}} = 1 
\end{align}
that is satisfied if and only if each term in the sum equals $1$. 
Indeed, this happens if and only if $2^{\ell}$ divides $(m - 2 w_H(v))$ for all $v \in C$.
\end{proof}
\end{lemma}

\section{A Quantum Reed-Muller (QRM) Family}

Finally, although we do not have the full extension of Theorem~\ref{thm:transversal_T} yet, we consider a family of quantum Reed-Muller codes QRM$(r,m)$ that supports $\pi/2^{\ell}$ $Z$-rotations from the Clifford hierarchy, and we also explicitly construct the logical operations induced by transversal $Z$-rotations on these codes.
The code QRM$(r,m)$ is a CSS code defined by $C_2 = \text{RM}(r-1,m)$ and $C_1 = \text{RM}(r,m)$.
Hence, we can identify the following relationships:
\begin{align}
\text{$X$-type\ stabilizers}\ \leftrightarrow c \in \text{RM}(r-1,m), \ \text{$Z$-type\ stabilizers}\ \leftrightarrow c \in \text{RM}(m-r-1,m), \nonumber \\
\text{$X$-type\ logical\ operators}\ \leftrightarrow c \in \text{RM}(r,m), \ \text{$Z$-type\ logical\ operators}\ \leftrightarrow c \in \text{RM}(m-r,m).
\end{align}
The parameters for QRM$(r,m)$ are given by $\llbr 2^m, \binom{m}{r}, 2^{\min\{r,m-r\}} \rrbr$.
Recollect that for $v_f \in \mathbb{Z}_2^k$, the CSS basis states are
\begin{align}
\label{eq:css_basis_states}
\dket{v_f}_L \equiv \frac{1}{|C_2|} \sum_{c \in C_2}  \dket{v_f \cdot G_{C_1/C_2} \oplus c} = \frac{1}{|C_2|} \sum_{y \in \mathbb{Z}_2^{k_2}}  \dket{v_f \cdot G_{C_1/C_2} \oplus y \cdot G_2},
\end{align}
where $G_{C_1/C_2}$ denotes the generator matrix for the linear subspace of coset representatives for $C_2$ in $C_1$, and $G_2$ denotes the generator matrix for the code $C_2$.
For QRM$(r,m)$, the rows of $G_{C_1/C_2}$ correspond to degree $r$ monomials, each identifying a logical qubit.
Hence, any polynomial $f$ comprised of these monomials corresponds to a distinct logical computational basis state $\dket{v_f}_L$.
So a non-trivial logical $X$ operator is described by a degree $r$ polynomial $f$, but only the degree $r$ terms will determine which logical qubits are acted upon.
Also, this implies that if a particular degree $r$ term is present in $f$, then the corresponding logical qubit is set to $\dket{1}_L$ in $\dket{v_f}_L$.

\addtocounter{example}{-3}

\begin{example}[contd.]
\normalfont
Before we state the general result, let us setup the notation through the $\llbr 64,15,4 \rrbr$ example from Section~\ref{sec:product_of_CCZs}.
Recollect that in this case we have $m = 6$ and $r = 2$, so the logical qubits can be identified with the degree $2$ monomials that define generators for logical $X$ operators.
Hence, the polynomial in~\eqref{eq:QRM26_poly} defining the logical gate realized by physical transversal $T$ can be represented in monomial subscripts as 
\begin{align} 
q(f) \equiv q(v_f) & = v_{x_1 x_2} v_{x_3 x_4} v_{x_5 x_6} + v_{x_1 x_2} v_{x_3 x_5} v_{x_4 x_6} + v_{x_1 x_2} v_{x_3 x_6} v_{x_4 x_5} \nonumber \\
     & \quad + v_{x_1 x_3} v_{x_2 x_4} v_{x_5 x_6} + v_{x_1 x_3} v_{x_2 x_5} v_{x_4 x_6} + v_{x_1 x_3} v_{x_2 x_6} v_{x_4 x_5}  \nonumber \\
     & \quad + v_{x_1 x_4} v_{x_2 x_3} v_{x_5 x_6} + v_{x_1 x_4} v_{x_2 x_5} v_{x_3 x_6} + v_{x_1 x_4} v_{x_2 x_6} v_{x_3 x_5}  \nonumber \\
     & \quad + v_{x_1 x_5} v_{x_2 x_3} v_{x_4 x_6} + v_{x_1 x_5} v_{x_2 x_4} v_{x_3 x_6} + v_{x_1 x_5} v_{x_2 x_6} v_{x_3 x_4}  \nonumber \\
     & \quad + v_{x_1 x_6} v_{x_2 x_3} v_{x_4 x_5} + v_{x_1 x_6} v_{x_2 x_4} v_{x_3 x_5} + v_{x_1 x_6} v_{x_2 x_5} v_{x_3 x_4},
\end{align}
where each term in the polynomial corresponds to a logical CCZ gate acting on the three logical qubits indexed by the three monomial subscripts, and the sum corresponds to a product of such gates (in the logical unitary space).
In the notation of~\eqref{eq:QRM26_poly}, $v_{x_1 x_2} v_{x_3 x_4} v_{x_5 x_6} \equiv v_1 v_{10} v_{15}$ and so on, which means $v_f = [v_{x_1 x_2}, v_{x_1 x_3}, \ldots, v_{x_5 x_6}]$, i.e., 
\begin{align}
\dket{v_f}_L = \dket{v_{x_1 x_2}}_L \otimes \dket{v_{x_1 x_3}}_L \otimes \cdots \otimes \dket{v_{x_5 x_6}}_L = \dket{v_1}_L \otimes \dket{v_2}_L \otimes \cdots \otimes \dket{v_{15}}_L.    
\end{align}
For this code, the rows of $G_{C_1/C_2}$ are evaluations of the $\binom{m}{m/3} = 15$ degree $r=2$ monomials, namely $x_1 x_2, x_1 x_3, x_1 x_4, \ldots, x_5 x_6$.
So, the polynomial $f \in \text{RM}(r,m)$ above is a linear combination of degree $r=2$ monomials, and possibly lower degree monomials that correspond to just $X$-type stabilizers. 
Hence, $v_f \in \mathbb{Z}_2^{15}$ exactly describes which corresponding rows of $G_{C_1/C_2}$ are chosen in this linear combination.
Therefore, if $f = x_1 x_2 + x_3 x_4 + x_5 x_6 + (\text{smaller\ degree\ terms})$, then $\dket{v_{x_1 x_2} v_{x_3 x_4} v_{x_5 x_6}}_L = \dket{111}_L$ and other qubits are set to $\dket{0}_L$, so $q(f) = 1$.
But if $f = x_1 x_2 + x_3 x_4 + x_5 x_6 + x_3 x_5 + x_4 x_6 + (\text{smaller\ degree\ terms})$, then $q(f) = 0$ as this polynomial corresponds to two CCZs applying the phase $-1$.

For stating the general result, it will be convenient to replace the monomial subscripts with binary vectors $p_1, p_2, p_3 = p_{m/r}$. 
So, for example, for the first term $v_{x_1 x_2} v_{x_3 x_4} v_{x_5 x_6}$ these index vectors are given by $p_1 = [1,1,0,0,0,0], p_2 = [0,0,1,1,0,0], p_3 = [0,0,0,0,1,1]$, which each have Hamming weight $r = 2$ and sum up to $\vecnot{1}$.
\end{example}

Define $\mu(x) \coloneqq (-1)^x$, and let $\nu_p(s)$ denote the largest integer $t$ such that $p^t$ divides $s$.

\begin{theorem} 
\label{thm:QRM_family}
Suppose that $1 \leq r \leq m/2$ and $r$ divides $m$.  
Then, the transversal $\exp\left( \frac{\imath\pi}{2^{m/r}} Z \right)$ gate is a logical operator for $\text{QRM}(r,m)$.  
Moreover, up to local corrections, the corresponding logical operator acts on a computational basis state by
\begin{align}
\label{eq:QRM_family_logical_action}
\dket{v_f}_L \mapsto \mu(q(f))\dket{v_f}_L,
\end{align} 
where $f = \sum_{\vecnot{d} \in \mathbb{Z}_2^m} a_{\vecnot{d}} x^{\vecnot{d}} \in \text{RM}(r,m), a_{\vecnot{d}} \in \{0,1\}$ with $a_{\vecnot{d}} = 0$ for all $w_H(\vecnot{d}) > r$, and 
\begin{align}
\label{eq:QRM_thm_poly}
q(f) \equiv q(v_f) = \sum_{(p_1,\ldots,p_{m/r}) \in P} \prod_{j = 1}^{m/r} v_{p_j} \ (\bmod\ 2),
\end{align}
where $P \coloneqq \{ (p_1,\ldots,p_{m/r}) \colon p_j \in \mathbb{Z}_2^m, \sum_{j=1}^{m/r} p_j = \vecnot{1}, w_H(p_j) = r \}$.
In particular, $deg(q) = m/r$ and so by~\cite{Cui-physreva17}, it is a gate from the $(m/r)$-th level of the Clifford hierarchy.
\end{theorem}

In general, the above theorem states that the logical gate polynomial $q(f)$ consists of all terms such that the monomials in the subscripts of each term form a unique partition of $m$ variables into $m/r$ groups of $r$ variables each.
Therefore, the number of terms in the polynomial $q(f)$, and hence the number of gates in the induced logical operator, is given by $\frac{m!}{(r!)^{m/r} \left( \frac{m}{r} \right)!}$.

\begin{proof} 
From the example above and~\eqref{eq:css_basis_states}, we realize that any $f \in \text{RM}(r,m)$ corresponds to a vector $u = v_f \cdot G_{C_1/C_2} \oplus y \cdot G_2 \in C_1$ in the CSS superposition for $\dket{v_f}_L$.
Define $\zeta_t \coloneqq e^{\frac{2\pi \imath}{t}}$ and note that on any state $\dket{u}$, transversal $\exp\left( \frac{\imath\pi}{2^{m/r}} Z \right)$ maps 
\begin{align}
\label{eq:u_zeta}
\dket{u} \mapsto \zeta_{2^{\frac{m}{r}}}^{w_H(u)} \dket{u}.
\end{align}
By fixing $v_f$ we fix the degree $r$ terms in $f$, and by sweeping over all $y \in \mathbb{Z}_2^{k_2}$ we exhaust all choices of degree at most $(r-1)$ terms in $f$, thereby examining all states $\dket{u}$ in the CSS superposition corresponding to $\dket{v_f}_L$.
For the logical operator to be well-defined as a diagonal gate acting as per~\eqref{eq:QRM_family_logical_action}, we need to show that $w_H(u)$ (mod $2^{m/r}$) depends only on the degree $r$ terms in $f$.
Thus, we are interested in $\nu_2(w_H(u))$ for different $u$ in a single coset of $\text{RM}(r-1,m)$ in $\text{RM}(r,m)$.

First, let us consider $\text{QRM}(1,m)$ separately for simplicity.  
Here, if $\dket{u} = \dket{00\cdots0}$, then for any $w \in u + \text{RM}(0,m)$, $\nu_2(w_H(w)) = m$ and so $\dket{w} \mapsto \dket{w}$.  
However, if $u = \text{ev}(f)$ with $\text{deg}(f) = 1$, then for any $w \in u + \text{RM}(0,m) = \{ u, u \oplus \vecnot{1} \}$, $w$ corresponds to a codimension-$1$ affine plane so that $\nu_2(w_H(w)) = m-1$, and so $\dket{w} \mapsto -\dket{w}$.
Hence, the logical diagonal unitary has diagonal entries $(1,-1,-1,\ldots,-1)$, which is equivalent to $(-1,1,1,\ldots,1)$ up to a global phase of $(-1)$.
Thus, up to local corrections (i.e., a logical transversal $X$ gate correction), transversal application of physical $Z$-rotation $\exp\left( \frac{\imath\pi}{2^{m/r}} Z \right)$ implements a logical $\text{C}^{m-1}Z$ gate.
This captures the $\llbr 8,3,2 \rrbr$ code we discussed previously in Section~\ref{sec:logical_ccz}.

Now consider the more general case where $r \geq 2$.  
We are interested in calculating $w_H(u) = 2^m - N(f)$ (mod $2^{m/r}$), where $N(f)$ denotes the number of zeros of $f$ over $\mathbb{F}_2$. 
Then, following the proofs of Ax's theorem~\cite{Ax-ajm64, McEliece-dm72, Hou-actaa16}, note that
\begin{align} 
N(f) & = \frac{1}{2}\sum_{\vecnot{x} = (x_0,x_1,\ldots,x_m) \in \mathbb{F}_2^{m+1}} \mu(x_0 f(x_1,\ldots,x_m)) \\
     & = \frac{1}{2}\sum_{\vecnot{x} \in \mathbb{F}_2^{m+1}} \mu\left( x_0 \sum_{\vecnot{d} \in \mathbb{F}_2^m} a_{\vecnot{d}} x^{\vecnot{d}} \right) \\ 
     & = \frac{1}{2}\sum_{\vecnot{x} \in \mathbb{F}_2^{m+1}} \prod_{\vecnot{d} \in \mathbb{F}_2^m} \mu\left( x_0 a_{\vecnot{d}} x^{\vecnot{d}} \right) \\
     & = \frac{1}{2}\sum_{\vecnot{x} \in \mathbb{F}_2^{m+1}} \prod_{\vecnot{d} \in \mathbb{F}_2^m} \left( 1 - 2 x_0 a_{\vecnot{d}} x^{\vecnot{d}} \right).
\end{align}
Define the function $t$ on $\mathbb{F}_2$ by $t(0) = 1$, $t(1) = -2$.  
Then distributing the product, we can express $N(f)$ as 
\begin{align}
N(f) = \frac{1}{2}\sum_i \sum_{\vecnot{x} \in \mathbb{F}_2^{m+1}} \prod_{\vecnot{d} \in \mathbb{F}_2^m} \left( t(i(\vecnot{d})) a_{\vecnot{d}}^{i(\vecnot{d})}(x_0x^{\vecnot{d}})^{i(\vecnot{d})} \right),
\end{align}
where the summation over indicators $i$ runs over all Boolean functions $i \colon \mathbb{F}_2^m \rightarrow \mathbb{F}_2$. 
We want to calculate $\nu_2(w_H(u))$, where $u = \text{ev}(f)$, and we observe that $\nu_2(w_H(u)) = \nu_2(2^m - N(f)) = \nu_2(N(f))$.
Hence, we are interested in the $2$-adic valuation of $N(f)$, so we group terms from this sum into products and rewrite this as 
\begin{align}
N(f) = \frac{1}{2}\sum_i \left(\prod_{\vecnot{d} \in \mathbb{F}_2^m} a_{\vecnot{d}}^{i(\vecnot{d})} \right) \left(\prod_{\vecnot{d} \in \mathbb{F}_2^m} t(i(\vecnot{d})) \right) \left(\sum_{\vecnot{x} \in \mathbb{F}_2^{m+1}} \prod_{\vecnot{d} \in \mathbb{F}_2^m} \left( x_0 x^{\vecnot{d}} \right)^{i(\vecnot{d})}\right).
\end{align}
Observe that for each function $i$, the first product is binary, and the remaining two terms are each powers of $2$, so the whole term is a power of $2$ and has a $2$-adic valuation.
The last term is a power of $2$ because it is precisely the Hamming weight of the monomial $\prod_{\vecnot{d} \in \mathbb{F}_2^m} \left( x_0 x^{\vecnot{d}} \right)^{i(\vecnot{d})}$.

Now we are interested in the quantity $\nu_2\left( \left[ \prod_{\vecnot{d} \in \mathbb{F}_2^m} t(i(\vecnot{d})) \right] \left[ \sum_{\vecnot{x} \in \mathbb{F}_2^{m+1}} \prod_{\vecnot{d}} (x_0x^{\vecnot{d}})^{i(\vecnot{d})} \right] \right)$, since the smallest value among all indicating functions $i$ will determine $\nu_2(N(f))$.
Observe that when $i$ is the zero function, this quantity takes the maximal value of $2^{m+1}$, and hence does not affect $\nu_2(N(f))$.
When $i$ is not the zero function, we can calculate
\begin{align}
\nu_2\left(\sum_{\vecnot{x} \in \mathbb{F}_2^{m+1}} \prod_{\vecnot{d} \in \mathbb{F}_2^m} \left( x_0 x^{\vecnot{d}} \right)^{i(\vecnot{d})}\right) & = m - w_H\left( \sum_{\vecnot{d} \in \mathbb{F}_2^m} i(\vecnot{d})\vecnot{d} \right) \text{ and} \\
\nu_2\left(\prod_{\vecnot{d} \in \mathbb{F}_2^m} t(i(\vecnot{d}))\right) & = \sum_{\vecnot{d} \in \mathbb{F}_2^m} i(\vecnot{d}).
\end{align}
So we conclude that 
\begin{align}
\nu_2\left( \left[ \prod_{\vecnot{d} \in \mathbb{F}_2^m} t(i(\vecnot{d})) \right] \left[ \sum_{\vecnot{x} \in \mathbb{F}_2^{m+1}} \prod_{\vecnot{d} \in \mathbb{F}_2^m} \left( x_0 x^{\vecnot{d}} \right)^{i(\vecnot{d})} \right] \right) = m - w_H\left( \sum_{\vecnot{d} \in \mathbb{F}_2^m} i(\vecnot{d}) \vecnot{d} \right) + \sum_{\vecnot{d} \in \mathbb{F}_2^m} i(\vecnot{d}).  
\end{align}
Now, because $f \in \text{RM}(r,m)$, $a_{\vecnot{d}}$ in the first term of each $i$ in $N(f)$ ensures that only terms with $\text{deg}(\vecnot{d}) \leq r$ survive. 
Hence, we have $w_H\left( \sum_{\vecnot{d} \in \mathbb{F}_2^m} i(\vecnot{d}) \vecnot{d} \right) \leq r \sum_{\vecnot{d} \in \mathbb{F}_2^m} i(\vecnot{d})$, with equality occurring only when all $\vecnot{d}$ with $i(\vecnot{d}) = 1$ are disjoint degree $r$ terms, i.e., weight $r$ vectors. 
From this, we can conclude
\begin{align}
& \nu_2\left( \left[ \prod_{\vecnot{d} \in \mathbb{F}_2^m} t(i(\vecnot{d})) \right] \left[ \sum_{\vecnot{x} \in \mathbb{F}_2^{m+1}} \prod_{\vecnot{d} \in \mathbb{F}_2^m} \left( x_0 x^{\vecnot{d}} \right)^{i(\vecnot{d})} \right] \right) \nonumber \\
  & \geq m - w_H\left( \sum_{\vecnot{d} \in \mathbb{F}_2^m} i(\vecnot{d}) \vecnot{d} \right) + \frac{1}{r} w_H\left( \sum_{\vecnot{d} \in \mathbb{F}_2^m} i(\vecnot{d})\vecnot{d} \right) \\
  & \geq \frac{m}{r}.
\end{align}
The second inequality holds because $m - t + \frac{t}{r} - \frac{m}{r} = (m-t) (1 - \frac{1}{r}) \geq 0$, since $t = w_H\left( \sum_{\vecnot{d} \in \mathbb{F}_2^m} i(\vecnot{d}) \vecnot{d} \right) \leq m$ and $r \geq 2$.

Furthermore, because $r|m$, we have equality if and only if $w_H\left( \sum_{\vecnot{d} \in \mathbb{F}_2^m} i(\vecnot{d}) \vecnot{d}\right) = m$ and $\sum_{\vecnot{d} \in \mathbb{F}_2^m} i(\vecnot{d}) = m/r$.  
In other words, the $2$-adic valuation of $N(f)$ is solely determined by those functions $i$ for which exactly $m/r$ \emph{disjoint} terms $\vecnot{d}$, each of weight $r$, have $i(\vecnot{d}) = 1$.
Put together, these conditions exactly define the coefficient products appearing in $q(f)$ in~\eqref{eq:QRM_thm_poly}.  
Let $P'$ denote the set of all such $i$ satisfying these conditions, so that this set has a bijective mapping to the set $P$ defined in the theorem statement.  
Then returning to $N(f)$, and noting that only those terms $i$ which contribute $2^{m/r}$ matter, we see that 
\begin{align}
w_H(u) & = 2^m - N(f) \\
  & \equiv - N(f) \ (\bmod\ 2^{\frac{m}{r}}) \\
  & \equiv 2^{\frac{m}{r}-1} \sum_{i \in P'} \prod_{\vecnot{d} \in \mathbb{F}_2^m} a_{\vecnot{d}}^{i(\vecnot{d})}\ (\bmod\ 2^{\frac{m}{r}}) \\
  & = 2^{\frac{m}{r}-1} \sum_{(p_1,\ldots,p_{m/r}) \in P} \prod_{j = 1}^{m/r} v_{p_j} \\
\label{eq:quasitransversal}
  & = 2^{\frac{m}{r}-1} q(f)\ (\bmod\ 2^{\frac{m}{r}}),
\end{align}
by construction of $P'$ and $P$.  
Here, $u$ determines which $a_{\vecnot{d}} = 1$, or equivalently which $v_{p_j} = 1$ ($u \leftrightarrow v_f$), since $u = \text{ev}(f)$ points to a specific coset of RM($r-1,m$) in RM($r,m$).
As $q(f)$ is oblivious to lower-order terms in $f$ (that correspond to $X$-type stabilizers), each coset indeed has a well-defined weight residue $(\text{mod}\ 2^{\frac{m}{r}})$, and thus the induced logical operation is also well-defined. 
Accordingly, by~\eqref{eq:u_zeta}, the logical action on (logical) computational basis vectors is defined by 
\begin{align}
\dket{v_f}_L \mapsto \zeta_{2^{m/r}}^{2^{m/r - 1} q(f)} \dket{v_f}_L = \mu(q(f)) \dket{v_f}_L.
\end{align}
This completes the proof. 
\end{proof}

\begin{remark}[Quasitransversality]
\normalfont
In~\cite{Campbell-pra17,Campbell-prl17}, Campbell and Howard considered diagonal gates $U_F \in \mathcal{C}^{(3)}$ that can be expressed as $U_F = \sum_{x \in \mathbb{Z}_2^k} \omega^{F(x)} \dketbra{x}$, where $\omega = e^{\imath\pi/4}$ and $F(x) = L(x) + 2Q(x) + 4C(x)$ (mod $8$) is a weighted polynomial with $L(x)$ linear (mod $8$), $Q(x)$ quadratic (mod $4$) and $C(x)$ cubic (mod $2$) polynomials.
So $L$ corresponds to single-qubit $Z$-rotations, $Q$ corresponds to controlled $Z$-rotations, and $C$ corresponds to CCZ gates.
They define a quantum code to be ``$F$-quasitransversal'' if there exists a Clifford $g$ such that $g T^{\otimes n}$ acting on the physical qubits realize the logical gate $U_F$.
In~\cite[Lemma 1]{Campbell-pra17}, they provided the following sufficient condition for a CSS code to be $F$-quasitransversal, which we rewrite in our notation (e.g., see~\eqref{eq:css_basis_states}):
\begin{align}
    w_H(x \cdot G_{C_1/C_2} \oplus y \cdot G_2) \sim_{c} F(x) \ (\bmod\ 8), \ y \in \mathbb{Z}_2^{k_2},
\end{align}
where the subscript ``$c$'' implies that the two sides are Clifford equivalent, i.e., there exists a weighted polynomial $\tilde{F}$ such that by replacing $F(x)$ with $F(x) + 2\tilde{F}(x)$ above, we can replace $\sim_{c}$ with equality.

Now, observe that $u = x \cdot G_{C_1/C_2} \oplus y \cdot G_2 \in C_1$ exactly corresponds to $u = \text{ev}(f)$ for some $f \in \text{RM}(r,m)$ above (with $x = v_f$).
Hence, we note that~\eqref{eq:quasitransversal} exactly matches the (quasi)transversality condition above (with equality and thereby no Clifford correction).
Therefore, QRM($m/3,m$) is $4 q(f)$-(quasi)transversal.
\end{remark}

\begin{remark}[Quantum Pin Codes]
\normalfont
Vuillot and Breuckmann~\cite{Vuillot-arxiv19} recently introduced ``Quantum Pin Codes'' as an abstract framework to synthesize stabilizer codes that support transversal, or partially transversal, physical $Z$-rotations.
These codes are inspired by topological constructions such as color codes~\cite{Kubica-pra15}, but the abstraction extends beyond algebraic topology while retaining transversality properties.
The authors produce several new codes using this formalism.
We note that the above result regarding QRM($r,m$) codes applies to a general family of quantum pin codes as discussed in~\cite[Section V-D]{Vuillot-arxiv19}.
\end{remark}

\section{Summary and Outlook}

In this chapter, we used the recent characterization of quadratic form diagonal (QFD) gates~\cite{Rengaswamy-pra19} to derive necessary and sufficient conditions for a stabilizer code to support a physical transversal $T$ gate.
Our Heisenberg approach allowed us to generalize all such existing constructions.
Using this, we showed that, for any non-degenerate stabilizer code with this property, there exists an equivalent CSS code that also possesses this property.
So for magic state distillation via transversal $T$ on non-degenerate stabilizer codes, CSS codes are essentially optimal.
We also showed that triorthogonal codes form the most general family of CSS codes that realize logical transversal $T$ via physical transversal $T$.
Among several examples, we constructed a $\llbr 16,3,2 \rrbr$ code using the decreasing monomial formalism, and demonstrated how to check that transversal $T$ realizes logical CCZ with the help of generalized triorthogonality conditions.
This points to a possibly general construction of CSS codes supporting transversal $T$ using this formalism.

We then extended the above results beyond $T$ gates, and derived trigonometric stabilizer conditions for the code to support a transversal $\pi/2^{\ell}$ $Z$-rotation.
However, we were only able to reduce this to finite geometric conditions under some assumptions. 
Finally, we considered a family of quantum Reed-Muller codes and determined the exact logical operation induced by transversal $Z$-rotations using Ax's theorem on residue weights of polynomials.
Although these logical operations involve products of overlapping many-controlled-$Z$ gates, it will be interesting to investigate their utility in magic state distillation and other proposals for universal quantum computation.
In certain systems, finer angle rotations often have better fidelity than coarser angle rotations. 
Hence, these native resources in the lab could be leveraged in combination with these codes to potentially obtain better circuit decompositions.

\section{Proofs for All Results}
\label{sec:all_proofs}

In all the proofs below we will use a few observations or identities repeatedly, so we mention them here.
\begin{enumerate}

\item[(O1)] As discussed in Chapter~\ref{sec:hermitian_paulis}, for $a,b,x \in \mathbb{Z}^n$ we have $E(a, b + 2x) = (-1)^{ax^T}$ and $E(a + 2x, b) = (-1)^{bx^T} E(a,b)$.
When multiplying two Pauli matrices we have the identities
\begin{align}
E(a,b) E(c,d) & = (-1)^{\syminn{[a,b]}{[c,d]}} E(c,d) E(a,b) \\
              & = \imath^{bc^T - ad^T \bmod 4} E(a+c, b+d) \\
              & = \imath^{bc^T - ad^T \bmod 4} E(a+c, (b \oplus d) + 2 (b \ast d) ) \\
              & = \imath^{bc^T - ad^T \bmod 4} (-1)^{(a + c) (b \ast d)^T} E( (a \oplus c) + 2 (a \ast c), b \oplus d) \\
              & =  \imath^{bc^T - ad^T \bmod 4} (-1)^{(a \oplus c) (b \ast d)^T + (b \oplus d) (a \ast c)^T} E(a \oplus c, b \oplus d).
\end{align}

\item[(O2)] For any binary subspace $A \subseteq \mathbb{Z}_2^n$, due to symmetry in binary subspaces we have $\sum_{x \in A} (-1)^{xv^T} = |A| \cdot \mathbbm{I}(v \in A^{\perp})$.
In other words, we have $|x \in A \colon xv^T \equiv 0| = |x \in A \colon xv^T \equiv 1|$ if and only if $v \notin A^{\perp}$.

\item[(O3)] For $a \in \mathbb{Z}_2^n$, a set such as $A \coloneqq \{ x \in \mathbb{Z}_2^n \colon x \preceq a \}$ is a subspace of $\mathbb{Z}_2^n$ since $x_1, x_2 \preceq a \Rightarrow (x_1 \oplus x_2) \preceq a$.
Hence, using (O2) we observe that $\sum_{x \in A} (-1)^{xv^T} = |A| \cdot \mathbbm{I}(v \in A^{\perp}) = 2^{w_H(a)} \cdot \mathbbm{I}(v \preceq \bar{a})$, where $\bar{a}$ is the ones' complement of $a$.

\item[(O4)] Let $a,b \in \mathbb{Z}_2^n$ have disjoint supports so that $a \ast b = 0$. Then we observe that
\begin{align}
\sum_{v \in \mathbb{Z}_2^n} \imath^{va^T} (-\imath)^{vb^T} & \overset{(i)}{=} \sum_{w \preceq \overline{(a \oplus b)}} \left( \sum_{v_a \preceq a} \imath^{v_a a^T} \right) \left( \sum_{v_b \preceq b} (-\imath)^{v_b b^T} \right) \\
  & = \sum_{w \preceq \overline{(a \oplus b)}} \left( \sum_{v_a \preceq a} \imath^{w_H(v_a)} \right) \left( \sum_{v_b \preceq b} (-\imath)^{w_H(v_b)} \right) \\
  & \overset{(ii)}{=} 2^{n - w_H(a) - w_H(b)} (1 + \imath)^{w_H(a)} (1 - \imath)^{w_H(b)}.
\end{align}
In step (i) we split any $v \in \mathbb{Z}_2^n$ uniquely as $v = v_a \oplus v_b \oplus w$, where $v_a$ is supported only on $a$, $v_b$ is supported only on $b$, and $w$ is supported outside the support of $a$ and $b$.
In step (ii), for each of the two sums over $v_a$ and $v_b$, we notice that the inner products $v_a a^T$ and $v_b b^T$ take values $q \in \{ 0,1,\ldots,w_H(v_a) \}$ and $q \in \{ 0,1,\ldots,w_H(v_b) \}$ according to the Hamming weights $w_H(v_a)$ and $w_H(v_b)$, respectively.
Hence, there are $\binom{n}{q}$ vectors $v_a \preceq a$ (resp. $v_b \preceq b$) that produce the inner product $q$, and this is captured in the binomial expansion of $(1 + \imath)^{w_H(a)}$ (resp. $(1 - \imath)^{w_H(b)}$).

\end{enumerate}

\subsection{Proof of Theorem~\ref{thm:transversal_T}}
\label{sec:proof_transversal_T}

We will make use of the above observations in the following proof.
We proved in~\cite{Rengaswamy-pra19} that given a tensor product of diagonal unitaries $\tau_{R_1}^{(\ell)} \otimes \tau_{R_2}^{(\ell)} \otimes \cdots \otimes \tau_{R_n}^{(\ell)}$, the result is also of the form $\tau_R^{(\ell)}$ with 
\begin{align}
R = 
\begin{bmatrix}
R_1 & 0 & \cdots & 0 \\
0 & R_2 & \cdots & 0 \\
\vdots & \vdots & \ddots & \vdots \\
0 & 0 & \cdots & R_n 
\end{bmatrix}.
\end{align}
Hence, for an $n$-qubit system, the transversal application of $T$ gate corresponds to $R = I_n$ and $\ell = 3$.
Based on the discussion in Section~\ref{sec:general_QFD}, for the case of transversal $T$, we need
\begin{align}
T^{\otimes n} \Pi_S (T^{\otimes n})^{\dagger} & = \frac{1}{2^r} \sum_{j = 1}^{2^r} \epsilon_j T^{\otimes n} E(a_j,b_j) (T^{\otimes n})^{\dagger} \\
  & = \frac{1}{2^r} \sum_{j = 1}^{2^r} \epsilon_j \cdot \frac{1}{2^{w_H(a_j)/2}} \sum_{y \preceq a_j} (-1)^{b_j y^T} E(a_j, b_j \oplus y) \\
  & = \frac{1}{2^r} \sum_{j = 1}^{2^r} \frac{\epsilon_j}{2^{w_H(a_j)/2}} \left[ E(a_j, b_j) + \sum_{\substack{y \preceq a_j\\y \neq 0}} (-1)^{b_j y^T} E(a_j, b_j \oplus y) \right] \\
\label{eq:transT_equality}
  & = \frac{1}{2^r} \sum_{j = 1}^{2^r} \epsilon_j E(a_j,b_j),
\end{align}
where only for the last step we have assumed that transversal $T$ preserves the code subspace.
Note that whenever $a_j = 0$, the denominator is $1$ and the inner summation is trivial since only $0 \preceq 0$.
Therefore, each such stabilizer $E(0,b_j)$ is retained unchanged (as we would expect since $T^{\otimes n}$ is diagonal and commutes with diagonal Paulis), and we only need to analyze the case $a_j \neq 0$.

For any index $j = h \in \{ 1,2,\ldots,2^r \}$, first we observe that we need $w_H(a_h)$ to be even in order to make the denominator an integer, which can be canceled by producing $2^{w_H(a_h)/2}$ copies of the stabilizer element $E(a_h,b_h)$ (with the appropriate sign $\epsilon_h \in \{ \pm 1 \}$) in the summation over all $2^r$ stabilizer elements.
This clearly shows that the first condition in the theorem is necessary for transversal $T$ to preserve the code space.
(We will call the sum over $j \in \{1,\ldots,2^r\}$ as the ``outer summation'' and the sum over $y \preceq a_j$ as the ``inner summation''; also, for a given $j = h$ we refer to the corresponding $E(a_h,b_h)$ as the ``outer summation term''.)
The only way to produce copies of $E(a_h,b_h)$ is through stabilizers $E(a_h,b_h')$ such that $b_h' \oplus y = b_h$ for some $y \preceq a_h$.
(Note: These two stabilizers correspond to two different outer summation terms for some indices $j$ and $j'$, where $a_j = a_{j'} = a_h$ but $b_j = b_h$ is distinct from $b_{j'} = b_h \oplus y$.)
If two such Paulis $E(a_h, b_h \oplus y), E(a_h, b_h \oplus z)$ must belong to $S$ then we need $E(a_h,b_h \oplus y)$ and $E(a_h,b_h \oplus z)$ to commute, which means we need $a_h y^T \equiv a_h z^T$ (mod $2$).
Since $y = 0$ must be included, we need $a_h z^T = zz^T = w_H(z) \equiv 0$ (mod $2$) for all such choices $z \preceq a_h$.

There are $n_h \coloneqq 2^{w_H(a_h)-1}$ such even weight vectors $z \preceq a_h$ but only a subset of them might correspond to stabilizers in the inner summation.
Hence, let us define
\begin{align}
Z_h \coloneqq \{ z \preceq a_h \colon \epsilon_z E(0,z) \in S\ \text{for\ some}\ \epsilon_z \in \{ \pm 1 \} \}
\end{align}
as in the statement of the theorem.
It is clear that $Z_h$ is a binary subspace of even weight vectors, so let its dimension be $t_h \leq w_H(a_h) - 1$.
As mentioned above, every $z \in Z_h$ provides an outer summation term $E(a_h, b_h \oplus z)$ that produces a copy of $E(a_h,b_h)$ in its corresponding inner summation.
So we need at least $2^{p_h}$ such $z$'s in order to cancel the factor in the denominator, where $p_h \coloneqq w_H(a_h)/2$, i.e., $t_h \geq p_h$.

In order to prove that the second and third conditions in the theorem are necessary for transversal $T$ to preserve the code space, we need to ensure two things based on the equality imposed in~\eqref{eq:transT_equality}:
\begin{itemize}
    \item[(a)] For any $h \in \{ 1,2,\ldots,2^r \}$, over all outer summation indices $j$ there are a net of $2^{w_H(a_h)/2}$ copies of $E(a_h, b_h)$ (with the right sign) so that this stabilizer element is preserved under conjugation of the code projector by transversal $T$.
    
    \item[(b)] For any $h \in \{ 1,2,\ldots,2^r \}$, for each $y \notin Z_h$ and $y \preceq a_h$, over all outer summation indices $j$ there are exactly the same number of positive and negative copies of the non-stabilizer element $E(a_h, b_h \oplus y)$, which cancel out.
\end{itemize}
Let us express these two conditions mathematically by collecting signs of the respective Pauli elements.
For notational clarity, if the outer summation term is $E(a_h, b_h)$ then we write its sign as $\epsilon_{(a_h, b_h)}$ and similarly if the term is $E(a_h, b_h \oplus z)$ then we write its sign as $\epsilon_{(a_h, b_h \oplus z)}$.
For conditions (a) and (b), we respectively require
\begin{align}
\sum_{z \in Z_h} \epsilon_{(a_h, b_h \oplus z)} (-1)^{(b_h \oplus z) z^T} & = \sum_{z \in Z_h} \epsilon_{(a_h, b_h \oplus z)} (-1)^{b_h z^T} = \epsilon_{(a_h, b_h)} 2^{w_H(a_h)/2}, \\
\sum_{z \in Z_h} \epsilon_{(a_h, b_h \oplus z)} (-1)^{(b_h \oplus z) (z \oplus y)^T} & = (-1)^{b_h y^T} \sum_{z \in Z_h} \epsilon_{(a_h, b_h \oplus z)} (-1)^{b_h z^T} (-1)^{zy^T} = 0.
\end{align}
If we ignore the overall sign $(-1)^{b_h y^T}$ in the second equation, then we see that it is impossible to have $(-1)^{z y^T} = 1$ for all $z \in Z_h$, because otherwise we have a contradiction between the two conditions.
Therefore, since $y \notin Z_h$, by the symmetry of binary vectors spaces (observation (O2)), we must have $|\{ z \in Z_h \colon zy^T \equiv 0 \}| = |\{ z \in Z_h \colon zy^T \equiv 1 \}|$.
To proceed further we will find the following lemma about binary subspaces to be useful.

\begin{lemma}
\label{lem:C_self_dual}
For even $n$, let $C$ be an $[n,k \geq \frac{n}{2}]$ binary linear code comprising only of even weight codewords.
Then the following are equivalent:
\begin{enumerate}
    \item $C$ contains its dual $C^{\perp}$.
    \item $C$ contains an $[n,\frac{n}{2}]$ self-dual subcode $C_s$.
    \item Any vector $y \in \mathbb{Z}_2^n \setminus C$ satisfies $|\{ x \in C \colon xy^T \equiv 0 \}| = |\{ x \in C \colon xy^T \equiv 1 \}|$.
\end{enumerate}
\end{lemma}
\begin{IEEEproof}
First we show that 2) implies 3).
Let $G \in \mathbb{Z}_2^{k \times n}$ be a generator matrix for $C$ with rows $g_1,g_2,\ldots,g_k$ such that $g_1,g_2,\ldots,g_{n/2}$ form a submatrix $G_s$ that generates the self-dual subcode $C_s$.
Then any $y \in \mathbb{Z}_2^n \setminus C$ cannot be orthogonal to all rows of $G_s$, since otherwise $y \in C_s$.
Let $yg_1^T = 1$ (mod $2$) without loss of generality.
Then for any other $g_i$ such that $yg_i^T = 1$ (mod $2$), one can replace $g_i$ with $g_i \oplus g_1$ to form a new matrix $G'$ with rows $g_1,g_2',g_3',\ldots,g_k'$ that still spans $C$ (i.e., some $g_i'$ can be equal to $g_i$).
Hence, now $y(g_i')^T = 0$ (mod $2$) for all $i=2,3,\ldots,k$ and $yx^T = 1$ (mod $2$) for all $x$ in the coset $g_1 \oplus \text{span}(g_2',g_3',\ldots,g_k')$.
This shows that $|\{ x \in C \colon xy^T \equiv 0 \}| = |\{ x \in C \colon xy^T \equiv 1 \}|$.


Next we show that 3) implies 1) and 1) implies 2).
Now we start by assuming that every $y \in \mathbb{Z}_2^n \setminus C$ satisfies $|\{ x \in C \colon xy^T \equiv 0 \}| = |\{ x \in C \colon xy^T \equiv 1 \}|$.
This means that all vectors orthogonal to $C$ are contained in $C$, i.e., $C^{\perp} \subseteq C$.
Without loss of generality, assume that we have a generator matrix $G$ for $C$ with rows $g_1,g_2,\ldots,g_k$ such that $g_1,g_2,\ldots,g_{n-k}$ form a submatrix $G^{\perp}$ that generates the dual code $C^{\perp}$.
Then we can enlarge $C^{\perp}$ into a dimension $(n-k+1)$ self-orthogonal code by adding $g_{n-k+1}$ to the rows of $G^{\perp}$, since $g_{n-k+1} \in C$ is dual to $C^{\perp}$ and to itself (as it has even weight).
Now notice that all vectors orthogonal to this new $C^{\perp}$ are still in the rowspace of $G$ since the dual of the span of the first $(n-k)$ rows are contained in the rowspace of $G$.
Hence, it is possible to find a row between $g_{n-k+2}$ and $g_k$ that can be used to enlarge $C^{\perp}$ into a dimension $(n-k+2)$ self-orthogonal code.
Intuitively, as $C^{\perp}$ is enlarged, its dual keeps shrinking starting from $C$ at the beginning.
We can continue this process and, since we assumed $k \geq n/2$, it stops only when $C^{\perp}$ is a dimension $n/2$ self-dual code, as required.
\end{IEEEproof}

Let $\tilde{Z}_h$ denote the subspace $Z_h$ where all the indices outside the support of $a_h$ are punctured, i.e., $\tilde{Z}_h \subseteq \{ 0,1 \}^{w_H(a_h)}$.
Setting $C = \tilde{Z}_h$ in Lemma~\ref{lem:C_self_dual}, we see that $\tilde{Z}_h$ must contain its dual $(\tilde{Z}_h)^{\perp}$ and also a dimension $w_H(a_h)/2$ self-dual code $\tilde{Z}_h^s$.
In other words, $Z_h$ must contain a dimension $w_H(a_h)/2$ self-dual code in the support of $a_h$.
This proves the necessity of the second condition in the theorem.

Let us compute the sign $\epsilon_{(a_h, b_h \oplus z)}$ of $E(a_h, b_h \oplus z)$ assuming that $\epsilon_h E(a_h, b_h), \epsilon_z E(0,z) \in S$.
Using~\eqref{eq:Eab_multiply},~\eqref{eq:Eab_2x} we calculate
\begin{align}
\epsilon_h E(a_h, b_h) \cdot \epsilon_z E(0,z) & = \epsilon_h \epsilon_z \imath^{-a_h z^T} E(a_h, b_h + z) \\
   & = \epsilon_h \epsilon_z \imath^{zz^T} E(a_h, (b_h \oplus z) + 2 (b_h \ast z)) \\
   & = \epsilon_h \epsilon_z \imath^{zz^T} (-1)^{a_h (b_h \ast z)^T} E(a_h, b_h \oplus z) \\
   & = \epsilon_h \epsilon_z \imath^{zz^T} (-1)^{b_h z^T} E(a_h, b_h \oplus z) \\
\Rightarrow \epsilon_{(a_h, b_h \oplus z)} & = \epsilon_{(a_h, b_h)} \epsilon_{(0,z)} \imath^{zz^T} (-1)^{b_h z^T},
\end{align}
where we have used observation (O1) and for the fourth equality we have assumed that $z \preceq a_h$.
Now we start analyzing the sum for condition (a) above by observing that $w_H(z \oplus v) = zz^T + vv^T - 2zv^T$.
\begin{align}
\Gamma & \coloneqq \sum_{z \in Z_h} \epsilon_{(a_h, b_h \oplus z)} (-1)^{b_h z^T} \\
   & = \sum_{z \in Z_h} \epsilon_{(a_h, b_h)} \epsilon_{(0,z)} \imath^{zz^T} (-1)^{b_h z^T} (-1)^{b_h z^T} \\
   & = \epsilon_{(a_h, b_h)} \sum_{z \in \tilde{Z}_h} \epsilon_{(0,z)} \imath^{zz^T} \\
\Rightarrow \Gamma^2 & = \sum_{z,v \in \tilde{Z}_h} \epsilon_{(0,z)} \epsilon_{(0,v)} \imath^{zz^T + vv^T} \\
   & = \sum_{z,v \in \tilde{Z}_h} \epsilon_{(0,z \oplus v)} \imath^{w_H(z \oplus v)} (-1)^{(z \oplus v) v^T} \\
   & = \sum_{u \in \tilde{Z}_h} \epsilon_{(0,u)} \imath^{uu^T} \left( \sum_{v \in \tilde{Z}_h} (-1)^{uv^T} \right)\ (u \coloneqq z \oplus v) \\
   & = |\tilde{Z}_h| \sum_{u \in (\tilde{Z}_h)^{\perp}} \epsilon_{(0,u)} \imath^{uu^T}\ (\text{by\ observation\ (O3)}).
\end{align}
We need this sum over $(\tilde{Z}_h)^{\perp}$ to be $|(\tilde{Z}_h)^{\perp}|$ so that $|\Gamma| = 2^{w_H(a_h)/2}$ as required.
Therefore, it is clear that we need $\epsilon_{(0,u)} \coloneqq \imath^{uu^T}$ for all $u \in (\tilde{Z}_h)^{\perp}$.
Again, note that these vectors $u$ essentially represent $n$-qubit pure $Z$-type stabilizers even though $(\tilde{Z}_h)^{\perp}$ is a space where all indices outside the support of $a_h$ have been punctured.
The subtlety is that $(\tilde{Z}_h)^{\perp} \neq \tilde{(Z_h^{\perp})}$ since the latter computes the dual over $\{ 0,1 \}^n$, which is not what we want here.
Let us also verify that condition (b) above is satisfied. 
\begin{align}
\Delta & \coloneqq \sum_{z \in Z_h} \epsilon_{(a_h, b_h \oplus z)} (-1)^{b_h z^T} (-1)^{zy^T} \\
  & = \sum_{z \in Z_h} \epsilon_{(a_h, b_h)} \epsilon_{(0,z)} \imath^{zz^T} (-1)^{b_h z^T} (-1)^{b_h z^T} (-1)^{zy^T} \\
  & = \epsilon_{(a_h, b_h)} \sum_{z \in \tilde{Z}_h} \epsilon_{(0,z)} \imath^{zz^T} (-1)^{zy^T} \\
  & = \epsilon_{(a_h, b_h)} \sum_{w \in \tilde{Z}_h/(\tilde{Z}_h)^{\perp}} \sum_{u \in (\tilde{Z}_h)^{\perp}} \epsilon_{(0, w \oplus u)} \imath^{w_H(w \oplus u)} (-1)^{(w \oplus u)y^T} \\
  & = \epsilon_{(a_h, b_h)} \sum_{w \in \tilde{Z}_h/(\tilde{Z}_h)^{\perp}} \epsilon_{(0,w)} \imath^{ww^T} (-1)^{wy^T} \left( \sum_{u \in (\tilde{Z}_h)^{\perp}} \epsilon_{(0,u)} \imath^{uu^T} (-1)^{(w \oplus y) u^T} \right) \\
  & = 0,
\end{align}
since $uw^T = 0$ for all $u \in (\tilde{Z}_h)^{\perp}$ and, again, by symmetry of binary vector spaces and because $y \notin \tilde{Z}_h$, we have $|u \in (\tilde{Z}_h)^{\perp} \colon uy^T \equiv 0| = |u \in (\tilde{Z}_h)^{\perp} \colon uy^T \equiv 1|$ (observation (O2)).
Thus, this establishes the necessity of the third condition in the theorem.

Now for the converse we assume that the first two conditions in the theorem hold true and that $\epsilon_{(0,u)} \coloneqq \imath^{uu^T}$ for all $u \in \tilde{Z}_h^{s}$, where $\tilde{Z}_h^{s}$ is the self-dual code present inside $\tilde{Z}_h$.
Once again, we just have to show that conditions (a) and (b) above are satisfied under these assumptions.
For (b), the above calculation for $\Delta$ itself suffices since $(\tilde{Z}_h)^{\perp} \subseteq \tilde{Z}_h^s$, so we are only left to show that $\Gamma = \epsilon_{(a_h, b_h)} 2^{w_H(a_h)/2}$.
We observe that
\begin{align}
\Gamma & = \epsilon_{(a_h, b_h)} \sum_{z \in \tilde{Z}_h} \epsilon_{(0,z)} \imath^{zz^T} \\
  & = \epsilon_{(a_h, b_h)} \sum_{w \in \tilde{Z}_h/\tilde{Z}_h^s} \sum_{u \in \tilde{Z}_h^s} \epsilon_{(0,w \oplus u)} \imath^{ww^T + uu^T - 2wu^T} \\
  & = \epsilon_{(a_h, b_h)} \sum_{w \in \tilde{Z}_h/\tilde{Z}_h^s} \epsilon_{(0,w)} \imath^{ww^T} 
  \left( \sum_{u \in \tilde{Z}_h^s} \epsilon_{(0,u)} \imath^{uu^T} (-1)^{wu^T} \right) \\
  & = \epsilon_{(a_h, b_h)} \sum_{w \in \tilde{Z}_h/\tilde{Z}_h^s} \epsilon_{(0,w)} \imath^{ww^T} 
  \left( \sum_{u \in \tilde{Z}_h^s} (-1)^{wu^T} \right) \\
  & = \epsilon_{(a_h, b_h)} |\tilde{Z}_h^s| \\
  & = \epsilon_{(a_h, b_h)} 2^{w_H(a_h)/2},
\end{align}
since only $w = 0 \in \tilde{Z}_h/\tilde{Z}_h^s$ is orthogonal to all $u \in \tilde{Z}_h^s$ (observation (O2)).
This completes the proof of the theorem\footnote{We would like to thank Jeongwan Haah for encouraging us to clarify the existence of the self-dual code; this motivated us to greatly improve this proof.}. \hfill \IEEEQEDhere

\subsection{Proof of Lemma~\ref{lem:conj_by_trans_T_Tinv}}
\label{sec:proof_conj_by_trans_T_Tinv}

We will use the observations listed at the beginning of Appendix~\ref{sec:all_proofs} to complete this proof.
As mentioned earlier in the proof of Theorem~\ref{thm:transversal_T}, we proved in~\cite{Rengaswamy-pra19} that given a tensor product of diagonal unitaries $\tau_{R_1}^{(\ell)} \otimes \tau_{R_2}^{(\ell)} \otimes \cdots \otimes \tau_{R_n}^{(\ell)}$, the result is also of the form $\tau_R^{(\ell)}$ with 
\begin{align}
R = 
\begin{bmatrix}
R_1 & 0 & \cdots & 0 \\
0 & R_2 & \cdots & 0 \\
\vdots & \vdots & \ddots & \vdots \\
0 & 0 & \cdots & R_n 
\end{bmatrix}.
\end{align}
Hence, for an $n$-qubit system, the transversal application of $T$ gate corresponds to $R = I_n$ and $\ell = 3$.
Similarly, it is easy to see that the symmetric matrix corresponding to $T^{\otimes t}$ is $R = D_t$, where $t = t_1 + 7 t_7$ and $D_t$ is a diagonal $n \times n$ matrix with the diagonal set to $t$.
Then, according to Corollary~\ref{cor:tauR_conjugate}, we see that
\begin{align}
\tilde{R}(D_t,a,3) & = 3 D_{a \ast t} - (D_{1-a} D_t D_a + D_a D_t D_{1-a} + 2 D_{a \ast t}) = D_{a \ast t} = D_{a \ast t_1} + 7 D_{a \ast t_7}, \\
\phi(D_t,a,b,3) & = - a D_t a^T = - at^T = -at_1^T - 7 at_7^T = - w_H(a \ast t_1) + w_H(a \ast t_7)\ (\text{mod}\ 8).
\end{align}
We need to calculate $c_{D_{a \ast t},x}^{(2)} = \frac{1}{\sqrt{2^n}} \sum_{v \in \mathbb{Z}_2^n} (-1)^{vx^T} \imath^{v D_{a \ast t_1} v^T} (-\imath)^{v D_{a \ast t_7} v^T}$ for all $x \in \mathbb{Z}_2^n$.
For $x = 0$, we observe
\begin{align}
c_{D_{a \ast t},0}^{(2)} & = \frac{1}{\sqrt{2^n}} \sum_{v \in \mathbb{Z}_2^n} \imath^{v D_{a \ast t_1} v^T} (-\imath)^{v D_{a \ast t_7} v^T} \\
  & = \frac{1}{\sqrt{2^n}} \sum_{v \in \mathbb{Z}_2^n} \imath^{v (a \ast t_1)^T} (-\imath)^{v (a \ast t_7)^T} \\
  & = \frac{1}{\sqrt{2^n}} 2^{n - w_H(a \ast t_1) - w_H(a \ast t_7)} (1 + \imath)^{w_H(a \ast t_1)} (1 - \imath)^{w_H(a \ast t_7)} \ (\text{observation\ (O4)}) \\
  & = e^{\frac{\imath \pi}{4} \left[ w_H(a \ast t_1) - w_H(a \ast t_7) \right]} 2^{\left( n - w_H(a \ast t') \right)/2} \ (\text{since}\ e^{\pm \frac{\imath \pi}{4}} = (1 \pm \imath)/\sqrt{2}).
\end{align}
For the case $x \neq 0$, we 
calculate as follows.
Recollect that we have defined $t' = t_1 + t_7 \in \mathbb{Z}_2^n$.
\begin{align}
c_{D_{a \ast t}, x}^{(2)} & = \frac{1}{\sqrt{2^n}} \sum_{v \in \mathbb{Z}_2^n} \imath^{v \left[ (a \ast t_1) + 3(a \ast t_7) + 2x \right]^T} \\
  & = \frac{1}{\sqrt{2^n}} \sum_{v \preceq (a \ast t')} \imath^{v \left[ (a \ast t_1) + 3(a \ast t_7) + 2x \right]^T} \sum_{w \preceq \overline{(a \ast t')}} (-1)^{w x^T} \ (\text{observation\ (O4)}) \\
  & = 2^{\frac{n}{2} - w_H(a \ast t')} \left( \sum_{v \preceq (a \ast t') \oplus x} \imath^{v \left[ (a \ast t_1) + 3(a \ast t_7) + 2x \right]^T} \right) \left( \sum_{w \preceq x} \imath^{w \left[ (a \ast t_1) + 3(a \ast t_7) + 2x \right]^T} \right),\ x \preceq (a \ast t').
\end{align}
In the last step, unless $x \preceq (a \ast t')$ it is easy to see that the inner summation in the second step vanishes (observation (O3)).
Furthermore, we have used this property of $x$ to split the sum into indices on the support of $x$ and the others (observation (O4)).
For convenience, let us denote by $A$ the support of $(a \ast t') \oplus x$, by $A_1$ the support of $(a \ast t_1) \oplus (x \ast a \ast t_1)$, and by $A_7$ the support of $(a \ast t_7) \oplus (x \ast a \ast t_7)$.
Note that $A = \text{supp}(a \ast t') \setminus \text{supp}(x)$ and hence $vx^T = 0$.
For simplicity, we will write $\{ v \in \mathbb{Z}_2^n \colon \text{supp}(v) \subseteq A \}$ as $v \preceq A$.
Then, using observation (O4), we can write the first sum above as
\begin{align}
\sum_{v \preceq A} \imath^{v (a \ast t_1)^T} (-\imath)^{v (a \ast t_7)^T} & = \sum_{z_1 \preceq A_1} \imath^{z_1 (a \ast t_1)^T} \sum_{z_2 \preceq A_7} (-\imath)^{z_2 (a \ast t_7)^T} \\
  & = (1 + \imath)^{w_H(a \ast t_1) - w_H(x \ast a \ast t_1)} (1 - \imath)^{w_H(a \ast t_7) - w_H(x \ast a \ast t_7)} \\
  & = e^{\frac{\imath\pi}{4} \left[ \left( w_H(a \ast t_1) - w_H(a \ast t_7) \right) - \left( w_H(x \ast a \ast t_1) - w_H(x \ast a \ast t_7) \right)  \right]} 2^{\left( w_H(a \ast t') - w_H(x) \right)/2}.
\end{align}
Now, again using observation (O4), we can calculate the second sum similarly as follows.
\begin{align}
\sum_{w \preceq x} \imath^{w \left[ (a \ast t_1) + 3(a \ast t_7) + 2x \right]^T} & = \sum_{w \preceq x \ast (a \ast t_1)} (-\imath)^{ww^T} \sum_{z \preceq x \ast (a \ast t_7)} \imath^{zz^T} \\
  & = (1 - \imath)^{w_H(x \ast a \ast t_1)} (1 + \imath)^{w_H(x \ast a \ast t_7)} \\
  & = e^{\frac{-\imath\pi}{4} \left[ w_H(x \ast a \ast t_1) - w_H(x \ast a \ast t_7) \right]} 2^{w_H(x)/2}.
\end{align}
Combing these two results and substituting back we get,
\begin{align}
c_{D_{a \ast t},x}^{(2)} = 
\begin{cases}
2^{\left( n - w_H(a \ast t') \right)/2} e^{\frac{\imath\pi}{4} \left[ \left( w_H(a \ast t_1) - w_H(a \ast t_7) \right) - 2 \left( w_H(x \ast a \ast t_1) - w_H(x \ast a \ast t_7) \right) \right]}  & \text{if}\ x \preceq (a \ast t'), \\
0 & \text{otherwise}.
\end{cases}
\end{align}
Recollect from~\eqref{eq:tau_Eab_expand} that the action of $\tau_R^{(\ell)}$ on a Pauli matrix $E(a,b)$ is given by
\begin{align}
\tau_R^{(\ell)} E(a,b) ( \tau_R^{(\ell)} )^{\dagger} & = \frac{1}{\sqrt{2^n}} \xi^{\phi(R,a,b,\ell)} \sum_{x \in \mathbb{Z}_2^n} c_{\tilde{R}(R,a,\ell),x}^{(\ell-1)} \imath^{-ax^T} E(a, b + aR + x).
\end{align}
Hence, using observation (O1), we can calculate the action of $T^{\otimes t}$ on $E(a,b)$ under conjugation to be
\begin{align}
& T^{\otimes t} E(a,b) \left( T^{\otimes t} \right)^{\dagger} \nonumber \\
  & = \frac{1}{\sqrt{2^n}} e^{\frac{-\imath\pi}{4} \left[ w_H(a \ast t_1) - w_H(a \ast t_7) \right]} \sum_{x \preceq (a \ast t')} c_{D_{a \ast t},x}^{(2)} \imath^{-a x^T} E(a, b + aD_t + x) \\
  & = \frac{2^{\left( n - w_H(a \ast t') \right)/2}}{\sqrt{2^n}} \sum_{x \preceq (a \ast t')} \imath^{-x \left[ (a \ast t_1) - (a \ast t_7) \right]^T - x \left[ (a \ast t_1) + (a \ast t_7) \right]^T} E(a, b + (a \ast t_1) + 7 (a \ast t_7) + x) \\
  & = \frac{1}{2^{w_H(a \ast t')/2}} \sum_{x \preceq (a \ast t')}  (-1)^{x (a \ast t_1)^T} E(a, b + ((a \ast t') + x) + 6 (a \ast t_7) ) \\
  & = \frac{1}{2^{w_H(a \ast t')/2}} \sum_{x \preceq (a \ast t')}  (-1)^{x (a \ast t_1)^T + a (a \ast t_7)^T} E\left( a, b + \left( (a \ast t') \oplus x + 2 (x \ast (a \ast t')) \right) \right) \\
  & = \frac{1}{2^{w_H(a \ast t')/2}} \sum_{x \preceq (a \ast t')}  (-1)^{x (a \ast t_1)^T + w_H(a \ast t_7)} E\left( a, b + \left( (a \ast t') \oplus x + 2 x \right) \right) \\
  & = \frac{1}{2^{w_H(a \ast t')/2}} \sum_{x \preceq (a \ast t')}  (-1)^{x (a \ast t_1)^T + w_H(a \ast t_7) + xa^T} E\left( a, b \oplus \left( (a \ast t') \oplus x \right) + 2 b \ast ((a \ast t') \oplus x) \right) \\
  & = \frac{1}{2^{w_H(a \ast t')/2}} \sum_{x \preceq (a \ast t')}  (-1)^{x (a \ast t_7)^T + (a \ast t_7) (a \ast t_7)^T + a \left[ b \ast ((a \ast t') \oplus x) \right]^T} E\left( a, b \oplus \left( (a \ast t') \oplus x \right) \right) \\
  & = \frac{1}{2^{w_H(a \ast t')/2}} \sum_{x \preceq (a \ast t')}  (-1)^{(a \ast t_7) \left[(a \ast t_7) \oplus x \right]^T + a \left[ b \ast ((a \ast t') \oplus x) \right]^T} E\left( a, b \oplus \left( (a \ast t') \oplus x \right) \right) \\
  & = \frac{1}{2^{w_H(a \ast t')/2}} \sum_{y \preceq (a \ast t')}  (-1)^{(a \ast t_7) \left[y \oplus (a \ast t_1) \right]^T + a ( b \ast y )^T} E\left( a, b \oplus y \right) \ (y \coloneqq (a \ast t') \oplus x) \\
  & = \frac{1}{2^{w_H(a \ast t')/2}} \sum_{y \preceq (a \ast t')}  (-1)^{y(a \ast t_7)^T + a ( b \ast y )^T} E\left( a, b \oplus y \right) \\
  & = \frac{1}{2^{w_H(a \ast t')/2}} \sum_{y \preceq (a \ast t')}  (-1)^{(b + t_7) y^T} E\left( a, b \oplus y \right).
\end{align}
The last step follows from $y \preceq (a \ast t') \Rightarrow y \preceq a \Rightarrow (b \ast y) \preceq a$ and $y(a \ast t_7)^T = w_H((y \ast a) \ast t_7) = w_H(y \ast t_7) = y t_7^T$.  \hfill \IEEEQEDhere

\subsection{Proof of Corollary~\ref{cor:conj_by_trans_T_Tinv}}
\label{sec:proof_cor_conj_by_trans_T_Tinv}

We have $t = \sum_{j=1}^{7} j t_j$ with $t_j \ast t_{j'} = 0$ for all $j \neq j'$.
Then we can write $T^{\otimes t} = T^{t_1 + t_3 + t_5 + t_7} P^{t_2 + t_3 + t_6 + t_7} Z^{t_4 + t_5 + t_6 + t_7}$, where $Z^{t_4 + t_5 + t_6 + t_7} = E(0, t_4 + t_5 + t_6 + t_7)$.
Now we can compute $T^{\otimes t} E(a,b) (T^{\otimes t})^{\dagger}$ as follows.
Firstly we observe
\begin{align}
E(0, t_4 + t_5 + t_6 + t_7) \cdot E(a,b) \cdot E(0, t_4 + t_5 + t_6 + t_7) = (-1)^{a (t_4 + t_5 + t_6 + t_7)^T} E(a,b).
\end{align}
Next, using the identity $u + v = u \oplus v + 2 (u \ast v)$, we can calculate
\begin{align}
P^{t_2 + t_3 + t_6 + t_7} \cdot & \, (-1)^{a (t_4 + t_5 + t_6 + t_7)^T} E(a,b) \cdot (P^{\dagger})^{t_2 + t_3 + t_6 + t_7} \nonumber \\
  & = (-1)^{a (t_4 + t_5 + t_6 + t_7)^T} E(a, b + (a \ast (t_2 + t_3 + t_6 + t_7)) ) \\
  & = (-1)^{a (t_4 + t_5 + t_6 + t_7)^T + a (b \ast (t_2 + t_3 + t_6 + t_7))^T} E(a, b \oplus (a \ast (t_2 + t_3 + t_6 + t_7)) )  \\
  & = (-1)^{a (t_4 + t_5 + t_6 + t_7)^T + a (b \ast (\tilde{t}_2 + \tilde{t}_3))^T} E(a, b \oplus (a \ast (\tilde{t}_2 + \tilde{t}_3)) ) \\
  & = (-1)^{a (t_4 + t_5 + t_6 + t_7)^T + w_H(a \ast b \ast \tilde{t}_2) + w_H(a \ast b \ast \tilde{t}_3)} E(a, c ),
\end{align}
where we have defined $c \coloneqq b \oplus (a \ast (\tilde{t}_2 + \tilde{t}_3))$ for convenience.
Finally we can invoke Lemma~\ref{lem:conj_by_trans_T_Tinv} with the $t_1, t_7$ in that lemma taken respectively to be $t_1 + t_3 + t_5 + t_7$ and $0$ here.
Then we have
\begin{align}
& T^{t_1 + t_3 + t_5 + t_7} \cdot (-1)^{a (t_4 + t_5 + t_6 + t_7)^T + w_H(a \ast b \ast \tilde{t}_2) + w_H(a \ast b \ast \tilde{t}_3)} E(a,c) \cdot (T^{\dagger})^{t_1 + t_3 + t_5 + t_7} \nonumber \\    
  & = \frac{(-1)^{a (t_4 + t_5 + t_6 + t_7)^T + w_H(a \ast b \ast \tilde{t}_2) + w_H(a \ast b \ast \tilde{t}_3)}}{2^{w_H(a \ast (\tilde{t}_1 + \tilde{t}_3))/2}} \sum_{y \preceq a \ast (\tilde{t}_1 + \tilde{t}_3)} (-1)^{cy^T} E(a, c \oplus y) \\
  & = \frac{(-1)^{a (t_4 + t_5 + t_6 + t_7)^T + w_H(a \ast b \ast \tilde{t}_2) + w_H(a \ast b \ast \tilde{t}_3)}}{2^{w_H(a \ast (\tilde{t}_1 + \tilde{t}_3))/2}} \nonumber \\
  & \hspace{1.5cm} \sum_{y \preceq a \ast (\tilde{t}_1 + \tilde{t}_3)} (-1)^{by^T + y (a \ast (\tilde{t}_2 + \tilde{t}_3))^T} E(a, b \oplus [(a \ast (\tilde{t}_2 + \tilde{t}_3)) \oplus y]) \\
  & = \frac{(-1)^{a (t_4 + t_5 + t_6 + t_7)^T + w_H(a \ast (\tilde{t}_2 + \tilde{t}_3)) }}{2^{w_H(a \ast (\tilde{t}_1 + \tilde{t}_3))/2}} \sum_{(a \ast \tilde{t}_2) \preceq z \preceq a \ast (\tilde{t}_1 + \tilde{t}_2 + \tilde{t}_3)} (-1)^{bz^T + z (a \ast (\tilde{t}_2 + \tilde{t}_3))^T} E(a, b \oplus z),
\end{align}
where we have defined $z \coloneqq (a \ast (\tilde{t}_2 + \tilde{t}_3)) \oplus y$.
In the last step, first we observe that $(a \ast \tilde{t}_2)$ is always present in $z$ and is unaffected by the range of $y$ above since $\tilde{t}_1 \ast \tilde{t}_2 = 0 = \tilde{t}_3 \ast \tilde{t}_2$.
Hence, we write that $z$ contains and is contained in $(a \ast \tilde{t}_2)$.
Next, since the change of variables does not change the support of $y$ inside $(a \ast \tilde{t}_1)$, the new variable $z$ also loops over all vectors contained in $(a \ast \tilde{t}_1)$.
Finally, since $z$ adds $(a \ast \tilde{t}_3)$ to $y$ (modulo 2), we see that $y$ undergoes a ones' complement in the support of $(a \ast \tilde{t}_3)$.
However, since $y$ takes all possible values in the support of $(a \ast \tilde{t}_3)$, we conclude that $z$ still retains that property.
Furthermore, under the change of variables, we have cancelled the new factors $(-1)^{b (a \ast \tilde{t}_2)^T}$ and $(-1)^{b (a \ast \tilde{t}_3)^T}$ correspondingly with the existing factors $(-1)^{w_H(b \ast a \ast \tilde{t}_2)}$ and $(-1)^{w_H(b \ast a \ast \tilde{t}_3)}$.
Now we observe that 
\begin{align}
z (a \ast (\tilde{t}_2 + \tilde{t}_3))^T & = w_H(z \ast (a \ast \tilde{t}_2)) + w_H((z \ast a) \ast \tilde{t}_3) \\
  & = w_H(a \ast \tilde{t}_2) + w_H(z \ast \tilde{t}_3) = w_H(a \ast \tilde{t}_2) + \tilde{t}_3 z^T,
\end{align}
since clearly $z \preceq a$.
Substituting this back, and noting that $\tilde{t}_3 = t_3 + t_7$, we obtain 
\begin{align}
T^{\otimes t} E(a,b) \left( T^{\otimes t} \right)^{\dagger} = \frac{ (-1)^{a (t_3 + t_4 + t_5 + t_6)^T} }{2^{w_H(a \ast (\tilde{t}_1 + \tilde{t}_3))/2}} \sum_{(a \ast \tilde{t}_2) \preceq z \preceq (a \ast (\tilde{t}_1 + \tilde{t}_2 + \tilde{t}_3))}  (-1)^{(b + \tilde{t}_3) z^T} E\left( a, b \oplus z \right),
\end{align}
which is the final expression given in the statement of the lemma. \hfill \IEEEQEDhere

\subsection{Proof of Theorem~\ref{thm:logical_identity}}
\label{sec:proof_logical_identity}

Let $G_1 = 
\begin{bmatrix}
G_{C_1/C_2} \\
G_2
\end{bmatrix}$ be a generator matrix for the code $C_1$.
Then by the CSS construction, the vectors $x = \bigoplus_{i = 1}^{k} c_i x_i$, where $c_i \in \{0,1\}$ and $x_i$ form the $k$ rows of the coset generator matrix $G_{C_1/C_2}$, determine all logical $X$ operators $E(x,0)$ for the code CSS($X, C_2; Z, C_1^{\perp}$).
Similarly, vectors $a \in C_2$ determine the $X$-type stabilizers $E(a,0)$ for the code. 
Therefore, $(x \oplus a) \in C_1$ represents all possible $X$-type representatives of all logical $X$ operators for the CSS-T code.
Recollect that by the CSS-T conditions, $C_2 \subset C_1^{\perp} \Rightarrow a \in C_1^{\perp}$ as well, and in fact $\tilde{a}$ belongs to the dual $\tilde{Z}_a^{\perp}$ of the (punctured) subspace $Z_a$ in $C_1^{\perp}$ that is supported on $a$ (see Theorem~\ref{thm:transversal_T} for notation), so that $\imath^{w_H(a)} E(0,a) \in S$.
By assumption, we have transversal $T$ acting trivially on the logical qubits, so transversal $P = T^2$ must also act trivially on the logical qubits.
Using this fact, and the identity $PXP^{\dagger} = Y$, let us observe the action of transversal $P$ on a logical $X$ representative $E(x \oplus a, 0)$.
We have
\begin{align}
P^{\otimes n} E(x \oplus a, 0) \left( P^{\otimes n} \right)^{\dagger} = E(x \oplus a, x \oplus a) = E(x \oplus a, 0) \cdot \imath^{w_H(x \oplus a)} E(0, x \oplus a),
\end{align}
where the second equality follows from the identity~\eqref{eq:Eab_multiply}.
Note that $w_H(x \oplus a) = w_H(x) + w_H(a) - 2 xa^T \equiv w_H(x) + w_H(a)$ (mod $4$), since $xa^T \equiv 0$ (mod $2$) due to the fact that logical $X$ operators must commute with $Z$-type stabilizers, and $\imath^{w_H(a)} E(0,a) \in S$.
(Recollect that $E(x,0)$ and $E(0,a)$ commute if and only if their symplectic inner product $\syminn{[x,0]}{[0,a]} = xa^T = 0$.)
Hence, $\imath^{w_H(x \oplus a)} E(0, x \oplus a) = \imath^{w_H(x)} E(0,x) \cdot \imath^{w_H(a)} E(0,a)$. 
Since $P^{\otimes n}$ must act trivially on logical qubits, we require that $P^{\otimes n} E(x \oplus a, 0) \left( P^{\otimes n} \right)^{\dagger} \equiv E(x \oplus a, 0)$, where the equivalence class is defined by multiplication with stabilizer elements.
For this equivalence to be true, we need $\imath^{w_H(x)} E(0,x) \in S$. 
This proves the first condition stated in the theorem.

The above ensures that $P^{\otimes n}$ acts like the logical identity.
Let us now examine $T^{\otimes n}$ along similar lines by using the identity $TXT^{\dagger} = e^{-\imath\pi/4} Y\, P$.
We require
\begin{align}
T^{\otimes n} E(x \oplus a, 0) \left( T^{\otimes n} \right)^{\dagger} & = e^{-\frac{\imath\pi}{4} w_H(x \oplus a)} E(x \oplus a, x \oplus a)\, P^{x \oplus a} \\ 
  & = e^{-\frac{\imath\pi}{4} w_H(x \oplus a)} E(x \oplus a, 0) \cdot \imath^{w_H(x \oplus a)} E(0, x \oplus a)\, P^{x \oplus a} \\
  & = e^{-\frac{\imath\pi}{4} w_H(x \oplus a)} E(x \oplus a, 0)\, P^{x \oplus a} \cdot \imath^{w_H(x \oplus a)} E(0, x \oplus a)\\
  & \equiv E(x \oplus a, 0).
\end{align}
Here the notation $P^{x \oplus a}$ means that the phase gate is applied to the qubits in the support of $(x \oplus a)$.
From the above calculation for $P^{\otimes n}$, we know that $\imath^{w_H(x \oplus a)} E(0, x \oplus a) \in S$, so for the last equivalence to be true, we need to ensure that $P^{x \oplus a}$ acts like the logical identity.
Let us examine its action on an arbitrary logical $X$ representative $E(y \oplus b, 0)$ for $y \in C_1/C_2, b \in C_2$.
We require
\begin{align}
P^{x \oplus a} E(y \oplus b, 0) \left( P^{x \oplus a} \right)^{\dagger} & = E(y \oplus b, (y \oplus b) \ast (x \oplus a)) \\
  & = E(y \oplus b, 0) \cdot \imath^{w_H((y \oplus b) \ast (x \oplus a))} E(0, (y \oplus b) \ast (x \oplus a)) \\
  & \equiv E(y \oplus b, 0).
\end{align}
Observe that this is satisfied for the case $y = x, b = a$ by the arguments above for $P^{\otimes n}$.
Clearly, the constraint we need is that $\imath^{w_H((y \oplus b) \ast (x \oplus a))} E(0, (y \oplus b) \ast (x \oplus a)) \in S$.
Since this must hold for all valid $x,y,a,b$, by setting $a = b = 0$ we obtain the third condition of the theorem.
Similarly, by setting $a = y = 0$ and $x = y = 0$ respectively, we obtain the second and fourth conditions of the theorem.
It can be verified that these alone ensure that $\imath^{w_H((y \oplus b) \ast (x \oplus a))} E(0, (y \oplus b) \ast (x \oplus a)) \in S$ for all combinations of $x,y,a,b$, since we can split $(y \oplus b) \ast (x \oplus a) = (y \ast x) \oplus (y \ast a) \oplus (b \ast x) \oplus (b \ast a)$ and using the Hamming weight identity we used above.
Finally, we will show that the last three conditions amount to triorthogonality.

Note that since we need $\imath^{w_H(y \ast x)} E(0, y \ast x) \in S$, it must be true that $(y \ast x) \in C_1^{\perp}$ since by the CSS construction pure $Z$-type stabilizers arise from the code $C_1^{\perp}$.
As logical $X$ operators and $X$-type stabilizers must each commute with $Z$-type stabilizers, by the symplectic inner product constraint we can see that this implies $z (y \ast x)^T = w_H(z \ast y \ast x) \equiv 0$ (mod $2$) for any $z \in C_1/C_2$ or $z \in C_2$.
Similarly, since we need $\imath^{w_H(b \ast a)} E(0, b \ast a) \in S$, we also have $w_H(z \ast b \ast a) \equiv 0$ for any $z \in C_1/C_2$ or $z \in C_2$.
These are exactly the triorthogonality conditions in Definition~\ref{def:triorthogonality}.
The first condition of Definition~\ref{def:triorthogonality} follows from the facts that $C_2 \subset C_1^{\perp}$, and since $\imath^{w_H(y \ast x)} E(0, y \ast x), \imath^{w_H(y \ast a)} E(0, y \ast a) \in S$ must be Hermitian, the phase has to be $\pm 1$, which implies $w_H(y \ast x) = xy^T \equiv 0, w_H(y \ast a) = ay^T \equiv 0$ (mod $2$) for any $x,y \in C_1/C_2$ and $a \in C_2$.
Hence, triorthogonality of $G_1$ is a necessary condition for transversal $T$ to realize the logical identity on a CSS-T code. \hfill \IEEEQEDhere

\subsection{Proof of Theorem~\ref{thm:logical_trans_T}}
\label{sec:proof_logical_trans_T}

The proof uses a very similar strategy as for Theorem~\ref{thm:logical_identity}.
As we observed there, $(x \oplus a) \in C_1$ for $x \in C_1, a \in C_2$ represents all possible $X$-type representatives of all logical $X$ operators for the CSS-T code.
Recollect that by the CSS-T conditions, $C_2 \subset C_1^{\perp} \Rightarrow a \in C_1^{\perp}$ as well, and in fact $\tilde{a}$ belongs to the dual $\tilde{Z}_a^{\perp}$ of the (punctured) subspace $Z_a$ in $C_1^{\perp}$ that is supported on $a$ (see Theorem~\ref{thm:transversal_T} for notation), so that $\imath^{w_H(a)} E(0,a) \in S$.
By assumption, we have physical transversal $T$ acting as transversal $T$ on the logical qubits, so physical transversal $P = T^2$ must also act as transversal $P$ on the logical qubits.
Using this fact, and the identity $PXP^{\dagger} = Y$, let us observe the action of transversal $P$ on a logical $X$ representative $E(x \oplus a, 0)$.
We require
\begin{align}
P^{\otimes n} E(x \oplus a, 0) \left( P^{\otimes n} \right)^{\dagger} & = E(x \oplus a, x \oplus a) \\
  & = E(x \oplus a, 0) \cdot \imath^{w_H(x \oplus a)} E(0, x \oplus a) \\
  & \equiv \imath^{w_H(c)} E(x \oplus a, 0) E(0,z),
\end{align}
where the second equality follows from~\eqref{eq:Eab_multiply}, and $z \in C_2^{\perp}/C_1^{\perp}$ is the corresponding logical $Z$ string for the logical $X$ string $x$%
\footnote{Using the notation $x = \bigoplus_{i=1}^{k} c_i x_i$ in the theorem statement, $E(x \oplus a, 0) = E(a,0) \prod_{i=1}^{k} \bar{X}_i^{c_i}$. Since $(P \otimes P) (X \otimes X) (P^{\dagger} \otimes P^{\dagger}) = Y \otimes Y = \imath^2 (X \otimes X) (Z \otimes Z)$ physically, the calculation above translates this to the logical level to produce $w_H(c)$, where $\bar{X}_i = E(x_i,0)$ and $\bar{Z}_i = E(0,z_i)$.}.
This also means $xz^T \equiv w_H(c)$ (mod $2$) since the respective pairs of logical $X$ and $Z$ must anti-commute.
Note that $w_H(x \oplus a) = w_H(x) + w_H(a) - 2 xa^T \equiv w_H(x) + w_H(a)$ (mod $4$) because $xa^T \equiv 0$ (mod $2$) due to the fact that logical $X$ operators must commute with $Z$-type stabilizers (and $\imath^{w_H(a)} E(0,a) \in S$).
(Recollect that $E(x,0)$ and $E(0,a)$ commute if and only if their symplectic inner product $\syminn{[x,0]}{[0,a]} = xa^T = 0$.)
Hence, $\imath^{w_H(x \oplus a)} E(0, x \oplus a) = \imath^{w_H(x)} E(0,x) \cdot \imath^{w_H(a)} E(0,a)$, and for the last equivalence to be true above, we need $\imath^{w_H(x) - w_H(c)} E(0,x \oplus z) \in S$. 
Therefore, $w_H(x) - w_H(c) \equiv 0$ (mod $2$) for this matrix to be Hermitian. 
Observing the case when $w_H(c) = 1$ and hence $x = x_i$ for some $i \in \{1,\ldots,k\}$, this implies that $w_H(x_i)$ must be odd. 
Moreover, since $a \in C_2 \subset C_1^{\perp}$, we see that $x_i a^T = 0$, and $z_i a^T = 0$ because $z_i \in C_2^{\perp}$.
Also, for any $x_j \in C_1/C_2$ that is distinct from $x_i$, we have $z_i x_j^T = 0$ by definition, and since the above condition means $x_i \oplus z_i \in C_1^{\perp}$, we also have $x_i x_j^T = 0$.
So we can assume $z_i = x_i$ to be the corresponding logical $Z$ string for $\bar{X}_i$ as well, in which case the above condition becomes trivial and the required equivalence is satisfied.
In this scenario, since $x_i \oplus z_i = 0 \Rightarrow E(0,x_i \oplus z_i) = I_N$, we need $w_H(x_i) \equiv 1$ (mod $4$) exactly as for a non-trivial stabilizer group we need to make sure that $-I_N \notin S$.

The above ensures that $P^{\otimes n}$ acts desirably and so $E(x \oplus a, x \oplus a)$ indeed corresponds to the logical $Y$ operator $\bar{Y}_c$ corresponding to the given logical $X$ string $x = \bigoplus_{i=1}^{k} c_i x_i \in C_1/C_2$.
Let us now examine $T^{\otimes n}$ along similar lines by using the identity $TXT^{\dagger} = e^{-\imath\pi/4} Y\, P$.
We require
\begin{align}
T^{\otimes n} \bar{X}_c \left( T^{\otimes n} \right)^{\dagger} & = T^{\otimes n} E(x \oplus a, 0) \left( T^{\otimes n} \right)^{\dagger} \\
  & = e^{-\frac{\imath\pi}{4} w_H(x \oplus a)} E(x \oplus a, x \oplus a)\, P^{x \oplus a} \\
%
  & \equiv e^{-\frac{\imath\pi}{4} w_H(c)} \bar{Y}_c\, \bar{P}_c,
\end{align}
where $\bar{P}_c$ denotes the logical phase gate corresponding to the given logical $X$ string $x = \bigoplus_{i=1}^{k} c_i x_i$, and the notation $P^{x \oplus a}$ means that the phase gate is applied to the qubits in the support of $(x \oplus a)$.
We have verified above that the $\bar{Y}_c$ condition is satisfied, and when $P^{x \oplus a}$ acts on $\bar{X}_c = E(x \oplus a, 0)$, it indeed acts desirably.
Therefore, to verify $P^{x \oplus a} \equiv \bar{P}_c$, we just need to ensure that $P^{x \oplus a}$ acts like the logical identity on all $E(y \oplus b, 0)$ for $y \neq x \in C_1/C_2$ and any $b \in C_2$.
This part of the proof is identical to the corresponding arguments in the proof of Theorem~\ref{thm:logical_identity}, and so this proves that triorthogonality of $G_1$ is a necessary condition.
Finally, we observe that we need the condition $w_H(x \oplus a) \equiv w_H(c)$ (mod $8$) for the above equivalence to hold, and this completes the proof of the theorem. \hfill \IEEEQEDhere

\subsection{Proof of Lemma~\ref{lem:conj_by_trans_Z_rot}}
\label{sec:proof_conj_by_trans_Z_rot}

In this proof, we will see that the ideas used to calculate the coefficients for the transversal $T$ gate generalize to the transversal application of power-of-$2$ roots of $T$, namely $\tau_{R}^{(\ell)} \coloneqq 
\begin{bmatrix}
1 & 0 \\
0 & e^{2\pi\imath/2^\ell}
\end{bmatrix} = 
\begin{bmatrix}
1 & 0 \\
0 & \xi
\end{bmatrix}
$ with $R = [\, 1\, ]$.
When this $Z$-rotation is applied transversally, we have $R = I_n$ as before for the $T$ gate.
Then from Corollary~\ref{cor:tauR_conjugate} we get $\phi(R,a,b,\ell) = (1 - 2^{\ell-2}) a I_n a^T = (1 - 2^{\ell-2}) w_H(a)$ and $\tilde{R} \coloneqq \tilde{R}(R,a,\ell) = (1 + 2^{\ell-2}) D_{a I_n} - (D_{1-a} I_n D_a + D_a I_n D_{1-a} + 2 D_{a I_n D_a}) = (1 + 2^{\ell-2}) D_a - (0 + 0 + 2 D_a) = (2^{\ell-2} - 1) D_a$.

First consider the case $a \neq [1,1,\ldots,1]$ and let $x \in \Fn$ such that $\text{supp}(x) \nsubseteq \text{supp}(a)$, so that there exists some $j \in \{1,\ldots,n\}$ satisfying $a_j = 0, x_j = 1$.
Once again, let $\tilde{x} = [x_1,x_2,\ldots,x_{j-1},x_{j+1},\ldots,x_n]$ and similarly for $\tilde{a}, \tilde{v}$.
Then we observe that
\begin{align}
c_{(2^{\ell-2} - 1) D_a,x}^{(\ell-1)} & = \frac{1}{\sqrt{2^n}} \sum_{v \in \Fn} (-1)^{vx^T} (\xi^2)^{v \tilde{R} v^T} \\
  & = \frac{1}{\sqrt{2^n}} \sum_{v \in \Fn} (-1)^{vx^T} \xi^{2(2^{\ell-2} - 1) va^T} \\
  & = \frac{1}{\sqrt{2^n}} \sum_{\substack{\tilde{v} \in \mathbb{F}_2^{n-1}\\(v_j = 0)}} (-1)^{\tilde{v} \tilde{x}^T} \xi^{(2^{\ell-1} - 2) \tilde{v} \tilde{a}^T} - \frac{1}{\sqrt{2^n}} \sum_{\substack{\tilde{v} \in \mathbb{F}_2^{n-1}\\(v_j = 1)}} (-1)^{\tilde{v} \tilde{x}^T} \xi^{(2^{\ell-1} - 2) \tilde{v} \tilde{a}^T} \\
  & = 0.
\end{align}
So we only need to consider coefficients corresponding to $x \preceq a$.
For the remaining calculations, it is convenient to note that $\xi^{2^{\ell-1} - 2} = - \xi^{-2}$.
Once again, for the case $x = [0,0,\ldots,0]$ and arbitrary $a \in \Fn$, we observe that
\begin{align}
\label{eq:trans_root_T_coeff_x_0}
c_{(2^{\ell-2} - 1) D_a,0}^{(\ell-1)} = \frac{1}{\sqrt{2^n}} \sum_{v \in \Fn} (-\xi^{-2})^{v a^T} = \frac{1}{\sqrt{2^n}} 2^{n - w_H(a)} (1 - \xi^{-2})^{w_H(a)} = \sqrt{2^n} \left( \frac{1 - \xi^{-2}}{2} \right)^{w_H(a)}.
\end{align}
For $x \neq [0,0,\ldots,0]$, and $x \preceq a$, let $j \in \text{supp}(x) \subseteq \text{supp}(a)$ so that $x_j = a_j = 1$ and $e_j x^T = 1$.
Then for each $v$ such that $vx^T \equiv 0$, we have $(v \oplus e_j) x^T \equiv 1$.
So we calculate
\begin{align}
& c_{(2^{\ell-2} - 1) D_a,x}^{(\ell-1)} \nonumber \\
  & = \frac{1}{\sqrt{2^n}} \sum_{v \in \Fn} (-1)^{vx^T} (-\xi^{-2})^{v a^T} \\
  & = \frac{1}{\sqrt{2^n}} \sum_{\substack{v \in \Fn\\vx^T \equiv 0}} (-\xi^{-2})^{v a^T} - \frac{1}{\sqrt{2^n}} \sum_{\substack{v \in \Fn\\vx^T \equiv 1}} (-\xi^{-2})^{v a^T} \\
  & = \frac{1}{\sqrt{2^n}} \sum_{\substack{v \in \Fn\\vx^T \equiv 0}} \left[ (-\xi^{-2})^{v a^T} - (-\xi^{-2})^{(v \oplus e_j) a^T} \right] \\
  & = \frac{1}{\sqrt{2^n}} \sum_{\substack{v \in \Fn\\vx^T \equiv 0}} \left[ (-\xi^{-2})^{v a^T} - (-\xi^{-2})^{(v + e_j - 2 (v \ast e_j)) a^T} \right] \\
  & = \frac{1}{\sqrt{2^n}} \sum_{\substack{v \in \Fn\\vx^T \equiv 0}} (-\xi^{-2})^{v a^T} \left[ 1 - (-\xi^{-2})^{a_j - 2 v_j a_j}  \right] \\
  & = \frac{(1 + \xi^{-2})}{\sqrt{2^n}} \sum_{\substack{\tilde{v} \colon \tilde{v} \tilde{x}^T \equiv 0\\(v_j = 0)}} (-\xi^{-2})^{\tilde{v} \tilde{a}^T} + \frac{(1 + \xi^2)}{\sqrt{2^n}} \sum_{\substack{\tilde{v} \colon \tilde{v} \tilde{x}^T \equiv 1\\(v_j = 1)}} (-\xi^{-2}) \cdot (-\xi^{-2})^{\tilde{v} \tilde{a}^T} \\
  & = 
\begin{dcases}
\left( \frac{1 + \xi^{-2}}{\sqrt{2}} \right) \frac{1}{\sqrt{2^{n-1}}} \sum_{\tilde{v} \in \mathbb{F}_2^{n-1}} (-\xi^{-2})^{\tilde{v} \tilde{a}^T} & \text{if}\ x = e_j, j \in \text{supp}(a), \\
\left( \frac{1 + \xi^{-2}}{\sqrt{2}} \right) \left[ \frac{1}{\sqrt{2^{n-1}}} \sum_{\substack{\tilde{v} \in \mathbb{F}_2^{n-1}\\\tilde{v} \tilde{x}^T \equiv 0}} (-\xi^{-2})^{\tilde{v} \tilde{a}^T} - \frac{1}{\sqrt{2^{n-1}}} \sum_{\substack{\tilde{v} \in \mathbb{F}_2^{n-1}\\\tilde{v} \tilde{x}^T \equiv 1}} (-\xi^{-2})^{\tilde{v} \tilde{a}^T} \right] & \text{if}\ w_H(x) \geq 2, x \preceq a
\end{dcases} \\
  & = \left( \frac{1 + \xi^{-2}}{\sqrt{2}} \right) \times 
\begin{dcases}
c_{(2^{\ell-2} - 1) D_{\tilde{a}},0}^{(\ell-1)} & \text{if}\ x = e_j, j \in \text{supp}(a), \\
c_{(2^{\ell-2} - 1) D_{\tilde{a}}, \tilde{x}}^{(\ell-1)} & \text{if}\ w_H(x) \geq 2, x \preceq a.
\end{dcases}
\end{align}
When $w_H(x) = 1$, i.e., $x = e_j$, we calculate 
\begin{align}
c_{(2^{\ell-2} - 1) D_a, e_j}^{(\ell-1)} = \frac{(1 + \xi^{-2})}{\sqrt{2}} \sqrt{2^{n-1}} \left( \frac{1 - \xi^{-2}}{2} \right)^{w_H(a) - 1} = \left( \frac{1 + \xi^{-2}}{1 - \xi^{-2}} \right) \sqrt{2^n} \left( \frac{1 - \xi^{-2}}{2} \right)^{w_H(a)}.
\end{align}
We can simplify the first factor as
\begin{align}
\frac{1 + \xi^{-2}}{1 - \xi^{-2}} \cdot \frac{1 - \xi^2}{1 - \xi^2} = \frac{1 + \xi^{-2} - \xi^2 - 1}{1 - \xi^{-2} - \xi^2 + 1} = \frac{-2\imath \sin\frac{2\pi}{2^{\ell-1}}}{2 \left( 1 - \cos\frac{2\pi}{2^{\ell-1}} \right)} = \frac{-2\imath \sin\frac{2\pi}{2^\ell} \cos\frac{2\pi}{2^\ell}}{2 \sin^2\frac{2\pi}{2^\ell}} = -\imath \cot\frac{2\pi}{2^\ell}.
\end{align}
Therefore, in general we have
\begin{align}
c_{(2^{\ell-2} - 1) D_a, x}^{(\ell-1)} & = 
\begin{dcases}
\left( \frac{1 + \xi^{-2}}{\sqrt{2}} \right)^{w_H(x)} \sqrt{2^{n-w_H(x)}} \left( \frac{1 - \xi^{-2}}{2} \right)^{w_H(a) - w_H(x)} & \text{if}\ x \preceq a, \\
0 & \text{otherwise}
\end{dcases} \\
\label{eq:D_ak_coeff_value}
  & = 
\begin{dcases}
\left( -\imath \cot\frac{2\pi}{2^\ell} \right)^{w_H(x)} \sqrt{2^n} \left( \frac{1 - \xi^{-2}}{2} \right)^{w_H(a)} & \text{if}\ x \preceq a, \\
0 & \text{otherwise}.
\end{dcases}
\end{align}
Using this expression, we proceed to calculate the action of $\tau_{I_n}^{(\ell)}$ on an arbitrary $E(a,b)$. 
\begin{align}
& \tau_{I_n}^{(\ell)} E(a,b) (\tau_{I_n}^{(\ell)})^{\dagger} \nonumber \\
  & = \frac{1}{\sqrt{2^n}} \xi^{\phi(I_n,a,b,\ell)} \sum_{x \in \Fn} c_{(2^{\ell-2} - 1) D_a, x}^{(\ell-1)} \imath^{-a x^T} E(a, b + a I_n + x) \\
  & = \frac{1}{\sqrt{2^n}} \xi^{(1 - 2^{\ell-2}) w_H(a)} \sum_{x \preceq a} \left( -\imath \cot\frac{2\pi}{2^\ell} \right)^{w_H(x)} \sqrt{2^n} \left( \frac{1 - \xi^{-2}}{2} \right)^{w_H(a)} \imath^{-w_H(x)} E(a, b + a + x) \\
  & = \left( \frac{\xi(1 - \xi^{-2})}{2 \imath} \right)^{w_H(a)} \sum_{x \preceq a} \left( - \cot\frac{2\pi}{2^\ell} \right)^{xx^T} E(a, b + a + x) \\
  & = \left( \frac{\xi - \xi^{-1}}{2 \imath} \right)^{w_H(a)} \sum_{x \preceq a} \left( - \cot\frac{2\pi}{2^\ell} \right)^{xx^T} (-1)^{a(b \ast (a \oplus x))^T + a x^T} E(a, b \oplus (a \oplus x)) \\
  & = \left( \frac{2\imath \sin\frac{2\pi}{2^\ell}}{2 \imath} \right)^{w_H(a)} \sum_{y \preceq a} \left( \cot\frac{2\pi}{2^\ell} \right)^{(a - y) (a - y)^T} (-1)^{a(b \ast y)^T} E(a, b \oplus y) \\
  & = \left( \sin\frac{2\pi}{2^\ell} \right)^{w_H(a)} \left( \cot\frac{2\pi}{2^\ell} \right)^{w_H(a)} \sum_{y \preceq a} \left( \tan\frac{2\pi}{2^\ell} \right)^{w_H(y)} (-1)^{a(b \ast y)^T} E(a, b \oplus y) \\
  & = \left( \cos\frac{2\pi}{2^\ell} \right)^{w_H(a)} \sum_{y \preceq a} \left( \tan\frac{2\pi}{2^\ell} \right)^{w_H(y)} (-1)^{a(b \ast y)^T} E(a, b \oplus y) \\
  & = \frac{1}{\left( \sec\frac{2\pi}{2^\ell} \right)^{w_H(a)}} \sum_{y \preceq a} \left( \tan\frac{2\pi}{2^\ell} \right)^{w_H(y)} (-1)^{by^T} E(a, b \oplus y).
\end{align}
Note that we have used the fact that $y \coloneqq a \oplus x = a - x$ since $x \preceq a$, and also that $a x^T = xx^T, a y^T = yy^T$.
\hfill\IEEEQEDhere

\subsection{Proof of Theorem~\ref{thm:transversal_Z_rot}}
\label{sec:proof_transversal_Z_rot}

As in the proof of Theorem~\ref{thm:transversal_T}, we need to determine necessary and sufficient conditions for the equality
\begin{align}
\tau_{I_n}^{(\ell)} \Pi_S \left( \tau_{I_n}^{(\ell)} \right)^{\dagger} & = \frac{1}{2^r} \sum_{j = 1}^{2^r} \epsilon_j \tau_{I_n}^{(\ell)} E(a_j,b_j) \left( \tau_{I_n}^{(\ell)} \right)^{\dagger} \\
  & = \frac{1}{2^r} \sum_{j = 1}^{2^r} \frac{\epsilon_j}{\left( \sec\frac{2\pi}{2^\ell} \right)^{w_H(a_j)}} \sum_{y \preceq a_j} \left( \tan\frac{2\pi}{2^\ell} \right)^{w_H(y)} (-1)^{b_j y^T} E(a_j, b_j \oplus y) \\
  & = \frac{1}{2^r} \sum_{j = 1}^{2^r} \epsilon_j E(a_j,b_j) \\
  & = \Pi_S.
\end{align}
Note that for $j \neq j'$, either $a_j \neq a_{j'}$ or at least $b_j \neq b_{j'}$ if $a_j = a_{j'}$.
By commutativity of stabilizers, for $v_1, v_2 \in Z_j$, we need $\syminn{[a_j, b_j \oplus v_1]}{[a_j, b_j \oplus v_2]} = 0$ which implies $a_j(v_1 \oplus v_2) = 0$ (mod $2$), or equivalently, $w_H(v_1 \oplus v_2) = 0$ (mod $2$).
Hence, all vectors in $Z_j$ must have even weight for any $a_j \neq 0$.

Let $\epsilon_v^{(j)} E(a_j, b_j \oplus v) \in S$ for $v \in Z_j$, for some $\epsilon_v^{(j)} \in \{ \pm 1 \}$.
Note that it is unclear if we need $\epsilon_v^{(j)} = \epsilon_j (-1)^{b_j v^T}$ always, as the expression above might suggest, and from the calculation in Theorem~\ref{thm:transversal_T}.
In general, if we collect all the coefficients for $E(a_j,b_j)$, then dividing it by $\left( \sec\frac{2\pi}{2^\ell} \right)^{w_H(a_j)}$ must leave $\epsilon_j$ alone in the numerator, i.e.,
\begin{align}
\epsilon_j + \sum_{v \in Z_j \setminus \{0\}} \epsilon_v^{(j)} (-1)^{(b_j \oplus v) v^T} \left( \tan\frac{2\pi}{2^\ell} \right)^{w_H(v)} & = \epsilon_j \left( \sec\frac{2\pi}{2^\ell} \right)^{w_H(a_j)} \\
\Rightarrow \sum_{v \in Z_j} \epsilon_v^{(j)} (-1)^{b_j v^T} \left( \tan\frac{2\pi}{2^\ell} \right)^{w_H(v)} & = \epsilon_j \left( \sec\frac{2\pi}{2^\ell} \right)^{w_H(a_j)} \ (\text{since}\ w_H(v)\ \text{even}).
\end{align}
Let $\epsilon_v E(0,v) \in S$ for $v \in Z_j$, for some $a_j$, and $\epsilon_v \in \{ \pm 1 \}$.
Then we calculate
\begin{align}
\epsilon_j E(a_j,b_j) \cdot \epsilon_v E(0,v) & = \epsilon_j \epsilon_v \imath^{-a_j v^T} E(a_j, b_j + v) \\
   & = \epsilon_j \epsilon_v \imath^{vv^T} E(a_j, b_j \oplus v + 2(b_j \ast v)) \\
   & = \epsilon_j \epsilon_v \imath^{vv^T} (-1)^{b_j v^T} E(a_j, b_j \oplus v) \\
   & \eqqcolon \epsilon_v^{(j)} E(a_j, b_j \oplus v).
\end{align}
Hence, $\epsilon_v^{(j)} = \epsilon_j \epsilon_v \imath^{vv^T} (-1)^{b_j v^T}$, where $w_H(v) = vv^T$ is even.
(Note that by the property of any non-trivial stabilizer, $E(0,0) = I_N$ has sign $+1$ always, so $\epsilon_0 = 1$.)
Substituting this value back, we get the condition
\begin{align}
\sum_{v \in Z_j} \epsilon_v \imath^{vv^T} \left( \tan\frac{2\pi}{2^\ell} \right)^{w_H(v)} & = \epsilon_j \left( \sec\frac{2\pi}{2^\ell} \right)^{w_H(a_j)} \\
\Rightarrow \sum_{v \in Z_j} \epsilon_v \left( \imath \tan\frac{2\pi}{2^\ell} \right)^{w_H(v)} & = \epsilon_j \left( \sec\frac{2\pi}{2^\ell} \right)^{w_H(a_j)},
\end{align}
for some choices of $\epsilon_v \in \{ \pm 1 \}$, except when $v = 0$ as commented above.
This proves the first condition in the theorem.

Now we need to ensure that all terms $E(a_j, b_j \oplus y)$ for $y \in W_j$ get cancelled properly, for all $a_j$, so that the code projector $\Pi_S$ is indeed preserved.
Observe that terms corresponding to the coset $\{ b_j \oplus y \colon y \preceq a_j \}$ can only appear in the inner summations corresponding to those $i \in \{ 1,2,\ldots,2^r \}$ where $\epsilon_i E(a_i, b_i) = \epsilon_v^{(j)} E(a_j, b_j \oplus v)$ with $v \in Z_j$.
Moreover, for every $y \in B_j$, the Pauli term $E(a_j, b_j \oplus y)$ appears exactly once in the inner summation corresponding to $\epsilon_v^{(j)} E(a_j, b_j \oplus v)$.
We just have to collect these coefficients and set their sum to zero.
Hence, for each $y \in B_j$ we need
\begin{align}
\sum_{v \in Z_j} \epsilon_v^{(j)} (-1)^{(b_j \oplus v) (v \oplus y)^T} \left( \tan\frac{2\pi}{2^\ell} \right)^{w_H(v \oplus y)} & = 0 \\
\Rightarrow \sum_{v \in Z_j} \epsilon_j \epsilon_v \imath^{vv^T} (-1)^{b_j v^T + (b_j \oplus v) (v \oplus y)^T} \left( \tan\frac{2\pi}{2^\ell} \right)^{w_H(v \oplus y)} & = 0 \\
\Rightarrow \sum_{v \in Z_j} \epsilon_v \left( \imath \tan\frac{2\pi}{2^\ell} \right)^{w_H(v \oplus y)} & = 0.
\end{align}
In the last equation, we can reduce the exponent $w_H(v \oplus y)$ to just $w_H(v) - 2vy^T$ since the additional term $w_H(y)$ does not affect the equality to zero.
This proves the second condition in the theorem. \hfill \IEEEQEDhere

%% file: conclusion.tex

In this dissertation we discussed our contributions to quantum communications and quantum computing.
For communication over the pure-state classical-quantum channel, we described the recently proposed quantum algorithm that performs belief-propagation with \emph{quantum} messages (BPQM) on the factor graph of a classical error-correcting code.
We performed extensive analysis to show that it provides optimal decisions for the value of each bit and appears to optimally decode the entire transmitted codeword.
Although the numerical simulations convincingly show that BPQM appears to be quantum optimal, we would like to develop a mathematical proof that establishes this analytically.
It will be interesting to investigate if an optical receiver can be built based on BPQM and used to decode messages optimally in deep-space optical communications.
Furthermore, it remains to be investigated if such optimality can be proven for a general family of binary linear codes, perhaps those with tree factor graphs.

For quantum computing, we described a systematic algorithm to synthesize logical Clifford gates for stabilizer codes.
However, further research is required to ensure that the algorithm produces fault-tolerant circuits, at least under certain conditions.
Then we generalized the binary symplectic formalism to a large set of diagonal (QFD) gates in the Clifford hierarchy via symmetric matrices over rings of integers.
We rigorously derived their action on Pauli matrices.
We discussed potential applications of these results in quantum information, and it would be interesting to further explore these preliminary ideas.
Finally, we used the QFD formalism to fully characterize all stabilizer codes whose code subspace is preserved under the application of transversal $T, T^{\dagger}$ and identity gates on the physical (code) qubits.
Using this result, we provided useful corollaries about triorthogonal codes and the optimality of CSS codes for magic state distillation. 
It remains to be explored if our general recipe can be used to characterize codes that support any given QFD gate.

%% file: ch5_appendix.tex

\section{{MATLAB\textsuperscript{\textregistered}} Code for Algorithm~\ref{alg:symp_lineq_all}}
\label{sec:alg2_matlab}

\begin{lstlisting}
function F_all = find_all_symp_mat(U, V, I, J)

I = I(:)';
J = J(:)';
Ibar = setdiff(1:m,I);
Jbar = setdiff(1:m,J);
alpha = length(Ibar) + length(Jbar); 
tot = 2^(alpha*(alpha+1)/2);
F_all = cell(tot,1);

% Find one solution using symplectic transvections (Algorithm 1)
F0 = find_symp_mat(U([I, m+J], :), V);

A = mod(U * F0, 2);
Ainv = gf2matinv(A);  
IbJb = union(Ibar,Jbar);
Basis = A([IbJb, m+IbJb],:);  % these rows span the subspace W^{\perp} in Theorem 23
Subspace = mod(de2bi((0:2^(2*length(IbJb))-1)',2*length(IbJb)) * Basis, 2);

% Collect indices of free vectors in the top and bottom halves of Basis
% Note: these are now row indices of Basis, not row indices of A!!
[~, Basis_fixed_I, ~] = intersect(IbJb,I);  % intersect(IbJb,I) = intersect(I,Jbar)
[~, Basis_fixed_J, ~] = intersect(IbJb,J);  % intersect(IbJb,J) = intersect(Ibar,J)
Basis_fixed = [Basis_fixed_I, length(IbJb) + Basis_fixed_J];
Basis_free = setdiff(1:2*length(IbJb), Basis_fixed);

Choices = cell(alpha,1);

% Calculate all choices for each free vector using just conditions imposed
% by the fixed vectors in Basis (or equivalently in A)
for i = 1:alpha
    ind = Basis_free(i);
    h = zeros(1,length(Basis_fixed));    
    % Impose symplectic inner product of 1 with the "fixed" symplectic pair
    if (i <= length(Ibar))
        h(Basis_fixed == length(IbJb) + ind) = 1;
    else
        h(Basis_fixed == ind - length(IbJb)) = 1;
    end    
    % Check the necessary conditions on the symplectic inner products
    Innpdts = mod(Subspace * fftshift(Basis(Basis_fixed,:), 2)', 2);
    Choices{i,1} = Subspace(bi2de(Innpdts) == bi2de(h), :);
end

% First free vector has 2^(alpha) choices, second has 2^(alpha-1) choices and so on
for l = 0:(tot - 1)
    Bl = A;
    W = zeros(alpha,2*m);   % Rows are choices made for free vectors
                                 % W(i,:) corresponds to Basis(Basis_free(i),:)
    lbin = de2bi(l,alpha*(alpha+1)/2,'left-msb');
    v1_ind = bi2de(lbin(1,1:alpha),'left-msb') + 1;
    W(1,:) = Choices{1,1}(v1_ind,:);
    for i = 2:alpha
        % vi_ind loops through the 2^(alpha-(i-1)) valid choices for the i-th free vector
        vi_ind = bi2de(lbin(1,sum(alpha:-1:alpha-(i-2)) + (1:(alpha-(i-1)))),'left-msb') + 1;
        Innprods = mod(Choices{i,1} * fftshift(W,2)', 2);        
        % Impose symplectic inner product of 0 with chosen free vectors
        h = zeros(1,alpha);
        % Handle case when Basis contains a symplectic pair of free vectors
        if (i > length(Ibar))
            h(Basis_free == Basis_free(i) - length(IbJb)) = 1;
        end
        % Check the necessary and sufficient conditions on the symplectic inner products
        Ch_i = Choices{i,1}(bi2de(Innprods) == bi2de(h), :);
        W(i,:) = Ch_i(vi_ind,:);  % use the vi_ind-th valid choice for the i-th free vector
    end
    Bl([Ibar, m+Jbar], :) = W;  % replace rows of free vectors with current choices
    F = mod(Ainv * Bl, 2);          % this is the matrix F' in Theorem 23
    F_all{l+1,1} = mod(F0 * F, 2);
end    

end
\end{lstlisting}

\section{Enumeration of Solutions for the $\llbr 6,4,2 \rrbr$ Code}
\label{sec:css642_all_phy_ops}

Using the algorithms described in Section~\ref{sec:discuss} we enumerate all symplectic solutions for each logical operator described in Section~\ref{sec:css642}.
The physical circuits corresponding to these matrices can be obtained by decomposing them into products of elementary symplectic transformations in Table~\ref{tab:std_symp} and using their circuits described in Section~\ref{sec:elem_symp}.
An algorithm for performing this decomposition is given in the proof of Theorem~\ref{thm:Trung} (from~\cite{Can-2017a}).
Note that this decomposition is not unique.
The \texttt{MATLAB\textsuperscript{\textregistered}} programs for reproducing the following results, along with their circuits obtained from the above decomposition, are available at \url{https://github.com/nrenga/symplectic-arxiv18a}.
These programs can perform this task for any stabilizer code.

For the $\llbr 6,4,2 \rrbr$ code, there are $2^{r(r+1)/2} = 8$ possible symplectic solutions that satisfy the linear constraints imposed by each of~\eqref{eq:phase_maps},\eqref{eq:cz_maps},\eqref{eq:lcnot21_maps}, and~\eqref{eq:hadamard_maps}.
For each of the $4$ logical operators below, $F_1$ is the solution discussed in Section~\ref{sec:css642}.


\subsection{Logical Phase Gate $(\lP_1)$}
\label{sec:css_phase_all}

~

{\scriptsize
\begin{align*}
F_1 = 
\left[
\begin{array}{cccccc|cccccc}
1 & 0 & 0 & 0 & 0 & 0 & 0 & 0 & 0 & 0 & 0 & 0 \\
0 & 1 & 0 & 0 & 0 & 0 & 0 & 1 & 0 & 0 & 0 & 1 \\
0 & 0 & 1 & 0 & 0 & 0 & 0 & 0 & 0 & 0 & 0 & 0 \\
0 & 0 & 0 & 1 & 0 & 0 & 0 & 0 & 0 & 0 & 0 & 0 \\
0 & 0 & 0 & 0 & 1 & 0 & 0 & 0 & 0 & 0 & 0 & 0 \\
0 & 0 & 0 & 0 & 0 & 1 & 0 & 1 & 0 & 0 & 0 & 1 \\
\hline
0 & 0 & 0 & 0 & 0 & 0 & 1 & 0 & 0 & 0 & 0 & 0 \\
0 & 0 & 0 & 0 & 0 & 0 & 0 & 1 & 0 & 0 & 0 & 0 \\
0 & 0 & 0 & 0 & 0 & 0 & 0 & 0 & 1 & 0 & 0 & 0 \\
0 & 0 & 0 & 0 & 0 & 0 & 0 & 0 & 0 & 1 & 0 & 0 \\
0 & 0 & 0 & 0 & 0 & 0 & 0 & 0 & 0 & 0 & 1 & 0 \\
0 & 0 & 0 & 0 & 0 & 0 & 0 & 0 & 0 & 0 & 0 & 1 \\
\end{array}
\right] & ,
F_2 = 
\left[
\begin{array}{cccccc|cccccc}
1 & 0 & 0 & 0 & 0 & 0 & 1 & 1 & 1 & 1 & 1 & 1 \\
0 & 1 & 0 & 0 & 0 & 0 & 1 & 0 & 1 & 1 & 1 & 0 \\
0 & 0 & 1 & 0 & 0 & 0 & 1 & 1 & 1 & 1 & 1 & 1 \\
0 & 0 & 0 & 1 & 0 & 0 & 1 & 1 & 1 & 1 & 1 & 1 \\
0 & 0 & 0 & 0 & 1 & 0 & 1 & 1 & 1 & 1 & 1 & 1 \\
0 & 0 & 0 & 0 & 0 & 1 & 1 & 0 & 1 & 1 & 1 & 0 \\
\hline
0 & 0 & 0 & 0 & 0 & 0 & 1 & 0 & 0 & 0 & 0 & 0 \\
0 & 0 & 0 & 0 & 0 & 0 & 0 & 1 & 0 & 0 & 0 & 0 \\
0 & 0 & 0 & 0 & 0 & 0 & 0 & 0 & 1 & 0 & 0 & 0 \\
0 & 0 & 0 & 0 & 0 & 0 & 0 & 0 & 0 & 1 & 0 & 0 \\
0 & 0 & 0 & 0 & 0 & 0 & 0 & 0 & 0 & 0 & 1 & 0 \\
0 & 0 & 0 & 0 & 0 & 0 & 0 & 0 & 0 & 0 & 0 & 1 \\
\end{array}
\right] , \\~\\
F_3 = 
\left[
\begin{array}{cccccc|cccccc}
1 & 0 & 0 & 0 & 0 & 0 & 0 & 0 & 0 & 0 & 0 & 0 \\
0 & 1 & 0 & 0 & 0 & 0 & 0 & 1 & 0 & 0 & 0 & 1 \\
0 & 0 & 1 & 0 & 0 & 0 & 0 & 0 & 0 & 0 & 0 & 0 \\
0 & 0 & 0 & 1 & 0 & 0 & 0 & 0 & 0 & 0 & 0 & 0 \\
0 & 0 & 0 & 0 & 1 & 0 & 0 & 0 & 0 & 0 & 0 & 0 \\
0 & 0 & 0 & 0 & 0 & 1 & 0 & 1 & 0 & 0 & 0 & 1 \\
\hline
1 & 1 & 1 & 1 & 1 & 1 & 1 & 0 & 0 & 0 & 0 & 0 \\
1 & 1 & 1 & 1 & 1 & 1 & 0 & 1 & 0 & 0 & 0 & 0 \\
1 & 1 & 1 & 1 & 1 & 1 & 0 & 0 & 1 & 0 & 0 & 0 \\
1 & 1 & 1 & 1 & 1 & 1 & 0 & 0 & 0 & 1 & 0 & 0 \\
1 & 1 & 1 & 1 & 1 & 1 & 0 & 0 & 0 & 0 & 1 & 0 \\
1 & 1 & 1 & 1 & 1 & 1 & 0 & 0 & 0 & 0 & 0 & 1 \\
\end{array}
\right] & ,
F_4 = 
\left[
\begin{array}{cccccc|cccccc}
1 & 0 & 0 & 0 & 0 & 0 & 1 & 1 & 1 & 1 & 1 & 1 \\
0 & 1 & 0 & 0 & 0 & 0 & 1 & 0 & 1 & 1 & 1 & 0 \\
0 & 0 & 1 & 0 & 0 & 0 & 1 & 1 & 1 & 1 & 1 & 1 \\
0 & 0 & 0 & 1 & 0 & 0 & 1 & 1 & 1 & 1 & 1 & 1 \\
0 & 0 & 0 & 0 & 1 & 0 & 1 & 1 & 1 & 1 & 1 & 1 \\
0 & 0 & 0 & 0 & 0 & 1 & 1 & 0 & 1 & 1 & 1 & 0 \\
\hline
1 & 1 & 1 & 1 & 1 & 1 & 1 & 0 & 0 & 0 & 0 & 0 \\
1 & 1 & 1 & 1 & 1 & 1 & 0 & 1 & 0 & 0 & 0 & 0 \\
1 & 1 & 1 & 1 & 1 & 1 & 0 & 0 & 1 & 0 & 0 & 0 \\
1 & 1 & 1 & 1 & 1 & 1 & 0 & 0 & 0 & 1 & 0 & 0 \\
1 & 1 & 1 & 1 & 1 & 1 & 0 & 0 & 0 & 0 & 1 & 0 \\
1 & 1 & 1 & 1 & 1 & 1 & 0 & 0 & 0 & 0 & 0 & 1 \\
\end{array}
\right] , \\~\\
F_5 = 
\left[
\begin{array}{cccccc|cccccc}
0 & 1 & 1 & 1 & 1 & 1 & 0 & 0 & 0 & 0 & 0 & 0 \\
1 & 0 & 1 & 1 & 1 & 1 & 0 & 1 & 0 & 0 & 0 & 1 \\
1 & 1 & 0 & 1 & 1 & 1 & 0 & 0 & 0 & 0 & 0 & 0 \\
1 & 1 & 1 & 0 & 1 & 1 & 0 & 0 & 0 & 0 & 0 & 0 \\
1 & 1 & 1 & 1 & 0 & 1 & 0 & 0 & 0 & 0 & 0 & 0 \\
1 & 1 & 1 & 1 & 1 & 0 & 0 & 1 & 0 & 0 & 0 & 1 \\
\hline
0 & 0 & 0 & 0 & 0 & 0 & 0 & 1 & 1 & 1 & 1 & 1 \\
0 & 0 & 0 & 0 & 0 & 0 & 1 & 0 & 1 & 1 & 1 & 1 \\
0 & 0 & 0 & 0 & 0 & 0 & 1 & 1 & 0 & 1 & 1 & 1 \\
0 & 0 & 0 & 0 & 0 & 0 & 1 & 1 & 1 & 0 & 1 & 1 \\
0 & 0 & 0 & 0 & 0 & 0 & 1 & 1 & 1 & 1 & 0 & 1 \\
0 & 0 & 0 & 0 & 0 & 0 & 1 & 1 & 1 & 1 & 1 & 0 \\
\end{array}
\right] & , 
F_6 = 
\left[
\begin{array}{cccccc|cccccc}
0 & 1 & 1 & 1 & 1 & 1 & 1 & 1 & 1 & 1 & 1 & 1 \\
1 & 0 & 1 & 1 & 1 & 1 & 1 & 0 & 1 & 1 & 1 & 0 \\
1 & 1 & 0 & 1 & 1 & 1 & 1 & 1 & 1 & 1 & 1 & 1 \\
1 & 1 & 1 & 0 & 1 & 1 & 1 & 1 & 1 & 1 & 1 & 1 \\
1 & 1 & 1 & 1 & 0 & 1 & 1 & 1 & 1 & 1 & 1 & 1 \\
1 & 1 & 1 & 1 & 1 & 0 & 1 & 0 & 1 & 1 & 1 & 0 \\
\hline
0 & 0 & 0 & 0 & 0 & 0 & 0 & 1 & 1 & 1 & 1 & 1 \\
0 & 0 & 0 & 0 & 0 & 0 & 1 & 0 & 1 & 1 & 1 & 1 \\
0 & 0 & 0 & 0 & 0 & 0 & 1 & 1 & 0 & 1 & 1 & 1 \\
0 & 0 & 0 & 0 & 0 & 0 & 1 & 1 & 1 & 0 & 1 & 1 \\
0 & 0 & 0 & 0 & 0 & 0 & 1 & 1 & 1 & 1 & 0 & 1 \\
0 & 0 & 0 & 0 & 0 & 0 & 1 & 1 & 1 & 1 & 1 & 0 \\
\end{array}
\right] , \\~\\
F_7 = 
\left[
\begin{array}{cccccc|cccccc}
0 & 1 & 1 & 1 & 1 & 1 & 0 & 0 & 0 & 0 & 0 & 0 \\
1 & 0 & 1 & 1 & 1 & 1 & 0 & 1 & 0 & 0 & 0 & 1 \\
1 & 1 & 0 & 1 & 1 & 1 & 0 & 0 & 0 & 0 & 0 & 0 \\
1 & 1 & 1 & 0 & 1 & 1 & 0 & 0 & 0 & 0 & 0 & 0 \\
1 & 1 & 1 & 1 & 0 & 1 & 0 & 0 & 0 & 0 & 0 & 0 \\
1 & 1 & 1 & 1 & 1 & 0 & 0 & 1 & 0 & 0 & 0 & 1 \\
\hline
1 & 1 & 1 & 1 & 1 & 1 & 0 & 1 & 1 & 1 & 1 & 1 \\
1 & 1 & 1 & 1 & 1 & 1 & 1 & 0 & 1 & 1 & 1 & 1 \\
1 & 1 & 1 & 1 & 1 & 1 & 1 & 1 & 0 & 1 & 1 & 1 \\
1 & 1 & 1 & 1 & 1 & 1 & 1 & 1 & 1 & 0 & 1 & 1 \\
1 & 1 & 1 & 1 & 1 & 1 & 1 & 1 & 1 & 1 & 0 & 1 \\
1 & 1 & 1 & 1 & 1 & 1 & 1 & 1 & 1 & 1 & 1 & 0 \\
\end{array}
\right] & , 
F_8 = 
\left[
\begin{array}{cccccc|cccccc}
0 & 1 & 1 & 1 & 1 & 1 & 1 & 1 & 1 & 1 & 1 & 1 \\
1 & 0 & 1 & 1 & 1 & 1 & 1 & 0 & 1 & 1 & 1 & 0 \\
1 & 1 & 0 & 1 & 1 & 1 & 1 & 1 & 1 & 1 & 1 & 1 \\
1 & 1 & 1 & 0 & 1 & 1 & 1 & 1 & 1 & 1 & 1 & 1 \\
1 & 1 & 1 & 1 & 0 & 1 & 1 & 1 & 1 & 1 & 1 & 1 \\
1 & 1 & 1 & 1 & 1 & 0 & 1 & 0 & 1 & 1 & 1 & 0 \\
\hline
1 & 1 & 1 & 1 & 1 & 1 & 0 & 1 & 1 & 1 & 1 & 1 \\
1 & 1 & 1 & 1 & 1 & 1 & 1 & 0 & 1 & 1 & 1 & 1 \\
1 & 1 & 1 & 1 & 1 & 1 & 1 & 1 & 0 & 1 & 1 & 1 \\
1 & 1 & 1 & 1 & 1 & 1 & 1 & 1 & 1 & 0 & 1 & 1 \\
1 & 1 & 1 & 1 & 1 & 1 & 1 & 1 & 1 & 1 & 0 & 1 \\
1 & 1 & 1 & 1 & 1 & 1 & 1 & 1 & 1 & 1 & 1 & 0 \\
\end{array}
\right] .
\end{align*}
}

\subsection{Logical Controlled-Z Gate $(\lcz{1}{2})$}
\label{sec:css_cz_all}


{\scriptsize
\begin{align*}
F_1 = 
\left[
\begin{array}{cccccc|cccccc}
1 & 0 & 0 & 0 & 0 & 0 & 0 & 0 & 0 & 0 & 0 & 0 \\
0 & 1 & 0 & 0 & 0 & 0 & 0 & 0 & 1 & 0 & 0 & 1 \\
0 & 0 & 1 & 0 & 0 & 0 & 0 & 1 & 0 & 0 & 0 & 1 \\
0 & 0 & 0 & 1 & 0 & 0 & 0 & 0 & 0 & 0 & 0 & 0 \\
0 & 0 & 0 & 0 & 1 & 0 & 0 & 0 & 0 & 0 & 0 & 0 \\
0 & 0 & 0 & 0 & 0 & 1 & 0 & 1 & 1 & 0 & 0 & 0 \\
\hline
0 & 0 & 0 & 0 & 0 & 0 & 1 & 0 & 0 & 0 & 0 & 0 \\
0 & 0 & 0 & 0 & 0 & 0 & 0 & 1 & 0 & 0 & 0 & 0 \\
0 & 0 & 0 & 0 & 0 & 0 & 0 & 0 & 1 & 0 & 0 & 0 \\
0 & 0 & 0 & 0 & 0 & 0 & 0 & 0 & 0 & 1 & 0 & 0 \\
0 & 0 & 0 & 0 & 0 & 0 & 0 & 0 & 0 & 0 & 1 & 0 \\
0 & 0 & 0 & 0 & 0 & 0 & 0 & 0 & 0 & 0 & 0 & 1 \\
\end{array}
\right] & , 
F_2 = 
\left[
\begin{array}{cccccc|cccccc}
1 & 0 & 0 & 0 & 0 & 0 & 1 & 1 & 1 & 1 & 1 & 1 \\
0 & 1 & 0 & 0 & 0 & 0 & 1 & 1 & 0 & 1 & 1 & 0 \\
0 & 0 & 1 & 0 & 0 & 0 & 1 & 0 & 1 & 1 & 1 & 0 \\
0 & 0 & 0 & 1 & 0 & 0 & 1 & 1 & 1 & 1 & 1 & 1 \\
0 & 0 & 0 & 0 & 1 & 0 & 1 & 1 & 1 & 1 & 1 & 1 \\
0 & 0 & 0 & 0 & 0 & 1 & 1 & 0 & 0 & 1 & 1 & 1 \\
\hline
0 & 0 & 0 & 0 & 0 & 0 & 1 & 0 & 0 & 0 & 0 & 0 \\
0 & 0 & 0 & 0 & 0 & 0 & 0 & 1 & 0 & 0 & 0 & 0 \\
0 & 0 & 0 & 0 & 0 & 0 & 0 & 0 & 1 & 0 & 0 & 0 \\
0 & 0 & 0 & 0 & 0 & 0 & 0 & 0 & 0 & 1 & 0 & 0 \\
0 & 0 & 0 & 0 & 0 & 0 & 0 & 0 & 0 & 0 & 1 & 0 \\
0 & 0 & 0 & 0 & 0 & 0 & 0 & 0 & 0 & 0 & 0 & 1 \\
\end{array}
\right] , \\~\\
F_3 = 
\left[
\begin{array}{cccccc|cccccc}
1 & 0 & 0 & 0 & 0 & 0 & 0 & 0 & 0 & 0 & 0 & 0 \\
0 & 1 & 0 & 0 & 0 & 0 & 0 & 0 & 1 & 0 & 0 & 1 \\
0 & 0 & 1 & 0 & 0 & 0 & 0 & 1 & 0 & 0 & 0 & 1 \\
0 & 0 & 0 & 1 & 0 & 0 & 0 & 0 & 0 & 0 & 0 & 0 \\
0 & 0 & 0 & 0 & 1 & 0 & 0 & 0 & 0 & 0 & 0 & 0 \\
0 & 0 & 0 & 0 & 0 & 1 & 0 & 1 & 1 & 0 & 0 & 0 \\
\hline
1 & 1 & 1 & 1 & 1 & 1 & 1 & 0 & 0 & 0 & 0 & 0 \\
1 & 1 & 1 & 1 & 1 & 1 & 0 & 1 & 0 & 0 & 0 & 0 \\
1 & 1 & 1 & 1 & 1 & 1 & 0 & 0 & 1 & 0 & 0 & 0 \\
1 & 1 & 1 & 1 & 1 & 1 & 0 & 0 & 0 & 1 & 0 & 0 \\
1 & 1 & 1 & 1 & 1 & 1 & 0 & 0 & 0 & 0 & 1 & 0 \\
1 & 1 & 1 & 1 & 1 & 1 & 0 & 0 & 0 & 0 & 0 & 1 \\
\end{array}
\right] & , 
F_4 = 
\left[
\begin{array}{cccccc|cccccc}
1 & 0 & 0 & 0 & 0 & 0 & 1 & 1 & 1 & 1 & 1 & 1 \\
0 & 1 & 0 & 0 & 0 & 0 & 1 & 1 & 0 & 1 & 1 & 0 \\
0 & 0 & 1 & 0 & 0 & 0 & 1 & 0 & 1 & 1 & 1 & 0 \\
0 & 0 & 0 & 1 & 0 & 0 & 1 & 1 & 1 & 1 & 1 & 1 \\
0 & 0 & 0 & 0 & 1 & 0 & 1 & 1 & 1 & 1 & 1 & 1 \\
0 & 0 & 0 & 0 & 0 & 1 & 1 & 0 & 0 & 1 & 1 & 1 \\
\hline
1 & 1 & 1 & 1 & 1 & 1 & 1 & 0 & 0 & 0 & 0 & 0 \\
1 & 1 & 1 & 1 & 1 & 1 & 0 & 1 & 0 & 0 & 0 & 0 \\
1 & 1 & 1 & 1 & 1 & 1 & 0 & 0 & 1 & 0 & 0 & 0 \\
1 & 1 & 1 & 1 & 1 & 1 & 0 & 0 & 0 & 1 & 0 & 0 \\
1 & 1 & 1 & 1 & 1 & 1 & 0 & 0 & 0 & 0 & 1 & 0 \\
1 & 1 & 1 & 1 & 1 & 1 & 0 & 0 & 0 & 0 & 0 & 1 \\
\end{array}
\right] , \\~\\
F_5 = 
\left[
\begin{array}{cccccc|cccccc}
0 & 1 & 1 & 1 & 1 & 1 & 0 & 0 & 0 & 0 & 0 & 0 \\
1 & 0 & 1 & 1 & 1 & 1 & 0 & 0 & 1 & 0 & 0 & 1 \\
1 & 1 & 0 & 1 & 1 & 1 & 0 & 1 & 0 & 0 & 0 & 1 \\
1 & 1 & 1 & 0 & 1 & 1 & 0 & 0 & 0 & 0 & 0 & 0 \\
1 & 1 & 1 & 1 & 0 & 1 & 0 & 0 & 0 & 0 & 0 & 0 \\
1 & 1 & 1 & 1 & 1 & 0 & 0 & 1 & 1 & 0 & 0 & 0 \\
\hline
0 & 0 & 0 & 0 & 0 & 0 & 0 & 1 & 1 & 1 & 1 & 1 \\
0 & 0 & 0 & 0 & 0 & 0 & 1 & 0 & 1 & 1 & 1 & 1 \\
0 & 0 & 0 & 0 & 0 & 0 & 1 & 1 & 0 & 1 & 1 & 1 \\
0 & 0 & 0 & 0 & 0 & 0 & 1 & 1 & 1 & 0 & 1 & 1 \\
0 & 0 & 0 & 0 & 0 & 0 & 1 & 1 & 1 & 1 & 0 & 1 \\
0 & 0 & 0 & 0 & 0 & 0 & 1 & 1 & 1 & 1 & 1 & 0 \\
\end{array}
\right] & , 
F_6 = 
\left[
\begin{array}{cccccc|cccccc}
0 & 1 & 1 & 1 & 1 & 1 & 1 & 1 & 1 & 1 & 1 & 1 \\
1 & 0 & 1 & 1 & 1 & 1 & 1 & 1 & 0 & 1 & 1 & 0 \\
1 & 1 & 0 & 1 & 1 & 1 & 1 & 0 & 1 & 1 & 1 & 0 \\
1 & 1 & 1 & 0 & 1 & 1 & 1 & 1 & 1 & 1 & 1 & 1 \\
1 & 1 & 1 & 1 & 0 & 1 & 1 & 1 & 1 & 1 & 1 & 1 \\
1 & 1 & 1 & 1 & 1 & 0 & 1 & 0 & 0 & 1 & 1 & 1 \\
\hline
0 & 0 & 0 & 0 & 0 & 0 & 0 & 1 & 1 & 1 & 1 & 1 \\
0 & 0 & 0 & 0 & 0 & 0 & 1 & 0 & 1 & 1 & 1 & 1 \\
0 & 0 & 0 & 0 & 0 & 0 & 1 & 1 & 0 & 1 & 1 & 1 \\
0 & 0 & 0 & 0 & 0 & 0 & 1 & 1 & 1 & 0 & 1 & 1 \\
0 & 0 & 0 & 0 & 0 & 0 & 1 & 1 & 1 & 1 & 0 & 1 \\
0 & 0 & 0 & 0 & 0 & 0 & 1 & 1 & 1 & 1 & 1 & 0 \\
\end{array}
\right] , \\~\\
F_7 = 
\left[
\begin{array}{cccccc|cccccc}
0 & 1 & 1 & 1 & 1 & 1 & 0 & 0 & 0 & 0 & 0 & 0 \\
1 & 0 & 1 & 1 & 1 & 1 & 0 & 0 & 1 & 0 & 0 & 1 \\
1 & 1 & 0 & 1 & 1 & 1 & 0 & 1 & 0 & 0 & 0 & 1 \\
1 & 1 & 1 & 0 & 1 & 1 & 0 & 0 & 0 & 0 & 0 & 0 \\
1 & 1 & 1 & 1 & 0 & 1 & 0 & 0 & 0 & 0 & 0 & 0 \\
1 & 1 & 1 & 1 & 1 & 0 & 0 & 1 & 1 & 0 & 0 & 0 \\
\hline
1 & 1 & 1 & 1 & 1 & 1 & 0 & 1 & 1 & 1 & 1 & 1 \\
1 & 1 & 1 & 1 & 1 & 1 & 1 & 0 & 1 & 1 & 1 & 1 \\
1 & 1 & 1 & 1 & 1 & 1 & 1 & 1 & 0 & 1 & 1 & 1 \\
1 & 1 & 1 & 1 & 1 & 1 & 1 & 1 & 1 & 0 & 1 & 1 \\
1 & 1 & 1 & 1 & 1 & 1 & 1 & 1 & 1 & 1 & 0 & 1 \\
1 & 1 & 1 & 1 & 1 & 1 & 1 & 1 & 1 & 1 & 1 & 0 \\
\end{array}
\right] & , 
F_8 = 
\left[
\begin{array}{cccccc|cccccc}
0 & 1 & 1 & 1 & 1 & 1 & 1 & 1 & 1 & 1 & 1 & 1 \\
1 & 0 & 1 & 1 & 1 & 1 & 1 & 1 & 0 & 1 & 1 & 0 \\
1 & 1 & 0 & 1 & 1 & 1 & 1 & 0 & 1 & 1 & 1 & 0 \\
1 & 1 & 1 & 0 & 1 & 1 & 1 & 1 & 1 & 1 & 1 & 1 \\
1 & 1 & 1 & 1 & 0 & 1 & 1 & 1 & 1 & 1 & 1 & 1 \\
1 & 1 & 1 & 1 & 1 & 0 & 1 & 0 & 0 & 1 & 1 & 1 \\
\hline
1 & 1 & 1 & 1 & 1 & 1 & 0 & 1 & 1 & 1 & 1 & 1 \\
1 & 1 & 1 & 1 & 1 & 1 & 1 & 0 & 1 & 1 & 1 & 1 \\
1 & 1 & 1 & 1 & 1 & 1 & 1 & 1 & 0 & 1 & 1 & 1 \\
1 & 1 & 1 & 1 & 1 & 1 & 1 & 1 & 1 & 0 & 1 & 1 \\
1 & 1 & 1 & 1 & 1 & 1 & 1 & 1 & 1 & 1 & 0 & 1 \\
1 & 1 & 1 & 1 & 1 & 1 & 1 & 1 & 1 & 1 & 1 & 0 \\
\end{array}
\right] .
\end{align*}
}

\subsection{Logical Controlled-NOT Gate $(\lcnot{2}{1})$}
\label{sec:css_cnot_all}


{\scriptsize
\begin{align*}
F_1 = 
\left[
\begin{array}{cccccc|cccccc}
1 & 0 & 0 & 0 & 0 & 0 & 0 & 0 & 0 & 0 & 0 & 0 \\
0 & 1 & 0 & 0 & 0 & 0 & 0 & 0 & 0 & 0 & 0 & 0 \\
1 & 1 & 1 & 0 & 0 & 0 & 0 & 0 & 0 & 0 & 0 & 0 \\
0 & 0 & 0 & 1 & 0 & 0 & 0 & 0 & 0 & 0 & 0 & 0 \\
0 & 0 & 0 & 0 & 1 & 0 & 0 & 0 & 0 & 0 & 0 & 0 \\
1 & 1 & 0 & 0 & 0 & 1 & 0 & 0 & 0 & 0 & 0 & 0 \\
\hline
0 & 0 & 0 & 0 & 0 & 0 & 1 & 0 & 1 & 0 & 0 & 1 \\
0 & 0 & 0 & 0 & 0 & 0 & 0 & 1 & 1 & 0 & 0 & 1 \\
0 & 0 & 0 & 0 & 0 & 0 & 0 & 0 & 1 & 0 & 0 & 0 \\
0 & 0 & 0 & 0 & 0 & 0 & 0 & 0 & 0 & 1 & 0 & 0 \\
0 & 0 & 0 & 0 & 0 & 0 & 0 & 0 & 0 & 0 & 1 & 0 \\
0 & 0 & 0 & 0 & 0 & 0 & 0 & 0 & 0 & 0 & 0 & 1 \\
\end{array}
\right] & , 
F_2 = 
\left[
\begin{array}{cccccc|cccccc}
1 & 0 & 0 & 0 & 0 & 0 & 1 & 1 & 1 & 1 & 1 & 1 \\
0 & 1 & 0 & 0 & 0 & 0 & 1 & 1 & 1 & 1 & 1 & 1 \\
1 & 1 & 1 & 0 & 0 & 0 & 1 & 1 & 1 & 1 & 1 & 1 \\
0 & 0 & 0 & 1 & 0 & 0 & 1 & 1 & 1 & 1 & 1 & 1 \\
0 & 0 & 0 & 0 & 1 & 0 & 1 & 1 & 1 & 1 & 1 & 1 \\
1 & 1 & 0 & 0 & 0 & 1 & 1 & 1 & 1 & 1 & 1 & 1 \\
\hline
0 & 0 & 0 & 0 & 0 & 0 & 1 & 0 & 1 & 0 & 0 & 1 \\
0 & 0 & 0 & 0 & 0 & 0 & 0 & 1 & 1 & 0 & 0 & 1 \\
0 & 0 & 0 & 0 & 0 & 0 & 0 & 0 & 1 & 0 & 0 & 0 \\
0 & 0 & 0 & 0 & 0 & 0 & 0 & 0 & 0 & 1 & 0 & 0 \\
0 & 0 & 0 & 0 & 0 & 0 & 0 & 0 & 0 & 0 & 1 & 0 \\
0 & 0 & 0 & 0 & 0 & 0 & 0 & 0 & 0 & 0 & 0 & 1 \\
\end{array}
\right] , \\~\\
F_3 = 
\left[
\begin{array}{cccccc|cccccc}
1 & 0 & 0 & 0 & 0 & 0 & 0 & 0 & 0 & 0 & 0 & 0 \\
0 & 1 & 0 & 0 & 0 & 0 & 0 & 0 & 0 & 0 & 0 & 0 \\
1 & 1 & 1 & 0 & 0 & 0 & 0 & 0 & 0 & 0 & 0 & 0 \\
0 & 0 & 0 & 1 & 0 & 0 & 0 & 0 & 0 & 0 & 0 & 0 \\
0 & 0 & 0 & 0 & 1 & 0 & 0 & 0 & 0 & 0 & 0 & 0 \\
1 & 1 & 0 & 0 & 0 & 1 & 0 & 0 & 0 & 0 & 0 & 0 \\
\hline
1 & 1 & 1 & 1 & 1 & 1 & 1 & 0 & 1 & 0 & 0 & 1 \\
1 & 1 & 1 & 1 & 1 & 1 & 0 & 1 & 1 & 0 & 0 & 1 \\
1 & 1 & 1 & 1 & 1 & 1 & 0 & 0 & 1 & 0 & 0 & 0 \\
1 & 1 & 1 & 1 & 1 & 1 & 0 & 0 & 0 & 1 & 0 & 0 \\
1 & 1 & 1 & 1 & 1 & 1 & 0 & 0 & 0 & 0 & 1 & 0 \\
1 & 1 & 1 & 1 & 1 & 1 & 0 & 0 & 0 & 0 & 0 & 1 \\
\end{array}
\right] & , 
F_4 = 
\left[
\begin{array}{cccccc|cccccc}
1 & 0 & 0 & 0 & 0 & 0 & 1 & 1 & 1 & 1 & 1 & 1 \\
0 & 1 & 0 & 0 & 0 & 0 & 1 & 1 & 1 & 1 & 1 & 1 \\
1 & 1 & 1 & 0 & 0 & 0 & 1 & 1 & 1 & 1 & 1 & 1 \\
0 & 0 & 0 & 1 & 0 & 0 & 1 & 1 & 1 & 1 & 1 & 1 \\
0 & 0 & 0 & 0 & 1 & 0 & 1 & 1 & 1 & 1 & 1 & 1 \\
1 & 1 & 0 & 0 & 0 & 1 & 1 & 1 & 1 & 1 & 1 & 1 \\
\hline
1 & 1 & 1 & 1 & 1 & 1 & 1 & 0 & 1 & 0 & 0 & 1 \\
1 & 1 & 1 & 1 & 1 & 1 & 0 & 1 & 1 & 0 & 0 & 1 \\
1 & 1 & 1 & 1 & 1 & 1 & 0 & 0 & 1 & 0 & 0 & 0 \\
1 & 1 & 1 & 1 & 1 & 1 & 0 & 0 & 0 & 1 & 0 & 0 \\
1 & 1 & 1 & 1 & 1 & 1 & 0 & 0 & 0 & 0 & 1 & 0 \\
1 & 1 & 1 & 1 & 1 & 1 & 0 & 0 & 0 & 0 & 0 & 1 \\
\end{array}
\right]  , \\~\\
F_5 = 
\left[
\begin{array}{cccccc|cccccc}
0 & 1 & 1 & 1 & 1 & 1 & 0 & 0 & 0 & 0 & 0 & 0 \\
1 & 0 & 1 & 1 & 1 & 1 & 0 & 0 & 0 & 0 & 0 & 0 \\
0 & 0 & 0 & 1 & 1 & 1 & 0 & 0 & 0 & 0 & 0 & 0 \\
1 & 1 & 1 & 0 & 1 & 1 & 0 & 0 & 0 & 0 & 0 & 0 \\
1 & 1 & 1 & 1 & 0 & 1 & 0 & 0 & 0 & 0 & 0 & 0 \\
0 & 0 & 1 & 1 & 1 & 0 & 0 & 0 & 0 & 0 & 0 & 0 \\
\hline
0 & 0 & 0 & 0 & 0 & 0 & 0 & 1 & 0 & 1 & 1 & 0 \\
0 & 0 & 0 & 0 & 0 & 0 & 1 & 0 & 0 & 1 & 1 & 0 \\
0 & 0 & 0 & 0 & 0 & 0 & 1 & 1 & 0 & 1 & 1 & 1 \\
0 & 0 & 0 & 0 & 0 & 0 & 1 & 1 & 1 & 0 & 1 & 1 \\
0 & 0 & 0 & 0 & 0 & 0 & 1 & 1 & 1 & 1 & 0 & 1 \\
0 & 0 & 0 & 0 & 0 & 0 & 1 & 1 & 1 & 1 & 1 & 0 \\
\end{array}
\right] & , 
F_6 = 
\left[
\begin{array}{cccccc|cccccc}
0 & 1 & 1 & 1 & 1 & 1 & 1 & 1 & 1 & 1 & 1 & 1 \\
1 & 0 & 1 & 1 & 1 & 1 & 1 & 1 & 1 & 1 & 1 & 1 \\
0 & 0 & 0 & 1 & 1 & 1 & 1 & 1 & 1 & 1 & 1 & 1 \\
1 & 1 & 1 & 0 & 1 & 1 & 1 & 1 & 1 & 1 & 1 & 1 \\
1 & 1 & 1 & 1 & 0 & 1 & 1 & 1 & 1 & 1 & 1 & 1 \\
0 & 0 & 1 & 1 & 1 & 0 & 1 & 1 & 1 & 1 & 1 & 1 \\
\hline
0 & 0 & 0 & 0 & 0 & 0 & 0 & 1 & 0 & 1 & 1 & 0 \\
0 & 0 & 0 & 0 & 0 & 0 & 1 & 0 & 0 & 1 & 1 & 0 \\
0 & 0 & 0 & 0 & 0 & 0 & 1 & 1 & 0 & 1 & 1 & 1 \\
0 & 0 & 0 & 0 & 0 & 0 & 1 & 1 & 1 & 0 & 1 & 1 \\
0 & 0 & 0 & 0 & 0 & 0 & 1 & 1 & 1 & 1 & 0 & 1 \\
0 & 0 & 0 & 0 & 0 & 0 & 1 & 1 & 1 & 1 & 1 & 0 \\
\end{array}
\right] , \\~\\
F_7 = 
\left[
\begin{array}{cccccc|cccccc}
0 & 1 & 1 & 1 & 1 & 1 & 0 & 0 & 0 & 0 & 0 & 0 \\
1 & 0 & 1 & 1 & 1 & 1 & 0 & 0 & 0 & 0 & 0 & 0 \\
0 & 0 & 0 & 1 & 1 & 1 & 0 & 0 & 0 & 0 & 0 & 0 \\
1 & 1 & 1 & 0 & 1 & 1 & 0 & 0 & 0 & 0 & 0 & 0 \\
1 & 1 & 1 & 1 & 0 & 1 & 0 & 0 & 0 & 0 & 0 & 0 \\
0 & 0 & 1 & 1 & 1 & 0 & 0 & 0 & 0 & 0 & 0 & 0 \\
\hline
1 & 1 & 1 & 1 & 1 & 1 & 0 & 1 & 0 & 1 & 1 & 0 \\
1 & 1 & 1 & 1 & 1 & 1 & 1 & 0 & 0 & 1 & 1 & 0 \\
1 & 1 & 1 & 1 & 1 & 1 & 1 & 1 & 0 & 1 & 1 & 1 \\
1 & 1 & 1 & 1 & 1 & 1 & 1 & 1 & 1 & 0 & 1 & 1 \\
1 & 1 & 1 & 1 & 1 & 1 & 1 & 1 & 1 & 1 & 0 & 1 \\
1 & 1 & 1 & 1 & 1 & 1 & 1 & 1 & 1 & 1 & 1 & 0 \\
\end{array}
\right] & , 
F_8 = 
\left[
\begin{array}{cccccc|cccccc}
0 & 1 & 1 & 1 & 1 & 1 & 1 & 1 & 1 & 1 & 1 & 1 \\
1 & 0 & 1 & 1 & 1 & 1 & 1 & 1 & 1 & 1 & 1 & 1 \\
0 & 0 & 0 & 1 & 1 & 1 & 1 & 1 & 1 & 1 & 1 & 1 \\
1 & 1 & 1 & 0 & 1 & 1 & 1 & 1 & 1 & 1 & 1 & 1 \\
1 & 1 & 1 & 1 & 0 & 1 & 1 & 1 & 1 & 1 & 1 & 1 \\
0 & 0 & 1 & 1 & 1 & 0 & 1 & 1 & 1 & 1 & 1 & 1 \\
\hline
1 & 1 & 1 & 1 & 1 & 1 & 0 & 1 & 0 & 1 & 1 & 0 \\
1 & 1 & 1 & 1 & 1 & 1 & 1 & 0 & 0 & 1 & 1 & 0 \\
1 & 1 & 1 & 1 & 1 & 1 & 1 & 1 & 0 & 1 & 1 & 1 \\
1 & 1 & 1 & 1 & 1 & 1 & 1 & 1 & 1 & 0 & 1 & 1 \\
1 & 1 & 1 & 1 & 1 & 1 & 1 & 1 & 1 & 1 & 0 & 1 \\
1 & 1 & 1 & 1 & 1 & 1 & 1 & 1 & 1 & 1 & 1 & 0 \\
\end{array}
\right] .
\end{align*}
}

\subsection{Logical Targeted Hadamard Gate $(\lH_1)$}
\label{sec:css_had_all}


{\scriptsize
\begin{align*}
F_1 = 
\left[
\begin{array}{cccccc|cccccc}
1 & 0 & 0 & 0 & 0 & 0 & 0 & 0 & 0 & 0 & 0 & 0 \\
1 & 0 & 0 & 0 & 0 & 0 & 0 & 1 & 0 & 0 & 0 & 1 \\
0 & 0 & 1 & 0 & 0 & 0 & 0 & 0 & 0 & 0 & 0 & 0 \\
0 & 0 & 0 & 1 & 0 & 0 & 0 & 0 & 0 & 0 & 0 & 0 \\
0 & 0 & 0 & 0 & 1 & 0 & 0 & 0 & 0 & 0 & 0 & 0 \\
1 & 1 & 0 & 0 & 0 & 1 & 0 & 1 & 0 & 0 & 0 & 1 \\
\hline
1 & 1 & 0 & 0 & 0 & 0 & 1 & 1 & 0 & 0 & 0 & 1 \\
1 & 1 & 0 & 0 & 0 & 0 & 0 & 0 & 0 & 0 & 0 & 1 \\
0 & 0 & 0 & 0 & 0 & 0 & 0 & 0 & 1 & 0 & 0 & 0 \\
0 & 0 & 0 & 0 & 0 & 0 & 0 & 0 & 0 & 1 & 0 & 0 \\
0 & 0 & 0 & 0 & 0 & 0 & 0 & 0 & 0 & 0 & 1 & 0 \\
0 & 0 & 0 & 0 & 0 & 0 & 0 & 0 & 0 & 0 & 0 & 1 \\
\end{array}
\right] & , 
F_2 = 
\left[
\begin{array}{cccccc|cccccc}
1 & 0 & 0 & 0 & 0 & 0 & 1 & 1 & 1 & 1 & 1 & 1 \\
1 & 0 & 0 & 0 & 0 & 0 & 1 & 0 & 1 & 1 & 1 & 0 \\
0 & 0 & 1 & 0 & 0 & 0 & 1 & 1 & 1 & 1 & 1 & 1 \\
0 & 0 & 0 & 1 & 0 & 0 & 1 & 1 & 1 & 1 & 1 & 1 \\
0 & 0 & 0 & 0 & 1 & 0 & 1 & 1 & 1 & 1 & 1 & 1 \\
1 & 1 & 0 & 0 & 0 & 1 & 1 & 0 & 1 & 1 & 1 & 0 \\
\hline
1 & 1 & 0 & 0 & 0 & 0 & 1 & 1 & 0 & 0 & 0 & 1 \\
1 & 1 & 0 & 0 & 0 & 0 & 0 & 0 & 0 & 0 & 0 & 1 \\
0 & 0 & 0 & 0 & 0 & 0 & 0 & 0 & 1 & 0 & 0 & 0 \\
0 & 0 & 0 & 0 & 0 & 0 & 0 & 0 & 0 & 1 & 0 & 0 \\
0 & 0 & 0 & 0 & 0 & 0 & 0 & 0 & 0 & 0 & 1 & 0 \\
0 & 0 & 0 & 0 & 0 & 0 & 0 & 0 & 0 & 0 & 0 & 1 \\
\end{array}
\right] , \\~\\
F_3 = 
\left[
\begin{array}{cccccc|cccccc}
1 & 0 & 0 & 0 & 0 & 0 & 0 & 0 & 0 & 0 & 0 & 0 \\
1 & 0 & 0 & 0 & 0 & 0 & 0 & 1 & 0 & 0 & 0 & 1 \\
0 & 0 & 1 & 0 & 0 & 0 & 0 & 0 & 0 & 0 & 0 & 0 \\
0 & 0 & 0 & 1 & 0 & 0 & 0 & 0 & 0 & 0 & 0 & 0 \\
0 & 0 & 0 & 0 & 1 & 0 & 0 & 0 & 0 & 0 & 0 & 0 \\
1 & 1 & 0 & 0 & 0 & 1 & 0 & 1 & 0 & 0 & 0 & 1 \\
\hline
0 & 0 & 1 & 1 & 1 & 1 & 1 & 1 & 0 & 0 & 0 & 1 \\
0 & 0 & 1 & 1 & 1 & 1 & 0 & 0 & 0 & 0 & 0 & 1 \\
1 & 1 & 1 & 1 & 1 & 1 & 0 & 0 & 1 & 0 & 0 & 0 \\
1 & 1 & 1 & 1 & 1 & 1 & 0 & 0 & 0 & 1 & 0 & 0 \\
1 & 1 & 1 & 1 & 1 & 1 & 0 & 0 & 0 & 0 & 1 & 0 \\
1 & 1 & 1 & 1 & 1 & 1 & 0 & 0 & 0 & 0 & 0 & 1 \\
\end{array}
\right] & , 
F_4 = 
\left[
\begin{array}{cccccc|cccccc}
1 & 0 & 0 & 0 & 0 & 0 & 1 & 1 & 1 & 1 & 1 & 1 \\
1 & 0 & 0 & 0 & 0 & 0 & 1 & 0 & 1 & 1 & 1 & 0 \\
0 & 0 & 1 & 0 & 0 & 0 & 1 & 1 & 1 & 1 & 1 & 1 \\
0 & 0 & 0 & 1 & 0 & 0 & 1 & 1 & 1 & 1 & 1 & 1 \\
0 & 0 & 0 & 0 & 1 & 0 & 1 & 1 & 1 & 1 & 1 & 1 \\
1 & 1 & 0 & 0 & 0 & 1 & 1 & 0 & 1 & 1 & 1 & 0 \\
\hline
0 & 0 & 1 & 1 & 1 & 1 & 1 & 1 & 0 & 0 & 0 & 1 \\
0 & 0 & 1 & 1 & 1 & 1 & 0 & 0 & 0 & 0 & 0 & 1 \\
1 & 1 & 1 & 1 & 1 & 1 & 0 & 0 & 1 & 0 & 0 & 0 \\
1 & 1 & 1 & 1 & 1 & 1 & 0 & 0 & 0 & 1 & 0 & 0 \\
1 & 1 & 1 & 1 & 1 & 1 & 0 & 0 & 0 & 0 & 1 & 0 \\
1 & 1 & 1 & 1 & 1 & 1 & 0 & 0 & 0 & 0 & 0 & 1 \\
\end{array}
\right]  , \\~\\
F_5 = 
\left[
\begin{array}{cccccc|cccccc}
0 & 1 & 1 & 1 & 1 & 1 & 0 & 0 & 0 & 0 & 0 & 0 \\
0 & 1 & 1 & 1 & 1 & 1 & 0 & 1 & 0 & 0 & 0 & 1 \\
1 & 1 & 0 & 1 & 1 & 1 & 0 & 0 & 0 & 0 & 0 & 0 \\
1 & 1 & 1 & 0 & 1 & 1 & 0 & 0 & 0 & 0 & 0 & 0 \\
1 & 1 & 1 & 1 & 0 & 1 & 0 & 0 & 0 & 0 & 0 & 0 \\
0 & 0 & 1 & 1 & 1 & 0 & 0 & 1 & 0 & 0 & 0 & 1 \\
\hline
1 & 1 & 0 & 0 & 0 & 0 & 0 & 0 & 1 & 1 & 1 & 0 \\
1 & 1 & 0 & 0 & 0 & 0 & 1 & 1 & 1 & 1 & 1 & 0 \\
0 & 0 & 0 & 0 & 0 & 0 & 1 & 1 & 0 & 1 & 1 & 1 \\
0 & 0 & 0 & 0 & 0 & 0 & 1 & 1 & 1 & 0 & 1 & 1 \\
0 & 0 & 0 & 0 & 0 & 0 & 1 & 1 & 1 & 1 & 0 & 1 \\
0 & 0 & 0 & 0 & 0 & 0 & 1 & 1 & 1 & 1 & 1 & 0 \\
\end{array}
\right] & , 
F_6 = 
\left[
\begin{array}{cccccc|cccccc}
0 & 1 & 1 & 1 & 1 & 1 & 1 & 1 & 1 & 1 & 1 & 1 \\
0 & 1 & 1 & 1 & 1 & 1 & 1 & 0 & 1 & 1 & 1 & 0 \\
1 & 1 & 0 & 1 & 1 & 1 & 1 & 1 & 1 & 1 & 1 & 1 \\
1 & 1 & 1 & 0 & 1 & 1 & 1 & 1 & 1 & 1 & 1 & 1 \\
1 & 1 & 1 & 1 & 0 & 1 & 1 & 1 & 1 & 1 & 1 & 1 \\
0 & 0 & 1 & 1 & 1 & 0 & 1 & 0 & 1 & 1 & 1 & 0 \\
\hline
1 & 1 & 0 & 0 & 0 & 0 & 0 & 0 & 1 & 1 & 1 & 0 \\
1 & 1 & 0 & 0 & 0 & 0 & 1 & 1 & 1 & 1 & 1 & 0 \\
0 & 0 & 0 & 0 & 0 & 0 & 1 & 1 & 0 & 1 & 1 & 1 \\
0 & 0 & 0 & 0 & 0 & 0 & 1 & 1 & 1 & 0 & 1 & 1 \\
0 & 0 & 0 & 0 & 0 & 0 & 1 & 1 & 1 & 1 & 0 & 1 \\
0 & 0 & 0 & 0 & 0 & 0 & 1 & 1 & 1 & 1 & 1 & 0 \\
\end{array}
\right] , \\~\\
F_7 = 
\left[
\begin{array}{cccccc|cccccc}
0 & 1 & 1 & 1 & 1 & 1 & 0 & 0 & 0 & 0 & 0 & 0 \\
0 & 1 & 1 & 1 & 1 & 1 & 0 & 1 & 0 & 0 & 0 & 1 \\
1 & 1 & 0 & 1 & 1 & 1 & 0 & 0 & 0 & 0 & 0 & 0 \\
1 & 1 & 1 & 0 & 1 & 1 & 0 & 0 & 0 & 0 & 0 & 0 \\
1 & 1 & 1 & 1 & 0 & 1 & 0 & 0 & 0 & 0 & 0 & 0 \\
0 & 0 & 1 & 1 & 1 & 0 & 0 & 1 & 0 & 0 & 0 & 1 \\
\hline
0 & 0 & 1 & 1 & 1 & 1 & 0 & 0 & 1 & 1 & 1 & 0 \\
0 & 0 & 1 & 1 & 1 & 1 & 1 & 1 & 1 & 1 & 1 & 0 \\
1 & 1 & 1 & 1 & 1 & 1 & 1 & 1 & 0 & 1 & 1 & 1 \\
1 & 1 & 1 & 1 & 1 & 1 & 1 & 1 & 1 & 0 & 1 & 1 \\
1 & 1 & 1 & 1 & 1 & 1 & 1 & 1 & 1 & 1 & 0 & 1 \\
1 & 1 & 1 & 1 & 1 & 1 & 1 & 1 & 1 & 1 & 1 & 0 \\
\end{array}
\right] & , 
F_8 = 
\left[
\begin{array}{cccccc|cccccc}
0 & 1 & 1 & 1 & 1 & 1 & 1 & 1 & 1 & 1 & 1 & 1 \\
0 & 1 & 1 & 1 & 1 & 1 & 1 & 0 & 1 & 1 & 1 & 0 \\
1 & 1 & 0 & 1 & 1 & 1 & 1 & 1 & 1 & 1 & 1 & 1 \\
1 & 1 & 1 & 0 & 1 & 1 & 1 & 1 & 1 & 1 & 1 & 1 \\
1 & 1 & 1 & 1 & 0 & 1 & 1 & 1 & 1 & 1 & 1 & 1 \\
0 & 0 & 1 & 1 & 1 & 0 & 1 & 0 & 1 & 1 & 1 & 0 \\
\hline
0 & 0 & 1 & 1 & 1 & 1 & 0 & 0 & 1 & 1 & 1 & 0 \\
0 & 0 & 1 & 1 & 1 & 1 & 1 & 1 & 1 & 1 & 1 & 0 \\
1 & 1 & 1 & 1 & 1 & 1 & 1 & 1 & 0 & 1 & 1 & 1 \\
1 & 1 & 1 & 1 & 1 & 1 & 1 & 1 & 1 & 0 & 1 & 1 \\
1 & 1 & 1 & 1 & 1 & 1 & 1 & 1 & 1 & 1 & 0 & 1 \\
1 & 1 & 1 & 1 & 1 & 1 & 1 & 1 & 1 & 1 & 1 & 0 \\
\end{array}
\right] .
\end{align*}
}